%% file: main.tex
\documentclass[11pt]{llncs}
\usepackage{geometry}
\usepackage{Xiao-llncs}
\input{Macro}

\usepackage{bm}

\makeatletter
\def\@fnsymbol#1{\ensuremath{\ifcase#1\or *\or \dagger\or \ddagger\or
   \mathsection\or \mathparagraph\or \|\or **\or \dagger\dagger
   \or \ddagger\ddagger \else\@ctrerr\fi}}
\makeatother

\begin{document}


\title{The Round Complexity of Black-Box Post-Quantum Secure Computation}

\author{
Rohit Chatterjee\inst{1}
\and
Xiao Liang\inst{2}
\and 
Omkant Pandey\inst{3}
\and
Takashi Yamakawa\inst{4}
}
\institute{
National University of Singapore, Singapore\\ \email{rochat@nus.edu.sg}
\and
The Chinese University of Hong Kong, Hong Kong\\ \email{xiaoliang@cuhk.edu.hk}
\and
Stony Brook University, USA\\ \email{omkant@cs.stonybrook.edu}
\and
NTT Social Informatics Laboratories, Japan\\ \email{takashi.yamakawa@ntt.com}
}

\let\oldaddcontentsline\addcontentsline
\def\addcontentsline#1#2#3{}
\maketitle
\def\addcontentsline#1#2#3{\oldaddcontentsline{#1}{#2}{#3}}

\input{sections/om-abstract}
	\pagenumbering{roman}
\tableofcontents
\addcontentsline{toc}{section}{Table of Contents}
\clearpage

\pagenumbering{arabic}

\input{sections/om-intro}

\input{sections/tech-overview}

\input{sections/simultaneous-overview}

\input{sections/preliminaries}
\input{sections/tct}

\input{sections/one-sided}

\input{sections/K-1-1}

\input{sections/SimExt-lemma}

\input{sections/SE-1-1}
\input{sections/two-sided}
\input{sections/simultaneous_extraction}

\input{sections/PQNMC-one-many}
\input{sections/parallelOT}
\input{sections/MPC}
\input{sections/ExtCom-from-OT}
\input{sections/full-MPC}

\newpage
\addcontentsline{toc}{section}{References}
\bibliographystyle{alpha}
\bibliography{main.bbl}

\newpage
\appendix

\input{sections/two-sided-formal}

\end{document}

%% file: Macro.tex
\newcommand{\msf}{\mathsf}
\newcommand{\mcal}{\mathcal}

\newcommand{\Adv}{\ensuremath{\mathcal{A}}\xspace}

\newcommand{\algo}[1]{\ensuremath{\mathsf{#1}}\xspace}

\renewcommand{\bar}{\overline}
\newcommand{\Bad}{\ensuremath{\mathsf{Bad}}\xspace}

\newcommand{\bits}{\ensuremath{\{0,1\}}\xspace}

\newcommand{\BPP}{\ensuremath{\mathbf{BPP}}\xspace}

\newcommand{\cind}{\ensuremath{\stackrel{\text{c}}{\approx}}\xspace}

\newcommand{\com}{\ensuremath{\mathsf{com}}\xspace}
\newcommand{\Com}{\ensuremath{\mathsf{Com}}\xspace}

\newcommand{\decom}{\ensuremath{\mathsf{decom}}\xspace}

\newcommand{\ENMC}{\ensuremath{\mathsf{ENMC}}\xspace}
\renewcommand{\epsilon}{\varepsilon}
\newcommand{\epsilonExtCom}{\ensuremath{\epsilon\text{-}\mathsf{ExtCom}}\xspace}
\newcommand{\EqCom}{\ensuremath{\mathsf{EqCom}}\xspace}

\newcommand{\ExtBCom}{\ensuremath{\mathsf{ExtBCom}}\xspace}
\newcommand{\ExtCom}{\ensuremath{\mathsf{ExtCom}}\xspace}

\newcommand{\Func}{\ensuremath{\mathcal{F}}\xspace}

\newcommand{\Good}{\ensuremath{\mathsf{Good}}\xspace}

\renewcommand{\hat}{\widehat}

\newcommand{\idind}{\ensuremath{\stackrel{\text{i.d.}}{=\mathrel{\mkern-3mu}=}}\xspace}
\newcommand{\IDEAL}{\ensuremath{\mathsf{IDEAL}}\xspace}

\newcommand{\inconsistent}{\ensuremath{\mathsf{Inconsistent}}\xspace}

\newcommand{\mlen}{\ensuremath{N}\xspace}

\newcommand{\Naturals}{\ensuremath{\mathbb{N}}\xspace}

\newcommand{\negl}{\ensuremath{\mathsf{negl}}\xspace}

\newcommand{\NP}{\ensuremath{\mathbf{NP}}\xspace}

\newcommand{\Output}{\ensuremath{\mathsf{Out}}\xspace}

\newcommand{\OT}{\ensuremath{\mathsf{OT}}\xspace}
\newcommand{\OUT}{\ensuremath{\mathsf{OUT}}\xspace}
\newcommand{\OverExtBCom}{\ensuremath{\mathsf{OverExtBCom}}\xspace}

\renewcommand{\paragraph}{\para}

\newcommand{\pick}{\ensuremath{\xleftarrow{\$}}\xspace}
\newcommand{\picks}{\pick}

\newcommand{\poly}{\ensuremath{\mathsf{poly}}\xspace}
\newcommand{\PPT}{\ensuremath{\mathrm{PPT}}\xspace}

\newcommand{\Prot}{\ensuremath{\mathrm{\Pi}}\xspace}

\newcommand{\REAL}{\ensuremath{\mathsf{REAL}}\xspace}
\newcommand{\Recon}{\ensuremath{\mathsf{Recon}}\xspace}

\newcommand{\Relation}{\ensuremath{\mathcal{R}}\xspace}

\newcommand{\SimExt}{\ensuremath{\mathcal{SE}}\xspace}
\newcommand{\SimExtO}{\ensuremath{\mathcal{SE}_{\mathrm{over}}}\xspace}
\newcommand{\SimEq}{\ensuremath{\mathcal{SQ}}\xspace}
\newcommand{\SecPar}{\ensuremath{\lambda}\xspace}

\newcommand{\Set}[1]{\ensuremath{\{#1\}}\xspace}

\newcommand{\Share}{\ensuremath{\mathsf{Share}}\xspace}

\newcommand{\Sim}{\ensuremath{\mathcal{S}}\xspace}

\newcommand{\sind}{\ensuremath{\stackrel{\text{s}}{\approx}}\xspace}
\newcommand{\ST}{\ensuremath{\mathsf{ST}}\xspace}
\newcommand{\state}{\ensuremath{\mathsf{st}}\xspace}

\renewcommand{\tilde}{\widetilde}

\newcommand{\Val}{\msf{Val}}

\renewcommand{\vec}{\vb}

\newcommand{\Verify}{\ensuremath{\mathsf{Verify}}\xspace}
\newcommand{\View}{\ensuremath{\mathsf{View}}\xspace}
\newcommand{\view}{\ensuremath{\mathsf{v}}\xspace}

\newcommand{\VSS}{\ensuremath{\mathsf{VSS}}\xspace}

\newcommand{\wEqCom}{\ensuremath{\mathsf{wEqCom}}\xspace}

\newcommand{\xor}{\ensuremath{\oplus}\xspace}

\newcommand{\sample}{\gets}

\newcommand{\ra}{\rightarrow}
\newcommand{\la}{\leftarrow}

\newcommand{\secpar}{\lambda}

\newcommand{\out}{\mathsf{out}}
\newcommand{\bit}{\{0,1\}}

\newcommand{\defeq}{:=}

\newcommand{\ext}{\mathsf{Ext}}

\newcommand{\BQP}{\mathbf{BQP}}

\newcommand{\TD}{\mathsf{TD}}

\newcommand{\reginp}{\mathsf{Inp}}

\newcommand*{\regX}{\mathbf{X}}
\newcommand*{\regY}{\mathbf{Y}}

\newcommand{\regW}{\mathbf{W}}

\newcommand*{\regS}{\mathbf{S}}

\newcommand{\compind}{\cind}
\newcommand{\statind}{\sind}

\newcommand{\val}{\mathsf{val}}

\newcommand{\execution}[2]{\langle #1, #2 \rangle}

\newcommand{\calM}{\mathcal{M}}

\newcommand{\sfQ}{\mathsf{Q}}
\newcommand{\sfR}{\mathsf{R}}

\newenvironment{boxfig}[2]{\begin{figure}[#1]\fbox{\begin{minipage}{0.97\linewidth}
                        \vspace{0.2em}
                        \makebox[0.025\linewidth]{}
                        \begin{minipage}{0.95\linewidth}
            {{
                        #2 }}
                        \end{minipage}
                        \vspace{0.2em}
                        \end{minipage}}}{\end{figure}}

%% file: sections/om-abstract.tex
\begin{abstract}\addcontentsline{toc}{section}{Abstract}

We study the round-complexity of secure multi-party computation (MPC) in the post-quantum regime where honest parties and communication channels are classical but the adversary can be a quantum machine. Our focus is on the {\em fully} black-box setting where both the construction as well as the security reduction are black-box in nature. In this context, Chia, Chung, Liu, and Yamakawa [FOCS'22] demonstrated the infeasibility of achieving standard simulation-based security within constant rounds, unless $\NP \subseteq \BQP$. This outcome leaves crucial feasibility questions unresolved. Specifically, it remains unknown whether black-box constructions are achievable within polynomial rounds; additionally, the existence of constant-round constructions with respect to {\em $\epsilon$-simulation}, a relaxed yet useful alternative to the standard simulation notion, remains unestablished.

\hspace{1em}
This work provides positive answers to the aforementioned questions. We introduce the first black-box construction for post-quantum MPC in polynomial rounds, from the minimal assumption of post-quantum semi-honest oblivious transfers. In the two-party scenario, our construction requires only $\omega(1)$ rounds. 
These results have already found application in the oracle separation between classical-communication quantum MPC and $\mathbf{P} = \NP$ in the recent work of Kretschmer, Qian, and Tal [STOC'25].

\hspace{1em} 
As for $\epsilon$-simulation, Chia, Chung, Liang, and Yamakawa [CRYPTO'22] resolved the issue for the two-party setting, leaving the general multi-party setting as an open question. We complete the picture by presenting the first black-box and constant-round construction in the multi-party setting. Our construction can be instantiated using various standard post-quantum primitives including lossy public-key encryption, linearly homomorphic public-key encryption, or dense cryptosystems.

\hspace{1em} 
En route, we obtain a black-box and constant-round post-quantum commitment that achieves a weaker version of the standard 1-many non-malleability, from the minimal assumption of post-quantum one-way functions. Besides its utility in our post-quantum MPC construction, this commitment scheme also reduces the assumption used in the lower bound of quantum parallel repetition recently established by Bostanci, Qian, Spooner, and Yuen [STOC'24]. We anticipate that it will find more applications in the future.

\keywords{Multi-Party Computation \and Post-Quantum \and Non-Malleability \and Black-Box \and Round Complexity}
\end{abstract}

%% file: sections/om-intro.tex



\section{Introduction}

{\em Secure multi-party computation} (MPC) allows two or more mutually distrustful parties to compute any functionality without compromising the privacy of their inputs \cite{FOCS:Yao86,STOC:GolMicWig87}. We study foundational questions pertaining to the efficiency of secure multiparty computation in the {\em post-quantum} regime where honest parties and communication channels are classical but the adversary can be a quantum machine. We focus on two specific efficiency criteria: (1) the round-complexity, and (2) the black-box nature of the protocols. We are only concerned with general-purpose protocols in this work, i.e., protocols that can compute any well-defined multiparty functionality.

The black-box nature of the protocols manifests itself in at least two ways. First, the MPC protocol is said to have a {\em black-box construction}, if it only relies on the input-output behaviour of the underlying cryptographic primitives/assumptions. That is, the description of the MPC protocol is independent of the {\em implementation} level details of the underlying cryptographic primitives. This ensures that the efficiency of the protocol does not change with the implementation details of the underlying primitives. Moreover, such constructions remain valid even if the building-block primitives are based on a physical object such as a noisy channel or tamper-proof hardware \cite{wyner1975wire,FOCS:CreKil88,TCC:GLMMR04}.

Second, the MPC protocol is said to have a {\em black-box security-proof (or reduction)} if the security proof uses the adversary only as a black-box (i.e., only relies on its input/output functionality). We are concerned with MPC protocols that are {\em fully} black-box \cite{STOC:ImpRud89,TCC:ReiTreVad04}, i.e., they have a black-box construction as well as a black-box reduction to the underlying cryptographic primitives. Protocols that admit black-box reductions are often simpler and tend to result in more efficient implementations.

The complexity of black-box MPC protocols is well understood in the classical setting, resulting in fully black-box constructions in a {\em constant} number of rounds under standard polynomial hardness assumptions \cite{STOC:Goyal11}, obtained after a long sequence of works in this direction \cite{STOC:IKLP06,STOC:IKOS07,TCC:Haitner08,C:IshPraSah08,TCC:PasWee09,TCC:CDMW09,FOCS:Wee10}.

However, these questions are wide open in the post-quantum MPC (PQ-MPC) setting where honest parties and communication channels are still classical but the adversary is allowed to be a quantum machine. This is in part due to the fact that classical techniques for performing simulation and extraction in MPC and Zero-Knowledge protocols rarely work when the adversary is a quantum machine. In fact, Chia, Chung, Liu, and Yamakawa \cite{chia2022impossibility} recently showed that standard (expected) polynomial-time black-box simulation is impossible to achieve by constant-round constructions in the post-quantum setting (unless $\NP \subseteq \BQP$). Indeed, this impossibility holds even in scenarios where honest parties have access to quantum capabilities \cite{arXiv:CCLL}. These strong results still leave  glaring feasibility questions unresolved, which we discuss next.

\para{Full simulation but non-constant rounds:} In this regime, all known constructions achieving full (or standard) simulation \cite{EC:ABGKM21,FOCS:LPY23,goyal2023concurrent}\footnote{Note that \cite{arXiv:LPY23} and \cite{FOCS:LPY23} refer to the same paper. We include two separate bibliography entries because certain lemmas appear exclusively in the arXiv version \cite{arXiv:LPY23} but not in the conference version \cite{FOCS:LPY23}, and we occasionally need to cite them specifically.} make extensive use of non-black-box techniques. Indeed, \cite{EC:ABGKM21,FOCS:LPY23} even achieve constant rounds by relying on non-black-box simulation. However, no results are known if we insist on black-box constructions, {\em even} for the two-party setting.\footnote{We note that fully black-box constructions exist if the honest parties are allowed to leverage quantum power (e.g., \cite{C:BCKM21b,EC:GLSV21}). However, this falls outside the scope of our focus on {\em post-quantum} protocols.} This raises the following question:
\begin{quote}
 {\bf Question 1:} Do there exist black-box constructions of post-quantum 2PC (and MPC) with full simulation (in more than constant number of rounds)?
\end{quote} 

\para{Relaxed simulation in constant rounds:} The everpresent desire for constant-round secure protocols has prompted exploration of alternative notions such as {\em with $\epsilon$-simulation}, which is a relaxed form of standard simulation-based security that allows for an arbitrarily small noticeable simulation error $\epsilon$. This is an extensively well-studied notion in the literature \cite{DBLP:conf/focs/DworkNRS99,C:JKKR17,STOC:BitKhuPan19} that implies other important security notions --- e.g., $\epsilon$-zero-knowledge protocols imply witness indistinguishability \cite{STOC:FeiSha90} and $\epsilon$-simulatable MPCs imply input-indistinguishable computation \cite{FOCS:MicPasRos06}. 
In this $\epsilon$-simulation regime, the recent work of \cite{C:CCLY22} made initial progress by presenting a constant-round fully black-box protocol for the two-party setting. However, obtaining similar results in the {\em multi-party setting} has remained an unsolved challenge, {\em even with stronger hardness assumptions than those in the classical setting}. This motivates our second question:
\begin{quote}
 {\bf Question 2:} Do there exist black-box, constant-round constructions of post-quantum MPC with $\epsilon$-simulation?
 \end{quote} 


We remark that the recent breakthrough by Lombardi, Ma, and Spooner \cite{lombardi2022post} proposed a new model for post-quantum simulation, called coherent-runtime expected quantum polynomial time simulation. In this model, a simulator is allowed to coherently run multiple computational branches with different runtime so that they can interfere with one another.   
They show a set of results in this model that bypass the impossibility result of \cite{chia2022impossibility}. 
We emphasize that in the current work, we focus on the traditional notion of quantum \emph{strict}, rather than \emph{expected}, polynomial-time simulation. It is also worth mentioning that although the \cite{lombardi2022post}'s coherent-runtime expected QPT simulation implies $\epsilon$-simulation, the round complexity of {\em fully} black-box PQ-MPC has not been resolved in their model either. We leave it as an interesting direction to investigate the implications of the \cite{lombardi2022post} model on the round complexity of black-box PQ-MPC.

\subsection{Our Results}
In this work, we give a positive resolution of these two questions.

\subsubsection{Black-Box PQ-2PC and PQ-MPC with Full Simulation}

We obtain the first fully black-box PQ-2PC protocol from minimal assumptions, in any super-constant number of rounds, which is (asymptotically) optimal for black-box simulation (due to the lower bound of \cite{chia2022impossibility}):
\begin{theorem}\label{thm:informal:full2PC}
There exists a $\omega(1)$-round,\footnote{While the term $\omega(1)$ is typically used for lower bounds, in our context, we use it to mean that ``any super-constant value suffices.''} black-box construction of PQ-2PC (with full simulation), from the minimal assumption of post-quantum, semi-honest oblivious transfers (OTs). 
\end{theorem}
To build this protocol, we follow the approach of \cite{C:CCLY22} originally designed for black-box PQ-2PC with {\em $\epsilon$-simulation}. Very roughly speaking, the most crucial component in their approach is a {\em post-quantum extractable commitment} with $\epsilon$-simulation. This primitive is similar to the standard notion of extractable commitments in the classical setting, but it additionally requires that the post-extraction state of $C^*$ (the malicious committer) should be $\epsilon$-indistinguishable from that in the real execution.\footnote{We remark that while simulating for $C^*$'s post-extraction state is trivial in the classical setting, this task is particularly challenging when $C^*$ is a quantum machine (see \cite{C:CCLY22}).} We observe that we can use \cite{C:CCLY22} template to also achieve the standard notion of fully simulatable PQ-2PC (instead of just $\epsilon$-simulatability) if we can just make the underlying extractable commitment fully-simulatable. 

While the goal is clear, achieving this turns out to be quite non-trivial. To the best of our knowledge, all existing black-box constructions for this task crucially utilize quantum communication in their protocol \cite{C:BCKM21b,EC:GLSV21}. Since our aim is to build a {\em post-quantum} protocol, this does not suit us. To address this issue, we introduce the first black-box construction of post-quantum extractable commitments with full simulation. Our construction makes use of post-quantum semi-honest OTs. We note that while semi-honest OTs may not be the minimal assumption for extractable commitments per se, it is however minimal for our eventual goal of PQ-2PC.
\begin{lemma}\label{lemma:informal:fullExtCom}
    Assuming the existence of post-quantum semi-honest OTs, there exists a $\omega(1)$-round, black-box construction of post-quantum extractable commitments with full simulation. 
\end{lemma}

Given our construction of black-box PQ-2PC, we can use it to get a construction for fully simulatable PQ-MPC. This is done by invoking the \cite{C:IshPraSah08} black-box compiler to get a polynomial round PQ-MPC --- the key thing to notice is that our 2PC construction can also serve as the kind of OT protocol that is required by this compiler, albeit necessitating sequential composition for multiple OT calls. We refer the reader to \Cref{sec:full-MPC} for further details.

\begin{theorem}\label{thm:informal:fullMPC}
There exists a black-box construction of PQ-MPC with full simulation, from the minimal assumption of post-quantum semi-honest OTs. 
\end{theorem}

\subsubsection{Application I: LOCC MPC without OWFs}

A recent breakthrough by Kretschmer, Qian, and Tal \cite{STOC:KreQiaTal25} constructed a classical oracle relative to which $\mathbf{P} = \NP$, yet $\BQP$-computable (and quantum-secure) trapdoor OWFs exist, making them impossible to ``de-quantize'' in a black-box manner. This relativized world is particularly surprising when contrasted with its classical counterpart, where $\BPP$-computable OWFs can be de-randomized in a black-box manner \cite{FOCS:ImpLub89}. 

\cite{STOC:KreQiaTal25} established their main theorem via a fully black-box reduction. Consequently, relative to the same classical oracle, their theorem extends to demonstrate the existence of any ``LOCC'' cryptographic object that admits a fully black-box reduction to trapdoor OWFs in the post-quantum setting.  Here, LOCC stands for ``local operations and classical communication,'' meaning that parties can perform local quantum operations, but all communication must be classical.

 By combining our \Cref{thm:informal:full2PC} (and \Cref{thm:informal:fullMPC}) above {\em and} the post-quantum fully black-box reduction from semi-honest OTs to trapdoor OWFs from \cite{FOCS:GKMRV00}\footnote{Although the original work \cite{FOCS:GKMRV00} was focused on the classical setting, it is straightforward to see that their reduction holds in the post-quantum setting as well.}, the authors of \cite{STOC:KreQiaTal25} were able to derive the following \Cref{cor:application-I} as a corollary of their main theorem. As explained in \cite{STOC:KreQiaTal25}, our \Cref{thm:informal:full2PC} (and \Cref{thm:informal:fullMPC}) are essential for this result, as previous 2PC/MPC constructions either make non-black-box use of semi-honest OTs or lack security proofs in the presence of a quantum attacker.

 \begin{corollary}[{\cite[Corollary 39]{STOC:KreQiaTal25}, strengthened\footnote{The original \cite[Corollary 39]{STOC:KreQiaTal25} relied only on our \Cref{thm:informal:full2PC} to obtain maliciously secure OTs (and thus 2PC). Here, we extend it to MPC using the stronger \Cref{thm:informal:fullMPC}.}}]
 \label{cor:application-I}
There exists a classical oracle relative to which classical-communication and quantum-secure MPC exist, yet $\mathbf{P} = \NP$. 
 \end{corollary}

\subsubsection{Constant-Round Black-Box PQ-MPC with $\epsilon$-Simulation}
As for $\epsilon$-simulation, we study the general multi-party setting, and obtain the first constant-round fully black-box construction for PQ-MPC by relying on the same (more accurately, the post-quantum analog of) hardness assumptions as for the state-of-the-art {\em classical} MPC protocols:
\begin{theorem}\label{thm:informal:main}
There exists a constant-round black-box construction of $\epsilon$-simulatable PQ-MPC from a variety of standard post-quantum cryptographic primitives, such as lossy public-key encryption, linearly homomorphic public-key encryption, or dense cryptosystems.\footnote{We did not mention post-quantum (enhanced) trapdoor permutations as they are not known from standard quantum hardness assumptions yet. But as long as they exist, they can be included in \Cref{thm:informal:main} as well.}
\end{theorem}
Our approach to \Cref{thm:informal:main} follows a pipeline established for classical constant-round black-box MPC, which has evolved through a series of prior work \cite{C:IshPraSah08,TCC:PasWee09,FOCS:Wee10,STOC:Goyal11,FOCS:GLOV12}. In broad terms, we demonstrate that if the building components used in this pipeline are properly instantiated using their post-quantum equivalents, the outcome can be extended to the post-quantum realm. Further insights into this process are elaborated upon in \Cref{sec:tech-oeverview:reduction-to-NMC}. For now, it is worth noting that a critical step in this framework is the development of a black-box {\em 1-many non-malleable} commitment scheme in constant rounds. This constitutes the primary technical challenge in the post-quantum setting.

\para{Post-Quantum 1-Many Non-Malleability.} Non-malleable commitments \cite{STOC:DolDwoNao91} are commitments secure in the so-called {\em man-in-the-middle} (MIM) setting: An adversary $\mcal{M}$ plays the role of a receiver in one instance of a commitment (referred to as the {\em left session}), while simultaneously acting as a committer in another session (referred to as the {\em right session}). During the execution, $\mcal{M}$ can potentially make the value committed in the right session depend on that in the left session, in a malicious manner that is to her advantage. Notice that this is not breaking the hiding property of the commitment scheme, as $\mcal{M}$ may be able to conduct the above attack without explicitly learning the value committed in the left session. Furthermore, a commitment is said to be {\em 1-many} non-malleable if it is secure in the MIM setting with one left session but {\em polynomially many} right sessions, i.e., the adversary $\mcal{M}$ cannot make the {\em joint distribution} of the values committed across all right sessions depend on the one committed in the left session.

In the classical setting, the existence of black-box constant-round 1-many non-malleable commitments was established under the minimal assumption of one-way functions \cite{STOC:Goyal11,FOCS:GLOV12}. Such commitments played a pivotal role in enabling black-box constant-round MPC. However, in the post-quantum context, achieving non-malleability (even in the 1-1 MIM setting) with constant rounds proves to be an exceptionally challenging task. A recent result by \cite{FOCS:LPY23} succeeded in obtaining a post-quantum 1-1 non-malleable commitment in constant rounds. Yet, their construction relies significantly on {\em non-black-box} usage of post-quantum one-way functions, and it remains uncertain if their scheme can maintain non-malleability in the more demanding 1-many scenario.

In this work, we obtain a black-box and constant-round construction for a {\em weak version} (explained shortly) of 1-many post-quantum non-malleable commitments, from the minimal assumption of post-quantum one-way functions. Compared to the standard notion of 1-many non-malleability, our construction is restricted in the following sense:
\begin{itemize}
 \item 
 It supports a polynomial {\em tag space}\footnote{Each execution of non-malleable commitments requires a unique tag; otherwise, it is impossible to protect against MIM attacks (see \cite{STOC:Pass04} for related discussions).}, instead of a exponential-size tag space as required by  the standard definition.
 \item 
It is non-malleable only in the {\em synchronous} setting, meaning that all the messages of the left session and the polynomially many right sessions are sent in parallel.
\item
It is non-malleable conditioned on the fact that the honest receiver {\em in every right session} accepts. That is, if there is some right session where the receiver rejects during the commit stage (the committed value for this session is then defined to be $\bot$), then our protocol does not provide any non-malleability guarantee. (We refer to \Cref{def:NMCom:weak:pq} for a formal treatment.)
 \end{itemize} 
We emphasize that while our construction may not be as powerful as the standard 1-many post-quantum non-malleable commitments, it already has non-trivial applications. Firstly, such a scheme suffices for our main focus of post-quantum MPC. Additionally, it also reduces the assumption utilized in a lower bound of quantum parallel repetition (as we will discuss shortly). We believe it will find more applications in the future.
\begin{theorem}\label{thm:informal:1-many-NMC}
Assuming the existence of post-quantum one-way functions, there exists a black-box and constant-round construction of {\em weak} (as explained above) post-quantum 1-many non-malleable commitments.
\end{theorem}

It is known that 1-many non-malleability implies the seemingly more demanding {\em many-many} non-malleability, using a standard hybrid argument. This reduction holds even in the post-quantum setting (see e.g., \cite[Lemma 7.3]{EC:ABGKM21}). This yields the following corollary of \Cref{thm:informal:1-many-NMC}. 

\begin{corollary}\label{cor:informal:many-many-NMC}
Assuming the existence of post-quantum one-way functions, there exists a black-box and constant-round construction of {\em weak} (as explained above) post-quantum many-many non-malleable commitments.
\end{corollary}

\subsubsection{Application II: Quantum Parallel Repetition Lower Bound} 

Interestingly, our many-many non-malleability commitments find further application in establishing the lower bound for parallel repetition of post-quantum arguments. The recent work by Bostanci, Qian, Spooner, and Yuen  \cite{bostanci2023efficient} shows that parallel repetition does not always reduce the soundness error of post-quantum interactive argument systems. In particular, for any polynomial $k(\secpar)$, the authors of \cite{bostanci2023efficient} constructed a {\em constant-round} interactive argument for which a $k$-fold parallel repetition does not reduce the (post-quantum) soundness at all. Their construction makes use of many-many post-quantum (synchronous) non-malleable commitments in constant rounds, which were not known previously. Now, the above \Cref{cor:informal:many-many-NMC} reduces the assumption used in \cite{bostanci2023efficient} to the existence of post-quantum one-way functions. We state the result in the following \Cref{cor:informal:parallel-rep} and refer the interested reader to \cite[Theorem 1.6 and Section 6]{bostanci2023efficient}\footnote{We remark that \cite[Theorem 1.6]{bostanci2023efficient} assumes `concurrent-secure' many-to-many non-malleable commitments. But as the authors have shown in \cite[Section 6]{bostanci2023efficient}, `parallel-secure' (i.e., synchronous) many-to-many non-malleable commitments suffice. Moreover, in their application, if the verifier in one session rejects, the entire execution is considered rejected. Thus, our weak many-many non-malleability notion suffices.} for more information.
\begin{corollary}\label{cor:informal:parallel-rep}
Assume the existence of post-quantum one-way functions. Then, for every polynomial $k(\secpar)$,
there is a constant-round post-quantum interactive argument such that a $k(\secpar)$-fold repetition does not decrease the soundness error compared to the original protocol.
\end{corollary}

\subsection{More Related Work on Non-Black-Box Constructions}

Besides the aforementioned works \cite{EC:ABGKM21,FOCS:LPY23,goyal2023concurrent}, other non-black-box constructions of PQ-2PC also exist, such as \cite{AFRICACRYPT:LunNie11,C:HalSmiSon11}. This naturally raises the question: how large is the gap between these non-black-box PQ-2PC protocols and our black-box PQ-2PC in \Cref{thm:informal:full2PC}? What are the key obstacles preventing the removal of non-black-box components in these constructions?

In fact, these works adopt a fundamentally different approach from ours, as we elaborate below.

\cite{AFRICACRYPT:LunNie11} primarily focused on feasibility results rather than the black-box nature of the construction. Indeed, it is unclear how to remove the non-black-box components from the \cite{AFRICACRYPT:LunNie11} approach. This is because \cite{AFRICACRYPT:LunNie11} builds PQ-2PC following the GMW approach \cite{STOC:GolMicWig87}: first constructing a semi-honest protocol and then achieving active security by adding ZK proofs on each message to enforce honest behavior from the parties. This GMW approach is inherently non-black-box due to its reliance on ZK proofs for cryptographic statements (i.e., the parties' next-message functions).

Even in the classical setting, black-box constructions of 2PC/MPC move away from the GMW approach and instead follow a very different path established by the line of works \cite{C:IshPraSah08,TCC:PasWee09,TCC:CDMW09,TCC:Haitner08,STOC:IKLP06,FOCS:Wee10,STOC:Goyal11}. Briefly, the key advantage of this line of work lies in the development of techniques that enforce honest behavior without requiring ZK proofs for cryptographic statements, while achieving a constant number of interactions. Our constructions follow this line of work in the post-quantum setting, and therefore have little overlap with the \cite{AFRICACRYPT:LunNie11} approach.

A similar situation applies to \cite{C:HalSmiSon11}. Essentially, the PQ-2PC from \cite{C:HalSmiSon11} follows the classical approach established by \cite{STOC:CLOS02}. This is another inherently non-black-box approach where a commit-and-prove protocol is executed on cryptographic languages to enforce honest behavior from the parties. This can be viewed as a variant of the GMW compiler in the Universal-Composable (UC) framework. As such, there is little common ground for further comparison.

%% file: sections/tech-overview.tex

\section{Technical Overview}
\label{sec:tech-overview}

This section provides an overview of our techniques. We first discuss our construction of $\epsilon$-simulatable PQ-MPC (i.e., \Cref{thm:informal:main}). This is covered in \Cref{sec:tech-oeverview:reduction-to-NMC,sec:overview:recall-LPY,sec:overview:NM:1-1,sec:QE-chal,sec:overview:NM:1-many,sec:simultaneous-ext:overview}. After that, we describe our approach to PQ-2PC and PQ-MPC {\em with full simulation} (i.e., \Cref{thm:informal:full2PC,thm:informal:fullMPC}). This is covered in \Cref{sec:overview:full-MPC,sec:overview:PQBExtCom}.

\subsection{Reduction to Post-Quantum 1-Many Non-Malleability}
\label{sec:tech-oeverview:reduction-to-NMC}
As mentioned earlier, our approach to black-box $\epsilon$-simulatable post-quantum MPC follows a pipeline established in the classical setting. In the following, we first recall it.

\para{Classical Framework.} In the classical setting, the aforementioned pipeline to obtain constant-round and black-box MPC proceeds as follows:
\begin{enumerate}
\item \label[Step]{item:classical:pipeline:1}
{\em Malicious-Sender OT:} First, build a 1-out-of-2 string OT with a weak property, namely, with security against malicious senders but only {\em semi-honest} receivers; Additionally, the associated simulator for proving security is required to be `straight-line' (i.e., not performing any rewindings). Such schemes are known from any of the following: certifiable enhanced trapdoor permutations, dense cryptosystems, linearly homomorphic PKE, or lossy PKE (see, e.g., \cite{TCC:CDMW09,FOCS:Wee10}). These schemes are black-box constructions and constant-round (indeed, two rounds suffice). 

\item \label[Step]{item:classical:pipeline:2}
{\em Multi-Party Parallel OT:} Next, a compiler is employed to transfer the malicious-sender OT to a fully-secure OT {\em in the $n$-party parallel setting}. This is the setting of $n$ parties where every pair of parties $(P_i, P_j)$ runs two executions of same OT protocol, one with $P_i$ as the sender and the other with $P_j$ as the sender. All of these $2\cdot \binom{n}{2}$ executions happen in parallel. Such a compiler was constructed in \cite{FOCS:Wee10,STOC:Goyal11}, which is constant-round and makes only black-box use of its building blocks. (We provide more details when describing our approach.)

\item \label[Step]{item:classical:pipeline:3}
{\em General-Purpose MPC:} Finally, another black-box compiler is employed to transfer the $n$-party parallel OT to a general-purpose $n$-party secure computation protocol. This compiler was introduced in \cite{C:IshPraSah08}. It blows up the round complexity only by a constant number.
\end{enumerate}

\para{Our Approach.} At a high-level, our approach is to replace all the primitives employed in the above pipeline with their post-quantum analog, preserving both the constant-round and black-box properties.

First, we notice that \Cref{item:classical:pipeline:1} extends to the post-quantum setting straightforwardly. That is, post-quantum malicious-sender OTs (with straight-line simulation) can be based on post-quantum dense cryptosystems, linearly homomorphic PKE, or lossy PKE, which can be in turn based on the quantum hardness of Learning with Errors (QLWE). 

Obtaining the post-quantum analog of \Cref{item:classical:pipeline:2} represents the main technical challenge. Let us first discuss about \Cref{item:classical:pipeline:3}, assuming the existence of post-quantum multi-party parallel OTs (with $\epsilon$-simulation). For that purpose, we notice that the same \cite{C:IshPraSah08} compiler (introduced in the classical setting) can be used to convert any {\em post-quantum} multi-party parallel OT (with $\epsilon$-simulation) to a {\em post-quantum} MPC (with $\epsilon$-simulation)\footnote{The same observation has been made in the two-party setting in \cite{C:CCLY22}.}. It adds at most constant rounds, makes only black-box use of the given OT protocol, and does not rely on any extra assumptions. Roughly, this is because the original security reduction in \cite{C:IshPraSah08} is in straight-line and does not copy (or `clone') the state of the adversary. Thus, the same proof can be migrated to the post-quantum setting. Although there are some caveats (e.g., how to handle the $\epsilon$-simulation error), we choose not to expand on them in this overview and refer the reader to \Cref{sec:MPC} for more details.

In the following, we focus on the post-quantum analog of \Cref{item:classical:pipeline:2}.

\para{Post-Quantum Multi-Party Parallel OT.} Our starting point is the constant-round, black-box compiler described in \cite{FOCS:Wee10,STOC:Goyal11}. The specific structure of this protocol is not the primary emphasis of this overview and is therefore omitted (see \Cref{sec:parallel-OT} for details). Our sole concern lies in the fact that this compiler relies on a distinct commitment scheme that enjoys the following properties:
\begin{itemize}
\item
{\em Constant-Round and Black-Box:} This is necessary because our ultimate goal is to obtain a constant-round OT (and MPC) that makes only black-box use of the building blocks.
\item
{\em Parallel-Extractable:} It considers the setting where a potentially malicious committer executes $n$ sessions of the scheme in parallel. It requires the existence of an extractor that can extract the committed values {\em in all the $n$ sessions simultaneously}.

\item
{\em 1-Many Non-Malleable:} As explained in the introduction, this notion considers a MIM adversary $\mcal{M}$ who plays the role of a receiver in one instance of the commitment (dubbed the left session), while simultaneously acting as a committer in polynomially many other instances (referred to as the right sessions). All the sessions happen in parallel. For security, we require that $\mcal{M}$ cannot correlate the {\em joint distribution} of the values committed in all of the right sessions with that in the (single) left session. (See \Cref{sec:nmcom} for a formal definition.)
\end{itemize}
It can be shown that as long as we have a post-quantum analog of the above commitment scheme, the same \cite{FOCS:Wee10,STOC:Goyal11} compiler can be used to convert a post-quantum malicious-sender OT (with $\epsilon$-simulation) to a post-quantum multi-party parallel OT (with $\epsilon$-simulation).

It is worth noting that the post-quantum equivalent of (parallel) extractability necessitates an additional requirement: the extractor must be capable of simulating the post-extraction state of the malicious committer. This aspect was not explicitly addressed in the classical setting because classical information can be `cloned,' allowing an extractor to create two copies of the committer---one for extraction and the other for simulating the post-extraction state, thereby mimicking a straight-line execution. However, achieving such a `simulatable' extraction becomes challenging in the post-quantum realm. In fact, there have been suggestions that achieving post-quantum extractable commitments with negligible simulation error may be impossible in constant rounds, if one insists on black-box simulation techniques \cite{chia2022impossibility,C:CCLY22,arXiv:CCLL}. This very challenge is the reason why we relax our security notion to $\epsilon$-simulatability. Looking forward, our objective is to aim for parallel-extractable commitments with $\epsilon$-simulation for post-extraction state in the post-quantum setting. We will demonstrate that this suffices for the \cite{FOCS:Wee10,STOC:Goyal11} compiler when our ultimate goal is $\epsilon$-simulatable PQ-MPC. To maintain our focus on the core topics of this overview, we will omit additional details in this regard and refer the reader to \Cref{sec:MPC}.

\para{Post-Quantum 1-Many Non-Malleability.} Next, our focus turns to the development of a commitment scheme that satisfies the post-quantum analog of the three properties mentioned earlier. To achieve this, we start with the constant-round post-quantum non-malleable commitment described in \cite{FOCS:LPY23}. First, we observe that the \cite{FOCS:LPY23} scheme is already post-quantum extractable (with $\epsilon$-simulation of the post-extraction state) {\em in the stand-alone setting}. Also, it is not hard to see that the techniques from \cite{C:CCLY22} can be used to prove that the \cite{FOCS:LPY23} scheme is post-quantum {\em parallel}-extractable as well. However, it is important to note that the \cite{FOCS:LPY23} scheme achieves non-malleability in the {\em 1-1 setting} only, as opposed to being {\em 1-many non-malleable}. Furthermore, it extensively relies on the use of its underlying primitive (i.e., a post-quantum one-way function) in a {\em non-black-box} manner. Indeed, the question of constructing constant-round post-quantum commitments that achieve {\em either of these two properties} remains an open challenge. In the following, we describe our ideas to achieves {\em both} properties, under the minimal assumption of post-quantum one-way functions.

\subsection{PQ-NMC from \cite{FOCS:LPY23}} 
\label{sec:overview:recall-LPY}

We first recall the \cite{FOCS:LPY23} construction and the salient features therein that help with the proof of non-malleability. For our purpose, it is sufficient to focus on the simplified scheme shown in the technical overview of \cite{FOCS:LPY23}. That construction achieves non-malleability in the synchronous 1-1 MIM setting, where the left-session tag $t$ is {\em strictly smaller than} the right-session tag $\tilde{t}$ (dubbed `one-sided' non-malleability). 

It works as follows: To commit to a message $m$ with tag $t \in [n]$, the committer $C$ first commits to $m$ using a statistically binding commitment scheme $\msf{com}=\Com(m;r)$ (e.g., Naor's commitment). Then, the receiver $R$ sends a hard puzzle that has {\em exactly} $t$ distinct solutions; $R$ also gives a witness-indistinguishable proof of knowledge (referred to as {\bf WIPoK-1}) to prove that it knows one of the $t$ solutions. Finally, $C$ is required to prove using another WIPoK (referred to as {\bf WIPoK-2}) that it knows {\em either} the value committed in $\msf{com}$ {\em or} one solution to $R$'s hard puzzle. 

\begin{figure}
\begin{subfigure}[h]{0.9\textwidth}
 \centering
         \fbox{
         \includegraphics[width=\textwidth,page=1]{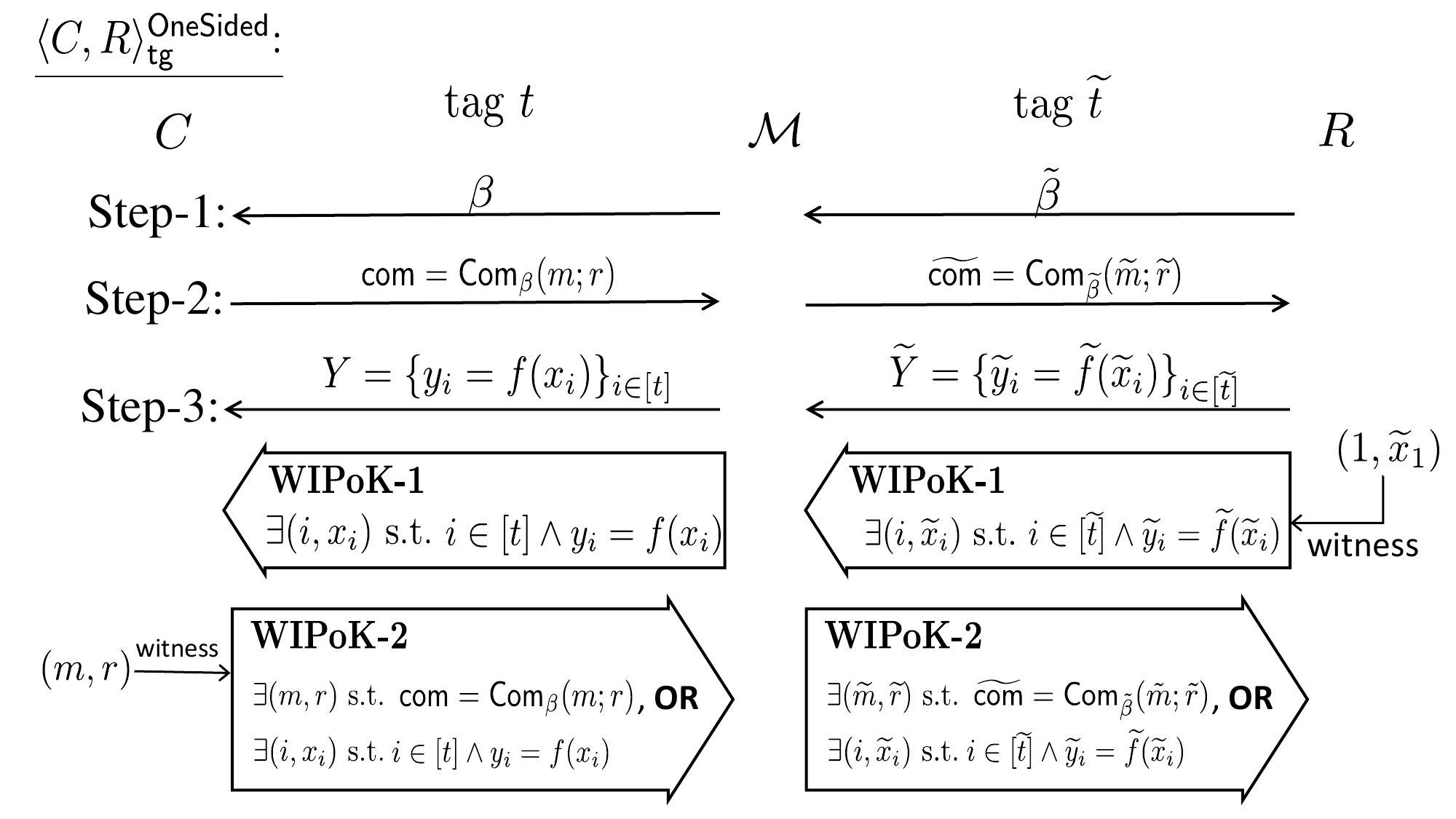}
         }
         \caption{Man-in-the-Middle Execution of $\langle C, R\rangle^{\msf{OneSided}}_{\msf{tg}}$}
         \label{figure:one-sided:tech-overview:real}
     \end{subfigure}
     \\~

     \begin{subfigure}[h]{0.9\textwidth}
      \vspace{3em}
		\centering
        \fbox{
         \includegraphics[width=\textwidth]{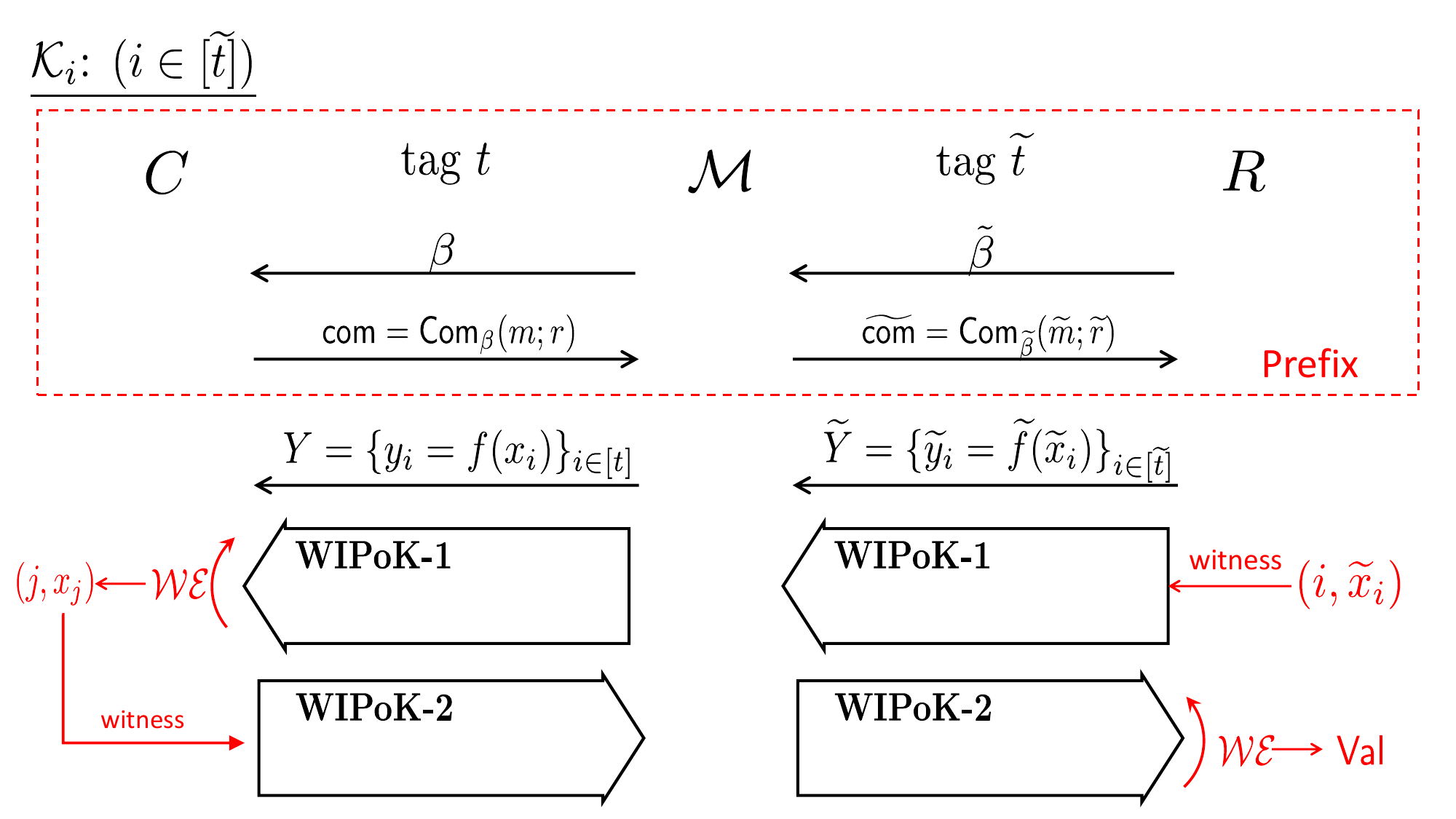}
         }
         \caption{The Simulation-less Extractor $\mcal{K}_i$}
         \label{figure:one-sided:tech-overview:Ki}
     \end{subfigure}
     \caption{}
\end{figure}

  We illustrate the 1-1 MIM execution of this protocol in \Cref{figure:one-sided:tech-overview:real} (borrowed from \cite{FOCS:LPY23}), where the $t$-solution hard puzzle is instantiated with $t$ images $(y_1, \ldots, y_t)$ of an {\em injective} OWF $f$, and the solutions are the preimage $x_i$'s satisfying $y_i = f(x_i)$ for all $i \in [t]$.

\para{A Pigeon-Hole Argument.} Proofs of non-malleability typically rely on the following intuitive claim: in the MIM interaction, we want the honest committer $C$ to be able to `cheat' on the left, while the MIM adversary $\mcal{M}$ should not be able to `cheat' similarly on the right. To show this, \cite{FOCS:LPY23} relies on a pigeon-hole based argument, which we sketch here. Note that in the MIM interaction depicted in \Cref{figure:one-sided:tech-overview:real}, there is an inherent asymmetry between the left and right side executions---there are more puzzle solutions on the right as compared to the number on the left (since $t < \tilde{t}$). This leads to the following intuitive observation: suppose the receiver $R$ switches the witness it uses in {\bf WIPoK-1} on the right. Due to this asymmetry, $\mcal{M}$ cannot switch its witness in every such case. Namely, by the pigeonhole principle, there {\em must} exist indices $i,j \in [\tilde{t}]$ on the right and $k \in [t]$ on the left such that no matter which of $\tilde{x}_i$ or $\tilde{x}_j$ is used as a witness, $\mcal{M}$ can only $x_k$ as witness in the left {\bf WIPoK-1}. 

For this `pigeon-hole tuple' $(\tilde{x}_i, \tilde{x}_j, x_k)$, we see that the following must also happen---suppose the left-session $C$ uses $x_k$ as witness in the left {\bf WIPoK-2}, we can then argue that $\mcal{M}$ must use $\tilde{m}$ in the right {\bf WIPoK-2} as follows: 
\begin{itemize}
\item
First, assuming $R$ uses $\tilde{x}_i$ in the right {\bf WIPoK-1} and $\mcal{M}$ uses $x_k$ in the left {\bf WIPoK-1}, if we extract from the right {\bf WIPoK-2}, the extracted value can only equal $\tilde{m}$ or $\tilde{x}_i$. That is because other $\tilde{x}_j$'s (with $j \ne i$) have not been used in the right {\bf WIPoK-1} and so we can appeal to the one-wayness of $f$ to say that $\mcal{M}$ cannot learn these values.
\item
Similarly, assuming $R$ uses $\tilde{x}_j$ in the right {\bf WIPoK-1}  and $\mcal{M}$ uses $x_k$ in the left {\bf WIPoK-1}, if we extract from the right {\bf WIPoK-2}, the extracted value can only take the values $\tilde{m}$ or $\tilde{x}_j$.  
\end{itemize}
Then, by the witness indistinguishability of the right {\bf WIPoK-1}, the extracted value should {\em not} change if $R$ switches between $\tilde{x}_i$ and $\tilde{x}_j$ (in the right {\bf WIPoK-1}). Thus the extracted value can only be $\tilde{m}$ if $R$ uses $\tilde{x}_i$ (or $\tilde{x}_j$) on the right and $C$ uses $x_k$ on the left.

It seems that the approach outlined above is promising and can help show the intuitive guarantee of allowing $C$ to ``cheat'' on the left while preventing such behavior from $\mcal{M}$ on the right. However, actually proving such a guarantee is quite challenging and is the core technical contribution of \cite{FOCS:LPY23}. In particular, they must address the following technical hurdles: (1) a mechanism is needed to efficiently identify the `magic' triples $(\tilde{x}_i,\tilde{x}_j,x_k)$---one cannot extract $x_k$ from the left {\bf WIPoK-1} simply by rewinding, because the above argument relies on the WI property of that stage, which may not hold if it is rewound. (2) More crucially, the above pigeon-hole argument assumed a one-to-one correspondence between the $\mcal{M}$'s witness and $R$'s witness used in {\bf WIPoK-1}. This is over simplified. Indeed, $\mcal{M}$ can switch its witness {\em probabilistically} when $R$ switches witnesses. 

To address these issues, \cite{FOCS:LPY23} develops an involved {\em distributional} pigeon-hole lemma to formally captures the intuition above. For the current overview, the details of this lemma is not crucial, and thus we do not dig it further. However, the structure of the proof in \cite{FOCS:LPY23} is crucial for understanding our new techniques later. Therefore, we briefly recall its structure below, focusing only on the aspects necessary to establish a foundation for the subsequent discussion of our techniques.

\para{\cite{FOCS:LPY23}'s Proof Structure.} At a high level, the \cite{FOCS:LPY23} approach involves a reduction from non-malleability to the hiding of the left Naor commitment $\Com$ performed initially. This is a rigorous formalization of the aforementioned intuition that `we can cheat in the left but $\mcal{M}$ cannot in the right.' In more detail, they first make the subsequent portion of the left execution after $\Com$ {\em independent} of message $m$, so that the reduction can go through. Next, the idea is to extract the $\tilde{m}$ committed initially by $\mcal{M}$ from the right session. If one can always extract the correct value $\tilde{m}$ and while not unduly disturbing $\mcal{M}$'s post-extraction state, the reduction to the hiding of $\Com$ is easily seen: If $\mcal{M}$'s $\tilde{m}$ changes according to $m$ (i.e., the message committed in $\Com$ on the left), then one can always use $\tilde{m}$ extracted from the right {\bf WIPoK-2} to detect the difference, compromising the hiding property of $\Com$. Thus, the {\em most challenging part} in this approach is to efficiently extract $\tilde{m}$, without disturbing $\mcal{M}$'s post-extraction state too much.

\cite{FOCS:LPY23} builds such an `extractor with simulation' in two steps. First, they build a `base' extractor $\mcal{K}$ without any simulation guarantee, whose job is only to extract $\tilde{m}$ correctly. $\mcal{K}$ works by sampling an uniform index $i \in [\tilde{t}]$ and running the machine $\mcal{K}_i$ depicted in \Cref{figure:one-sided:tech-overview:Ki}. In particular $\mcal{K}_i$ differs from the real MIM execution in the following way: (1) it uses $\tilde{x}_i$ (instead of $\tilde{x}_1$) as the witness in the right {\bf WIPoK-1}; (2) it uses the witness extractor $\mcal{WE}$ to obtain the witness $j\|x_j$ used by $\mcal{M}$ in the left {\bf WIPoK-1}, and then uses this extracted $j\|x_j$ to finish the left {\bf WIPoK-2}; (3) it uses the witness extractor $\mcal{WE}$ to extract the witness $\msf{Val}$ used by $\mcal{M}$ in the right {\bf WIPoK-2} and hopes that $\msf{Val} = \tilde{m}$.

 \cite{FOCS:LPY23} uses the aforementioned distributional pigeon-hole lemma to prove that such a machine $\mcal{K}$ will indeed extract $\msf{Val} =\tilde{m}$ with noticeable probability, {\em conditioned on a `good' prefix (i.e., Steps 1 and 2) from which $\mcal{M}$ will indeed finish the execution with noticeable probability} (this condition is necessary: one cannot hope to extract $\tilde{m}$ with noticeable probability if, say, $\mcal{M}$ always aborts in the real MIM execution).

Next, \cite{FOCS:LPY23} develops a {\em simulation-extraction} lemma. Using this lemma, they are able to convert the simulation-less $K$ into a new machine $\SimExt$ that extracts $\tilde{m}$ while {\em also} being able to simulate the post-extraction state. 
  As remarked before, this gets them most of the way through the proof---\cite{FOCS:LPY23} show that one can use $\SimExt$ to complete the outlined reduction to hiding of $\Com$, and thus demonstrate non-malleability.

\subsection{Our Black-Box Construction: 1-1 Setting}
\label{sec:overview:NM:1-1}
 While the \cite{FOCS:LPY23} commitment is non-malleable and works in constant rounds, it does not suffice for our application because the construction makes heavy non-black-box use of its cryptographic components, and it is unclear if their security proof holds in the more demanding 1-many MIM setting. Now, we first introduce new ideas to obtain a black-box construction.

Observe that there are two sources of non-black-box usage in \Cref{figure:one-sided:tech-overview:real}. First, $R$'s proof in {\bf WIPoK-1} needs the code of the hard puzzle (i.e., the OWF $f$); Second, $C$'s consistency proof {\bf WIPoK-2} makes non-black-box use of {\em both} the hard puzzle {\em and} Naor's commitment in {\bf Step-1}.

We first notice that it is not hard to make $R$'s behavior black-box. Essentially, what $R$ does in the hard-puzzle set-up stage is first `committing' to $t$ solutions and then proving that it knows one solution. This is actually a classical task called {\em witness indistinguishable commit-and-prove (of knowledge)}. It is not hard to modify existing black-box witness indistinguishable commit-and-prove protocols (e.g., \cite{C:CCLY22}) to make $R$'s hard puzzle and the {\bf WIPoK-1} steps black-box. In the following, we only focus on the non-black-box usage on $C$'s side. 


Making $C$'s consistency proof {\bf WIPoK-2} black-box turns out to be quite challenging. One may hope to re-use the aforementioned black-box commit-and-prove technique to resolve the non-black-box use of the {\bf Step-1} Naor's commitment. However, the real difficulty lies in its dependency on the puzzle (i.e., the alternate clause in {\bf WIPoK-2} in \Cref{figure:one-sided:tech-overview:real}). Notice that the puzzle solutions are {\em only known to $R$}! This means that the statement becomes one about the {\em preimages} of $f$, for which the committer/prover (i.e., party $C$) does not have a witness.  In this scenario, it is unclear how the black-box commit-and-prove techniques could help. 

We develop new ideas to tackle this challenge. Our guiding principle is to modify {\bf WIPoK-2} so that it proves only {\em non-cryptographic} relations. That is, we try to make {\bf WIPoK-2} depend only on values that are {\em either} committed by $C$ itself (so that it can be handled by black-box commit-and-prove), {\em or} otherwise made available to both parties in the course of the protocol. 

To do this, we start with careful scrutiny of \cite{FOCS:LPY23}'s simulation-less extractor $\mcal{K}_i$ shown in \Cref{figure:one-sided:tech-overview:Ki}. We observe that the {\bf WIPoK-2} stage there can be interpreted as serving a {\em dual} function in their security proof: (1) It is used to ensure honest behavior of the committer (or $\mcal{M}$ in the MIM setting); This is what essentially helps them perform the distributional pigeon hole argument. (2) The proof of knowledge property of this stage provides {\em extraction opportunities} to efficiently  learn $\tilde{m}$. 

We find that these two purposes can in fact be `decoupled', leading to a more modular security proof as follows: first, one can imagine a $\mcal{K}'_i$ that is identical to the original $\mcal{K}_i$ but does not invoke $\mcal{WE}$ for the right {\bf WIPoK-2}. For this new $\mcal{K}'_i$, we could hope to re-use \cite{FOCS:LPY23}'s distributional pigeon-hole lemma to argue that {\em the witness `used' by $\mcal{K}'_i$ in the right {\bf WIPoK-2} should be $\tilde{m}$ with good probability}. Second, it is a simple application of the proof of knowledge property to extract the witness used in this {\bf WIPoK-2}, {\em which is guaranteed to be $\tilde{m}$ (with good probability) by the previous step}.

On the other hand, notice that {\em once the protocol is completed}, it is safe for $R$ to disclose its hard puzzle solutions. This seems vacuously true and useless. However, it becomes very useful once combined with the observation in the last paragraph, which yields our key idea---we propose to replace the {\bf WIPoK-2} stage by the following three steps:
\begin{enumerate}
\item 
{\bf ExtCom:} $C$ commits to $m$ again using an extractable commitment;
\item
{\bf Solution Reveal:} $R$ reveals all the hard puzzle solutions $(x_1, \ldots, x_t)$;
\item
{\bf WIP-2:} $C$ proves the same relation as in \cite{FOCS:LPY23} {\bf WIPoK-2}, using a WI proof. (Note that we do not require the proof of knowledge property anymore.) In particular, $C$ proves that the value committed in $\ExtCom$ is equal to {\em either} the value committed in {\bf Step-1} Naor's commitment, {\em or} one of the puzzle solutions among the (already revealed) $x_i$'s.
\end{enumerate}
This structure exercises our previous observations as follows. We first `decouple' the two functions of the original {\bf WIPoK-2} (as explained above) by delegating the extractability to {\bf ExtCom} and the consistency proof to {\bf WIP-2}. Then, we can ask $R$ to reveal the puzzle solutions right after {\bf ExtCom}, because at that moment $C$ (or $\mcal{M}$ in the MIM setting) has already `fixed' the witness for consistency proof (i.e., the current {\bf WIP-2}) in {\bf ExtCom}, and cannot change its mind anymore even if the hard puzzle solutions are revealed to it. 

Indeed, we can show this protocol is non-malleable as follows. Consider a $\mcal{K}''_i$ that is similar to $\mcal{K}$ in \Cref{figure:one-sided:tech-overview:Ki}, but instead extract $\msf{Val}$ from $\mcal{M}$'s {\bf ExtCom} on the right. Then, a similar distributional pigeon-hole argument can be established, proving that the $\msf{Val}$ extracted from {\bf ExtCom} indeed equals to $\tilde{m}$ with good probability. Then, using the same technique as in \cite{FOCS:LPY23}, we can build a simulation-less extractor $\mcal{K}$ from this new $\mcal{K}''_i$ and convert $\mcal{K}$ to a simulation extractor to finish the final reduction to the hiding property of the left Naor's commitment.

It seems we are already done---To make $C$'s behavior black-box, note that all $C$ does now is to commit to two values, one in the original {\bf Step-1} and the other in the new {\bf ExtCom}, and then proves in the new {\bf WIP-2} a non-cryptographic predicate (since the puzzle solutions are revealed) over the two committed values and the revealed puzzle solutions. As mentioned earlier, this task can be made black-box by a simple application of known black-box commit-and-prove techniques. This indeed works {\em if our goal were to build a classically secure scheme}. Unfortunately, this step turns out to be challenging in the presence of a quantum $\mcal{M}$, due to reasons exclusive to the quantum setting. We describe these in \Cref{sec:QE-chal}.

\subsection{Quantum-Exclusive Challenges}
\label{sec:QE-chal}

To explain these challenges, we first briefly recall how canonical black-box commit-and-prove protocols broadly work. At a very cursory level, such protocols have (as indicated by their nomenclature) well-defined {\em commit} and {\em prove} stages. The {\bf commit stage} has the prover commit to values involved later in the proof statement, but not directly. Instead, one commits to shares obtained from a {\em Verifiable Secret Sharing} (VSS) of the intended value. These are a strengthening of standard secret secret sharing schemes and allow for reconstruction of the secret even if some shares are adversarially tampered with---and like standard secret sharing schemes, hide the secret completely if not enough shares are collected. 

This is done for compatibility with the subsequent (black-box) {\bf prove stage}, where the prover follows the {\em MPC-in-the-head} \cite{STOC:IKOS07,FOCS:GLOV12} approach: First the committer (or prover) $C$ emulates in her head a MPC execution with $n$ parties (where the inputs are the VSS shares from the commit stage), and commits to the views of each party during this virtual execution. This is followed by a `cut-and-choose' interaction, where $C$ and the receiver (or verifier) $R$ agree on some subset $\eta \subset [n]$ (of size $k$) of the views from the virtual MPC execution (which includes the initial VSS shares)---typically via a coin-tossing step which helps with extablishing zero knowledge for this protocol, but we gloss over this for now---and the prover reveals the corresponding views. The verifier then checks {\em consistency} of the views; by the design of the protocol, this allows the verifier to catch a cheating prover out with a fairly high probability (establishing soundness), but reveals nothing about the value committed by the prover in the commit stage (leading to zero knowledge).  


As indicated, we want to use such a scheme to supply the commmitter's proof of consistency. Accordingly, we will have {\bf Step-1} and the new {\bf ExtCom} correspond to the commit stage, and the {\bf WIP-2} will correspond to the prove stage of the commit-and-prove protocol. In particular, this makes the initial commitment in {\bf Step-1} of our protocol no longer a straightforward Naor commitment to $m$---instead, this is now a commitment (in parallel) to {\em VSS shares} of $m$, as is necessary for the commit-and-prove technique. Since we aim to overall reduce non-malleability to the hiding of the initial commitment, we must now consider a reduction directly to the hiding of the commit stage of the commit-and-prove protocol.  

This is deceptively tricky. The reason lies in the operation of the black-box commit-and-prove protocol; recall that this entails that some of the VSS shares from the {\bf commit phase} be revealed during the {\bf prove stage}. To realize this with standard commitments is hard because the hiding guarantee does not cover such `partial' revealing of information about the committed value. The solution turns out to involve modifying the standard hiding game to incorporate VSS shares (namely, to perform the commitment as described earlier by committing to VSS shares of the message) and allowing for a partial reveal of a certain subset shares of the adversary's choice later in the challenge (this `mimics' the subsequent proof stage interaction, allowing for a subset of shares corresponding to $\eta$ to be revealed by the VSS-based hiding challenger). We articulate such a `VSS-based hiding' game in the course of our proof and show that this is hard to win given standard (computational) hiding of commitments and the secrecy property of VSS schemes.

We now turn to how the reduction itself works. As a first attempt, consider a reduction $R^*$ that runs the MIM game internally and forwards the {\bf Step-1} commitment to the external VSS hiding challenger. Note that $R^*$ needs to specify a challenge set $\eta$. $R^*$ could simply wait until the MIM $\mcal{M}$ sends $\eta$ in the left {\bf WIP-2} execution, and forward this $\eta$ to the external challenger. This seems to work, assuming $R^*$ ran machine $\mcal{K}''_i$ (or the $\mcal{K}$ built out of $\mcal{K}''_i$) to thereby win the VSS hiding game. However, note that in the final reduction to hiding, $R^*$ instead needs to run the simulation extractor $\mcal{SE}$ that is built out of $\mcal{K}$ using the \cite{FOCS:LPY23} simulation-extraction lemma. In more detail, $\mcal{SE}$ involves {\em coherently rewinding} the machine $\mcal{K}$. This is problematic: recall that $\mcal{M}$ sends set $\eta$ in {\bf WIP-2}, and this $\eta$ needs to be forwarded to the external VSS hiding challenger, which we can of course not rewind \footnote{This is of course not an issue in the classical setting, wherein the machine $\mcal{K}$ does not need enhancement, because simulation can be easily added on by making two copies of the adversary: one for extraction and the other for simulation.}. 

Our heuristic towards solving this is to make sure that $R^*$'s communication with the external challenger should {\em end before the post-prefix phase of its internal MIM execution starts}. As a first try, we let $R^*$ sample the set $\eta$ by itself. To make sure that $\mcal{M}$ indeed uses the same $\eta$ during the left {\bf WIP-2} stage, we add a coin-flipping protocol to generate the challenge set. The hope is to let $R^*$ use the simulator for the coin-flipping interaction to `force' $\mcal{M}$'s challenge set to the $\eta$ it sampled beforehand.

Unfortunately this does not work. Recall that we are in the MIM setting: if we force $\mcal{M}$'s coin-flip protocol on the left, it is possible that $\mcal{M}$ could in turn force the coin-flipping result in the right session with $R$, where the soundness guarantee of the right {\bf WIP-2} may not hold anymore. It thus seems what we need is a {\em non-malleable} coin-flipping protocol; where we can force the result in the left session but $\mcal{M}$ cannot. This however puts the cart before the horse in that it is known that {\em non-malleable} coin-flipping implies non-malleable commitments, which is our object. Thus it seems this approach is a dead end. 

\para{Trapdoor Coin-Flipping}. Fortunately, we manage to resuscitate this approach with the following new ideas. To understand that, let us first delve deeper into the \cite{FOCS:LPY23} proof of non-malleability.  We have already mentioned that for the machine $\mcal{K}''_i$ described before, even if we {\em do not} extract from the right {\bf ExtCom}, we can use a similar pigeon-hole argument to show that the value committed there is indeed $\tilde{m}$ with good probability. This is argued in two steps:
\begin{itemize}
      \item
      First,  we can use the \cite{FOCS:LPY23} pigeon-hole argument to show that even if the left $C$ commits in the left {\bf ExtCom} to some $j\|x_j$ it extracted from the hard puzzle, $\mcal{M}$ cannot in turn commit to some puzzle solution in the right {\bf ExtCom}. We emphasize that this step {\em does not} rely on the soundness of the right {\bf WIP-2}. (This is a feature inherited from \cite{FOCS:LPY23} design, though there this observation was superfluous given that there was no call to `decouple' the {\bf WIPoK-2} into an {\bf ExtCom} and a {\bf WIP-2} as we do).
      \item 
      Second, we can now invoke the soundness of the left {\bf WIP-2}---since $\tilde{m}$ and the right puzzle solutions are the only witnesses for the relation {\bf WIP-2} proves, if the committed value in the right $\ExtCom$ is not any of the puzzle solutions (as argued above), then it must be $\tilde{m}$. 
\end{itemize}

It follows that even if we were to remove the {\bf WIP-2} step in our protocol, it would still enjoy a limited form of `non-malleability'---$\mcal{M}$ cannot commit to a left hard puzzle solution {\em even if} the left $C$ does so. Our idea is to leverage this limited `non-malleability' by constructing a {\em limited} version of non-malleable coin-flipping that suffices for our purposes. Specifically, we construct what we call a {\em trapdoor coin-flipping} protocol. This is an augmented coin-flipping protocol between two parties $C$ and $R$ where $C$ additionally commits to some string $x$ before coin-flipping starts. The security guarantee of the actual coin-flipping stage is `controlled' by the committed string $x$ and a predicate $\phi(\cdot)$ given to both parties at the beginning of the coin-flipping stage. In particular, if the committed $x$ satisfies the predicate $\phi(\cdot)$, then by design a committer can `force' the coin-flipping result to a pre-sampled random string $\eta$; but if the committed $x$ does not satisfy the predicate, then no QPT $C^*$ can bias the coin-flipping result. (see \Cref{sec:td-cf} for a formal treatment).

We will use this trapdoor coin-flipping to determine the challenge in {\bf WIP-2}, setting the trapdoor predicate to `{\bf ExtCom} commits to one of the puzzle solutions.' Then for machine $\mcal{K}''_i$, we can enforce the {\bf WIP-2} challenge to $\eta$ in the left execution using the trapdoor predicate (because $C$ does commit to some $j\|x_j$ in the left {\bf ExtCom}). Then, we use the aforementioned `limited non-malleability' guarantee to argue that $\mcal{M}$ cannot force the coin-flipping for the right {\bf WIP-2}, and thus the soundness guarantee of the right {\bf WIP-2} still holds. This obviates the issues raised above. 

Putting everything together, our proof works as follows: 
\begin{itemize}
      \item
      We first use the \cite{FOCS:LPY23} pigeon-hole argument to argue that $\mcal{M}$ cannot commit to any puzzle solutions in the right {\bf ExtCom}. As mentioned above, this step does not make use of the soundness of {\bf WIP-2}.
      \item 
      Since the right {\bf ExtCom} does not commit to any puzzle solution, the soundness of the right {\bf WIP-2} must hold, {\em even if we enforce the coin-flipping step in the left {\bf WIP-2}}. This comes as a guarantee of our trapdoor coin-flipping protocol. Therefore, by the soundness of the right {\bf WIP-2}, the value committed in the right  {\bf ExtCom} must be $\tilde{m}$ (and will be extracted by $\mcal{K}''_i$).
\end{itemize}
As mentioned before, since we pre-sample $\eta$, $R^*$'s communication with the external VSS hiding challenger can be pushed entirely to the prefix phase. So $R^*$ can make use of a `full-fledged' simulation-extractor $\mcal{SE}$ which is built from the new $\mcal{K}''_i$ (indeed, from $\mcal{K}$ that picks a random $i$ and runs $\mcal{K}''_i$ as in \cite{FOCS:LPY23}), to complete the reduction.

\para{Noisy Simulation-Extraction Lemma.} We emphasize that the overview above forms only an intuitive explanation of our ideas. To implement them formally is more challenging as the aforementioned issues appear in a more subtle and technical manner. Owing to the paucity of space, we refer the reader to \Cref{sec:PQNMC:1-1} for fuller details. However, there is a particularly subtle issue unique to our protocol (i.e., not appearing in \cite{FOCS:LPY23}) that we would like to highlight. The above discussion pays much attention to the value committed in $\ExtCom$. In certain steps of our proof, it becomes important to extract this value {\em efficiently}, in order to reduce the security to some {\em falsifiable}\footnote{An assumption is falsifiable \cite{C:Naor03,STOC:GenWic11} if it can be modeled as an interactive game between an {\em efficient} challenger and an adversary, at the conclusion of which the challenger can {\em efficiently} decide whether the adversary won the game.} assumptions. For that we often need to consider the extracted value and take it to be the committed message if it is {\em not} any of the puzzle solutions on the right. As reasoned above, this is the case in all but a noticeable fraction of cases.

However, this indeed starts affecting the conversion from $\mcal{K}$ to $\mcal{SE}$: the simulation-extraction lemma given in \cite{FOCS:LPY23} crucially relies on the fact that the simulation-less extractor $\mcal{K}$ will, if it does not abort, extract a {\em unique} string with good (technically, {\em noticeable}) probability. Put another way, $\mcal{K}$ needs to know that what it extracts is indeed $\tilde{m}$, and if not, it needs to output $\bot$. However, our new $\mcal{K}$ cannot perform such checks---we only argued that $\mcal{K}$ extracts $\tilde{m}$ with noticeable probability. However, it could still be the case with noticeable probability that the extracted value is simply garbage or `noise.' Even worse, $\mcal{K}$ cannot detect this case because the {\bf Step-1} commitment is performed in a black-box commit-and-prove format.\footnote{This issue does not happen in \cite{FOCS:LPY23} because their {\bf Step-1} is Naor's commitment and $\mcal{K}$ extracts the committed value together with the randomness from {\bf WIPok-2}, so that it can check validity using Naor's decommitment algorithm.} Fortunately, we can upgrade the simulation-extraction lemma from \cite{FOCS:LPY23} to tolerate such noise, which suffices for our purpose. We refer to \Cref{sec:simext:1-1} for details.

This finishes the description of our 1-1 non-malleable commitments. The above discussion is based on the `one-sided' (i.e., $t < \tilde{t}$) scheme in the technical overview of \cite{FOCS:LPY23} and thus our protocol inherits this restriction. Fortunately, we can remove this restriction using exactly the same `two-slot' trick (initiated by Pass and Rosen \cite{STOC:PasRos05}) as in \cite{FOCS:LPY23}. We refer to \Cref{sec:two-sided:main-body} for details.

\subsection{One-Many Non-Malleability}
\label{sec:overview:NM:1-many}

So far, we have restricted our attention to the 1-1 setting for non-malleability. We now discuss our approach to obtain 1-many non-malleable commitments. Recall that this involves a man-in-the-middle adversary $\mcal{M}$ that runs a single commitment session with an honest committer $C$ on the left, and runs up to polynomially many commitment sessions on the right with honest receivers $R^{(1)},\dots,R^{(N)}$ (where we have used $N$ to denote the total number of right sessions). We stress that 1-1 non-malleability does not directly imply 1-many non-malleability; this is a known barrier even in the classical setting and we elide further explanation for the sake of conciseness.  

In spite of this, we are able to demonstrate that the same black-box construction we described above also enjoys 1-many non-malleability. In fact, we rely for the most part on very similar strategies to those we employed in the 1-1 case. The key similarity we exploit is that in the reduction to hiding that we outlined in the 1-1 case (that formed the base of our proof), the majority of the modifications we make are with respect to the {\em left} session---the only changes made on the right are changing which puzzle solution is committed to in {\bf ExtCom} by the honest receiver, and {\em extracting} the committed value. We can thus hope to translate our technique to the 1-many setting as well. 
At a high level, we accomplish this in three steps: 

\para{Step 1.} First, we design analogs of the simulation-less extractor $\mcal{K}$ from earlier, that we call {\em localized} simulation-less extractors $\mcal{K}^{(j)}$ (for $j \in [N]$). $\mcal{K}^{(j)}$ performs extraction only in the $j$-th right session, and acts as an honest receiver in all other right sessions. It is not hard to establish the same guarantees for {\em each} $\mcal{K}^{(j)}$ from the guarantees for $\mcal{K}$ in the 1-1 setting, by means of a `wrapper' reduction: namely, since the other right sessions of $\mcal{M}$ are run honestly, we can run the 1-many interaction internally while treating the $j$-th session as the sole right session in a 1-1 MIM interaction, and handling all the other internal right sessions by itself. 

\para{Step 2.}  Next, we show how to build a {\em simultaneous} simulation-less extractor that is then able to extract the committed values in {\em all} right sessions with high success (but without any simulation guarantees). This step turns out to be particularly challenging due to the quantum nature of the MIM adversary $\mcal{M}$. Note that the instanced extractors $\mcal{K}^{(j)}$'s described above enjoys extractability only for a single run. If we want to perform simultaneous extraction, the natural attempt of running them one-by-one does not work---after the execution of, say, $\mcal{K}^{(1)}$, the internal quantum state of $\mcal{M}$ may have already been disturbed too much to support the execution of $\mcal{K}^{(2)}$. 

    Another natural idea is to first convert $\mcal{K}^{(1)}$ to a simulatable extractor, using the above noisy simulation-extraction lemma, and hope that the simulation guarantees of that lemma can `protect' the state of $\mcal{M}$ after the execution of $\mcal{K}^{(1)}$ so that we can keep running $\mcal{K}^{(2)}$ (and also convert $\mcal{K}^{(2)}$ to a simulatable extractror to support $\mcal{K}^{(3)}$ and so on). Unfortunately, this idea does not work either due to a subtle technical reason: The simulation guarantees of the noisy simulation-extraction lemma is for the {\em post-extraction} extraction state of $\mcal{M}$, which means it simulates $\mcal{M}$'s state at the end (or `bottom') of the protocol; However, to be able to run $\mcal{K}^{(2)}$ right after $\mcal{K}^{(1)}$, we have to simulates $\mcal{M}$'s state at the beginning (or `top') of the protocol! Thus, new ideas are needed to resolve this problem.

    We tackle this problem by crafting a novel {\em simultaneous} post-quantum extraction lemma, drawing upon the measure-and-repair technique introduced in \cite{TCC:Zhandry20,FOCS:CMSZ21}. This represents another main technical contribution of this work and it may find future applications where simultaneous post-quantum extraction is needed. We provide an overview of it in \Cref{sec:simultaneous-ext:overview}.

\para{Step 3.}  Finally, we show that the same noisy simulation-extraction lemma described above can be used to upgrade this simultaneous simulation-less extractor to one {\em with simulation}, and thus finish the proof of non-malleability with a similar reduction to the VSS hiding game as in the 1-1 case.

%% file: sections/simultaneous-overview.tex
\subsection{Simultaneous Extraction}
\label{sec:simultaneous-ext:overview}

We now provide a brief overview of our simultaneous extraction lemma as mentioned above. We will use slightly different notation. In the following description, machines $\mcal{K}_1, \ldots,\mcal{K}_n$ play the role of our instanced simulation-less extractor $\mcal{K}^{(1)}, \ldots, \mcal{K}^{(N)}$ mentioned above (setting $n = N$). And $\mcal{V}$ is a machine that should be treated as enforcing the condition that we start from a `good' prefix (i.e., the {\bf Step-1} Naor's commitment in the VSS form). It is necessary because as we mentioned earlier, we cannot hope to extract $\tilde{m}$'s with good probability if $\mcal{M}$ always aborts the execution before it naturally ends. The machine $\mcal{K}$ is our desired simultaneous simulation-less extractor.


\begin{lemma}[Simultaneous extraction lemma (informal)]
Let $\mcal{V}$ and $\mcal{K}_1, \ldots,\mcal{K}_n$ be QPT algorithms for a polynomial $n$ that satisfy the following: 
\begin{itemize}
    \item {\bf $\mcal{V}$'s syntax:} $\mcal{V}$ takes a quantum state in Hilbert space $\mcal{H}$ and outputs $\top$ or $\bot$; 
    \item {\bf $\mcal{K}_i$'s syntax:} $\mcal{K}_i$ takes a quantum state in Hilbert space $\mcal{H}$ and outputs a classical string $s_i$ or $\bot$;
    \item {\bf Uniqueness of $\mcal{K}_i$'s output:} 
    For each $i\in [n]$, there is a classical string $s_i^*$ such that $\mcal{K}_i$'s output is either $s_i^*$ or $\bot$ on any input;\footnote{The formal version of this lemma (see \Cref{lem:Simultaneous-SimExt}) permits the outputs of $\mathcal{K}_i$ to be other `noise' values, provided that the probability of this occurrence can be bounded by a noticeable function. This is essential for compatibility with the previously described noisy simulation-extraction lemma. However, for this overview, we overlook this detail to maintain focus on the main idea.} 
    \item {\bf ``Good" states for $\mcal{V}$ is also ``good" for $\mcal{K}_i$:}  
    For any noticeable $\gamma$, there is noticeable $\delta$ such that for any quantum state $\rho$, if 
    $$
    \Pr[\mcal{V}(\rho)=\top]\ge \gamma,
    $$
    then 
     $$
    \Pr[\mcal{K}_i(\rho)=s_i^*]\ge \delta.
    $$
\end{itemize}
Then there is a QPT algorithm $\mcal{K}$ (called a simultaneous extractor) satisfying the following:
\begin{itemize}
\item {\bf $\mcal{K}$'s syntax:} $\mcal{K}$ takes a quantum state in Hilbert space $\mcal{H}$ and outputs $n$ classical strings $(s_1,s_2, \ldots,s_n)$ or $\bot$; 
\item  {\bf Uniqueness of $\mcal{K}$'s output:}  $\mcal{K}$'s output is either $(s_1^*,s_2^*, \ldots,s_n^*)$ or $\bot$ on any input; 
\item {\bf ``Good" states for $\mcal{V}$ is also ``good" for $\mcal{K}$:}  
For any noticeable $\gamma$, there is noticeable $\delta'$ such that for any quantum state $\rho$, if 
  $$
    \Pr[\mcal{V}(\rho)=\top]\ge 8\gamma,\footnote{An arbitrary constant factor larger than $1$ suffices, but we choose $8$ to match the formal version of the lemma.}
    $$
    then 
     $$
    \Pr[\mcal{K}(\rho)=(s_1^*,s_2^*, \ldots,s_n^*)]\ge \delta'.
    $$
\end{itemize} 
\end{lemma}

If the input $\rho$ is classical, then the above lemma trivially holds: $\mcal{K}$ can simply run each $\mcal{K}_i$ many times until it succeeds. However, if $\rho$ is quantum, the state may collapse once we run $\mcal{K}_i$ for some $i$, which prevents us from running it on the same state again. To resolve the issue, our idea is to use the ``state repairing" technique introduced in \cite{FOCS:CMSZ21}. 
To explain their technique, we first review the concept of (approximate) projective implementation introduced by Zhandry \cite{TCC:Zhandry20}. 
Let $\{(\Pi_i,I-\Pi_i)\}_i$ be a family of binary-outcome projective measurements 
indexed by a classical index $i$ of a certain length. 
Consider the ``mixture" $M$ of  $\{(\Pi_i,I-\Pi_i)\}_i$, i.e., the procedure that first randomly samples $i$ and then applies the projective measurement  $(\Pi_i,I-\Pi_i)$.  
Zhandry showed the existence of ``projective implementation" $\mathsf{ProjImp}$ of $M$, which is a real-valued projective measurement that ``measures" the success probability of $M$, i.e., for any state $\rho$, the distribution of applying $M$ on $\rho$ is identical to first applying $\mathsf{ProjImp}$ on $\rho$ to obtain $p\in [0,1]$ and then outputting $1$ with probability $p$. 
Though it is unknown how to efficiently implement $\mathsf{ProjImp}$, Zhandry gave an efficient procedure called $\mathsf{API}$ (``Approximate Projective Implementation") that approximates $\mathsf{ProjImp}$ in an appropriate sense. 
Now, we are ready to describe the state repairing technique of \cite{FOCS:CMSZ21}. 
Suppose that we apply $\mathsf{API}$ on some state, which yields an outcome $p$, and then apply $M$.  
At this point, there is no guarantee on the outcome if we apply $\mathsf{API}$ again.  
The work \cite{FOCS:CMSZ21} constructed an efficient state repairing procedure $\mathsf{Repair}$ which acts on the post-execution state so that the output of $\mathsf{API}$ on the resulting state is at least $p-\epsilon$ with overwhelming probability for an arbitrary noticeable function $\epsilon$. 

Our idea is to apply their technique in our context as follows. We consider a family $\{(\Pi_i,I-\Pi_i)\}_{i\in[n]}$ where $\Pi_i$ corresponds to the event that $\mcal{K}_i$ successfully extracts $s_i^*$ and $\mathsf{API}$ that approximates the probability that $\mcal{V}$ returns $\top$. 
If the initial state is accepted with probability sufficiently larger than $\gamma$ and $\epsilon$ is set to be sufficiently small,  
then if we alternately apply $(\Pi_i,I-\Pi_i)$ (i.e., run $\mcal{K}_i$)  and the state repair procedure, we can guarantee that each application of  $(\Pi_i,I-\Pi_i)$ results in the first outcome, which corresponds to successfully extracting $s_i^*$, with probability at least $\delta$. 
Thus, we can simultaneously extract $s_1^*, \ldots,s_n^*$ if we repeat the above sufficiently many times.  
Though we eventually prove that this idea works, this is not a direct application of the result of \cite{FOCS:CMSZ21} since the situation is somewhat different. 
In particular, we have to make sure that 
\begin{itemize}
\item we can construct $\mathsf{API}$ for any binary-outcome POVMs (that correspond to the success of $\mcal{V}$), and 
\item the state repairing procedure still works even if $\mathsf{API}$ is defined for a binary-outcome POVM that is irrelevant to the projections $\{(\Pi_i,I-\Pi_i)\}_{i\in[n]}$. 
\end{itemize}
First, we observe that the second point is actually not an issue since this is techncially already proven in \cite{FOCS:CMSZ21}. 
That is, even though they only apply their technique in the setting where $\mathsf{API}$ is defined for the mixture of projections, their core technical lemma~\cite[Lemma 4.10]{FOCS:CMSZ21} already captures the situation where $\mathsf{API}$ is irrelevant to those projections. 
For the first point, though Zhandry showed that an (inefficient) projective implementation can be defined for any binary-outcome POVMs, he did not show how to efficiently approximate it.  
Thus, we give a construction of $\mathsf{API}$ for any binary-outcome POVMs (that correspond to the success of $\mcal{V}$), which generalizes Zhandry's construction. 
The construction and its analysis are similar to Zhandry's original one for the case of mixtures of projective measurements 
while we rely on Jordan's lemma as an additional tool.\footnote{\cite{FOCS:CMSZ21} also gives a variant of Zhandry's $\mathsf{API}$ by using Jordan's lemma, but they also only consider mixtures of projective measurements.}


\subsection{Black-Box Post-Quantum 2PC and MPC with Full Simulation}
\label{sec:overview:full-MPC}

We begin by recalling the framework established in \cite{C:CCLY22}, which was devised originally for constant-round black-box PQ-2PC {\em with $\epsilon$-simulation}.

A key component of the \cite{C:CCLY22} framework is a black-box extractable commit-and-prove protocol with $\epsilon$-simulation, which we refer to as ``$\epsilon$-ExtCom-n-Prove'' henceforth. This primitive enables a committer to commit to a message $m$ during the {\em Commit Stage} and subsequently prove, with $\epsilon$-zero-knowledge, that the committed $m$ satisfies a predicate $\phi$ during the {\em Prove Stage}. Furthermore, the Commit Stage itself functions as a post-quantum extractable commitment with $\epsilon$-simulation, meaning that the post-extraction state of the malicious committer is $\epsilon$-indistinguishable from that in the real execution. It is worth noting that the symbol $\epsilon$ in the name ``$\epsilon$-ExtCom-n-Prove'' indicates that both the zero-knowledge property of the Prove Stage and the post-extraction simulation of the Commit Stage are defined with $\epsilon$-simulation.

We note that \cite{C:CCLY22} can be interpreted as a compiler that transforms a $O(k)$-round black-box ExtCom-n-Prove protocol into a $O(k)$-round black-box PQ-2PC protocol; moreover, if the ExtCom-n-Prove protocol is defined with $\epsilon$-simulation (resp.\  full simulation), then the resulting PQ-2PC protocol would be $\epsilon$-simulatable (resp.\ fully simulatable). This observation is formalized in \Cref{sec:full-MPC:2PC}. Consequently, our goal of constructing $\omega(1)$-round black-box PQ-2PC can be simplified to the task of building $\omega(1)$-round black-box ExtCom-n-Prove protocols with full simulation.

\para{Post-Quantum Extractable Batch Commitments.} We further observe that in order to build the desired ExtCom-n-Prove protocols, it (almost) \takashi{I added "almost". I'm not sure if this is nice, but without this, this sentence is a little bit inaccurate.} suffices to develop black-box $\omega(1)$-round post-quantum extractable commitments. In this overview, we do not delve into the explanation of why this is true (refer to \Cref{sec:bb-extcom-n-prove} for details). 
More precisely, what we require and refer to as {\em post-quantum extractable batch commitments} is as follows: The committer is able to commit to a vector of messages $\vb{m} = (m_1, \ldots, m_n)$ {\em collectively}. The commitment can be decommited locally on each index $i\in [n]$. 
We require the following security to hold: 

\begin{description}
\item{\bf Hiding} 
For any index $i^* \in [n]$, $m_{i^*}$ remains concealed even if the adversary is given $m_i$ and the corresponding decommitment information for all $i\neq i^*$. 
In particular, this implies that for any subset $I\subseteq [n]$, the messages corresponding to indices in $I$ remain concealed even if the adversary is given $m_i$ and the corresponding decommitment information for all $i\notin I$.\footnote{This may look similar to {\em selective-opening security} \cite{JC:Hofheinz11}, but we remark that we only require the hiding when $I$ is fixed at the beginning 
whereas selective-opening security in \cite{JC:Hofheinz11} allows the adversary to adaptively choose $I$ depending on the commitment.} 

\item{\bf Extractability}
There exists a QPT machine $\mathcal{SE}$ (dubbed the {\em simulation extractor}) capable of extracting the committed vector message $\vb{m}^* = (m^*_1, \ldots, m^*_n)$ from the malicious committer $C^*$, while simultaneously (fully) simulating $C^*$'s post-extraction state to be negligibly close to that in the real execution between $C^*$ and the honest receiver. 
\end{description} 
We remark that if we only consider the hiding and binding (rather than extractability), then a simple parallel composition of a stand-alone commitment scheme would suffice. 
However, extractability may not be preserved under parallel composition, and this is why we need to introduce the above notion of extractable batch commitments.

We manage to build a black-box $\omega(1)$-round construction for such a post-quantum extractable batch commitment, assuming the existence of post-quantum semi-honest oblivious transfer. We provide an overview of this construction in \Cref{sec:overview:PQBExtCom}. For now, we simply remark that this commitment scheme leads to a black-box $\omega(1)$-round construction of ExtCom-n-prove (with full simulation), which, as discussed earlier, results in the first black-box $\omega(1)$-round PQ-2PC (with full simulation) from the minimal assumption of post-quantum semi-honest oblivious transfers.

\para{Extension to the Multi-Party Setting.} Note that the above results for PQ-2PC imply, in particular, a black-box $\omega(1)$-round construction of post-quantum oblivious transfers (maliciously secure, with full simulation). Utilizing a known compiler from \cite{C:IshPraSah08}, such an oblivious transfer protocol can be converted into a black-box PQ-MPC (with full simulation), where the round complexity is polynomial in the number of parties (and consequently polynomial in the security parameter $\secpar$). We refer to \Cref{sec:full-MPC} for details.

\subsection{Post-Quantum Extractable Batch Commitments}
\label{sec:overview:PQBExtCom}

We now provide an overview of our $\omega(1)$-round construction of post-quantum extractable batch commitments (with full simulation), which only makes black-box use of a constant-round post-quantum semi-honest OT. Since all the primitives in the sequel is post-quantum, we henceforth drop the the quantifier ``post-quantum'' for succinctness.

We start by describing a protocol that only supports committing to vectors $\vb{m}$ of length 1. In this case, the notion of extractable batch commitment degenerates to standard extractable commitments (with full simulation). We emphasize that even such a commitment is previously unknown, if one insists on black-box constructions. Our construction can be divided into the following three steps. 
\begin{enumerate}
\item Construct $\omega(1)$-round \emph{equivocal} commitments with full simulation based on OWFs. Here, equivocality means that one can simulate the commit stage for malicious receivers in such a way that the commitment can be opened to an arbitrary message in the reveal stage. There are well-known classical black-box constructions of equivocal commitments from OWFs \cite{STOC:Kilian88,FOCS:Kilian94,TCC:PasWee09}, which are later adapted into the post-quantum setting \cite{C:BCKM21b}. Though those constructions are $O(\secpar)$-round, we observe that they can be easily optimized to $\omega(1)$ rounds.
\item 
Convert equivocal commitments into extractable commitments with a weaker security guarantee which we call \emph{extractability with over-extraction}. It is similar to the standard extractability (with full simulation) except that we allow the extractor to extract a non-$\bot$ message even if the commitment is ill-formed (i.e., there is no valid opening to any message).  In fact, we show that our protocol supports (a certain form of) \emph{parallel} extraction with over-extraction, which we elaborate on later. 
The conversion incurs a constant-round overhead and makes black-box use of constant-round $\epsilon$-simulatable parallel OT, which in turn is constructed from a constant-round \emph{semi-honest} OT in a black-box manner in \cite{C:CCLY22}. We stress that the resulting (parallel) extractable commitment protocol with over-extraction supports full simulation even though the base OT only supports $\epsilon$-simulation. Since this step is the technical core of our construction of extractable commitments, we will provide more details shortly.  
\item Eliminate over-extraction with a $\omega(1)$-round overhead based on OWFs using a standard cut-and-choose technique. Roughly, the commit stage of the protocol works as follows: The committer generates VSS shares of the message $m$ and commits to each share using an extractable commitment scheme with over-extraction in parallel. Then the committer and receiver  execute coin-flipping to agree on a subset on which the committer reveals the committed shares along with the corresponding decommitment information. 
If all of the decommitments are valid, then the receiver is convinced that a large-fraction of the unrevealed commitments is likely to be well-formed (i.e., has a valid decommitment). Then the transcript of the commit stage is well-formed whenever the receiver accepts (except for a negligible probability). In this case, the simulation-extractor can figure out whether the commitment is well-formed by itself and thus over-extraction never occurs. We remark that we only need the coin-flipping protocol to be simulatable against one malicious party (i.e., the role played by the receiver in the commitment protocol), which can be constructed from any equivocal commitments with a constant-round overhead. Since the first step gives $\omega(1)$-round equivocal commitments from OWFs, the overall overhead of round-complexity in this step is just $\omega(1)$. 
\end{enumerate}

Below, we give more details of the second step. 
First, we explain how to achieve extractability with over-extraction in the stand-alone setting (where there is no parallel execution). Our construction works as follows: 

\noindent\textbf{Commit stage.} 
\begin{enumerate}
\item \label[Step]{abstract_ExtCom_1}
The committer $C$ commits to the message $m$ using Naor's commitment. 
\item \label[Step]{abstract_ExtCom_2}
$C$ generates $2k$-out-of-$2k$ secret sharing $\{s_{j}^{b}\}_{j\in [k],b\in\bit}$ of $m$ where $k=\omega(\log \secpar)$. That is, they are uniformly random under the constraint that $\bigoplus_{j\in[k],b\in\bit} s_j^b=m$.  
\item \label[Step]{abstract_ExtCom_3}
$C$ and the receiver $R$ execute $k$-parallel execution of $\epsilon$-simulatable OT where in the $j$-th execution $C$ uses $(s_j^0,s_j^1)$ as input and $R$ uses an independently random bit $r_j$ as input.  
\item \label[Step]{abstract_ExtCom_4}
$C$  and $R$ engage in the following coin-flipping subprotocol to agree on $t\in \bit^k$:
\begin{enumerate}
\item \label[Step]{abstract_ExtCom_4a}
$R$ samples a random string $t_R\gets \bit^k$ and commits to it using the equivocal commitment scheme. 
\item \label[Step]{abstract_ExtCom_4b}
$C$ samples a random string $t_C\gets \bit^k$ and sends it to $C$.
\item \label[Step]{abstract_ExtCom_4c}
$R$ sends to $C$ the value $t_R$ together with the corresponding decommitment information. 
At this point, $C$ and $R$ agree on $t\coloneqq t_R \oplus t_C$. 
\end{enumerate}
\item \label[Step]{abstract_ExtCom_5}
$C$ sends $s_j^{t_j}$ for $j\in [k]$ where $t_j$ is the $j$-th bit of $t$. 
\end{enumerate}

\noindent\textbf{Decommit stage.} 
$C$ sends all the randomness used during the commitment stage as decommitment information, and $R$ accepts if it is consistent to the transcript. 

\smallskip
The statistical binding property of the above protocol follows straightforwardly from that of Naor's commitment. 
The computational hiding property can be shown as follows: Since the OT satisfies $\epsilon$-simulatable security, for any malicious receiver $R^*$, we can simulate the execution of the OT in \Cref{abstract_ExtCom_3} only using $\{s_j^{r^*_j}\}_{j\in [k]}$ for some sequence $\{r^*_j\}_{j\in [k]}$ of bits 
with a noticeable simulation error $\epsilon$. 
Since $\{s_j^b\}_{j\in [k],b\in\bit}$ is $2k$-out-of-$2k$ secret sharing of $m$, $R^*$ cannot learn any information of $m$ unless $t_j$ happens to be $1-r^*_j$ for all $j\in [k]$. However, by the binding property of the equivocal commitment scheme, $R^*$ can cause only a negligible bias on the distribution of $t$. Thus, the probability that $t_j=1-r^*_j$  for all $j\in [k]$ is $2^{-k}+\negl(\secpar)=\negl(\secpar)$.  
The above argument implies that $R^*$ can distinguish commitments to different messages with advantage at most $\epsilon+\negl(\secpar)$. Here, $\epsilon$ can be any noticeable function in $\secpar$, and thus this actually implies the standard computational hiding. 

Below, we give a proof sketch for extractability with over-extraction. We construct the simulation extractor as follows:
\begin{itemize}
\item Execute  \Cref{abstract_ExtCom_1,abstract_ExtCom_2,abstract_ExtCom_3} of the commit stage with the malicious committer $C^*$ while playing the role of the honest receiver. At this point, the simulation extractor obtains $\{s_j^{r_j}\}_{j\in[k]}$ for some random bits $r_j$. 
\item Use equivocality to simulate the commit stage of the equivocal commitment scheme in \Cref{abstract_ExtCom_4a}.  
\item Receive $t_C$ from $C^*$ in \Cref{abstract_ExtCom_4b}.  
\item Set $t\coloneqq (1-r_1)||(1-r_2)||...||(1-r_k)$ and $t_R \coloneqq t_C\oplus t$. Then open the equvocal commitment in \Cref{abstract_ExtCom_4a} to $t_R$. 
\item Receive $\{s_j^{t_j}=s_j^{1-r_j}\}_{j\in [k]}$ from $C^*$ in \Cref{abstract_ExtCom_5}. 
\item Output the final state of $C^*$ along with the extracted message $m\coloneqq \bigoplus_{j\in[k],b\in\bit} s_j^b$.
\end{itemize}

It is straightforward to see that the extracted message is equal to the committed message assuming that the transcript is well-formed, i.e., it has a valid opening to some message. Indeed, since the committer is required to reveal all the randomness in the decommit stage, to generate a well-formed transcript, a malicious sender has to follow the protocol albeit with a possibly skewed randomness distribution. In this case, the simulation extractor obtains half of the shares as the output of the OT by the perfect completeness of the OT, and the rest of the shares in the final step. 
Since the simulation extractor obtains all the secret shares of $m$, it recovers the correct committed message. 

Moreover, we can see that the simulated state of $C^*$ is computationally indistinguishable from the real one (regardless of whether the transcript is well-formed) as follows. We observe that the only difference between the real and simulated execution is that the result $t$ of the coin-flipping is programmed to be $(1-r_1)||(1-r_2)||...||(1-r_k)$ by using equivocality. Though the bits $r_1,...,r_k$ are also used as the receiver's inputs of the OT,  $\epsilon$-simulatable security of the OT against malicious senders ensure that they are computationally hidden from the view of the malicious committer.\footnote{We rely on a well-known fact that $\epsilon$-simulatable security implies indistinguishability-based security.} Thus, $t_R=t_C\oplus t$ is indistinguishable from uniformly random from the view of the malicious committer. Then we can reduce the indistinguishability between the real and simulated execution to equivocality of the equivocal commitment scheme.  

\para{Extractable Batch Commitments.} The remaining issue is how to achieve parallel extraction. If we have parallel equivocal commitments, then the above simulation extractor readily extends to the parallel setting. However, the problem is that we do not know how to achieve parallel equivocality in $\omega(1)$ rounds. To circumvent this issue, we change the syntax of the commitment protocol in the parallel setting. That is, instead of considering parallel execution of many copies of the same protocol, we consider a protocol where the committer commits to multiple messages at once, and we require the simulation extractor to extract all the committed messages. In this setting, we can use a single execution of coin-flipping subprotocol to generate the coins (i.e., $t$ in the above protocol) for all the sessions at once. This completely resolves the problem since now there is no parallel execution of the equivocal commitment scheme. In the actual proof, we formalize commitments with such modified syntax as \emph{batch} commitments (see \Cref{def:bcom}) and show that all the remaining steps work with this definition.

%% file: sections/preliminaries.tex

\section{Preliminaries}

\subsection{Basic Notations} 
Let $\secpar \in \Naturals$ denote security parameter. 
For a positive integer $n$, let $[n]$ denote the set $\{1,2,...,n\}$.
For a finite set $\mcal{X}$, $x\sample \mcal{X}$ means that $x$ is uniformly chosen from $\mcal{X}$.

A function $f:\mathbb{N}\ra [0,1]$ is said to be \emph{negligible} if for all polynomial $p$ and sufficiently large $\secpar \in \mathbb{N}$, we have $f(\secpar)< 1/p(\secpar)$; it is said to be \emph{overwhelming} if $1-f$ is negligible, and said to be \emph{noticeable} if there is a polynomial $p$ such that $f(\secpar)\geq  1/p(\secpar)$ for sufficiently large $\secpar\in \mathbb{N}$. We denote by $\poly$ an unspecified polynomial and by $\negl$ an unspecified negligible function. For two functions $f_1(\secpar)$ and $f_2(\secpar)$, we will often use $f_1(\secpar) = f_2(\secpar) \pm \negl(\secpar)$ as a shorthand for $|f_1(\secpar) - f_2(\secpar)| \le  \negl(\secpar)$.

Honest (classical) parties are modeled as interactive Turing machines (ITMs). We use PPT and QPT to denote (classical) probabilistic polynomial time and quantum polynomial time, respectively.
For a classical probabilistic or quantum algorithm $\mcal{A}$, $y\sample \mcal{A}(x)$ means that $\mcal{A}$ is run on input $x$ and outputs $y$. When we consider a non-uniform QPT adversary, we specify it by a sequence of polynomial-size quantum circuits with quantum advice $\{\mcal{A}_\secpar, \rho_\secpar\}_{\secpar\in\mathbb{N}}$. In an execution with the security parameter $\secpar$, $\mcal{A}$ runs $\mcal{A}_{\secpar}$ taking $\rho_\secpar$ as the advice.  
For simplicity, we often omit the index $\secpar$ and just write $\mcal{A}(\rho)$ to mean a non-uniform QPT algorithm specified by  $\{\mcal{A}_\secpar, \rho_\secpar\}_{\secpar\in \mathbb{N}}$.

\para{Notations for Indistinguishability.}
We may consider random variables over bit strings or over quantum states. 
This will be clear from the context.
We use the same notations for classical and quantum computational indistinguishability, but there should be no fear of confusion; It means computational indistinguishability against PPT (resp.\ QPT) distinguishers whenever we consider classical (resp.\ post-quantum) security.
For ensembles of random variables $\mathcal{X}=\{X_i\}_{\secpar\in \mathbb{N},i\in I_\secpar}$ and $\mathcal{Y}=\{Y_i\}_{\secpar\in \mathbb{N},i\in I_\secpar}$ over the same set of indices $I=\bigcup_{\secpar\in\mathbb{N}}I_\secpar$ and a function $\delta$,       
we use $\mathcal{X}\compind_{\delta}\mathcal{Y}$ to mean that for any non-uniform PPT (resp.\ QPT) algorithm $\mcal{A}$, there exists a negligible function $\negl(\cdot)$ such that for all $\secpar\in\mathbb{N}$, $i\in I_\secpar$, we have
\begin{equation}\label[Inequality]{eq:def:ind}
|\Pr[\mcal{A}(X_i)]-\Pr[\mcal{A}(Y_i)]|\leq \delta(\secpar) + \negl(\secpar).
\end{equation}
We say that $\mathcal{X}$ and $\mathcal{Y}$ are $\delta$-computationally indistinguishable if the above holds. 
In particular, when the above holds for $\delta=0$, we say that $\mcal{X}$ and $\mcal{Y}$ are computationally indistinguishable, and simply write $\mcal{X}\compind \mcal{Y}$.

Similarly, we use $\mcal{X}\sind_{\delta}\mcal{Y}$ to mean that for any unbounded time  algorithm $\mcal{A}$, there exists a negligible function $\negl(\cdot)$ such that for all $\secpar\in\mathbb{N}$, $i\in I_\secpar$, \Cref{eq:def:ind} holds. In particular, when the above hold for $\delta=0$, we say that $\mcal{X}$ and $\mcal{Y}$ are statistically indistinguishable, and simply write $\mcal{X}\sind \mcal{Y}$.
Moreover,  
we write $\mcal{X} \idind \mcal{Y}$ to mean
that $X_i$ and $Y_i$ are distributed identically for all $i\in I$. 

When we consider an ensemble $\mcal{X}$ that is only indexed by $\secpar$ (i.e., $I_\secpar=\{\secpar\}$), we write $\mcal{X}=\{X_\secpar\}_\secpar$ for simplicity.

\subsection{Post-Quantum Commitments}
We define (classically-secure and post-quantum) commitments. The following definitions are based on those in \cite{C:CCLY22}. 
\begin{definition}[Post-Quantum Commitments]\label{def:com}
A {\em post-quantum commitment scheme}  $\langle C, R \rangle$ is a classical interactive protocol between interactive \PPT machines $C$ and $R$. Let $m\in \bits^{\ell(\secpar)}$ (where $\ell(\cdot)$ is some polynomial) is a message that $C$ wants to commit to. The protocol consists of the following stages:
\begin{itemize}
\item
{\bf Commit Stage:} $C(m)$ and $R$ interact with each other to generate a transcript (which is also called a commitment) denoted by $\tau$,\footnote{That is, we regard the whole transcript as a commitment.} 
$C$'s state $\ST_{C}$, and 
$R$'s output $b_{\mathrm{com}}\in\Set{\bot,\top}$ indicating acceptance $(i.e., b_{\mathrm{com}}=\top)$ 
or rejection $(i.e., b_{\mathrm{com}}=\bot)$.
 We denote this execution by $(\tau,\ST_{C},b_{\mathrm{com}}) \gets \langle C(m), R \rangle(1^\secpar)$. When $C$ is honest, $\ST_C$ is classical, but when we consider a malicious quantum committer $C^*(\rho)$, we allow it to generate any quantum state $\ST_{C^*}$. 
 Similarly, a malicious quantum receiver $R^*(\rho)$ can output any quantum state, which we denote by $\OUT_{R^*}$ instead of $b_{\mathrm{com}}$. 
\item
{\bf Decommit Stage:}
$C$ generates a decommitment $\decom$ from $\ST_C$.
We denote this procedure by $\decom \gets C(\ST_C)$. 
Then it sends a message $m$ and decommitment $\decom$ to $R$, 
and $R$ outputs a bit
$b_{\mathrm{dec}}\in\Set{\bot, \top}$ indicating acceptance $(i.e., b_{\mathrm{dec}}=\top)$ 
or rejection $(i.e., b_{\mathrm{dec}}=\bot)$.
We assume that $R$'s verification procedure is deterministic and denote it
by $\Verify(\tau,m,\decom)$.\footnote{Note that $\Verify$ is well-defined since our syntax does not allow $R$ to keep a state from the commit stage.}
W.l.o.g., we assume that $R$ always rejects (i.e., $\Verify(\tau,\cdot,\cdot) = \bot$) whenever $b_\mathrm{com} = \bot$. (Note that w.l.o.g., $\tau$ can include $b_\mathrm{com}$ because we can always modify the protocol to ask $R$ to send $b_\mathrm{com}$ as the last round message.)
\end{itemize}

The scheme satisfies the following requirements:
\begin{enumerate}
\item
{\bf (Completeness.)} For any polynomial $\ell:\mathbb{N} \rightarrow \mathbb{N}$ and any $m \in \bits^{\ell(\secpar)}$, it holds that
\begin{equation*}
\Pr[b_{\mathrm{com}}=b_{\mathrm{dec}}=\top : 
\begin{array}{l}
(\tau, \ST_{C}, b_{\mathrm{com}}) \gets \langle C(m),R \rangle(1^\secpar) \\
\decom \gets C(\ST_C)\\
b_{\mathrm{dec}}\gets \Verify(\tau,m,\decom)
\end{array}
] = 1.
\end{equation*}

\item
{\bf (Statistically binding.)} For any unbounded-time committer $C^*$, the following holds: 
\begin{align*}
    \Pr[
    \begin{array}{l}
    \exists~m_0,m_1,\decom_0,\decom_1,~s.t.~m_0\neq m_1 ~\land\\
     \Verify(\tau,m_0,\decom_0)=\Verify(\tau,m_1,\decom_1)=\top
    \end{array}
    :(\tau,\ST_{C^*},b_{\mathrm{com}}) \gets \langle C^*, R \rangle(1^\secpar)]=\negl(\secpar).
\end{align*}

\item
{\bf (Computationally Hiding.)} For any non-uniform QPT receiver $R^*$ and any polynomial $\ell : \mathbb{N} \rightarrow \mathbb{N}$, the following holds:
$$\big\{ \OUT_{R^*}\langle C(m_0),R^* \rangle(1^\secpar)\big\}_{\secpar \in \mathbb{N}, ~m_0, m_1 \in \bits^{\ell(\secpar)}} ~\cind~ \big\{ \OUT_{R^*}\langle C(m_1),R^* \rangle(1^\secpar)\big\}_{\secpar \in \mathbb{N}, ~m_0, m_1 \in \bits^{\ell(\secpar)}},$$
where $\OUT_{R^*}\langle C(m_b),R^* \rangle(1^\secpar)$ $(b \in \bits)$ denotes the output of $R^*$ at the end of the commit stage. 
\end{enumerate}
\end{definition}

For a statistically binding commitment scheme (e.g., the one defined in \Cref{def:com}), we often need to talk about the actual value that is ``committed'' by the committer at the end of the commit stage. For that purpose. we develop a notion in \Cref{def:com-val} for such a value. Note that this value is not efficiently computable (before the starting of the decommit stage) due to the hiding property of the commitment scheme. Rather, it is simply defined in a information-thoeretical sense.
\begin{definition}[Committed Values]\label{def:com-val}
For a statistically binding commitment scheme $\langle C, R \rangle$ (as per \Cref{def:com}), we define the value function as follows:
\begin{equation*}
    \val(\tau)\defeq 
    \begin{cases}
    m&\text{~if~}\exists\text{~unique~}m\text{~s.t.~}\exists~\decom, \Verify(\tau,m,\decom)=1\\
    \bot &\text{otherwise}
    \end{cases}. 
\end{equation*}
where $\Verify$ is as defined in \Cref{def:com}.
\end{definition}

\subsection{Post-Quantum Extractable Commitments}
We define the post-quantum analog of extractable commitments, which we denote as PQ-ExtCom. As mentioned in the introduction, in the post-quantum setting, we need to explicitly require that the extractor (almost) does not disturb the (potentially malicious) committer's state during the extraction. However, it is not known black-box constructions of such post-quantum extractable commitments exist from (polynomially hard) post-quantum OWFs \cite{chia2022impossibility}.  
Fortunately, a recent work \cite{C:CCLY22} showed that a constant-round construction from post-quantum OWFs is possible if we relax the extractability to allow an (arbitrarily small) noticeable simulation error. The following definitions are taken from \cite{C:CCLY22}.

\begin{definition}[PQ-ExtCom with $\epsilon$-Simulation]\label{def:epsilon-sim-ext-com:strong}
A post-quantum commitment scheme $\langle C, R\rangle$ (as per \Cref{def:com}) is {\em extractable with $\epsilon$-simulation} if there exists a QPT algorithm $\SimExt$ (called the $\epsilon$-simulation extractor) such that for any noticeable $\epsilon(\secpar)$ and any non-uniform QPT $C^*(\rho)$, 
\begin{equation*}
\big\{ \SimExt^{C^*(\rho)}(1^\secpar,1^{\epsilon^{-1}}) \big\}_\secpar
\sind_\epsilon
\big\{(\val(\tau), \ST_{C^*}):(\tau,\ST_{C^*},b_{\mathrm{com}}) \gets \langle C^*(\rho), R \rangle(1^\secpar)\big\}_\secpar,  
\end{equation*}
where $\val(\tau)$ is the value committed by $C^*$ as defined in \Cref{def:com-val}.
\end{definition}


\para{Parallel Extractability.} We also define in \Cref{def:epsilon-sim-ext-com:parallel} the parallel version of \Cref{def:epsilon-sim-ext-com:strong}. This definition considers polynomially many instances of a commitment {\em in parallel}, where the committers are malicious. It requires the existence of a simulation-extractor $\SimExt$ that simultaneously extract the values committed in all the sessions, while $\epsilon$-simulating the post-extraction state of the malicious committers. This definition is a little weak in the sense that it only requires $\SimExt$ to succeed when $R$ accepts in all the parallel sessions. In particular, when $R$ accepts in some sessions but not in others, the $\SimExt$ does not need to extract (or simulate) anything. As remarked in \cite{C:CCLY22}, an alternative stronger (and more natural) definition would require the $\SimExt$ to extract the committed values in all the sessions where $R$ accepts in the $j$-th session, and $\epsilon$-simulate the post-extraction state of the malicious committers across all the sessions (even for those where $R$ rejects at the end of commit stage). However, such a construction {\em in constant rounds} remains an open challenge, even with non-black-box techniques.

Fortunately, this weak parallel version as per \Cref{def:epsilon-sim-ext-com:parallel} suffices for our purpose.

\begin{definition}[Parallel Extractability with $\epsilon$-Simulation]\label{def:epsilon-sim-ext-com:parallel}
A post-quantum commitment scheme $\langle C, R\rangle$ (as per \Cref{def:com}) is {\em parallelly extractable with $\epsilon$-simulation} if 
for any integer $n=\poly(\secpar)$, there exists a QPT algorithm $\SimExt$ (called the parallel $\epsilon$-simulation extractor) such that for any noticeable $\epsilon(\secpar)$ and any non-uniform QPT $C^*(\rho)$, 
\begin{align*}
&\big\{ \SimExt^{C^*(\rho)}(1^\secpar,1^{\epsilon^{-1}}) \big\}_\secpar\\
\sind_\epsilon
&\big\{\big(\Gamma_{\{b_{\mathrm{com},j}\}_{j=1}^{n}}(\{\val(\tau_j)\}_{j=1}^{n}), \ST_{C^*}\big):(\{\tau_j\}_{j=1}^{n},\ST_{C^*},\{b_{\mathrm{com},j}\}_{j=1}^{n})\sample\execution{C^*(\rho)}{R^n}(1^\SecPar)\big\}_\secpar
\end{align*}
where 
$(\{\tau_j\}_{j=1}^{n},\ST_{C^*},\{b_{\mathrm{com},j}\}_{j=1}^{n})\sample\execution{C^*(\rho)}{R^n}(1^\SecPar)$
means that $C^*(\rho)$ interacts with $n$ copies of the honest receiver $R$ in parallel and the execution results in transcripts  $\{\tau_j\}_{j=1}^{n}$, the final state $\ST_{C^*}$, and outputs $\{b_{\mathrm{com},j}\}_{j=1}^{n}$ of each copy of $R$ and 
$$\Gamma_{\{b_{\mathrm{com},j}\}_{j=1}^{n}}(\{\val(\tau_j)\}_{j=1}^{n}) \coloneqq 
\begin{cases}
\{\val_\Prot(\tau_j)\}_{j=1}^{n} & \text{if}~\forall~j\in[n]~ b_{\mathrm{com},j} = \top \\
\bot & \text{otherwise}
\end{cases}.
$$
\end{definition}

Constant-round and black-box constructions are know for the above versions of post-quantum extractable commitments.
\begin{lemma}[\cite{C:CCLY22}]
\label{lem:parallel-extcom:CCLY}
Assume the existence of post-quantum one-way functions, there exist constant-round black-box constructions of:
\begin{itemize}
\item
post-quantum extractable commitments with $\epsilon$-simulation (as per \Cref{def:epsilon-sim-ext-com:strong})
\item
post-quatnum parallelly extractable commitments with $\epsilon$-simulation (as per \Cref{def:epsilon-sim-ext-com:parallel}).
\end{itemize}
\end{lemma}

\subsection{Post-Quantum Non-Malleable Commitments}\label{sec:nmcom}

We define post-quantum non-malleable commitments (PQ-NMC). Our definition follows the one in \cite{FOCS:LPY23}, but here we define {\em 1-many} non-malleability directly. We only state the definition in the synchronous setting, supporting polynomially many tags. As mentioned in the introduction, this version suffices for all applications herein. 

In fact, we will define and rely on a form of 1-many non-malleability that is weaker than the standard notion, which we accordingly title {\em weak} (1-many) non-malleability.  This bears resemblance to the weak parallel extractability defined above for extractable commitments, in that we will only expect the non-malleability condition to hold in the parallel execution {\em provided every session of this execution is completed successfully} (i.e., the receiver in each session accepts the corresponding interaction).

We will start by defining the notion of a man-in-the-middle execution in the 1-many setting, and then present the standard and our weak definition of non-malleability in this setting.

\para{1-Many Man-in-the-Middle Execution.} Let $\langle C, R\rangle$ be a statistically binding and computationally hiding post-quantum commitment scheme. We use a {\em tag-based} specification so that every execution of $\langle C, R\rangle$ is associated with a tag $t\in [T]$, where $T$ is an integer. Consider a non-uniform QPT adversary $\mcal{M} = \Set{\mcal{M}_\secpar, \rho_\secpar}_\secpar$ participating in $(k+1)$ instances of $\langle C, R\rangle$ as follows: $\mcal{M}_\secpar(\rho_\secpar)$ plays the role of the receiver in one instance (referred to as the {\em left session}), while simultaneously acting as a committer in the other $k$ sessions (referred to as the {\em right sessions}). {\em All the $(k+1)$ sessions are execute in parallel}, and we refer to this setting as the {\em synchronous 1-$k$ MIM execution}, where ``MIM'' is the acronym for ``man-in-the-middle.'' 

Notation-wise, we denote the relevant entities used in the right interaction as the ``tilde'd'' version of the corresponding entities on the left. In particular, let $t$ denote the tag associated with the left session and $(\tilde{t}_1, \ldots, \tilde{t}_k)$ denote the tags for the $k$ right sessions respectively; let $m$ denote the value committed by the honest $C$ in the left session, and $(\tilde{m}_1, \ldots, \tilde{m}_k)$ the values committed by $\mcal{M}_{\secpar}(\rho_\secpar)$ in the $k$ right sessions respectively, i.e., we set $\tilde{m}_i=\val(\tilde{\tau}_i)$ where $\tilde{\tau}_i$ is the transcript of the $i$-th right session (see \Cref{def:com-val}). 

For this 1-$k$ MIM execution, let $\msf{mim}[k]^{\mcal{M}_\secpar}_{\langle C, R \rangle}(m, \rho_\secpar)$ denote concatenation of the final output of $\mcal{M}_\secpar(\rho_\secpar)$ and the values committed in all the $k$ right sessions, {\em when the honest $C$ in the left session commits to value $m$}. That is, 
$$\msf{mim}[k]^{\mcal{M}_\secpar}_{\langle C, R \rangle}(m, \rho_\secpar)\coloneqq \big(\OUT_{\mcal{M}}, (\tilde{m}_1, \ldots, \tilde{m}_k) \big).$$

\begin{definition}[Standard Synchronous 1-Many PQ-NMC]\label{def:NMCom:pq}
A post-quantum statistically binding commitment $\langle C, R \rangle$ is said to be 1-$k$ non-malleable if for all polynomial $\ell(\cdot)$ and all non-uniform QPT adversaries $\mcal{M} = \Set{\mcal{M}_\secpar, \rho_\secpar}_\secpar$ participating the above synchronous 1-$k$ MIM execution with $t \ne \tilde{t}_i$ for all $i \in [k]$, it holds that
$$\big\{\msf{mim}[k]^{\mcal{M}_\secpar}_{\langle C, R \rangle}(m_0, \rho_\secpar)\big\}_{\secpar \in \Naturals, m_0, m_1 \in \bits^{\ell(\secpar)}} ~\cind~ \big\{\msf{mim}[k]^{\mcal{M}_\secpar}_{\langle C, R \rangle}( m_1, \rho_\secpar)\big\}_{\secpar \in \Naturals, m_0, m_1 \in \bits^{\ell(\secpar)}}.$$
\end{definition}
Some remarks follow:
\begin{enumerate}
\item 
\Cref{def:NMCom:pq} requires that the left-session tag $t$ is different from those for all right sessions. This is standard practice when defining non-malleability, with the purpose of ruling out the uninteresting case when $\mcal{M}$ is simply acting as a channel, forwarding messages from $C$ on the left to the $R$ on some right session. 
\item 
 \Cref{def:NMCom:pq} does not consider entanglement between $\mcal{M}$'s auxiliary input and distinguisher's auxiliary input. However, \cite[Claim 3.1]{EC:BitLinShm22} shows that the above definition implies the version that considers such entanglement.  
 \item 
 {\bf (1-1 Non-Malleability.)}
 When $k=1$, \Cref{def:NMCom:pq} degenerates to the standard definition of post-quantum non-malleability (in the synchronous setting).
 \end{enumerate} 

Next we turn to our weaker definition of 1-many non-malleability. 

\begin{definition}[Weak 1-Many PQ-NMC]\label{def:NMCom:weak:pq}
Let $\msf{mim}[k]^{\mcal{M}_\secpar}_{\langle C, R \rangle}(m, \rho_\secpar)$ denote the output of the synchronous 1-$k$ man-in-the-middle execution as above, and let $d_j \in \Set{\top, \bot}$ denote the decision of the receiver in each session for $j \in [k]$. Define the function $\Gamma_{\{d_j\}_{j=1}^{k}}(\cdot)$ as follows: 
$$\Gamma_{\{d_j\}_{j=1}^{k}}(\msf{mim}[k]^{\mcal{M}_\secpar}_{\langle C, R \rangle}(m, \rho_\secpar)) \coloneqq 
\begin{cases}
\msf{mim}[k]^{\mcal{M}_\secpar}_{\langle C, R \rangle}(m, \rho_\secpar) & \text{if}~\forall~j\in [k], ~d_j = \top \\
(\OUT_{\mcal{M}},\bot^k) & \text{otherwise}
\end{cases},
$$
where $\OUT_{\mcal{M}}$ is the first component of $\msf{mim}[k]^{\mcal{M}_\secpar}_{\langle C, R \rangle}(m, \rho_\secpar)$.

\xiao{need to check: if we really need to simulate for $\OUT_{\mcal{M}}$ in the bad case.   }

A post-quantum statistically binding commitment $\langle C, R \rangle$ is said to be {\em weakly} 1-$k$ non-malleable if for all polynomial $\ell(\cdot)$ and all non-uniform QPT adversaries $\mcal{M} = \Set{\mcal{M}_\secpar, \rho_\secpar}_\secpar$ participating the above synchronous 1-$k$ MIM execution with $t \ne \tilde{t}_i$ for all $i \in [k]$, it holds that 

\begin{align*}
&\big\{\Gamma_{\{d_j\}_{j=0}^{k}}(\msf{mim}[k]^{\mcal{M}_\secpar}_{\langle C, R \rangle}(m_0, \rho_\secpar))\big\}_{\secpar \in \Naturals, m_0, m_1 \in \bits^{\ell(\secpar)}}\\
\cind~
&\big\{\Gamma_{\{d_j\}_{j=0}^{k}}(\msf{mim}[k]^{\mcal{M}_\secpar}_{\langle C, R \rangle}(m_1, \rho_\secpar))\big\}_{\secpar \in \Naturals, m_0, m_1 \in \bits^{\ell(\secpar)}}
.\end{align*}

\end{definition}

\begin{remark}
    Note that \Cref{def:NMCom:pq} and \Cref{def:NMCom:weak:pq} are in fact {\em equivalent} in the basic 1-1 setting (i.e., the man-in-the-middle runs exactly one left and right session each). This is easily verified by observing the definition of the output in the MIM experiment: it is clear to see that in cases where the right interaction is completed successfully, the definitions are identical. On the other hand, when the right interaction is not completed successfully, the output according to both \Cref{def:NMCom:pq} and \Cref{def:NMCom:weak:pq} consists of the output state of the MIM and the $\bot$ symbol. 
\end{remark}

\begin{remark}
    We further observe that in analogy with the standard definition, weak one-many non-malleability also implies weak {\em many-many} non-malleability (which can be defined analogously). As in the standard case, this can be easily inferred from a standard hybrid argument (where the input in each left session is switched in turn). 
\end{remark}






\subsection{Post-Quantum MPC with $\epsilon$-Simulation}
We present the formal definition for PQ-MPC with $\epsilon$-simulation. It is identical to the standard MPC definition in the classical setting except that:
\begin{enumerate}
\item
The malicious party can be a QPT machine;
\item
The indistinguishability between the real-world execution and the simulated one is parameterized by a noticeable function $\epsilon(\secpar)$. 
\end{enumerate}
Consider $n$ parties $P_1, \ldots, P_n$ who wish to interact in a protocol $\Prot$ to evaluate a $n$-party classical functionality $f$ on their joint inputs. They communicate via authenticated point-to-point channels as well as broadcast channels, where everyone can send messages in the same round. The network is assumed to be synchronous with rushing adversaries, i.e. adversaries may generate their messages for any round after observing the messages of all honest parties in that round, but before observing the messages of honest parties in the next round. 

In this work, we consider a {\em static} adversary, namely, at the
beginning of the execution the adversary specifies a set $I$ of corrupted parties which she controls, and through the execution she will not change the set $I$. The ideal and real executions follow the standard description as in, e.g., \cite{Goldreich04,EPRINT:Lindell16}.

In the real world, a non-uniformal QPT adversary $\Adv= \Set{\Adv_\secpar, \rho_\secpar}_\secpar$ corrupting $\Set{P_i}_{i\in I}$ interacts with $\Set{P_i}_{i\in [n] \setminus I}$. Let $\vb{x} = (x_1, \ldots, x_n)$ denote the respective initial input to each party. Let $\REAL_{\Prot, \Adv_\secpar, I}(\secpar, \vb{x}, \rho_\secpar)$ denote the random variable consisting of the output of the adversary (which may be an arbitrary function of its view and in particular may be a quantum state) {\em and} the outputs of the uncorrupted parties $\Set{P_i}_{i\in [n] \setminus I}$. 

In the ideal world, a QPT machine $\Sim$ controls the same parties in $I$ as $\Adv_\secpar$. It gets $\Set{x_i}_{i \in I}$ as input and is granted black-box access to $\Adv_\secpar(\rho_\secpar)$. Similar as in the $\epsilon$-ZK definition \cite{C:ChiChuYam21,C:CCLY22}, $\Sim$ additionally takes as input a ``slackness parameter'' $\epsilon(\secpar)$, which is a noticeable function on $\secpar$. Henceforth, we always require that $\Sim$'s running time is a polynomial on both $\secpar$ and $\epsilon^{-1}$. In this ideal-world execution, let $\IDEAL_{f,\Sim,I}(\secpar, \epsilon, \vb{x}, \rho_\secpar)$ denote the outputs of $\Sim$ (with slackness $\epsilon$) and the outputs of the uncorrupted parties $\Set{P_i}_{i\in [n] \setminus I}$. 

\begin{definition}[Post-Quantum MPC with  $\epsilon$-Simulation] \label{def:mpc}
 Let $f$ be a classical $n$-party functionality, and $\Prot$ be a classical $n$-party protocol. We say that $\Prot$ is a post-quantum MPC protocol for $f$ with $\epsilon$-simulation if there exists a QPT simulator $\Sim$ such that for any non-uniform QPT adversary $\Adv = \Set{\Adv_\secpar, \rho_\secpar}_{\secpar\in\Naturals}$, any $I \subset [n]$, any $\vb{x}\in (\bits^*)^n$, and any noticeable function $\epsilon(\secpar)$, it holds that:
$$\Set{\REAL_{\Prot, \Adv_\secpar, I}(\secpar, \vb{x}, \rho_\secpar)}_{\secpar \in \Naturals} ~\cind_\epsilon~ \Set{\IDEAL_{f,\Sim,I}(\secpar, \epsilon, \vb{x}, \rho_\secpar)}_{\secpar \in \Naturals}.$$
\end{definition}

\subsection{Verifiable Secret Sharing and Information-Theoretic MPC}

\para{Verifiable Secret Sharing.} We present in \Cref{def:VSS} the definition of verifiable secret sharing (VSS) schemes \cite{FOCS:CGMA85}. We remark that \cite{STOC:BenGolWig88,EC:CDDHR99} implemented $(n+1, \lfloor n/3 \rfloor)$-perfectly secure VSS schemes. These constructions suffice for all the applications in the current paper.
\begin{definition}[Verifiable Secret Sharing]\label{def:VSS}
An $(n + 1, t)$-perfectly secure VSS scheme $\Prot_\VSS$ consists of a pair of protocols $(\VSS_\Share, \VSS_\Recon)$ that implement respectively the sharing and reconstruction phases as follows.
\begin{itemize}
\item {\bf Sharing Phase $\VSS_\Share$:} 
Player $P_{n+1}$ (referred to as dealer) runs on input a secret $s$ and randomness $r_{n+1}$, while any other player $P_i$ $(i \in [n])$ runs on input a randomness $r_i$. During this phase players can send (both
private and broadcast) messages in multiple rounds.
\item {\bf Reconstruction Phase $\VSS_\Recon$:}
Each shareholder sends its view $v_i$ $(i \in [n])$ of the Sharing Phase to each other player, and on input the views of all players (that can include bad or empty views) each player outputs a reconstruction of the secret $s$.
\end{itemize}
All computations performed by honest players are efficient. The computationally unbounded adversary can corrupt up to t players that can deviate from the above procedures. The following security properties
hold.
\begin{enumerate}
\item {\bf Perfectly Verifiable-Committing:} \label[Property]{item:def:VSS:vc}
if the dealer is dishonest, then one of the following two cases happen (i.e., with probability 1): 
\begin{enumerate}
\item \label[Case]{item:def:VSS:vc:case:1}
During the Sharing Phase, honest players disqualify the dealer, therefore they output a special value $\bot$ and will refuse to play the reconstruction phase; 
\item \label[Case]{item:def:VSS:vc:case:2}
During the Sharing Phase, honest players do not disqualify the dealer. Therefore such a phase determines a unique value $s^*$
that belongs to the set of possible legal values that does not include $\bot$, which will be reconstructed by the honest players during the
reconstruction phase.
\end{enumerate} 
\item \label[Property]{item:def:VSS:secrecy}
{\bf Secrecy:} 
if the dealer is honest, then the adversary obtains no information about the shared secret before running the protocol $\Recon$. More accurately, there exists a PPT oracle machine $\Sim^{(\cdot)}$ such that for any message $m$, and every (potentially inefficient) adversary $\Adv$ corrupting a set $T$ of parties with $|T|\le t$ during the Sharing Phase $\VSS_\Share(m)$ (denote $\Adv$'s view in this execution as $\View_{\Adv, T}(1^\secpar, m)$), the following holds: 
$\Set{\View_{\Adv, T}(1^\secpar, m)} \idind \Set{\Sim^\Adv(1^\secpar, T)}$.
 


\item {\bf Correctness:}
 if the dealer is honest throughout the protocols, then each honest player will output the shared secret $s$ at the end of protocol $\Recon$.
\end{enumerate}
\end{definition}

\para{Information-Theoretically Secure MPC.} We first recall {\em information-theoretically secure} MPC and relevant notions that will be employed in the MPC-in-the-head paradigm shown later.

\para{Information-Theoretic MPC.} We now define MPC in the information-theoretic setting (i.e., secure against unbounded adversaries). 


\begin{definition}[Perfectly/Statistically-Secure MPC]\label{def:MPC}
Let $f:(\Set{0,1}^*)^n \mapsto (\Set{0,1}^*)^n$ be an n-ary functionality, and let $\Prot$ be a protocol. We say that $\Prot$ {\em $(n,t)$-perfectly (resp., statistically) securely} computes $f$ if for every static, malicious, and (possibly-inefficient) probabilistic adversary $\Adv$ in the real model, there exists a probabilistic adversary $\Sim$ of comparable complexity (i.e., with running time polynomial in that of $\Adv$) in the ideal model, such that for every $I \subset [n]$ of cardinality at most $t$, every $\vec{x} = (x_1,\dots,x_n) \in (\Set{0,1}^*)^n$ (where $|x_1|=\dots=|x_n|$), and every $z \in \Set{0,1}^*$, it holds that: 
$$\Set{\REAL_{\Prot,\Adv(z),I}(\vec{x})} \idind \Set{\IDEAL_{f,\Sim(z),I}(\vec{x})} ~~~~\big(\text{resp.,}~\Set{\REAL_{\Prot,\Adv(z),I}(\vec{x})} \sind \Set{\IDEAL_{f,\Sim(z),I}(\vec{x})}\big).$$   
\end{definition}
Recall that the MPC protocol from \cite{STOC:BenGolWig88} achieves $(n,t)$-perfect security (against static and malicious adversaries) with $t$ being a constant fraction of $n$.
\begin{theorem}[\cite{STOC:BenGolWig88}]\label{thm:BGW88}
Consider a synchronous network  with  pairwise  private  channels. Then,  for every $n$-ary functionality $f$, there exists a protocol that $(n,t)$-perfectly securely computes $f$ in the presence of a static malicious adversary for any $t < n/3$.
\end{theorem}

\para{Consistency, Privacy, and Robustness.} We now define some {notation} related to MPC protocols. Their roles will become clear when we discuss the MPC-in-the-head technique later. 
\begin{definition}[View Consistency]
\label{def:view-consistency}
A view $\View_i$ of an honest player $P_i$ during an MPC computation $\Prot$ contains input and randomness used in the computation, and all messages received from and sent to the communication tapes. A pair of views $(\View_i,\View_j)$ is {\em consistent} with each other if 
\begin{enumerate}
  \item Both corresponding players $P_i$ and $P_j$ individually computed each outgoing message honestly by using the random tapes, inputs and incoming messages specified in $\View_i$ and $\View_j$ respectively, and:
  \item All output messages of $P_i$ to $P_j$ appearing in $\View_i$ are consistent with incoming messages of $P_j$ received from $P_i$ appearing in $\View_j$ (and vice versa).  
\end{enumerate}   
\end{definition}
\begin{remark}[View Consistency of VSS]\label{rmk:VSS:view-consistency}
Although \Cref{def:view-consistency} defines view consistency for MPC protocols, we will also refer to the view consistency for the execution of verifiable secret sharing schemes (\Cref{def:VSS}). The views $(\view_i, \view_j)$ of players $i$ and $j$ (excluding the dealer) during the execution of $\VSS_\Share$ is said to be consistent if any only if $(\view_i, \view_j)$ satisfies the two requirements in \Cref{def:view-consistency}.
\end{remark}

We further define the notions of correctness, privacy, and robustness for multi-party protocols. 

\begin{definition}[Semi-Honest Computational Privacy]\label{def:t-privacy}
Let $1\leq t<n$, let $\Prot$ be an MPC protocol, and let $\Adv$ be any {\em static, PPT, and semi-honest} adversary. We say that $\Prot$ realizes a function $f:(\Set{0,1}^*)^n \mapsto (\Set{0,1}^*)^n$ with {\em semi-honest $(n,t)$-computational privacy} if there is a PPT simulator $\Sim$ such that for any inputs $x,w_1,\dots,w_n$, every subset $T \subset [n]$ $(|T| \leq t)$ of players corrupted by $\Adv$, and every $D$ with circuit size at most $\mathsf{poly(\SecPar)}$, it holds that 
{\fontsize{10.5pt}{0pt}\selectfont
\begin{equation}
\big|\Pr[D(\View_T(x,w_1,\dots,w_n)) =1]- \Pr[D(\Sim(T,x,\Set{w_i}_{i \in T},f_T(x,w_1,\dots,w_n)))=1]\big| \leq \negl(\SecPar),
\end{equation}}
where $\View_T(x,w_1,\dots,w_n)$ is the joint view of all players.
\end{definition}

\begin{definition}[Statistical/Perfect Correctness]\label{def:MPC-correctness}
Let $\Prot$ be an MPC protocol. We say that $\Prot$ realizes a deterministic n-party functionality $f(x,w_1,\dots,w_n)$ with {\em perfect (resp., statistical)} correctness if for all inputs $x,w_1,\dots,w_n$, the probability that the output of some party is different from the output of $f$ is 0 (resp., negligible in $k$), where the probability is over the independent choices of the random inputs $r_1,\dots,r_n$ of these parties.   
\end{definition}

\begin{definition}[Perfect/Statistical Robustness] \label{def:t-robustness}
Assume the same setting as the previous definition. We say that $\Prot$ realizes $f$ with {\em $(n,t)$-perfect (resp., statistical) robustness} if in addition to being perfectly (resp., statistical) correct in the presence of a semi-honest adversary as above, it enjoys the following {\em robustness} property against any computationally unbounded malicious adversary corrupting a set $T$ of at most $t$ parties, and for any inputs $(x,w_1,\dots,w_n)$: if there is no $(w_1',\dots,w_n')$ such that $f(x,w_1',\dots,w_n')=1$, then the probability that some uncorrupted player outputs 1 in an execution of $\Prot$ in which the inputs of the honest parties are consistent with $(x,w_1,\dots,w_n)$ is 0 (resp., negligible in \SecPar).    
\end{definition}

\subsection{MPC-in-the-Head}
\label{sec:prelim:MitH}

 MPC-in-the-head (MitH) is a technique originally developed for constructing  black-box ZK protocols from MPC protocols \cite{STOC:IKOS07}. Intuitively, the MPC-in-the-head idea works as follows.  Let $\Func_\textsc{zk}$ be the zero-knowledge functionality for an \NP language. Assume there are $n$ parties holding a witness in a secret-sharing form. $\Func_\textsc{zk}$ takes as public input $x$ and one share from each party, and outputs 1 iff the secret reconstructed from the shares is a valid witness.
To build a ZK protocol, the prover runs in his head an execution of MPC w.r.t.\ $\Func_\textsc{zk}$ among $n$ imaginary parties, each one participating in the protocol with a share of the witness. Then, it commits to the view of each party separately. The verifier obtains $t$ randomly chosen views, checks that such views are ``consistent'' (see \Cref{def:view-consistency}), and accepts if the output of every party is 1. The idea is that, by selecting the $t$ views at random, $V$ will catch inconsistent views if the prover cheats.

We emphasize that, in this paradigm, a malicious prover decides the randomness of each virtual party, including those not checked by the verifier (corresponding to honest parties in the MPC execution). Therefore, MPC protocols with standard computational security may fail to protect against such attacks. We need to ensure that the adversary cannot force a wrong output even if it additionally controls the honest parties' random tapes. The $(n,\lfloor n/3 \rfloor)$-perfectly secure MPC protocol in \Cref{thm:BGW88} suffices for this purpose (see also \Cref{rmk:exact-mpc-requriements:mpc-in-the-head}).

One can extend this technique further (as in \cite{FOCS:GLOV12}), to prove a general predicate $\phi$ about an arbitrary value $\alpha$.  Namely, one can consider the functionality $\Func_\phi$ in which party $i$ participates with input a VSS share $[\alpha]_i$. $\Func_\phi$ collects all such shares, and outputs 1 iff $\phi(\VSS_\Recon([\alpha]_1,\ldots,[\alpha]_n)) = 1$.

\begin{remark}[Exact Security Requirements on the Underlying MPC.] 
\label{rmk:exact-mpc-requriements:mpc-in-the-head}
To be more accurate, any MPC protocol that achieves {\em semi-honest $(n,t)$-computational privacy (as per \Cref{def:t-privacy}) and $(n,t)$-perfect robustness (as per \Cref{def:t-robustness})} will suffice for the MPC-in-the-head application.\footnote{It is also worth noting that the $(n,t)$-perfect robustness could be replaced with {\em adaptive $(n,t)$-statistical robustness}. See \cite[Section 4.2]{STOC:IKOS07} for more details.} These two requirements are satisfied by any $(n,t)$-perfectly secure MPC (and, in particular, the one from \Cref{thm:BGW88}).
\end{remark}

\para{MPC-in-the-Head Commitments.} We present a hiding game that relies on the (MitH) technique, and show that no QPT adversary can win this game with non-negligible probability. This game is essentially a black-box post-quantum commitment protocol due to \cite{FOCS:GLOV12,C:CCLY22}. We choose to present it as the following hiding game because this game will be of direct use later when proving the security of our protocol in \Cref{sec:pq-nmc:1-1:proof:reduction-to-hiding} (particularlly in \Cref{lem:bb-nmc:similarity:g:til}).

\begin{ExperimentBox}[label={chall:vss:hide}]{VSS Hiding Game}
{\bf Parameters:} Let $n(\secpar)$ be a polynomial in $\secpar$. Let $k$ be a constant-fraction of $n$ such that $k \le \frac{n}{3}$.

This involves an (efficient) challenger $\algo{Ch}$ interacting with the adversary $\Adv$. The interaction proceeds as follows: 
\begin{enumerate}
    \item \label[Step]{chall:vss:hide:step:1}
    $\Adv$ selects messages $m_0,m_1 \in \bits^\secpar$ and sends these to $\algo{Ch}$.
    
    \item \label[Step]{chall:vss:hide:step:2}
    Next, $\Adv$ samples a random size-$k$  subset $\eta$ of $[n]$. It then runs an interaction of $\ExtCom$ (as per \Cref{def:epsilon-sim-ext-com:strong}) with $\algo{Ch}$, where it commits to $\eta$. 

    \item \label[Step]{chall:vss:hide:step:3}

    $\algo{Ch}$ prepares $n$ views $\Set{\msf{v}_i}_{i \in [n]}$, corresponding to an MitH execution for the $(n+1, k)$-$\VSS_\Share$ of the message $m_b$ (see \Cref{rmk:mpc-in-the-head-vss} for details). $\algo{Ch}$ commits to each $\msf{v}_i$ ($i \in [n]$) independently in parallel, using Naor's commitment.

\begin{remark}\label{rmk:mpc-in-the-head-vss}
 We describe this step more explicitly. $\algo{Ch}$ emulates $n+1$ virtual parties $\Set{P_i}_{i \in [n+1]}$ `in its head.' Party $P_{n+1}$ is the dealer, possessing the string $x$. Other parties do not have any input. These parties execute the $\VSS_\Share$ stage of the $(n+1,k)$-VSS scheme to compute the functionality $\VSS_\Share$. At the end of the execution, $P_i$ ($i\in [n]$) obtains the $i$-th $\VSS$ share of $m_b$ as the output, and $P_{n+1}$ does not receive any output. The $\Set{\msf{v}_i}_{i \in [n]}$ corresponds to the views of $\Set{P_i}_{i \in [n]}$ from this execution (emulated in $\algo{Ch}$'s head).
 \end{remark} 

    \item \label[Step]{chall:vss:hide:step:4}
    $\Adv$ sends $\eta$ togther with the decommitment informaiotn w.r.t.\ the $\ExtCom$ in \Cref{chall:vss:hide:step:2}.

    \item \label[Step]{chall:vss:hide:step:5}
    $\algo{Ch}$ then decommits to the VSS shares in the set $\eta$, i.e. it sends $\Set{\msf{v}_i}_{i \in \eta}$ along with the corresponding decommitment information w.r.t.\ the commitment in \Cref{chall:vss:hide:step:3}. 

    \item \label[Step]{chall:vss:hide:step:6}
    Finally, $\Adv$ submits a guess bit $b'$ corresponding to its estimate of which message was committed to by $\algo{Ch}$.
\end{enumerate}

\para{Output:} We use $\VSS_{\msf{hd}}(1^\secpar, \Adv)$ to denote the output of this game, where $\VSS_{\msf{hd}}(\Adv) = 1$ iff $b' = b$.
\end{ExperimentBox}

\begin{lemma}\label{lem:game:VSS:hiding}
For any QPT adversary $\Adv$, it holds that
$\Pr[ \VSS_{\msf{hd}}(1^\secpar, \Adv) = 1] = \frac{1}{2} \pm \negl(\secpar)$, 
where $\VSS_{\msf{hd}}(1^\secpar, \Adv)$ is defined in \Cref{chall:vss:hide}.
\end{lemma}

\begin{proof}[Proof Sketch]
The proof of this lemma already appears in \cite[Section 6.5]{C:CCLY22}. Here, we only recall the high-level idea. We will show the when $\algo{Ch}$ changes the committed value from $m_0$ to $m_1$, $\Adv$ cannot tell the difference. For this, we assume for contradiction that $\Adv$ can tell the different with some inverse-polynomial advantage $\delta(\secpar)$ for infinitely many $\secpar \in \Naturals$. Then, $\algo{Ch}$ can extract the subset $\eta$ from \Cref{chall:vss:hide:step:2}, using the simulation-extractor $\SimExt$ guaranteed by \Cref{def:epsilon-sim-ext-com:strong}, setting the error parameter $\epsilon \coloneqq \frac{\delta}{3}$.

With the $\eta$ in hand, $\algo{Ch}$ does not need to generate the views $\Set{\msf{v}_i}_{i \in [n]}$ in \Cref{chall:vss:hide:step:3} honestly. Instead, it can invoke the $(n,k)$-MitH simulator to simulate the views in set $\eta$, and set other views  $\Set{\msf{v}_i}_{i \in [n]\setminus \eta}$ to all-0 strings with proper length. By the security of the underlying $(n,k)$-MPC, the views in $\eta$ does not contain any information of the committed value $m_b$. This helps $\algo{Ch}$ to change the committed value from $m_0$ to $m_1$.

By our parameter setting $\epsilon \coloneqq \frac{\delta}{3}$, after $\algo{Ch}$ changes the committed value from $m_0$ to $m_1$, $\Adv$ can tell the different with probability at most $\frac{2\delta}{3}$, which is still smaller than $\delta$, reaching the desired contradiction. 
    
\end{proof}

\para{MPC-in-the-Head Interactive Arguments.} As a proof of concept, we show in the following a constant-round black-box interactive argument built from the MitH technique. This protocol is taken from \cite{STOC:IKOS07,C:CCLY22}. A prover $C$ first commits to a string $x$, and then starts to interact with the verifier $R$ for the statement that the committed $x$ satisfies a predicate $\phi(\cdot)$. The soundness requirement is: if $\phi(x) = 0$, then the verifier will reject the proof except for negligible probability. 

\begin{ProtocolBox}[label={protocol:BB-ZK}]{MPC-in-the-Head Interactive Argument}
{\bf Parameters:} Let $n$ be a polynomial in $\SecPar$, and $k$ be a constant fraction of $n$ such that $k \le n/3$. We will employ a $(n+1, k)$ $\VSS$ scheme and a $(n,k)$-secure MPC scheme. 

{\bf Inputs:} Both parties receive $\secpar$ as the common input. The prover in addition gets a string $x$ as its private input.  

\para{Commit Stage:} In this stage, $C$ commits to the string $x$ (using the MitH approach).
\begin{itemize}
\item $C$ prepares $n$ views $\Set{\msf{v}_i}_{i \in [n]}$, corresponding to an MitH execution for the $(n+1, k)$-$\VSS_\Share$ of the string $x$ (see \Cref{rmk:mpc-in-the-head-vss} for details). $C$ commits to each $\msf{v}_i$ ($i \in [n]$) independently in parallel, using Naor's commitment.
\end{itemize}


{\bf Proof Stage:} Both parties learn an efficiently computable predicate $\phi(\cdot)$. 
\begin{enumerate}
   
    \item 
    $C$ then prepares $n$ views $\Set{\msf{v'}_i}_{i \in [n]}$ corresponding to an $(n, k)$-MitH execution for the functionality $F_\phi$ described below, where party $P_i$ uses $\msf{v}_i$ as input. It then commits to each of these views $\msf{v'}_i$ independently in parallel using Naor's commitment. 
    \begin{itemize}
        \item {\bf Functionality $F_\phi$:} This collects inputs $\msf{v}_i$ from party $i$, runs $\VSS_\Recon$ on these inputs to recover a value $x$, and outputs $\phi(x)$. 
    \end{itemize} 

    \item \label[Step]{protocol:BB-ZK:R-challenge}
    $R$ then samples a size-$k$ random subset $\eta \subset [n]$ and sends it to $C$.
    
    \item 
    $C$ now decommits to the views $\Set{\msf{v}_i}_{i \in \eta}$ and $\Set{\msf{v'}_i}_{i \in \eta}$. 
    \item 
    Finally, $R$ checks these decommitments, and also checks if these revealed views are {\em consistent} w.r.t.\ the VSS and MPC executions, here by `consistent' we refer to the consistency requirements as per \Cref{def:view-consistency} and \Cref{rmk:VSS:view-consistency}. It also checks for each $i\in \eta$ the final output of $P_i$ contained in $\msf{v}'_i$ is 1. It aborts immediately if any of the checks fail.
\end{enumerate}
\end{ProtocolBox}
We know state the properties \Cref{protocol:BB-ZK} satisfies as \Cref{protocol:BB-ZK:property}.

\begin{lemma}[\cite{STOC:IKOS07,C:CCLY22}.]\label{protocol:BB-ZK:property}
\Cref{protocol:BB-ZK} satisfies the following properties:
\begin{enumerate}
    \item
    The {\bf Commit Stage} is a statistically binding commitment.
\item
If the $x$ committed in the {\bf Commit Stage} satisfies $\phi(x) = 1$, then $R$ accepts with probability 1.
\item
If the  $x$ committed in the {\bf Commit Stage} satisfies $\phi(x) = 0$, then $R$ rejects except for with probability $O(2^k)$, even if $C^*$ is a malicious QPT machine and $\phi$ is picked by $C^*$.
\end{enumerate}
\end{lemma}

\para{Extension of \Cref{protocol:BB-ZK}.} We remark that \cite{STOC:IKOS07,C:CCLY22} indeed proved that \Cref{protocol:BB-ZK} can be converted into a $\epsilon$-simulatable zero-knowledge protocol if \Cref{protocol:BB-ZK:R-challenge} is replaced with a coin-flipping protocol that is $\epsilon$-simultable against any QPT $R^*$. Such a coin-flipping protocol can be constructed as follows: (1) $R^*$ to commit to a share $\eta_R$ using a extractable commitment with $\epsilon$-simulation (as per \Cref{def:epsilon-sim-ext-com:strong}); (2) $C$ sends a random share $\eta_C$; (3) $R$ decommits to $\eta_R$. The coin-flipping result will be $\eta \coloneqq \eta_C \xor \eta_R$.

\subsection{Watrous' Rewinding Lemma} \takashi{I guess this is only used in the construction of equivocal commitments. We may move it to a later section.}
The following is Watrous' rewinding lemma \cite{SIAM:Watrous09} in the form of  \cite[Lemma 2.1]{STOC:BitShm20}. 
\begin{lemma}[Watrous' Rewinding Lemma \cite{SIAM:Watrous09}]\label{lem:Watrous}
There is a quantum algorithm $\sfR$ that gets as input the following:
\begin{itemize}
\item A quantum circuit $\sfQ$ that takes $n$-input qubits in register $\reginp$ and outputs a classical bit $b$ and an $m$-qubit output.  
\item An $n$-qubit state $\rho$ in register $\reginp$.
\item A number $T\in \mathbb{N}$ in unary.
\end{itemize}

$\sfR(1^T,\sfQ,\rho)$ executes in time $T\cdot|\sfQ|$  and outputs a distribution over $m$-qubit states  $D_{\rho}\defeq \sfR(1^T,\sfQ,\rho)$  with the following guarantees.

For an $n$-qubit state $\rho$, denote by $\sfQ_{\rho}^0$ the conditional distribution of the output distribution $\sfQ(\rho)$,
conditioned on $b = 0$, and denote by $p(\rho)$ the probability that $b = 0$. If there exist $p_0, q \in (0,1)$, $\gamma \in (0,\frac{1}{2})$
such that:
\begin{itemize}
    \item  Amplification executes for enough time: $T\geq \frac{\log (1/\gamma)}{4p_0(1-p_0)}$,
    \item  There is some minimal probability that $b = 0$: For every $n$-qubit state $\rho$, $p_0\leq p(\rho)$,
    \item  $p(\rho)$ is input-independent, up to $\gamma$ distance: For every $n$-qubit state $\rho$, $|p(\rho)-q|<\gamma$, and
    \item  $q$ is closer to $\frac{1}{2}$: $p_0(1-p_0)\leq q(1-q)$,
\end{itemize}
then for every $n$-qubit state $\rho$,
\begin{align*}
    \TD(\sfQ_{\rho}^0,D_{\rho})\leq 4\sqrt{\gamma}\frac{\log(1/\gamma)}{p_0(1-p_0)}.
\end{align*}

Moreover, $\sfR(1^T,\sfQ,\rho)$ only makes black-box use of $\sfQ(\rho)$.\footnote{The black-boxness is observed in \cite{C:ChiChuYam21,C:CCLY22}} 
\end{lemma} 

%% file: sections/tct.tex

\section{Post-Quantum Trapdoor Coin-Tossing with $\epsilon$-Simulation}
\label{sec:td-cf}

In this section, we build a {\em trapdoor coin-flipping} protocol. This is a coin-flipping protocol between two parties $C$ and $R$ where $C$ additionally commits to some string $x$ before the coin-flipping starts. The security guarantee of the actual coin-flipping stage is `controlled' by the committed string $x$ and a predicate $\phi(\cdot)$ that is given to both party at the beginning of the coin-flipping stage. In particular, if the committed $x$ does not satisfy the predicate $\phi(\cdot)$, then even a cheating $C^*$ cannot gain any advantage over $R$ in the coin-flipping stage. This is formalized as the standard simulation-based requirement, namely, we require the existence of a simulator that can `enforce' the coin-tossing result to a given random string if $C^*$ does not abort the execution.

On the other hand, if the committed $x$ satisfies the predicate, then $C$ can enforce the coin-tossing result to a given random string against any efficient $R^*$ that does not abort the execution (this is what we call `trapdoor'). We require that this `enforcing procedure' should happen in straight-line, i.e., it is not allowed to rewind $R^*$. Indeed, we actually require a stronger version of this property as follows.

We consider a real-world execution where $C$ commits honestly to a string $x$ to the malicious $R^*$; Here, we do not require that $x$ satisfies $\phi$. Then, in the simulation world, we provide a potentially different $x'$ to a `straight-line' simulator $\Sim$ such that $\phi(x')=1$. With this valid witness $x'$, $\Sim$ can enforce the coin-flipping result to a pre-sampled random string, interacting in straight-line with a potentially malicious $R^*$.

In the following, we first present the definition in \Cref{sec:com-n-prove:def} and then show our construction with security proof in \Cref{sec:com-n-prove:constr}. We remark that this construction is a simple application of the MPC-in-the-Head technique. Indeed, the main contribution of this work does not lie in this construction. We abstract out the notion of `trapdoor coin-flipping' simply because it helps us present our post-quantum non-malleable commitments (in \Cref{sec:PQNMC:1-1}) in a modular manner.  

\subsection{Definition}
\label{sec:com-n-prove:def}

\begin{definition}
\label{def:com-n-prove}
A post-quantum trapdoor coin-tossing with $\epsilon$-simulation protocol consists of a pair of \PPT ITM $\langle C, R \rangle$. Let $x\in \bits^{\ell(\secpar)}$ (where $\ell(\cdot)$ is some polynomial) is a message that $C$ wants to commit to. The protocol consists of the following stages (we omit the input $1^\secpar$ to $C$ and $R$): 
\begin{itemize}
\item 
{\bf Commit Stage:} 
$C(x)$ and $R$ interacts to generates a transcript (commitment) $\com$, 
$C$'s state $\ST_C$, and $R$'s decision bit $b\in \Set{\top, \bot}$ indicating acceptance (i.e., $b=\top$) or rejection (i.e., $b=\bot$). We denote this execution as $(\com, \ST_C, b) \gets \langle C(x), R \rangle_\textsc{com}$. 
Note that a malicious receiver is allowed to output any quantum state, which we denote by $\ST_{R^*}$ instead of $b$, and to keep the state for the following stages.

\item 
{\bf Decommit Stage:}\footnote{This stage is rarely executed in applications.} $C(\ST_C)$ generates a decommitment $\decom$ and sends it to $R$ along with a message $x$. $R$ accepts or rejects. 

\item 
{\bf Coin-Flipping Stage:} 
Let $\phi(\cdot)$ be any predicate. $C(\ST_C, \phi)$ and $R(\com,\phi)$ interacts, after which $R$ outputs $\top$ (accept) or $\bot$ (reject). We denote the execution of this stage as $(\eta_1, \eta_2) \gets \langle C(\ST_C), R(\com) \rangle^\phi_\textsc{cf}$, where $\eta_1, \eta_2 \in \bits^*$ are the respective output of $C$ and $R$. Note that a malicious receiver is allowed to output any quantum state, which we denote by $\OUT_{R^*}$ instead of $\eta_2$. 

\end{itemize}
The following requirements are satisfied:
\begin{enumerate}
	\item \label[Property]{def:com-n-prove:property:1}
{\bf Statistically Binding.} The Commit Stage and Decommit Stage together constitute a post-quantum commitment scheme that is statistically binding and (post-quantum) computational hiding.

\item \label[Property]{def:com-n-prove:property:2}
{\bf Completeness.} For any $x \in \bits^{\ell(\secpar)}$ and any predicate $\phi$, it holds that
\begin{equation}
\Pr[\eta_1 = \eta_2 ~:~ 
\begin{array}{l}
(\com,\ST_C,b) \gets \langle C(x),R \rangle_\textsc{com} \\
(\eta_1, \eta_2) \gets \langle C(\ST_C), R(\com) \rangle^\phi_\textsc{cf}
\end{array}
] = 1.
\end{equation}

\item \label[Property]{def:com-n-prove:property:3}
{\bf Security against Malicious $R^*$.} For any QPT $R^*$, there exists a `straight-line' QPT simulator $\Sim$ such that for any efficiently computable predicate $\phi(\cdot)$, any $x, x'\in \bits^{\ell(\secpar)}$ such that $\phi(x') = 1$, it holds that
\begingroup
 \fontsize{8pt}{10pt}\selectfont
$$
\bigg\{
(\eta_1, \OUT_{R^*})
~:~
\begin{array}{l}
(\com,\ST_C, \ST_{R^*}) \gets \langle C(x),R^* \rangle_\textsc{com} \\
(\eta_1, \OUT_{R^*}) \gets \langle C(\ST_C), R^*(\ST_{R^*}) \rangle^\phi_\textsc{cf}
\end{array}
\bigg\}_\secpar 
\cind
\bigg\{
\big(\Gamma_{d}(\eta), \OUT_{R^*}\big)
~:~
\begin{array}{l}
(\ST_S, \ST_{R^*}) \gets \langle \Sim(x') ,R^* \rangle_\textsc{com} \\
\eta \picks \bits^{k(\secpar)}\\
(d, \OUT_{R^*}) \gets \langle \Sim(\ST_S,\eta), R^*(\ST_{R^*}) \rangle^\phi_\textsc{cf}
\end{array}
\bigg\}_\secpar, 
$$
\endgroup
where $\Gamma_d(\eta) \coloneqq \begin{cases}
\eta & \text{if}~d = \top\\
\bot & \text{if}~d = \bot
\end{cases}$.

\item \label[Property]{def:com-n-prove:property:4}
{\bf Security against Malicious $C^*$.} For any QPT $C^*$ with $\ST_{C^*}$  there exists a QPT simulator $\Sim$ such that for any noticeable $\epsilon(\secpar)$,  any predicate $\phi(\cdot)$ and and any $\com^*$such that $\phi\big(\msf{val}(\com^*)\big) = 0$,\footnote{Recall from \Cref{def:com-val} that $\msf{val}(\com^*)$ is the value statistically bound in transcript $\com^*$.} it holds that
\begingroup
 \fontsize{9pt}{10pt}\selectfont
$$
\bigg\{
(\OUT_{C^*}, \eta_2)
~:~
(\OUT_{C^*}, \eta_2) \gets \langle C^*(\ST_{C^*}), R(\com^*) \rangle^\phi_\textsc{cf}
\bigg\}_\secpar 
\cind_\epsilon
\bigg\{
\big(\OUT_{C^*}, \Gamma_{d}(\eta) \big)
~:~
\begin{array}{l}
\eta \picks \bits^{k(\secpar)}\\
(\OUT_{C^*}, d) \gets  \Sim(1^\secpar, \eta, \phi, \com^*)
\end{array}
\bigg\}_\secpar, 
$$
\endgroup
where $\Gamma_d(\eta)$ is defined as above.

\end{enumerate}
\end{definition}
Some remarks regarding \Cref{def:com-n-prove} follows:
\begin{itemize}
\item
Note that the above completeness condition (\Cref{def:com-n-prove:property:2}) holds regardless of whether $x$ satisfies the predicate $\phi(\cdot)$ or not.

\item
When defining \Cref{def:com-n-prove:property:3,def:com-n-prove:property:4}, we essentially follow the standard simulation-based notion for two-party coin-flipping {\em with aborting}. That is, we require the simulator to successfully `enforce' the coin-flipping result to the pre-sampled $\eta$ only if the malicious party does not abort the protocol. Indeed, if the malicious party aborts before the protocol ends, the honest party cannot receive any output, and thus of course we cannot enforce its output to $\eta$ while being consistent with the real-world execution (where the honest party simply outputs $\bot$).  
\end{itemize}

\subsection{Construction}
\label{sec:com-n-prove:constr}

We present the construction in \Cref{prot:td:ct}. It makes use of (the post-quantum version of) Naor's commitment and the post-quantum extractable commitment $\ExtCom$ with $\epsilon$-simulation (as per \Cref{def:epsilon-sim-ext-com:strong}). Note that both of these building blocks are known in black-box and constant-round from post-quantum OWFs.

\begin{ProtocolBox}[label={prot:td:ct}]{Post-Quantum Trapdoor Coin-Flipping with $\epsilon$-Simulation}

\para{Input:}
Both the  $C$ and $R$ get the security parameter $1^\secpar$ as the common input. $C$ gets a string $x \in \bits^{\ell(\secpar)}$ as its private input.

\para{Commit Stage:}
\begin{enumerate}
\item \label[Step]{prot:td:ct:step:1}
$C$ prepares $n$ views $\Set{\msf{v}_i}_{i \in [n]}$, corresponding to an MitH execution for the $(n+1, k)$-$\VSS_\Share$ of the string $x$ (as detailed in \Cref{rmk:mpc-in-the-head-vss}). $C$ commits to each $\msf{v}_i$ ($i \in [n]$) independently in parallel, using Naor's commitment.
\end{enumerate}

\para{Decommit Stage:}
\begin{enumerate}
\item
$C$ sends $\Set{\msf{v}_i}_{i\in [n]}$ together with the decommitment information w.r.t.\ the commitments in \Cref{prot:td:ct:step:1}. 
\item
$R$ checks the validity of the decommitment information and the consistency among $\Set{\msf{v}_i}_{i\in [n]}$ (as per \Cref{rmk:VSS:view-consistency}). If these checks are successful, $R$ recovers $x$ as $x \coloneqq \VSS_\Recon(\msf{v}_1, \ldots, \msf{v}_n)$; otherwise, $R$ rejects and output $\bot$. 
\end{enumerate}

\para{Coin-Flipping Stage:} At the beginning of this stage, both parties learn  the description of a predicate $\phi(\cdot)$. They proceed as follows.
\begin{enumerate}
\item \label[Step]{prot:td:ct:step:2}

$C$ picks $\eta_C \picks \bits^{k(\secpar)}$ and commits to it with MitH. Namely, $C$ prepares $n$ views $\Set{\msf{cfv}^{(1)}_i}_{i \in [n]}$ corresponding to an MitH execution for the $(n+1, k)$-$\VSS_\Share$ of the value $\eta_C$. $C$ commits to each $\msf{cfv}^{(1)}_i$ ($i \in [n]$) independently in parallel, using $\ExtCom$. 

\item \label[Step]{prot:td:ct:step:3}
$R$ picks $\eta_R \picks \bits^{k(\secpar)}$ and sends it to $C$.

\item \label[Step]{prot:td:ct:step:4}
$C$ now sets $\eta'_C \coloneqq \eta_C$ and sends $\eta'_C$ to $R$.

\item \label[Step]{prot:td:ct:step:5}
$C$ prepares $n$ views $\Set{\msf{cfv}^{(2)}_i}_{i \in [n]}$, corresponding to an $(n,k)$-MitH execution of the $n$-party functionality $F_\phi$ described below, where party $P_i$ ($i\in [n]$) uses $\msf{v}_i \| \msf{cfv}^{(1)}_i$ as input. (Note that $\msf{v}_i \| \msf{cfv}^{(1)}_i$ will be the prefix of $\msf{cfv}^{(2)}_i$.)  $C$ commits to each $\msf{cfv}^{(2)}_i$ ($i \in [n]$) independently in parallel, using Naor's commitment. Here, the honest committer will use as an `effective witness' the value $\eta_C$ reconstructed from $\Set{\msf{cfv}^{(2)}_i}_{i \in [n]}$, and hence only evaluates the `first clause' of $F_\phi$ (described below) in the virtual MPC execution. 
\begin{itemize}
\item {\bf Functionality  $F_\phi$:} It has $\eta'_C$ hare-wired. It collects input (and parses it as) $\msf{v}_i \| \msf{cfv}^{(1)}_i$ from party $i$ for each $i \in [n]$. It then runs the recovery algorithm of $\VSS$ to obtain $a \coloneqq \VSS_\Recon(\msf{v}_1, \ldots, \msf{v}_n)$ and $b \coloneqq \VSS_\Recon(\msf{cfv}^{(1)}_1, \ldots, \msf{cfv}^{(1)}_n)$. It outputs 1 to each party if {\em either}
\begin{itemize}
    \item
    (First clause.) $b$ equals $\eta'_C$ sent in \Cref{prot:td:ct:step:4}, {\em or}
    \item
    (Second clause.) $\phi(a) = 1$.
\end{itemize} 
Otherwise, it outputs 0 to each party.
\end{itemize}

\item \label[Step]{prot:td:ct:step:6}
$C$ and $R$ now engage in the following coin-flipping subprotocol as detailed below.  
\begin{enumerate}
\item \label[Step]{prot:td:ct:step:6:1}
$R$ samples a random string $\theta_R$ of proper length and commits to it using $\ExtCom$.   
\item
$C$ samples a random string $\theta_C$ of proper length and sends it to $R$.  
\item
$R$ sends to $C$ the value $\theta_R$ together with the corresponding decommitment information w.r.t.\
the $\ExtCom$ in \Cref{prot:td:ct:step:6:1}. Now, $C$ and $R$ agree on a random string $\theta \coloneqq \theta_R \xor \theta_C$. By a proper choice of length, the string $\theta$ it can be interpreted as specifying a size-$k$ random subset of $[n]$. In the following, we abuse notation by using $\theta$ to denote the corresponding size-$k$ random subset.
\end{enumerate}

\item \label[Step]{prot:td:ct:step:7}
$C$ sends back the list $\Set{(\msf{v}_i,\msf{cfv}^{(1)}_i,\msf{cfv}^{(2)}_i)}_{i \in \theta}$ 
together with the corresponding decommitment information w.r.t.\ the commitment made in \Cref{prot:td:ct:step:1,prot:td:ct:step:2,prot:td:ct:step:5}.

\item
$R$ now checks the validity of the decommitments provided by $C$ and the consistency among the revealed views $\Set{(\msf{v}_i, \msf{cfv}^{(1)}_i, \msf{cfv}^{(2)}_i}_{i \in \theta}$. It also checks for each $i\in \theta$ the final output of $P_i$ contained in $\msf{cfv}^{(2)}_i$ is 1. If any of these checks fail, $R$ aborts immediately.
\end{enumerate}

\para{Output:} $C$ and $R$ output $\eta \coloneqq \eta'_C \xor \eta_R$.
\end{ProtocolBox}

\para{Security.} The security of \Cref{prot:td:ct} is stated as the following lemma.
\begin{lemma}\label{lem:td-cf}
\Cref{prot:td:ct} is a black-box, constant-round trapdoor coin-flipping protocol (as per \Cref{def:com-n-prove}).
\end{lemma}
\begin{proof}

It is straightforward to see that \Cref{prot:td:ct} is constant-round and makes only black-box use of its underlying cryptographic components. Also, the completeness (i.e., \Cref{def:com-n-prove:property:2}) and statistically binding property (i.e., \Cref{def:com-n-prove:property:1}) of \Cref{prot:td:ct} is straightforward from the protocol description. Thus, to prove \Cref{lem:td-cf}, we only need to show that \Cref{prot:td:ct} satisfies \Cref{def:com-n-prove:property:3,def:com-n-prove:property:4} in \Cref{def:com-n-prove}.

As we mentioned earlier, \Cref{prot:td:ct} is a simple application of the MPC-in-the-Head technique. Indeed, similar constructions have appeared previously in, e.g., \cite{FOCS:GLOV12,C:CCLY22}. Our proof for \Cref{def:com-n-prove:property:3,def:com-n-prove:property:4} follows almost immediately from these known techniques. In the following, we only provide a proof sketch and refer the interested readers to \cite[Section 6.5]{C:CCLY22} for similar proofs.

\para{Proving \Cref{def:com-n-prove:property:3}.} Note that \Cref{prot:td:ct} can be viewed as a black-box commit-and-prove protocol where $C$ commits to two values, one being the $x$ committed in \Cref{prot:td:ct:step:1} of the {\bf Commit Stage} and the other being $\eta_C$ committed in \Cref{prot:td:ct:step:2} of the {\bf Coin-Flipping Stage}; $C$ then proves that these two values satisfies the predicate $F_\phi$ defined in \Cref{prot:td:ct:step:5}. Note that such a black-box commit-and-prove protocol is {\em witness indistinguishable} (see \cite[Section 6.5]{C:CCLY22} for a proof of this fact\footnote{Indeed, \cite[Section 6.5]{C:CCLY22} proves a stronger claim that such a protocol is $\epsilon$-zero-knowledge, which implies witness indistinguishability.}), namely, even if $C$ changes the committed values to another $x'$ and $\eta''_C$, {\em as long as $x'$ and $\eta''_C$ together satisfy the predicate $F_\phi$}, then no (potentially malicious) $R^*$ could tell the difference.

With this observation, a simulator $\Sim$ can be constructed as follows. $\Sim$ commits to $x'$ in \Cref{prot:td:ct:step:1} of the {\bf Commit Stage} and emulates the honest $C$, where recall from the condition of \Cref{def:com-n-prove:property:3} that $x'$ is a valid witness for $\phi$ and is given to $\Sim$ as input. In \Cref{prot:td:ct:step:4}, instead of setting $\eta'_C$ to $\eta_C$ honestly, $\Sim$ sets $\eta'_C = \eta \xor \eta_R$, where recall that $\eta$ is the input to $\Sim$ that it wants to enforce. Now, since $\eta'_C \ne \eta_C$, the `first clause' of $F_\phi$ does not hold. Fortunately, since $\phi(x') = 1$, the `second clause' of $F_\phi$ still holds. Thus, the new $x'$ together with $\eta_C$ constitute a valid witness for $F_\phi$. Then, the witness indistinguishability of this commit-and-prove protocol implies \Cref{def:com-n-prove:property:3}. 

\para{Proving \Cref{def:com-n-prove:property:4}.} This proof is even simpler. $\Sim$ works by extracting the $\eta^*_C$ committed by $C^*$ in \Cref{prot:td:ct:step:2} of the {\bf Coin-Flipping Stage}, using the (parallel) extractability of $\ExtCom$ with $\epsilon$-simulation. In more details, recall that the shares $\Set{\msf{cfv}^{(1)}_i}_{i \in [n]}$  is committed in parallel using $\ExtCom$ in \Cref{prot:td:ct:step:2}. $\Sim$ will extract all of these shares using the parallel extractability of $\ExtCom$ with $\epsilon$-simulation (as per \Cref{def:epsilon-sim-ext-com:parallel}), and compute $\eta^*_C\coloneqq \VSS_{\Recon}(\msf{cfv}^{(1)}_1, \ldots, \msf{cfv}^{(1)}_n)$.

 Then, $\Sim$ sends $\eta_R \coloneqq \eta \xor \eta^*_C$ in \Cref{prot:td:ct:step:3}, and finishes the remaining execution emulating the honest $R$.

First, note that in a real-world execution between $C^*$ and $R$, the value $\eta^*_C$ (committed by $C^*$ in \Cref{prot:td:ct:step:2} of the {\bf Coin-Flipping Stage}) must equal to the $\eta'_C$ sent by $C^*$ in \Cref{prot:td:ct:step:4}. This is because we know that the value committed by $C^*$ in \Cref{prot:td:ct:step:1} of the {\bf Commit Stage} does not satisfies the predicate $\phi(\cdot)$ (this is state as $\phi\big(\msf{val}(\com^*)\big) = 0$ in the condition of  \Cref{def:com-n-prove:property:4}). Thus, if $C^*$ sends $\eta'_C \ne \eta^*_C$ in \Cref{prot:td:ct:step:4}, then both clauses of $F_\phi$ are unsatisfied. In this situation, $C^*$ cannot provide a convincing proof to $R$; Otherwise, she breaks the soundness of the commit-and-prove protocol. Again, see \cite[Section 6.5]{C:CCLY22} for a similar proof.

 By the extractability of $\ExtCom$, $\Sim$ is able to extract the same $\eta^*_C$ while simulating the post-extraction state of $C^*$ up to a arbitrarily small noticeable $\epsilon(\secpar)$ error, and successfully enforce the coin-flipping result to $\eta^*_C \xor \eta^*_C\xor\eta= \eta$. This establishes \Cref{def:com-n-prove:property:4}.

This concludes the proof of \Cref{lem:td-cf}.

\end{proof}

%% file: sections/one-sided.tex

\section{Post-Quantum Non-Malleable Commitments: One-to-One and One-sided}
\label{sec:PQNMC:1-1}
We present a black-box, constant-round construction of post-quantum non-malleable commitment in the synchronous setting and supporting a polynomial number of tags. The construction shown in this section is non-malleable under the `one-sided' assumption, i.e., its non-malleability holds in the 1-1 MIM execution where the left-session tag $t$ is strictly smaller than the right-session tag $\tilde{t}$. This constraint will be subsequently alleviated in \Cref{sec:two-sided:main-body}. 

\subsection{Construction}
\label{sec:BB-NMC:construction}

The construction is described in \Cref{protocol:BB-NMCom}. It makes black-box use of the following building blocks:
\begin{enumerate}
\item
The post-quantum parallelly extractable commitment scheme $\ExtCom$ with $\epsilon$-simulation \Cref{def:epsilon-sim-ext-com:parallel}, which can be built in black-box from any post-quantum OWFs (see \Cref{lem:parallel-extcom:CCLY}).

\item
    A post-quantum commitment scheme that is statistically-binding and computationally-hiding (against QPT adversaries). This is also known assuming only black-box access to post-quantum secure OWFs. In particular, we will make use of Naor's commitment which can be built in black-box from any post-quantum OWFs as well.
\item
     A perfectly secure verifiable secret sharing scheme $\VSS = (\VSS_{\Share}, \VSS_{\Recon})$ (as per \Cref{def:VSS}); We will set the parameters to make use of an $(n+1, k)$ scheme and an $(n+1, 2k)$ scheme. See \Cref{protocol:BB-NMCom} for details.
\item
    A semi-honest computationally private and perfectly robust MPC protocol (see \Cref{rmk:exact-mpc-requriements:mpc-in-the-head}); We will set the parameters to make use of an $(n, k)$ scheme and an $(n, 2k)$ scheme. See \Cref{protocol:BB-NMCom} for details. 
\end{enumerate}
We remark that the MPC and VSS are used to implement the MPC-in-the-Head (MitH) technique as explained in \Cref{sec:prelim:MitH}.

\begin{ProtocolBox}[label={protocol:BB-NMCom}]{One-Sided PQ-NMC: Black-Box and Constant-Round}
{\bf Parameter Setting:} The tag space is defined to be $[T]$, where $T$ is a polynomial in the security parameter $\secpar$. Let $n$ be a polynomial in $\SecPar$, and $k$ be a constant fraction of $n$ such that $2k \le n/3$. 

\para{Input:}
Both the committer $C$ and the receiver $R$ get the security parameter $1^\secpar$ and a tag $t \in [T]$ as the common input; $C$ gets a string $m \in \bits^{\ell(\SecPar)}$ as its private input, where $\ell(\cdot)$ is a polynomial. 

\para{Commit Phase:}
\begin{enumerate}
\item\label[Step]{bbnmc:init-com}
{\bf (Initial Com to $m$.)} In this stage, $C$ commits to the message (using the MPC-in-the-head approach).
\begin{itemize}
\item $C$ prepares $n$ views $\Set{\msf{cv}^{(1)}_i}_{i \in [n]}$, corresponding to an MitH execution for the $(n+1, 2k)$-$\VSS_\Share$ of the message $m$ (as detailed in \Cref{rmk:mpc-in-the-head-vss}). $C$ commits to each $\msf{cv}^{(1)}_i$ ($i \in [n]$) independently in parallel, using Naor's commitment.
\end{itemize}


\item \label[Step]{bbnmc:hard-puzzle:puzzle-setup}
{\bf (Hard Puzzle Setup.)} In this stage, $R$ sets up a $t$-solution hard puzzle. It then commits to one solution of the puzzle and proves in zero-knowledge the consistency (using the MPC-in-the-head approach).
\begin{enumerate}

\item \label[Step]{bbnmc:hard-puzzle:com-ch}
$C$ samples a size-$k$ random subset $\msf{ch} \subseteq [n]$, and commits to it using $\ExtCom$. 

\item \label[Step]{bbnmc:hard-puzzle:rv-1}
$R$ samples $t$ random strings $x_1,\dots,x_t \pick \bits^\secpar$. $R$ prepares $n$ views $\Set{\msf{rv}^{(1)}_i}_{i \in [n]}$, corresponding to an MitH execution for the $(n+1, k)$-$\VSS_\Share$ of the string $x_1 \| \ldots \|x_t$ (as detailed in \Cref{rmk:mpc-in-the-head-vss}). $R$ commits to each $\msf{rv}^{(1)}_i$ ($i \in [n]$) independently in parallel, using Naor's commitment.

\item \label[Step]{bbnmc:hard-puzzle:rv-2}
$R$ prepares another $n$ views $\Set{\msf{rv}^{(2)}_i}_{i \in [n]}$, corresponding to an MitH execution for the $(n+1, k)$-$\VSS_\Share$ of the string $1\|x_1$ (as detailed in \Cref{rmk:mpc-in-the-head-vss}).  $R$ commits to each $\msf{rv}^{(2)}_i$ ($i \in [n]$) independently in parallel, using $\ExtCom$.

\item \label[Step]{bbnmc:hard-puzzle:rv-3}
$R$ then prepares another $n$ views $\Set{\msf{rv}^{(3)}_i}_{i \in [n]}$, corresponding to an  $(n, k)$-MitH execution of the $n$-party functionality $F^R_{\msf{consis}}$ described below (intuitively, $F^R_{\msf{consis}}$ checks the consistency between \Cref{bbnmc:hard-puzzle:rv-1} and \Cref{bbnmc:hard-puzzle:rv-2}), where party $P_i$ ($i\in [n]$) uses $\msf{rv}^{(1)}_i \| \msf{rv}^{(2)}_i$ as input. (Note that $\msf{rv}^{(1)}_i \| \msf{rv}^{(2)}_i$ will be a prefix of $\msf{rv}^{(3)}_i$.)  $R$ commits to each $\msf{rv}^{(3)}_i$ ($i \in [n]$) independently in parallel, using Naor's commitment.
\begin{itemize}
\item {\bf Functionality  $F^R_{\msf{consis}}$:} It collects  input (and parses it as) $\msf{rv}^{(1)}_i \| \msf{rv}^{(2)}_i$ from party $i$ for each $i \in [n]$. It then runs the recovery algorithm of $\VSS$ to obtain $a_1\|\ldots\|a_t \coloneqq \VSS_\Recon(\msf{rv}^{(1)}_1, \ldots, \msf{rv}^{(1)}_n)$ and $j\|b_{j} \coloneqq \VSS_\Recon(\msf{rv}^{(2)}_1, \ldots, \msf{rv}^{(2)}_n)$. If $j \in [t]$ and $b_{j} = a_{j}$, it outputs 1 to each party; otherwise, it outputs 0 to each party.
\end{itemize}

\item \label[Step]{bbnmc:hard-puzzle:decom-ch}
$C$ sends $\msf{ch}$ together with the decommitment information (w.r.t.\ \Cref{bbnmc:hard-puzzle:com-ch}).
\item \label[Step]{bbnmc:hard-puzzle:open}
$R$ sends $\Set{(\msf{rv}^{(1)}_i, \msf{rv}^{(2)}_i, \msf{rv}^{(3)}_i)}_{i \in \msf{ch}}$ together with the decommitment information (w.r.t.\ their respective commitments in \Cref{bbnmc:hard-puzzle:rv-1,bbnmc:hard-puzzle:rv-2,bbnmc:hard-puzzle:rv-3}).
\item
$C$ checks the validity of the decommitment information and the consistency among the revealed views $\Set{(\msf{rv}^{(1)}_i, \msf{rv}^{(2)}_i, \msf{rv}^{(3)}_i)}_{i \in \msf{ch}}$. In particular, it checks for each $i \in \msf{ch}$ that $\msf{rv}^{(1)}_i\| \msf{rv}^{(2)}_i$ is the prefix of $\msf{rv}^{(3)}_i$. It also checks for each distinct pair $i, j\in \msf{ch}$ that $(\msf{rv}^{(1)}_i, \msf{rv}^{(1)}_j)$ are consistent, $(\msf{rv}^{(2)}_i, \msf{rv}^{(2)}_j)$ are consistent, and $(\msf{rv}^{(3)}_i, \msf{rv}^{(3)}_j)$ are consistent, where by `consistent' we refer to the consistency requirements as per \Cref{def:view-consistency} and \Cref{rmk:VSS:view-consistency}. It also checks for each $i\in \msf{ch}$ the final output of $P_i$ contained in $\msf{rv}^{(3)}_i$ is 1. It aborts immediately if any of the checks fail.
\end{enumerate}

\item \label[Step]{prot:bbnmc:extcom}
{\bf (ExtCom to $m$.)} $C$ commits to $m$ once again, using $\ExtCom$ in the MitH format. In more detail:
\begin{itemize}
\item
$C$ prepares $n$ views $\Set{\msf{cv}^{(2)}_i}_{i \in [n]}$, corresponding to an MitH execution for the $(n+1, 2k)$-$\VSS_\Share$ of the message $m$ (as detailed in \Cref{rmk:mpc-in-the-head-vss}). $C$ commits to each $\msf{cv}^{(2)}_i$ ($i \in [n]$) independently in parallel, using $\ExtCom$.
\end{itemize}
We remark that this step can be viewed as the {\bf Commit Stage} of the trapdoor coin-flipping protocol shown in \Cref{prot:td:ct}. Here are two key points we want to emphasize: (1) The original \Cref{prot:td:ct} uses Naor's commitment for this stage, but here we instead use $\ExtCom$. This is to make this commitment extractable. (2) The original \Cref{prot:td:ct} uses a $(n+1, k)$-VSS scheme, but here we instead use a $(n+1, 2k)$ VSS scheme. That is because we need to open two subsets of size $k$, one in \Cref{prot:bbnmc:cointoss}, which is the {\bf Coin-Flipping Stage}, and the other in \Cref{prot:bbnmc:mpc:reveal}, which is for the proof of consistency of the current PQ-NMC protocol.

These modifications will be important later when we prove non-malleability.

\item\label[Step]{prot:bbnmc:puzzle-sol-reveal}
{\bf (Puzzle Solution Reveal.)} $R$ reveals $(x_1, \ldots, x_t)$ by decommitting to $\Set{\msf{rv}^{(1)}_i}_{i \in [n]}$. In more detail,
\begin{enumerate}
\item
$R$ sends $\Set{\msf{rv}^{(1)}_i}_{i \in [n]}$ together with the decommitment information w.r.t.\ the commitments in \Cref{bbnmc:hard-puzzle:rv-1}.
\item
$C$ checks the validity of the decommitment information and the consistency among $\Set{\msf{rv}^{(1)}_i}_{i\in [n]}$ (as per \Cref{rmk:VSS:view-consistency}). If these checks are successful, $C$ recovers $x_1 \| \ldots \|x_t \coloneqq \VSS_\Recon(\msf{rv}^{(1)}_1, \ldots, \msf{rv}^{(1)}_n)$; otherwise, $C$ rejects and output $\bot$. 
\end{enumerate}

\item \label[Step]{prot:bbnmc:PoC}
{\bf (Committer's Consistency Proof.)} This stage should be interpreted as $C$ proving consistency between its actions in \Cref{bbnmc:init-com,prot:bbnmc:extcom} (i.e., these two steps commit to the same value) using a witness indistinguishable argument, where the trapdoor statement is that $C$ manages to commit to a puzzle solution in \Cref{prot:bbnmc:extcom}. This step is again conducted in the MitH format. Note that for the honest committer, the `effective witness' in MitH is the message $m$ reconstructed from both $\Set{\msf{cv}^{(1)}_i}_{i \in [n]}$ and $\Set{\msf{cv}^{(2)}_i}_{i \in [n]}$, and so the virtual MPC execution in reality evaluates the `first clause' of $F^C_{\msf{consis}}$ as defined below. In more detail, this stage proceeds as follows.
\begin{enumerate}
\item\label[Step]{prot:bbnmc:C-consis-com}
$C$ prepares $n$ views $\Set{\msf{cv}^{(3)}_i}_{i \in [n]}$, corresponding to an $(n,2k)$-MitH execution of the $n$-party functionality $F^C_{\msf{consis}}$ described below, where party $P_i$ ($i\in [n]$) uses $\msf{cv}^{(1)}_i \| \msf{cv}^{(2)}_i$ as input. (Note that $\msf{cv}^{(1)}_i \| \msf{cv}^{(2)}_i$ will be the prefix of $\msf{cv}^{(3)}_i$.)  $C$ commits to each $\msf{cv}^{(3)}_i$ ($i \in [n]$) independently in parallel, using Naor's commitment.
\begin{itemize}
\item {\bf Functionality  $F^C_{\msf{consis}}$:} It collects input (and parses it as) $\msf{cv}^{(1)}_i \| \msf{cv}^{(2)}_i$ from party $i$ for each $i \in [n]$. It then runs the recovery algorithm of $\VSS$ to obtain $a \coloneqq \VSS_\Recon(\msf{cv}^{(1)}_1, \ldots, \msf{cv}^{(1)}_n)$ and $b \coloneqq \VSS_\Recon(\msf{cv}^{(2)}_1, \ldots, \msf{cv}^{(2)}_n)$. It outputs 1 to each party if {\em either}
\begin{itemize}
\item
(First clause.) $b = a$, {\em or}
\item
(Second clause.) $b$ can be parsed as $j\|x'$ such that $j\in [t]$ and $x' = x_j$ (recall that $x_j$ is among the puzzle solutions revealed by $R$ in \Cref{prot:bbnmc:puzzle-sol-reveal}.
\end{itemize} 
Otherwise, it outputs 0 to each party.
\end{itemize}

\item\label[Step]{prot:bbnmc:cointoss}
{\bf (Trapdoor Coin-Flipping)} $C$ and $R$ then execute the {\bf Coin-Flipping Stage} of the trapdoor coin-flipping protocol shown in \Cref{prot:td:ct}, with the trapdoor predicate $\phi(\cdot)$ defined as follows
\begin{itemize}
\item
{\bf Predicate $\phi(\cdot)$:} It has the values $(x_1, \ldots, x_t)$ hard-wired (recall that these values are revealed in \Cref{prot:bbnmc:puzzle-sol-reveal}). On input $i\|a$, it outputs 1 iff  $i\in [t]$ and $a = x_i$.
\end{itemize}
By the completeness of the trapdoor coin-flipping protocol (i.e., \Cref{def:com-n-prove:property:2} in \Cref{def:com-n-prove}), at the end of this step, $C$ and $R$ agree on a string $\eta$. By a proper choice of length, the string $\eta$ can be interpreted as specifying a size-$k$ random subset of $[n]$. In the following, we abuse notation by using $\eta$ to denote the corresponding size-$k$ random subset.

\item\label[Step]{prot:bbnmc:mpc:reveal}
$C$ sends $\Set{(\msf{cv}^{(1)}_i, \msf{cv}^{(2)}_i, \msf{cv}^{(3)}_i)}_{i \in \eta}$ together with the decommitment information (w.r.t.\ their respective commitments in \Cref{bbnmc:init-com,prot:bbnmc:extcom,prot:bbnmc:C-consis-com}).

\item
$R$ checks the validity of the decommitment information and the consistency among the revealed views $\Set{(\msf{cv}^{(1)}_i, \msf{cv}^{(2)}_i, \msf{cv}^{(3)}_i)}_{i \in \eta}$.  
It also checks for each $i\in \eta$ the final output of $P_i$ contained in $\msf{cv}^{(3)}_i$ is 1. $R$ aborts if any of these checks fail.
\end{enumerate}

\end{enumerate}

\para{Decommit Stage:} 
\begin{enumerate}
\item
$C$ sends $\Set{\msf{cv}^{(1)}_i}_{i\in [n]}$ together with the decommitment information w.r.t.\ the commitments in \Cref{bbnmc:init-com}. 
\item
$R$ checks the validity of the decommitment information and the consistency among $\Set{\msf{cv}^{(1)}_i}_{i\in [n]}$ (as per \Cref{rmk:VSS:view-consistency}). If these checks are successful, $R$ recovers $m$ as $m \coloneqq \VSS_\Recon(\msf{cv}^{(1)}_1, \ldots, \msf{cv}^{(1)}_n)$; otherwise, $R$ rejects and output $\bot$. 
\end{enumerate}

\end{ProtocolBox}

\para{Security.} The security of \Cref{protocol:BB-NMCom} is stated as the following theorem.
\begin{theorem}\label{thm:one-sided:non-malleability}
 Assuming the existence of post-quantum one-way functions, there exists (i.e., \Cref{protocol:BB-NMCom}) a black-box, constant-round construction of 1-1 post-quantum non-malleable commitments (as per \Cref{def:NMCom:pq} with $k=1$) in synchronous setting, with one-sided security, and supporting tag space $[T]$ with $T(\secpar)$ being any polynomial in the security parameter $\secpar$.
\end{theorem}   
It is straightforward to see that \Cref{protocol:BB-NMCom} is constant-round and makes only black-box use of its underlying cryptographic components. Completeness of \Cref{protocol:BB-NMCom} is also straightforward from the protocol description. The statistical binding property follows from that of Naor's commitment in \Cref{bbnmc:init-com}. Computational-hiding property of any non-malleable commitment scheme follows directly from its non-malleability. So, to prove \Cref{thm:one-sided:non-malleability}, we only need to prove the post-quantum non-malleability of \Cref{protocol:BB-NMCom}, which we prove in subsequent subsections.

\subsection{Outline for the Proof of Non-Malleability}
\label{sec:BB-PQNMC:1-1:proof:outline}
The proof for the non-malleability of \Cref{protocol:BB-NMCom} is very involved and lengthy. To help the reader understand it better, we provide an outline delving into it.

Recall from \Cref{def:NMCom:pq} (with $k =1$) that to prove non-malleability of \Cref{protocol:BB-NMCom}, we consider the synchronous 1-1 MIM execution of \Cref{protocol:BB-NMCom} where the left session uses tag $t$ and the right session uses tag $\tilde{t}$. Also recall that $\tilde{t} \ge t+1$ since we focus on the `one-sided' setting. We need to show that for any QPT MIM adversary $\mcal{M}_\secpar(\rho_\secpar)$, it holds that
\begin{equation}\label{thm:pqnmc:eq:target}
\big\{\msf{mim}^{\mcal{M}_\secpar}(m_0, \rho_\secpar)\big\}_{\secpar \in \Naturals, m_0, m_1 \in \bits^{\ell(\secpar)}} ~\cind~ \big\{\msf{mim}^{\mcal{M}_\secpar}( m_1, \rho_\secpar)\big\}_{\secpar \in \Naturals, m_0, m_1 \in \bits^{\ell(\secpar)}},
\end{equation}
where $\msf{mim}^{\mcal{M}_\secpar}(m_b, \rho_\secpar)$ denotes the {\em joint} distribution of the output of $\mcal{M}_\secpar$ {\em and} the value $\tilde{m}$ committed in the right session, where the left (honest) committer $C$ commits to message $m_b$.

At a high level, our proof follows the template from \cite{FOCS:LPY23}. We reduce non-malleability to the computational-hiding property of the Naor's commitment in \Cref{bbnmc:init-com} of the left session. That is, we consider in the MIM execution that the Naor's commitment in \Cref{bbnmc:init-com} of the left session comes from an external challenger, committing to an underlying message $m_b$ with $b$ picked uniformly from $\bits$. Our goal is to construct a machine $\SimExt$ (dubbed simulation-extractor) that can efficiently extract the value $\tilde{m}$ committed by $\mcal{M}_\secpar$ in the left session, while simulating $\mcal{M}_\secpar$'s post-extraction state (possibly with an arbitrarily small simulation error $\epsilon$). Note that if such an $\SimExt$ exists, it indeed {\em efficiently} outputs the value $\msf{mim}^{\mcal{M}_\secpar}(m_b, \rho_\secpar)$ (which was not efficiently computable since the $\tilde{m}$ value is hidden in the transcript). Then, if \Cref{thm:pqnmc:eq:target} does not hold, $\SimExt$'s output will be distinguishable when the external challenger changes the committed value between $m_0$ and $m_1$. This breaks the computational hiding property of Naor's commitment.

However, there are some challenges in implementing this template. 
\begin{enumerate}
    \item
    First, note that the message $m$ committed in \Cref{bbnmc:init-com} of the left session is used again in \Cref{prot:bbnmc:extcom} of the left session. Thus, to be able to forward the left \Cref{bbnmc:init-com} to the external challenger, we have to come up with some method to avoid using the underlying $m$ in \Cref{prot:bbnmc:extcom} of the left session (as this $m$ is known only to the external challenger in the reduction we outlined above).
    \item
    In sharp contrast to the original \cite{FOCS:LPY23} construction, \Cref{bbnmc:init-com} of our \Cref{protocol:BB-NMCom} is not a Naor's commitment to the underlying message $m$. Rather, it is a parallel execution of Naor's scheme committing to $n$ VSS shares of the message $m$.     Moreover, note that \Cref{bbnmc:init-com,prot:bbnmc:extcom,prot:bbnmc:PoC} of \Cref{protocol:BB-NMCom} essentially consist of a black-box commit-and-prove protocol, where $C$ first commits to two messages in \Cref{bbnmc:init-com,prot:bbnmc:extcom} respectively, and then proves a predicate $F^C_{\msf{consis}}$ over the two committed values. This structure induces new challenges: Even if we can make \Cref{prot:bbnmc:extcom} independent of the $m$ committed in \Cref{bbnmc:init-com}, it still need to reveal some of the shares committed in \Cref{bbnmc:init-com} when proving consistency in \Cref{prot:bbnmc:PoC} (in particular, in \Cref{prot:bbnmc:mpc:reveal}). 
\end{enumerate}
Resolving these issues requires us to use new ideas described in \Cref{sec:tech-overview}. In more detail, we will first show that {\em assume the existence of the $\SimExt$}, we can still follow the same template but rather reduce non-malleability to the VSS hiding game shown in \Cref{chall:vss:hide}, instead of to the vanilla Naor's commitment. We then show that the desired $\SimExt$ can indeed be constructed. In the following, we present an outline for these two steps.

\para{Reducing Non-Malleability to VSS Hiding.} This step is performed in \Cref{sec:pq-nmc:1-1:proof:reduction-to-hiding}. It is organized as follows.
\begin{enumerate}
\item
We the first `decouple’ the the committed message in \Cref{bbnmc:init-com} of the left session from the remaining steps. To do that, we design a game $\tilde{H}^{\mcal{M}_\secpar}$ in \Cref{pq:game:Htil}. This $\tilde{H}^{\mcal{M}_\secpar}$ is essentially identical to the 1-1 MIM execution, but makes use of a new machine $\mcal{G}_1$ (in \Cref{pq:machine:g1}) to finish the steps after \Cref{bbnmc:init-com}. A key feature of $\mcal{G}_1$ is that it does not need to know the $m$ committed in \Cref{bbnmc:init-com} of the left session.

\item
Next, we define another game $\tilde{G}^{\mcal{M}_\secpar}$ (in \Cref{pq:game:Gtil}), which performs \Cref{bbnmc:init-com} in the same manner as $\tilde{H}^{\mcal{M}_\secpar}$, but replace the $\mcal{G}_1$ machine by a simulation extractor $\SimExt$ {\em that we assume to exist}. We will show (in \Cref{lem:simext:closeness}) that this $\SimExt$ can extract from $\mcal{G}_1$ the $\tilde{m}$ committed by $\mcal{M}_\secpar$ in the right session, while simulating the post-extraction state of $\mcal{G}_1$ (with a noticeable error $\epsilon$ that can be made arbitrarily small).
\item
Finally, we show that the machine $\tilde{G}^{\mcal{M}_\secpar}$ is designed on purpose so that we can reduce non-malleability to the VSS hiding game. This is done in \Cref{bbnmc:proof:fin}.
\end{enumerate}

\para{Building Simulation-Extractor $\mcal{SE}$.} To build the desired $\mcal{SE}$, we first build a machine $\mcal{K}$ that is able to extract the correct $\tilde{m}$ from the machine $\mcal{G}_1$ mentioned above. But $\mcal{K}$ is not capable of simulating the post-extraction state of $\mcal{G}_1$. We will also show that $\mcal{K}$ satisfies some extra requirements so that we can later convert it to an extractor {\em with simulation}. The description of $\mcal{K}$ and the proof for its properties are the focus of \Cref{sec:simless-ext:1-1}.

We then show how to equip $\mcal{K}$ with simulation in \Cref{sec:simext:1-1}. To do that, we first need a generalization of the counterpart lemma from \cite[Lemma 20]{arXiv:LPY23}. This is because our construction makes heavy use of black-box commit-and-prove techniques so that the simulation-less extractor $\mcal{K}$ satisfies only weaker properties than its counterpart in \cite[Lemma 31]{arXiv:LPY23}. In particular, our $\mcal{K}$ cannot check if the value it extracted is indeed the correct $\tilde{m}$. Thus, we need to generalize the simulation-extraction lemma in \cite[Lemma 20]{arXiv:LPY23} to take care of related issues. This is handled in \Cref{sec:noisy-sim-ext}.

Finally, with all the preparatory work before, we can eventually convert our $\mcal{K}$ to the desired $\mcal{SE}$ using the a `noisy' simulation-extraction lemma developed in \Cref{sec:noisy-sim-ext}. This is done in \Cref{sec:sim-less-to-sim:1-1}.

This finishes the outline for our proof of non-malleability.

\subsection{Reduction to VSS Hiding Game}
\label{sec:pq-nmc:1-1:proof:reduction-to-hiding}

We first define some notion related to the MIM execution of \Cref{protocol:BB-NMCom}.
\begin{AlgorithmBox}[label={pq:gameH:description}]{Machine \textnormal{$H^{\mcal{M}_\secpar}(\secpar,m,\rho_\secpar)$}}

{\bf Machine $H^{\mcal{M}_\secpar}(\secpar,m,\rho_\secpar)$:} This is the man-in-the-middle execution of the commit stage of \Cref{protocol:BB-NMCom}, where the left committer commits to $m$ and $\mcal{M}_\secpar$'s non-uniform advice is $\rho_\secpar$. The output of this game is denoted by $\Output_{H^{\mcal{M}_\secpar}}(\secpar, m, \rho_\secpar)$ and consists of the following three parts:
\begin{enumerate}
\item
$\OUT$: This is the (quantum) output of $\mcal{M}$ at the end of this game;
\item
$\tilde{\tau}$: This is the commitment transcript sent by $\mcal{M}$ in the \Cref{bbnmc:init-com} of the right session;
\item
$\tilde{d} \in \Set{\top, \bot}$: This is the output of the honest receiver $R$ in the right session, indicating if the man-in-the-middle's commitment (i.e., the right session) is accepted ($\tilde{d} = \top$) or not ($\tilde{d} = \bot$).
\end{enumerate} 
\end{AlgorithmBox}


Also, to prove non-malleability, we need to talk about the value committed in the right session. Toward that, we define the following function:
$$
\msf{val}_{\tilde{d}}(\tilde{\tau}) \coloneqq 
\begin{cases}
\msf{val}(\tilde{\tau}) & \tilde{d} = \top\\
\bot & \tilde{d} = \bot
\end{cases},
$$
where $\msf{val}(\tilde{\tau})$ denote the value statistically-bound $\tilde{\tau}$ (i.e., the value that can be re-constructable from $\Set{\msf{cv}^{(1)}_i}_{i \in [n]}$). Note that $\msf{val}_{\tilde{d}}(\tilde{\tau})$ is exactly the value committed in the right session by $\mcal{M}$. Thus, to prove satisfies \Cref{def:NMCom:pq} (when $k = 1$), we only need to establish the following equation:
\begin{align*}
&\big\{\big(\OUT^0, \msf{val}_{\tilde{d}^0}(\tilde{\tau}^0)\big): (\OUT^0, \tilde{\tau}^0, \tilde{d}^0) \gets H^{\mcal{M}_\secpar}(\secpar, m_0, \rho_\secpar) \big\} \\
\cind ~& 
\big\{\big(\OUT^1, \msf{val}_{\tilde{d}^1}(\tilde{\tau}^1)\big): (\OUT^1, \tilde{\tau}^1, \tilde{d}^1) \gets H^{\mcal{M}_\secpar}(\secpar, m_1, \rho_\secpar) \big\} \numberthis \label{pq:eq:classical:H:m-0:m-1},
\end{align*}
where both ensembles are indexed by $\secpar \in \Naturals$ and $(m_0, m_1) \in \bits^{\ell(\secpar)} \times \bits^{\ell(\secpar)}$.

Next, we describe a new game $\tilde{H}^{\mcal{M}_\secpar}$, which is a slight modification of the game  $\tilde{H}^{\mcal{M}_\secpar}$. It will help us switch out different commitments in the left interaction. 

\begin{AlgorithmBox}[label={pq:game:Htil}]{Game \textnormal{$\tilde{H}^{\mcal{M}_\secpar}(\secpar, \epsilon,m,\rho_\secpar)$}}
{\bf Input:} It takes as input the same parameters $\secpar$, $\rho_\secpar$, and $m$ as for $H^{\mcal{M}_\secpar}$. It additionally takes as input a noticeable function $\secpar(\cdot)$. 

It proceeds as follows:
\begin{enumerate}
    \item \label[Step]{pq:game:Htil:pref}
    {\bf (Prefix phase.)} 
    This proceeds as follows. 
    \begin{enumerate}
        \item \label[Step]{pq:game:Htil:pref:1}
        Sample a random size-$k$ subset $\eta \subset [n]$.
        \item \label[Step]{pq:game:Htil:pref:2}
       	Execute $H^{\mcal{M}_\secpar}(m_1,\secpar,\rho_\secpar)$ until the end of \Cref{bbnmc:init-com}. At the moment, it already receives the \Cref{bbnmc:init-com} commitment made by the left-session honest committer $C$. It performs brute-force computation to obtain from $C$'s commitment the committed shares $\msf{cv}_i$ and their decommitment information for $i \in \eta$. We denote these values as $\msf{VI}_\eta \coloneqq \Set{(\msf{cv}_i, \msf{decom}_i}_{i \in \eta}$. 
    \end{enumerate}
    \subpara{Notation:}  Let $\msf{st}_{\mcal{M}}$ denote the state of $\mcal{M}$ at the end of \Cref{bbnmc:init-com}; Let $\msf{st}_C$  (resp.\ $\msf{st}_R$) denote the state of the honest committer (resp.\ receiver) at the end of \Cref{bbnmc:init-com}; Let $\tilde{\tau}$ denote the commitment sent by $\mcal{M}$ in \Cref{bbnmc:init-com} of the right session. We denote the tuple $(\msf{st}_{\mcal{M}}, \msf{st}_R, \tau,\tilde{\tau})$ as $\msf{pref}$. We use the following nation to express the execution of this {\bf Prefix phase}:
    \begin{equation}
    (\msf{pref}, \eta, \msf{VI}_\eta) \la \tilde{H}_{\msf{pref}}^{\mcal{M}_\secpar}(\secpar,  m, \rho_\secpar).
    \end{equation}
    We will often use $\msf{pref}' \coloneqq (\msf{pref}, \eta, \msf{VI}_\eta)$ to refer to the concatenation of $\msf{pref}$ and the $\eta, \msf{VI}_\eta$. We remark that this prefix generation step is independent of the error parameter $\epsilon$.
  
    \item \label[Step]{pq:game:Htil:body} 
    {\bf (Remainder phase.)} This involves the following steps: 
    \begin{enumerate}
        \item 
        $\tilde{H}^{\mcal{M}_\secpar}$ now invokes $\mcal{G}_1(1^\secpar, 1^{\epsilon^{-1}}, \msf{pref}, \eta, \msf{VI}_\eta))$ (as described in \Cref{pq:machine:g1}), which outputs a tuple $(\OUT, \tilde{d})$. 
        \item 
        $\tilde{H}^{\mcal{M}_\secpar}$ outputs $(\OUT, \tilde{\tau},\tilde{d})$. 
    \end{enumerate}
\end{enumerate}

\end{AlgorithmBox}


We now describe the subprocedure $\mcal{G}_1(\cdot)$. 

\begin{AlgorithmBox}[label={pq:machine:g1}]{Machine \textnormal{$\mcal{G}_1(1^\secpar, 1^{\epsilon^{-1}}, \msf{pref}, \eta, \msf{VI}_\eta)$}}
Game $\mcal{G}_1(1^\secpar, 1^{\epsilon^{-1}}, \msf{pref}, \eta, \msf{VI}_\eta)$ continues the execution using $\msf{pref}$ just as in $H^{\mcal{M}_\secpar}(\secpar, m, \rho_\secpar)$, apart from the following differences:
\begin{enumerate}
    \item  \label[Step]{machine:G1:step:2}
    In the left \Cref{bbnmc:hard-puzzle:rv-2} against $\mcal{M}_\secpar$, instead of following the honest receiver algorithm for $\ExtCom$, it instead uses the extractor $\SimExt_\ExtCom^{\mcal{M}_\secpar}(1^\secpar,1^{\epsilon^{-1}})$ to obtain an extracted value of the form $j'||x'_{j'}$ (and continuing the execution with the simulated state). 

    In more detail, the shares $\Set{\msf{rv}^{(2)}_i}_{i \in [n]}$ are committed by $\mcal{M}_\secpar$ using independent $\ExtCom$ in parallel. $\mcal{G}_1$ will extract all of these shares using the parallel extractability of $\ExtCom$ with $\epsilon$-simulation (as per \Cref{def:epsilon-sim-ext-com:parallel}), and compute $j'||x'_{j'} \coloneqq \VSS_{\Recon}(\msf{rv}^{(2)}_1, \ldots, \msf{rv}^{(2)}_n)$.

    \item \label[Step]{machine:G1:step:2}
    In the $\ExtCom$ execution for \Cref{prot:bbnmc:extcom} on the left, it commits to the extracted value $j'\|x'_{j'}$.

    In more detail, $\mcal{G}_1$ first prepares $n$ views $\Set{\msf{cv}^{(2)}_i}_{i \in [n]}$, corresponding to an MitH execution for the $(n+1, 2k)$-$\VSS_\Share$ of the value $j'\|x'_{j'}$ extracted in \Cref{machine:G1:step:2}, and then commits to each $\msf{cv}^{(2)}_i$ ($i \in [n]$) independently in parallel, using $\ExtCom$.

    \item \label[Step]{machine:G1:step:3}
    In \Cref{prot:bbnmc:puzzle-sol-reveal} of the left session, after reconstructing the receiver's puzzle solutions $x_1\|\dots\|x_t$, it checks whether $x_{j'} = x'_{j'}$, i.e., the value that it extracted earlier. If not, it aborts the execution and outputs $\bot$. 
    
    \item \label[Step]{machine:G1:step:3}
     In \Cref{prot:bbnmc:C-consis-com} of the left session, instead of generating the views $\Set{\msf{cv}^{(3)}_i}_{i\in [n]}$ using the `first clause' of $F^C_{\msf{consist}}$, generate these views using the `second clause'. We remark that this is possible because in \Cref{machine:G1:step:2} above, we already commit to the extracted $j'\|x'_{j'}$, which satisfies the `second clause' of $F^C_{\msf{consist}}$. 

     Recall that each $\msf{cv}^{(3)}_i$ has $\msf{cv}^{(1)}_i$ as a prefix. However, $\mcal{G}_1$ only knows the $\msf{cv}^{(1)}_i$ for $i \in \eta$. For that, $\mcal{G}_1$ simply set $\msf{cv}^{(1)}_i$ for $i \in [n] \setminus \eta$ to all-$0$ strings. This does not affect this step as $\mcal{G}_1$ now is proving the `second clause' of $F^C_{\msf{consist}}$, which is independent of the value determined by the real $\Set{\msf{cv}^{(1)}_i}_{i \in [n]}$. (Also note that this problem does not occur for $\Set{\msf{cv}^{(2)}_i}_{i \in [n]}$, which is generated by $\mcal{G}_1$ itself in \Cref{machine:G1:step:2}.)

    \item \label[Step]{machine:G1:step:4}
    In \Cref{prot:bbnmc:cointoss} of the left session, instead of executing the trapdoor coin-flipping protocol honestly, use the `straight-line' simulator $\Sim(\ST_C, \eta, j'\|x'_{j'}, \phi)$ that is guaranteed to exist by \Cref{def:com-n-prove:property:3} in \Cref{def:com-n-prove} (where $\ST_C$ is the classical state of the committer at the end of \Cref{prot:bbnmc:extcom}, emulated by $\mcal{G}_1$). Note that this is possible because we current have $\phi(j'\|x'_{j'}) = 1$. This effectively `enforce' the coin-flipping result to the $\eta$ contained in the input to $\Sim$.
    
    \item 
    It concludes the execution by performing the remaining steps as in $H^{\mcal{M}_\secpar}(\secpar, m, \rho_\secpar)$. One caveat is: In \Cref{prot:bbnmc:mpc:reveal} of the left session, $\mcal{G}_1$ will be asked to decommitment to $\Set{\msf{cv}^{(1)}_i}_{i \in [n]\setminus \eta}$. Note that the $\eta$ is already `enforced' to the $\eta$ in \Cref{machine:G1:step:4}, and $\mcal{G}_1$ does know the $\msf{cv}^{(i)}_i$ shares and decommitment information for $i \in \eta$ (contained in $\msf{VI}_\eta$).

    \item It finally outputs the values $(\OUT, \tilde{d})$, where again $\OUT$ is $\mcal{M}$'s final output and $\tilde{d}$ is the honest $R$'s final decision in the right session.
\end{enumerate}

\end{AlgorithmBox}

\begin{lemma}\label{lem:Htil:similarity}
For all $m \in \bits^{\ell(\secpar)}$ and all noticeable $\epsilon(\cdot)$, it holds that  
    \begin{equation*}
    \big\{\big(\OUT, \msf{val}_{\tilde{d}}(\tilde{\tau})\big): (\OUT, \tilde{\tau}, \tilde{d}) \gets H^{\mcal{M}_\secpar}(\secpar, m, \rho_\secpar) \big\}_\secpar 
    \cind_\epsilon  
    \big\{\big(\OUT, \msf{val}_{\tilde{d}}(\tilde{\tau})\big): (\OUT, \tilde{\tau}, \tilde{d}) \gets \tilde{H}^{\mcal{M}_\secpar}(\secpar, \epsilon, m, \rho_\secpar) \big\}_\secpar.
    \end{equation*}
\end{lemma}

\begin{proof}
We prove this lemma by a hybrid argument. Some of the modifications are pretty standard within black-box MPC literature and we sketch them for brevity. In the following, we fix arbitrary an $m$ and an $\epsilon$. For each hybrid $H_i$, we use $\OUT_{H_i}$ to denote its output.

\para{Hybrid $H_0$:} This is simply the game $H^{\mcal{M}_\secpar}(m,\rho_\secpar)$, renamed for convenience.

\para{Hybrid $H_1$:} This hybrid is identical to the previous one, except for the following changes. In \Cref{bbnmc:hard-puzzle:rv-2} of the left session, instead of following the honest $C$'s algorithm, it uses the extractor $\SimExt_\ExtCom^{\mcal{M}_\secpar}(1^\secpar,1^{\epsilon^{-1}})$ to obtain an extracted a value of the format $j'\|x'_{j'}$ and a (simulated) state $\state'_{\mcal{M}}$ (as described in \Cref{machine:G1:step:2} of $\mcal{G}_1$). It records $j'\|x'_{j'}$ and continues the execution of the MIM interaction with $\state'_{\mcal{M}}$ up till \Cref{prot:bbnmc:puzzle-sol-reveal}, where it obtains $x_1,\dots,x_t$ and checks if $x_{j'} = x'_{j'}$. If not, it aborts; otherwise it finishes the execution. All other steps are carried out as in the previous hybrid. Note that $H_1$ now requires the additional input $1^{\epsilon^{-1}}$. 

\subpara{$\Output_{H_0} \sind_\epsilon \Output_{H_1}$:} This follows directly from the parallel extractability with $\epsilon$-simulation of $\ExtCom$ (as per \Cref{def:epsilon-sim-ext-com:parallel}).

\para{Hybrid $H_2$:} This hybrid is identical to the previous one, except for the following changes. In \Cref{prot:bbnmc:extcom} of the left session, it uses the extracted $j'\|x'_{j'}$ as the committed message; And in \Cref{prot:bbnmc:C-consis-com} of the left session, instead of generating the views $\Set{\msf{cv}^{(3)}_i}_{i\in [n]}$ using the `first clause' of $F^C_{\msf{consist}}$, generate these views using the `second clause'.

\begin{remark}\label{rmk:known-shares}
We remark that this step is slight different from \Cref{machine:G1:step:3} of $\mcal{G}_1$, where the shares $\Set{\msf{cv}^{(1)}_i}_{i \in [n] \setminus \eta}$ (as prefix of $\msf{cv}^{(3)}_i$'s) is set to $0$-strings. In the current hybrid, $H_2$ still uses the honest $\Set{\msf{cv}^{(1)}_i}_{i \in [n] \setminus \eta}$ shares, because these shares are generated by itself and thus it knows them. We will change these shares to $0$-strings in a later hybrid (i.e., $H_4$). 
\end{remark}

\subpara{$\Output_{H_1} \cind \Output_{H_2}$:} Note that \Cref{prot:bbnmc:extcom} and \Cref{prot:bbnmc:PoC} constitutes a black-box commit-and-prove protocol. What we did in this step is to switch from one witness for the target predicate to another witness. This switch is computationally indistinguishable. Since this argument is standard, we omit the details (see \cite[Section 6.5]{C:CCLY22} for an example).

\para{Hybrid $H_3$:} This hybrid is identical to the previous one, except for the following changes. At the very beginning, it samples a random size-$k$ subset $\eta\subset [n]$; and In \Cref{prot:bbnmc:cointoss} of the left session, instead of executing the trapdoor coin-flipping protocol honestly, it uses the `straight-line' simulator $\Sim(\ST_C, \eta, j'\|x'_{j'}, \phi)$ to `enforce' the coin-flipping result (in the same manner as described in \Cref{machine:G1:step:4} of $\mcal{G}_1$).

\subpara{$\Output_{H_2} \cind \Output_{H_3}$:} This follows directly from the security guarantee of $\Sim$ (i.e., \Cref{def:com-n-prove:property:3} in \Cref{def:com-n-prove}).

\para{Hybrid $H_4$:} This hybrid is identical to the previous one, except for the following changes. In \Cref{prot:bbnmc:C-consis-com} of the left session, when generating $\msf{cv}^{(3)}_i$, it sets the shares $\Set{\msf{cv}^{(1)}_i}_{i \in [n] \setminus \eta}$ to $0$-strings. This is to compensate the concern in \Cref{rmk:known-shares}. 

\subpara{$\Output_{H_3} \cind \Output_{H_4}$:} First, note that changing the shares  $\Set{\msf{cv}^{(1)}_i}_{i \in [n] \setminus \eta}$ does not affect \Cref{prot:bbnmc:C-consis-com} of the left session, as we already switch to using the witness for the `second clause' of it in $H_3$. Also note that these shares will never be revealed because in $H_2$ we already `enforce' the challenge set to $\eta$. Thus, the indistinguishability of $H_3$ and $H_4$ follows directly from the computational hiding property of Naor's commitment in \Cref{prot:bbnmc:C-consis-com} of the left session.

\para{Hybrid $H_5$:} Note that in $H_4$, we already do not need to use any information about the shares $\Set{\msf{cv}^{(1)}_i}_{i \in [n] \setminus \eta}$. In this hybrid, we can think that after \Cref{bbnmc:init-com} of the execution, $H_5$ performs brute-force computation to learn the shares $\Set{\msf{cv}^{(1)}_i}_{i \in \eta}$ and their corresponding decommitment information (and put them together as $\msf{VI}_\eta$). Indeed, these are the only information that is need to finish the remaining execution. 

\subpara{$\Output_{H_4} \idind \Output_{H_5}$:} There is no real change in $H_5$ other than a change of perspective. These two hybrids are thus identical.

Finally, note that $H_5$ is exactly $\tilde{H}^{\mcal{M}_\secpar}(\secpar, \epsilon, m, \rho_\secpar)$. This concludes the proof of \Cref{lem:Htil:similarity}.

\end{proof}

Next, we define a further modified game $\tilde{G}$ that instead invokes a particular simulator-extractor $\SimExt$ that helps obtain the value committed to by $\mcal{M}_\secpar$ on the right (which then helps to demonstrate non-malleability). 

\begin{AlgorithmBox}[label={pq:game:Gtil}]{Game \textnormal{$\tilde{G}^{\mcal{M}_\secpar}(\secpar, \epsilon, m,\rho_\secpar)$}}
This proceeds in two phases as well: 
\begin{enumerate}

    \item \label[Step]{pq:game:Gtil:pref}
    {\bf (Prefix phase.)} This is identical to {the prefix phase of \Cref{pq:game:Htil}, i.e., it computes $(\msf{pref}, \eta, \msf{VI}_\eta) \la \tilde{H}_{\msf{pref}}^{\mcal{M}_\secpar}(\secpar, m, \rho_\secpar)$.}

    \item {\bf Remainder phase:}\label[Step]{pq:game:Gtil:body} This involves the following steps: 
    \begin{itemize} 
        \item It invokes a machine $\SimExt$, {which is guranteed to exist by the following \Cref{lem:simext:closeness}}: $\SimExt$ takes in as input a tuple $(1^\secpar, 1^{\epsilon^{-1}}, \msf{pref}, \eta, \msf{VI}_\eta)$ and outputs $(\msf{OUT}, \msf{Val})$. 
        \item  $\tilde{G}^{\mcal{M}_\secpar}$ outputs $(\msf{OUT}, \msf{Val})$ as its own output.
    \end{itemize}
\end{enumerate}
\end{AlgorithmBox}

The following \Cref{lem:simext:closeness} serves as assurance that we can build such a machine $\SimExt$ so that the games $\tilde{H}^{\mcal{M}_\secpar}$ and $\tilde{G}^{\mcal{M}_\secpar}$ present $\epsilon$-close views to the adversary (where we control the closeness parameter). \Cref{lem:simext:closeness} represents the most challenging task in the current proof of non-malleability. We will prove it in \Cref{sec:simless-ext:1-1,sec:simext:1-1}
\begin{lemma}[1-1 Simulation-Extractor]\label{lem:simext:closeness}
    Let $\mcal{G}_1(\cdot)$ be the efficient procedure defined in \Cref{pq:machine:g1}. There exists a simulation-extractor $\SimExt$ such that for any $(\msf{pref},\eta,\msf{VI}_\eta)$ in the support of $\tilde{H}^{\mcal{M}_\secpar}_\msf{pre}$, and for any noticeable $\epsilon(\secpar)$, there is a noticeable $\epsilon'(\secpar) \le 8\epsilon(\secpar)$ that is efficiently computable form $\epsilon(\secpar)$ such that the following holds:
    \begin{align*}
        &
        \big\{(\msf{OUT}, \msf{Val}): (\msf{OUT}, \msf{Val}) \gets \SimExt(1^\secpar, 1^{\epsilon^{-1}},\msf{pref},\eta,\msf{VI}_\eta) \big\}
        \\
        \cind_{\epsilon} 
        ~&
        \big\{\big(\OUT, \msf{val}_{\tilde{d}}(\tilde{\tau})\big): (\OUT,  \tilde{d}) \gets \mcal{G}_1(1^\secpar, 1^{\epsilon'^{-1}}, \msf{pref},\eta,\msf{VI}_\eta) \big\}.
    \end{align*}
\end{lemma} 
The following \Cref{lem:bbnmc:compare:Htil:Gtil} is an immediate consequence of \Cref{lem:simext:closeness}. 
\begin{corollary}\label{lem:bbnmc:compare:Htil:Gtil}
    Let $\tilde{H}^{\mcal{M}_\secpar}$ and $\tilde{G}^{\mcal{M}_\secpar}$ be as defined in \Cref{pq:game:Htil} and \Cref{pq:game:Gtil} respectively. For any QPT adversary $\mcal{M}_\secpar(\rho_\secpar)$, any $m \in \bits^{\ell(\secpar)}$, 
    and any noticeable $\epsilon(\secpar)$, there is a noticeable $\epsilon'(\secpar) \le 8\epsilon(\secpar)$ that is efficiently computable form $\epsilon(\secpar)$ such that the following holds: 
    \begin{align*}
        &
        \big\{(\msf{OUT}, \msf{Val}): (\msf{OUT}, \msf{Val}) \gets \tilde{G}^{\mcal{M}_\secpar}(\secpar, \epsilon, m, \rho_\secpar) \big\} 
        \\
        \cind_{\epsilon} 
        ~&
        \big\{\big(\OUT, \msf{val}_{\tilde{d}}(\tilde{\tau})\big): (\OUT, \tilde{\tau}, \tilde{d}) \gets \tilde{H}^{\mcal{M}_\secpar}(\secpar, \epsilon', m, \rho_\secpar) \big\}.
    \end{align*}
\end{corollary}

\begin{proof}
    Note that the prefix stages of $\tilde{H}^{\mcal{M}_\secpar}$ (see \Cref{pq:game:Htil:pref} in \Cref{pq:game:Htil}) and $\tilde{G}^{\mcal{M}_\secpar}$ (see \Cref{pq:game:Gtil:pref} in \Cref{pq:game:Gtil}) are identical, and therefore \Cref{lem:simext:closeness} immediately applies to show that $(\OUT, \msf{val}_{\tilde{d}}(\tilde{\tau}))$ obtained by running $\tilde{H}^{\mcal{M}_\secpar}$ and $(\msf{OUT}, \msf{Val})$ obtained by running $\tilde{G}^{\mcal{M}_\secpar}$ are $\epsilon$-close.

\end{proof}

\subsection{Finishing the Proof of Non-Malleability}
\label{bbnmc:proof:fin} 

With the helper machines defined in \Cref{sec:pq-nmc:1-1:proof:reduction-to-hiding}, we can now finish the proof of non-malleability. This is a proof by contradiction. We will show that if the non-malleability of \Cref{protocol:BB-NMCom} does not hold, then the machine $\tilde{G}^{\mcal{M}_\secpar}$ can be used to break the VSS hiding game defined in \Cref{chall:vss:hide}. We present the formal argument in the following.

We first show a lemma that relates machine $\tilde{G}^{\mcal{M}_\secpar}$ to the VSS hiding game defined in \Cref{chall:vss:hide}.
\begin{lemma}\label{lem:bb-nmc:similarity:g:til}
Let $\tilde{G}^{\mcal{M}_\secpar}$ be defined as in \Cref{pq:game:Gtil}. For For any QPT adversary $\mcal{M}_\secpar(\rho_\secpar)$ and any noticeable $\epsilon(\secpar)$, it holds that
    \begin{align*}
        &\big\{(\OUT^0, \Val^0):(\OUT^0, \Val^0) \gets \tilde{G}^{\mcal{M}_\secpar}(\secpar, \epsilon, m_0, \rho_\secpar)\big\} \\
        \cind ~&
        \big\{(\OUT^1, \Val^1):(\OUT^1, \Val^1) \gets \tilde{G}^{\mcal{M}_\secpar}(\secpar, \epsilon, m_1, \rho_\secpar)\big\}, \numberthis
    \end{align*}    
    where both ensembles are indexed by $\secpar \in \Naturals$ and $(m_0,m_1) \in \bits^{\ell(\secpar)} \times \bits^{\ell(\secpar)}$. 
\end{lemma}

\begin{proof}
    We show this by means of a reduction to the VSS hiding game defined in \Cref{chall:vss:hide}. We assume for contradiction that there exist a machine $\mcal{M}_\secpar(\rho_\secpar)$, a distinguisher $\mcal{D}_\secpar$, and a pair of messages $(m_0, m_1)$ so that \Cref{lem:bb-nmc:similarity:g:til} does not hold. We build a malicious $\Adv$ that wins the VSS hiding game (i.e., breaking \Cref{lem:game:VSS:hiding}).

    The $\Adv$ works as follows:
    \begin{enumerate}
    \item
    It sends $(m_0, m_1)$ to the external $\algo{Ch}$ for the VSS hiding game, as the \Cref{chall:vss:hide:step:1} message of \Cref{chall:vss:hide}.
    \item
    It internally samples a random size-$k$ subset $\eta \subset [n]$. 
    \item
    It commits to $\eta$ to the external $\algo{Ch}$ using $\ExtCom$, as the \Cref{chall:vss:hide:step:2} message of \Cref{chall:vss:hide}.
    \item
    When the external $\algo{Ch}$ sends the \Cref{chall:vss:hide:step:3} message of \Cref{chall:vss:hide}, it uses this message as the \Cref{bbnmc:init-com} message of the left session in its internal emulation of $\tilde{G}^{\mcal{M}_\secpar}$ with $\mcal{M}_\secpar$.
    \item
    It then decommits to $\eta$ to the external $\algo{Ch}$, as the \Cref{chall:vss:hide:step:4} message of \Cref{chall:vss:hide}.
    \item
    When it receives the revealed shares and their corresponding decommitment information form the external $\algo{Ch}$ (i.e., the \Cref{chall:vss:hide:step:5} message of \Cref{chall:vss:hide}), it puts them together to define $\msf{VI}_\eta$.
    \item
    It executes the machine $\SimExt(1^\secpar, 1^{\epsilon^{-1}}, \msf{pref}, \eta, \msf{VI}_\eta)$ as in the {\bf Remainder phase} of $\tilde{G}^{\mcal{M}_\secpar}$ (see \Cref{pq:game:Gtil}).
    \item
    It finally invokes the distinguisher $\mcal{D}_\secpar$ on $\SimExt$'s output, and outputs whatever $\mcal{D}_\secpar$ outputs.
    \end{enumerate}
Note that this $\Adv$ simulates perfectly emulates the execution of $\tilde{G}^{\mcal{M}_\secpar}(\secpar, \epsilon, m_b, \rho_\secpar)$ (when the external $\algo{Ch}$ uses $m_b$). Note that in game $\tilde{G}^{\mcal{M}_\secpar}$, we define the variable $\msf{VI}_\eta$ by brute force (see \Cref{pq:game:Htil:pref:2} in \Cref{pq:game:Htil}). But in the above VSS hiding game, $\Adv$ learns the values in $\msf{VI}_\eta$ from the external $\algo{Ch}$. This is only a syntax change as these two way leads to the same $\msf{VI}_\eta$. 

Therefore, if \Cref{lem:bb-nmc:similarity:g:til} does not hold, the above $\Adv$ will win the VSS hiding game with advantage non-negligibly greater than $1/2$.

\end{proof}

\para{Derivation of Contradiction.} We assume for contradiction that \Cref{pq:eq:classical:H:m-0:m-1} does not hold (i.e.,  the non-malleability of \Cref{protocol:BB-NMCom} does not hold). This means that there must be  a (possibly non-uniform) QPT distinguisher $\mcal{D} = \Set{\mcal{D}_\secpar, \sigma_\secpar}_{\secpar \in \Naturals}$, an ensemble of messages $\Set{(m_0, m_1)}_{\secpar \in \Naturals}$ and a function $\delta(\secpar) = 1/\poly(\secpar)$ such that for infinitely many $\secpar \in \Naturals$, it holds that
\begin{equation}\label[Inequality]{pq:eq:one-sided:proof:contra-assump}
\bigg|\Pr[\mcal{D}_\secpar\big(\OUT^0, \msf{val}_{\tilde{d}^0}(\tilde{\tau}^0); \sigma_\secpar\big)=1] - \Pr[\mcal{D}_\secpar\big(\OUT^1, \msf{val}_{\tilde{d}^1}(\tilde{\tau}^1); \sigma_\secpar\big)=1] \bigg|\ge  \delta(\secpar),
\end{equation}
where the first probability is taken over the random procedure $(\OUT^0, \tilde{\tau}^0, \tilde{d}^0) \gets H^{\mcal{M}_\secpar}(\secpar,m_0,\rho_\secpar)$, and the second probability is taken over the random procedure $(\OUT^1, \tilde{\tau}^1, \tilde{d}^1) \gets H^{\mcal{M}_\secpar}(\secpar,m_1,\rho_\secpar)$ (and the randomness due to the measurements performed by $\mcal{D}_\secpar$).

Now recall that by \Cref{lem:Htil:similarity}, we have that for any $m$ and any noticeable $\epsilon_1(\secpar)$, it holds that
\begin{align*}
& 
\big\{\big(\OUT^H, \msf{val}_{\tilde{d}^H}(\tilde{\tau}^H)\big)~:~(\OUT^{{H}}, \tilde{\tau}^{{H}}, \tilde{d}^{{H}}) \gets {H}^{\mcal{M}_\secpar}(\secpar, m, \rho_\secpar)\big\} \\ 
\cind_{\epsilon_1}~ &
\big\{\big(\OUT^{\tilde{H}}, \msf{val}_{\tilde{d}^{\tilde{H}}}(\tilde{\tau}^{\tilde{H}})\big)~:~(\OUT^{\tilde{H}}, \tilde{\tau}^{\tilde{H}}, \tilde{d}^{\tilde{H}}) \gets \tilde{H}^{\mcal{M}_\secpar}(\secpar, \epsilon_1, m, \rho_\secpar\big\}.
\end{align*} 
Using this to replace terms on both sides of \Cref{pq:eq:one-sided:proof:contra-assump}, we get  
\begin{equation}\label[Inequality]{pq:eq:contra:Htil}
    \bigg|\Pr[\mcal{D}_\secpar\big(\OUT^0, \msf{val}_{\tilde{d}^0}(\tilde{\tau}^0); \sigma_\secpar\big)=1] - \Pr[\mcal{D}_\secpar\big(\OUT^1, \msf{val}_{\tilde{d}^1}(\tilde{\tau}^1); \sigma_\secpar\big)=1] \bigg|\ge \delta(\secpar) - 2\epsilon_1(\secpar),
\end{equation}
where the inputs to $\mcal{D}_\secpar$ in the above are sampled as $(\OUT^0, \tilde{\tau}^0, \tilde{d}^0) \gets \tilde{H}^{\mcal{M}_\secpar}(\secpar, \epsilon_1, m_0,\rho_\secpar)$ and $(\OUT^1, \tilde{\tau}^1, \tilde{d}^1) \gets \tilde{H}^{\mcal{M}_\secpar}(\secpar, \epsilon_1, m_1,\rho_\secpar)$. 

Further, we have from \Cref{lem:bbnmc:compare:Htil:Gtil} that for any $m$ and any noticeable $\epsilon_2(\secpar)$, there exists a noticeable $\epsilon_1(\secpar) \le 8\epsilon_2(\secpar)$ that is efficiently computable from $\epsilon_2(\secpar)$ such that
\begin{align*}
    &\big\{\big(\OUT^{\tilde{H}}, \msf{val}_{\tilde{d}^{\tilde{H}}}(\tilde{\tau}^{\tilde{H}})\big): (\OUT^{\tilde{H}}, \tilde{\tau}^{\tilde{H}}, \tilde{d}^{\tilde{H}}) \gets \tilde{H}^{\mcal{M}_\secpar}(\secpar, \epsilon_1, m,\secpar,\rho_\secpar) \big\} \\
    \cind_{\epsilon_2} ~& 
    \big\{\big(\msf{OUT}_{\mcal{SE}}, \msf{Val}_{\mcal{SE}}\big): (\msf{OUT}_{\mcal{SE}}, \msf{Val}_{\mcal{SE}} \gets \tilde{G}^{\mcal{M}_\secpar}(\secpar, \epsilon_2, m,\secpar,\rho_\secpar) \big\}.
\end{align*}
 Again, replacing terms on both sides of \Cref{pq:eq:contra:Htil}, we have 
\begin{equation}\label[Inequality]{pq:eq:contra:Gtil}
    \bigg|\Pr[\mcal{D}_\secpar\big(\OUT^0_{\mcal{SE}}, \msf{Val}_{\mcal{SE}}^0; \sigma_\secpar\big)=1] - \Pr[\mcal{D}_\secpar\big(\msf{OUT}_{\mcal{SE}}^1, \msf{Val}_{\mcal{SE}}^1; \sigma_\secpar\big)=1] \bigg|\ge \delta(\secpar) -\epsilon_1(\secpar) - \epsilon_2(\secpar),
\end{equation}
where the inputs to $\mcal{D}_\secpar$ in the above are sampled as $(\OUT^0_{\mcal{SE}}, \msf{Val}_{\mcal{SE}}^0) \gets \tilde{G}^{\mcal{M}_\secpar}(\secpar, \epsilon_2, m_0,\secpar,\rho_\secpar)$ and $(\OUT^0_{\mcal{SE}}, \msf{Val}_{\mcal{SE}}^0) \gets \tilde{G}^{\mcal{M}_\secpar}(\secpar, \epsilon_2, m_1,\secpar,\rho_\secpar)$.

If we set $\epsilon_2 \coloneqq \frac{\delta}{18}$, then the lower-bound in \Cref{pq:eq:contra:Gtil} becomes 
$$\delta(\secpar) -\epsilon_1(\secpar) - \epsilon_2(\secpar) \ge \delta(\secpar) -8 \epsilon_2(\secpar) - \epsilon_2(\secpar) = \delta(\secpar) -9  \epsilon_2(\secpar) = \delta(\secpar) -9 \cdot \frac{\delta(\secpar)}{18}= \frac{\delta(\secpar)}{2}.$$

Since $\frac{\delta(\secpar)}{2}$ is noticeable, this is in direct contradiction to \Cref{lem:bb-nmc:similarity:g:til}, which says that the LHS of \Cref{pq:eq:contra:Gtil} can at most be negligible. We conclude that \Cref{pq:eq:one-sided:proof:contra-assump} is false. 

This establishes that our protocol described in \Cref{protocol:BB-NMCom} is indeed non-malleable as per \Cref{def:NMCom:pq} with $k =1$.

%% file: sections/K-1-1.tex

\section{Simulation-less Extractor \textnormal{$\mcal{K}$}: 1-1 Settings}
\label{sec:simless-ext:1-1}

In this section we introduce a `basic' extractor machine $\mcal{K}$. We then describe its operation and show that the stated properties hold. 

We will rely on some of the notation from the previous section. In particular, recall from \Cref{pq:game:Htil} that the procedure $\tilde{H}^{\mcal{M}_\secpar}_{\msf{pre}}(\secpar, m, \rho_\secpar)$ generates $\eta$, $\msf{VI}_\eta$, and $\msf{pref}=(\msf{st}_{\mcal{M}}, \msf{st}_R, \tau,\tilde{\tau})$. We will define $\msf{pref'} \coloneqq (\msf{pref},\eta,\msf{VI}_{\eta})$. Also, we define the following quantity $p^{\msf{Sim}}_{\msf{pref'}}[\epsilon_1]$ that will be important in the statement of $\mcal{K}$: 
\begin{equation}\label{eq:def:p-pref}
p^{\msf{Sim}}_{\msf{pref'}}[\epsilon_1] \coloneqq \Pr[\tilde{d} = \top : (\OUT,\tilde{d}) \gets \mcal{G}_1(1^\secpar, 1^{\epsilon_1^{-1}}, \msf{pref}')].
\end{equation}

\begin{lemma}[Simulation-less Extraction]\label{pq:lem:small-tag:proof:se:proof:K}
Let $\tilde{H}^{\mcal{M}_\secpar}_{\msf{pre}}(\secpar, m, \rho_\secpar)$ be as defined in \Cref{pq:game:Htil}. There exists a QPT machine $\mcal{K}$ such that for any noticeable $\epsilon(\secpar)$, there is a noticeable $\epsilon_1(\secpar) \le \epsilon(\secpar)$ that can be efficiently computed form $\epsilon$, such that for any noticeable $\epsilon_2(\secpar)$ and any tuple $\msf{pref'} = (\msf{st}_{\mcal{M}}, \msf{st}_R, \tau,\tilde{\tau},\eta,\msf{VI}_{\eta})$ in the support of $\tilde{H}^{\mcal{M}_\secpar}_{\msf{pre}}(\secpar, m, \rho_\secpar)$, the following holds
\begin{enumerate}
\item \label[Property]{pq:property:small-tag:proof:se:proof:K:syntax}
{\bf (Almost Uniqueness:)} $\mcal{K}$ takes as input $(1^\secpar, 1^{\epsilon_1^{-1}}, 1^{\epsilon_2^{-1}}, \msf{pref}')$. It outputs a value $\msf{Val} \in \bits^{\ell(\secpar)} \cup \Set{\bot}$ such that 
$$\Pr[\msf{Val} \notin \Set{\msf{val}(\tilde{\tau}) , \bot} ~:~\msf{Val} \ra \mcal{K}{(1^\secpar, 1^{\epsilon_1^{-1}}, 1^{\epsilon_2^{-1}}, \msf{pref}')}] \leq {\epsilon_2(\secpar) + \negl(\secpar)}.$$

\item \label[Property]{pq:property:small-tag:proof:se:proof:K}
{\bf (Extraction:)} If $p^{\msf{Sim}}_{\msf{pref'}}[\epsilon_1] \ge \epsilon(\secpar)$, then it holds that
$$\Pr[\Val = \msf{val}(\tilde{\tau}) : \Val \gets \mcal{K}{(1^\secpar, 1^{\epsilon_1^{-1}},1^{\epsilon_2^{-1}}, \msf{pref}')}] \ge {\frac{\epsilon'(\secpar)-\epsilon_2(\secpar)}{\tilde{t}}},$$
where $p^{\msf{Sim}}_{\msf{pref}'}[\epsilon_1]$ is defined in \Cref{eq:def:p-pref} and  $\epsilon'(\secpar) \coloneqq \frac{\epsilon(\secpar)}{10t^2}$. 
\end{enumerate}
\end{lemma}

In the following, we fix a noticeable function $\epsilon(\secpar)$ for which we want to prove \Cref{pq:lem:small-tag:proof:se:proof:K}. We show that it suffices to set $\epsilon_1(\secpar)\defeq \frac{t+1}{t^2+4t+2}\cdot \epsilon'(\secpar)$. 




\subsection{Description of $\mcal{K}$}
Before describing the extractor $\mcal{K}$, we first need to introduce some new machines related to $\mcal{G}_1$. 
\begin{AlgorithmBox}[label={algo:G:i}]{Machine \textnormal{$\mcal{G}_i(1^\secpar, 1^{\epsilon_1^{-1}}, \msf{pref}')$}}
\para{Machine $\mcal{G}_i(1^\secpar, 1^{\epsilon_1^{-1}}, \msf{pref}')$:} Recall that we have already defined $\mcal{G}_1$ in \Cref{pq:machine:g1}. For $i \in [\tilde{t}]\setminus \Set{1}$, the machine $\mcal{G}_i(1^\secpar, 1^{\epsilon_1^{-1}}, \msf{pref}')$ works identically to $\mcal{G}_1(1^\secpar, 1^{\epsilon_1^{-1}}, \msf{pref}')$, apart from the following changes: 
\begin{enumerate}
\item It commits to the value $i||\tilde{x}_i$ instead of the value $1||\tilde{x}_1$ in \Cref{bbnmc:hard-puzzle:rv-2} of the right session. 

In more detail, $\mcal{G}_i$ first prepares $n$ views $\Set{\msf{rv}^{(2)}_j}_{j \in [n]}$, corresponding to an MitH execution for the $(n+1, k)$-$\VSS_\Share$ of the string $i\|\tilde{x}_i$, and then commits to each $\msf{rv}^{(2)}_j$  independently in parallel, using $\ExtCom$. 

\item Additionally, now the string $i||\tilde{x}_i$ is used as the `effective input' in the (virtual) MPC execution computing $F^R_{\msf{consis}}$ in \Cref{bbnmc:hard-puzzle:rv-3}. (Indeed, this is an implicit change and occurs automatically when the first change is made.) 
\end{enumerate} 
\end{AlgorithmBox}

Next we define another machine $\mcal{K}_i$ for each $i \in [\tilde{t}]$ in \Cref{algo:K:i}. These $\mcal{K}_i$'s sever as the basic component for the eventual extractor $\mcal{K}$ we are going to build.
\begin{AlgorithmBox}[label={algo:K:i}]{Machine \textnormal{$\mcal{K}_i{(1^\secpar, 1^{\epsilon_1^{-1}}, 1^{\epsilon_2^{-1}}, \msf{pref}')}$}}
{\bf Machine $\mcal{K}_i{(1^\secpar, 1^{\epsilon_1^{-1}}, 1^{\epsilon_2^{-1}}, \msf{pref}')}$:} For each $i \in [\tilde{t}]$, the machine $\mcal{K}_i{(1^\secpar, 1^{\epsilon_1^{-1}}, 1^{\epsilon_2^{-1}}, \msf{pref}')}$ works identically to machine $\mcal{G}_i{(1^\secpar, 1^{\epsilon_1^{-1}}, \msf{pref}')}$ except that
\begin{itemize}
\item
In \Cref{prot:bbnmc:extcom} of the right session, instead of following the honest receiver's algorithm, it invokes $\SimExt_\ExtCom(1^\secpar, 1^{\epsilon_2^{-1}})$ to obtain an extracted value $\tilde{v}$.

 In more detail, the shares $\Set{\msf{cv}^{(2)}_j}_{j \in [n]}$ are committed by $\mcal{M}_\secpar$ in \Cref{prot:bbnmc:extcom} of the right session using independent $\ExtCom$ in parallel. $\mcal{K}_i$ will extract all of these shares using the parallel extractability of $\ExtCom$ with $\epsilon_2$-simulation (as per \Cref{def:epsilon-sim-ext-com:parallel}), and compute $\tilde{v} \coloneqq \VSS_{\Recon}(\msf{cv}^{(2)}_1, \ldots, \msf{cv}^{(2)}_n)$.
\end{itemize}

{\bf Outputs of $\mcal{K}_i$:} To aid in our proof, we define the output of the machines $\mcal{K}_i$ differently from the outputs of the machines described so far. 

Let $\tilde{v}$ denote the value extracted and recorded by $\mcal{K}_i$ in \Cref{prot:bbnmc:extcom}. As described, $\mcal{K}_i$ will complete the execution of both left and right sessions (just as in $\mcal{G}_i$). Recall that we use $\tilde{d}$ to denote the acceptance or rejection of the right-session honest receiver (i.e., its verdict). The output of $\mcal{K}_i$ is denoted as $\msf{Val} \in \bits^{\ell(\secpar)} \cup \Set{\bot_{\tilde{Y}}, \bot_{\msf{invalid}}}$ (where $(\bot_{\tilde{Y}}, \bot_{\msf{invalid}})$ are two specialized abort symbols), and is computed as follows: 
\begin{enumerate}
\item \label[Case]{pq:K-i:output:case:1}
If $\tilde{d} = \top$ . Then, there are two sub-cases:
\begin{enumerate}
\item \label[Case]{pq:K-i:output:case:1a}
$\tilde{v} \notin \Set{\tilde{x}_i}_{i \in [\tilde{t}]}$ : In this case, we set $\msf{Val} \coloneqq \tilde{v}$. 
\item \label[Case]{pq:K-i:output:case:1b}
$\tilde{v} \in \Set{\tilde{x}_i}_{i \in [\tilde{t}]}$ : In this case, we set $\msf{Val} \coloneqq \bot_{\tilde{Y}}$.
\end{enumerate}

\item \label[Case]{pq:K-i:output:case:2}
Otherwise, if $\tilde{d} = \bot$, set $\msf{Val} \coloneqq \bot_{\msf{invalid}}$.
\end{enumerate}
We emphasize that such a $\msf{Val}$ satisfies the {\em syntactic} requirement in \Cref{pq:property:small-tag:proof:se:proof:K:syntax} of \Cref{pq:lem:small-tag:proof:se:proof:K} \footnote{Note that here we defined two types of abortion: $\bot_{\tilde{Y}}$ and $\bot_{\msf{invalid}}$, while \Cref{pq:property:small-tag:proof:se:proof:K:syntax} of \Cref{pq:lem:small-tag:proof:se:proof:K} only allows a single abortion symbol $\bot$. We remark that this is only a cosmetic difference---It can be made consistent using the following rules: $\bot = \bot_{\tilde{Y}}$ {\em and} $\bot = \bot_{\msf{invalid}}$ (i.e., $\msf{Val} = \bot  \Leftrightarrow (\msf{Val} = \bot_{\tilde{Y}} \vee \msf{Val} = \bot_{\msf{invalid}})$).} (but does not imply the actual probabilistic condition, which we will show separately).
\end{AlgorithmBox}


Finally, we are ready to define the extractor $\mcal{K}$. Intuitively, $\mcal{K}$ can be thought of as an average-case version of $\Set{\mcal{K}_i}_{i\in[\tilde{t}]}$:
\begin{itemize}
\item
{\bf Extractor $\mcal{K}$:} On input ${(1^\secpar, 1^{\epsilon_1^{-1}}, 1^{\epsilon_2^{-1}}, \msf{pref}')}$, $\mcal{K}$ samples an index $i \pick [\tilde{t}]$ uniformly and runs $\mcal{K}_i{(1^\secpar, 1^{\epsilon_1^{-1}}, 1^{\epsilon_2^{-1}}, \msf{pref}')}$, and outputs the resulting output of $\mcal{K}_i$.  
\end{itemize}
It is easy to see that the extractor $\mcal{K}$ runs in polytime, and hence is a QPT machine. We now show the other properties in \Cref{pq:lem:small-tag:proof:se:proof:K} are satisfied as well. 


\subsection{Almost Uniqueness of $\mcal{K}$}
\label{sec:proof:K:1-1:almost-uniqueness}
In this part, we prove \Cref{pq:property:small-tag:proof:se:proof:K:syntax} of \Cref{pq:lem:small-tag:proof:se:proof:K}. 

First, note that $\mcal{K}$ by definition samples a random $i$ and run $\mcal{K}_i$. Thus, to prove \Cref{pq:property:small-tag:proof:se:proof:K:syntax}, it suffices to prove the inequality shown in  \Cref{pq:property:small-tag:proof:se:proof:K:syntax} for all $\mcal{K}_i$'s. That is, to prove \Cref{pq:property:small-tag:proof:se:proof:K:syntax}, it suffices to show the following: Under the same parameter settings as in \Cref{pq:lem:small-tag:proof:se:proof:K}, it holds that
\begin{equation}\label[Inequality]{eq:reduce:K:Ki}
\forall i\in[\tilde{t}],~
\Pr[\msf{Val} \notin \Set{\msf{val}(\tilde{\tau}), \bot}~:~\msf{Val} \la \mcal{K}_i{(1^\secpar, 1^{\epsilon_1^{-1}}, 1^{\epsilon_2^{-1}}, \msf{pref}')}] \leq {\epsilon_2(\secpar)+\negl(\secpar)}.
\end{equation}

In the following, we focus on establishing \Cref{eq:reduce:K:Ki}.

First, note that the following holds for any $i\in [\tilde{t}]$ (all the provabilities below is taken over the execution $\msf{Val} \la \mcal{K}_i{(1^\secpar, 1^{\epsilon_1^{-1}}, 1^{\epsilon_2^{-1}}, \msf{pref}')}$):
\begin{align*}
\Pr[\msf{Val} \notin \Set{\msf{val}(\tilde{\tau}), \bot}] 
& =
\Pr[\msf{Val} \notin \Set{\msf{val}(\tilde{\tau}),\bot_\msf{invalid},\bot_{\tilde{Y}}}] \numberthis \label{eq:uniqueness:Ki:1} \\
& =
\Pr[\big(\msf{Val} \notin \Set{\msf{val}(\tilde{\tau}),\bot_\msf{invalid},\bot_{\tilde{Y}}}\big) \wedge \big( \tilde{d} = \top \big)] \numberthis \label{eq:uniqueness:Ki:2} \\
& \le
\Pr[\big(\msf{Val} \notin \Set{\msf{val}(\tilde{\tau}),\bot_{\tilde{Y}}}\big) \wedge \big( \tilde{d} = \top \big)] \numberthis \label[Inequality]{eq:uniqueness:Ki:3} 
,\end{align*}
where \Cref{eq:uniqueness:Ki:1} follows from the fact that the symbol $\bot$ corresponds to both $\bot_{\msf{invalid}}$ and $\bot_{\tilde{Y}}$ (recall it from \Cref{algo:K:i}), \Cref{eq:uniqueness:Ki:2} follows from the fact that 
$\msf{Val}$ is set to $\bot_\msf{invalid}$ on the right whenever $\tilde{d}=\bot$ (see \Cref{algo:K:i}).

Thus, to prove \Cref{eq:reduce:K:Ki}, it suffices to upper-bound the RHS of \Cref{eq:uniqueness:Ki:3}  by {$\epsilon_2(\secpar)+\negl(\secpar)$}. Towards that, we now compare the RHS of \Cref{eq:uniqueness:Ki:3} with the corresponding condition on the {\em committed} value in \Cref{prot:bbnmc:extcom} on the right in machine $\mcal{G}_i$ (see \Cref{algo:G:i}). Recall that $\mcal{K}_i$ differs from $\mcal{G}_i$ only by its invocation of the $\SimExt_\ExtCom$ with error parameter {$\epsilon_2$} in the right \Cref{prot:bbnmc:extcom}. Thus, it follows from the extractability with $\epsilon_2$-simulation of the right \Cref{prot:bbnmc:extcom} that
\begin{align*}
    \forall i \in [\tilde{t}],&~\Pr[\big(\msf{Val} \notin \Set{\msf{val}(\tilde{\tau}),\bot_{\tilde{Y}}}\big) \wedge \big( \tilde{d} = \top \big): \msf{Val} \gets \mcal{K}_i{(1^\secpar, 1^{\epsilon_1^{-1}}, 1^{\epsilon_2^{-1}}, \msf{pref}')}] \\
    & \leq \Pr[\big(\tilde{\alpha} \notin \Set{\msf{val}(\tilde{\tau})} \cup \Set{\tilde{x}_j}_{j \in [\tilde{t}]}\big) \wedge \big(\tilde{d} = \top \big): (\OUT,\tilde{d})\gets \mcal{G}_i{(1^\secpar, 1^{\epsilon_1^{-1}},  \msf{pref}')}] + \epsilon_2, \numberthis \label[Inequality]{pq:eq:bound:Ki:invalid:to-Gi}
\end{align*}
where $\tilde{\alpha}$ denotes the value statistically bound (i.e., the committed value) in \Cref{prot:bbnmc:extcom} of the right in machine $\mcal{G}_i$.

\Cref{pq:eq:bound:Ki:invalid:to-Gi} essentially reduces the almost uniqueness of $\mcal{K}_i$ to that of machine $\mcal{G}_i$. That is, we claim that to prove \Cref{eq:reduce:K:Ki}, it suffices to prove the following \Cref{pq:claim:Gi:lb:soundness}.
\begin{lemma}[Almost Uniqueness of $\mcal{G}_i$]\label{pq:claim:Gi:lb:soundness}
 For $\mcal{G}_i$ as defined, we have that
$$\forall i\in [\tilde{t}],~ \Pr[\big(\tilde{\alpha} \notin \Set{\msf{val}(\tilde{\tau})} \cup \Set{\tilde{x}_j}_{j \in [\tilde{t}]}\big) \wedge \big( \tilde{d} = \top\big):(\OUT,\tilde{d}) \gets \mcal{G}_i{(1^\secpar, 1^{\epsilon_1^{-1}},  \msf{pref}')}] \leq  \negl(\secpar).$$
\end{lemma}
\begin{proof}[Proof of \Cref{pq:claim:Gi:lb:soundness} (Sketch)]
This proof follows from standard techniques. Thus, we only provide a sketch.

At a high level, we prove this lemma by a reduction to the soundness of the commit-and-prove protocol shown in \Cref{protocol:BB-ZK}. Assuming \Cref{pq:claim:Gi:lb:soundness} is false, we can build a malicious $C^*$ that first commit to the value $\tilde{\alpha}$ (by forwarding $\mcal{M}$'s commitment in \Cref{prot:bbnmc:extcom} of the right session to the external honest receiver), and then convince the external receiver that $F^C_{\msf{consis}}$ is satisfied with non-negligible probability. However, this should not happen because, by the condition in \Cref{pq:claim:Gi:lb:soundness}, $\tilde{\alpha} \notin \Set{\msf{val}(\tilde{\tau})} \cup \Set{\tilde{x}_j}_{j \in [\tilde{t}]}$ and thus $F^C_{\msf{consis}}$ is {\em not} satisfied. 

The only caveat is: the external receiver will specify a random challenge set $\eta$, but in machine $\mcal{G}_i$ this set $\eta$ is determined by the trapdoor coin-flipping in \Cref{prot:bbnmc:cointoss}. To make sure that we can indeed employ the internal right session with $\mcal{M}$ to convince the external receiver, we have to make sure that the internal right session uses the $\eta$ sampled by the external receiver. To do that, first notice that if $\tilde{\alpha} \notin \Set{\tilde{x}_j}_{j \in [\tilde{t}]}$, then the trapdoor predicate $\phi(\cdot)$ in \Cref{prot:bbnmc:cointoss} in the right session will not be satisfied. By the security of the trapdoor coin-flipping protocol (in particular, \Cref{def:com-n-prove:property:4} in \Cref{def:com-n-prove}), there must exist a simulator $\Sim$ that can `enforce' the coin-flipping result to a random $\eta$ sampled independently. Using this $\Sim$, we can make sure that the internal $\mcal{M}$ indeed generates a proof of consistency using the external receiver's $\eta$.

\end{proof}

With \Cref{pq:claim:Gi:lb:soundness} in hand, it is straightforward to see that \Cref{eq:uniqueness:Ki:3,pq:eq:bound:Ki:invalid:to-Gi} and \Cref{pq:claim:Gi:lb:soundness} together immediately implies \Cref{eq:reduce:K:Ki}. 

This finishes the proof for \Cref{pq:property:small-tag:proof:se:proof:K:syntax} in \Cref{pq:lem:small-tag:proof:se:proof:K}.

\subsection{Extraction Property of $\mcal{K}$}

In this part, we prove \Cref{pq:property:small-tag:proof:se:proof:K} of \Cref{pq:lem:small-tag:proof:se:proof:K}.

By definition, $\mcal{K}$ simply picks an $i$ uniformly from $[\tilde{t}]$ and runs machine $\mcal{K}_i$. Thus, to establish \Cref{pq:property:small-tag:proof:se:proof:K} in \Cref{pq:lem:small-tag:proof:se:proof:K}, it suffices to show the following: 
\begin{lemma}\label{pq:lem:bound:Ki}
    For the same parameter settings as in \Cref{pq:lem:small-tag:proof:se:proof:K}, it holds that 
    $$\exists i \in [\tilde{t}], ~\Pr[\Val = \msf{val}(\tilde{\tau}) :\Val \gets \mcal{K}_i{(1^\secpar, 1^{\epsilon_1^{-1}}, 1^{\epsilon_2^{-1}}, \msf{pref}')}] \ge {\epsilon'(\secpar) - \epsilon_2(\secpar)}.$$ 
\end{lemma}
We claim that \Cref{pq:lem:bound:Ki} follows as a result of the following \Cref{pq:lem:bound:Gi} regarding machine $\mcal{G}_i$'s defined in \Cref{algo:G:i}.
\begin{lemma}[Validity of $\mcal{G}_i$]\label{pq:lem:bound:Gi}
    For the same parameter settings as in \Cref{pq:lem:small-tag:proof:se:proof:K}, it holds that 
$$
    \exists i \in [\tilde{t}],\ \Pr[\big(\tilde{\alpha} = \msf{val}( \tilde{\tau}) \big) \wedge \big(\tilde{d} = \top\big) ~:~(\OUT,\tilde{d}) \gets \mcal{G}_i{(1^\secpar, 1^{\epsilon_1^{-1}},  \msf{pref}')}] \ge \epsilon'(\secpar).
$$
\end{lemma}
In the following, we first prove \Cref{pq:lem:bound:Ki}, assuming \Cref{pq:lem:bound:Gi} holds. Then, we will show the proof of \Cref{pq:lem:bound:Gi} in \Cref{sec:proof:pq:lem:bound:Gi}, which represents the main technical task we perform in this section.

\begin{proof}(Proving \Cref{pq:lem:bound:Ki})
We will show this via contradiction. For the sake of contradiction, assume that \Cref{pq:lem:bound:Ki} does not hold. That is, we assume that under the parameter conditions in \Cref{pq:lem:bound:Ki}, it holds that 
\begin{equation}\label[Inequality]{pq:eq:Ki:contra-assump}
    \forall i \in [\tilde{t}],\ \Pr[\Val = \msf{val}(\tilde{\tau}) :\Val \gets \mcal{K}_i{(1^\secpar, 1^{\epsilon_1^{-1}}, 1^{\epsilon_2^{-1}}, \msf{pref}')}] < {\epsilon'(\secpar) - \epsilon_2(\secpar)}.
\end{equation} 

First, recall from the description of machine $\mcal{K}_i$ (\Cref{algo:K:i}) that $\msf{Val}$ is set to $\bot_{\msf{invalid}}$ if $\tilde{d} = \bot$. Thus, it must hold that
\begin{align*}
& \Pr[\Val = \msf{val}(\tilde{\tau}) :\Val \gets \mcal{K}_i{(1^\secpar, 1^{\epsilon_1^{-1}}, 1^{\epsilon_2^{-1}}, \msf{pref}')}] \\ 
=~& 
\Pr[\big(\tilde{v} = \msf{val}(\tilde{\tau})\big) \wedge \big( \tilde{d} = \top \big):\Val \gets \mcal{K}_i{(1^\secpar, 1^{\epsilon_1^{-1}}, 1^{\epsilon_2^{-1}}, \msf{pref}')}]. \numberthis \label{pq:lem:bound:Gi:proof:diff:Ki:Ki:middle}
\end{align*} 

Next, recall that the only different between $\mcal{K}_i$ and $\mcal{G}_i$ lies in that $\mcal{K})_i$ additionally invoke the extractor with error parameter $\epsilon_2$ in \Cref{prot:bbnmc:extcom} of the right session. By the $\epsilon_2$-simulatable extractability of $\ExtCom$ (as per \Cref{def:epsilon-sim-ext-com:parallel}), it holds hat




\begin{align*}
&  \bigg| \Pr[\big(\tilde{\alpha} = \msf{val}(\tilde{\tau})\big) \wedge \big(\tilde{d} = \top \big) :(\OUT,\tilde{d}) \gets \mcal{G}_i(1^\secpar, 1^{\epsilon_1^{-1}},  \msf{pref}')] \\ 
 &\hspace{3em} - 
 \Pr[\big(\Val = \msf{val}(\tilde{\tau})\big) \wedge \big(\tilde{d} = \top \big) :\Val \gets \mcal{K}_i(1^\secpar, 1^{\epsilon_1^{-1}}, 1^{\epsilon_2^{-1}}, \msf{pref}')] \bigg| \leq \epsilon_2(\secpar) \numberthis \label[Inequality]{pq:lem:bound:Gi:proof:diff:Ki:Gi}
\end{align*}

\Cref{pq:lem:bound:Gi:proof:diff:Ki:Gi} \Cref{pq:lem:bound:Gi:proof:diff:Ki:Ki:middle}, and \Cref{pq:eq:Ki:contra-assump} together imply the following
$$
    \forall i \in [\tilde{t}],\ \Pr[\big(\tilde{\alpha} = \msf{val}( \tilde{\tau}) \big) \wedge \big(\tilde{d} = \top\big) :(\OUT,\tilde{d}) \gets \mcal{G}_i(\msf{pref'})] <  {\epsilon'(\epsilon)},
$$
which contradicts \Cref{pq:lem:bound:Gi}.

This completes the proof of \Cref{pq:lem:bound:Ki}.

\end{proof} 

This completes the proof for \Cref{pq:property:small-tag:proof:se:proof:K} of \Cref{pq:lem:small-tag:proof:se:proof:K}, modulo the proof of \Cref{pq:lem:bound:Gi} that we will present in \Cref{sec:proof:pq:lem:bound:Gi}.

\subsection{Validity of $\mcal{G}_i$}
\label{sec:proof:pq:lem:bound:Gi}
In this part, we present the proof for \Cref{pq:lem:bound:Gi}.

We first need to define (in \Cref{algo:G'i:G''i}) two helper machines $\mcal{G}'_i$ and $\mcal{G}''_i$ ($\forall i \in [\tilde{t}]$). They are machines very similar to the $\mcal{G}_i$.  
\begin{AlgorithmBox}[label={algo:G'i:G''i}]{Machines $\mcal{G}'_i$ and $\mcal{G}''_i$}
\para{Machine $\mcal{G}_i'{(1^\secpar, 1^{\epsilon_1^{-1}},  \msf{pref}')}$:} For each $i \in [\tilde{t}]$, machine $\mcal{G}_i'{(1^\secpar, 1^{\epsilon_1^{-1}},  \msf{pref}')}$ proceeds as follows: 
\begin{enumerate}
    \item \label[Step]{algo:G'i:G''i:step:1}
    It behaves identically as $\mcal{G}_i{(1^\secpar, 1^{\epsilon_1^{-1}},  \msf{pref}')}$ (see \Cref{algo:G:i}) until the end of \Cref{bbnmc:hard-puzzle:rv-1}.

    \item \label[Step]{algo:G'i:G''i:step:2}
    It then performs {\em brute-force computation} to obtain the VSS shares $\Set{\msf{rv}^{(1)}_i}_{i \in [\tilde{t}]}$ committed by $\mcal{M}_\secpar$ in the \Cref{bbnmc:hard-puzzle:rv-1} Naor's commitment of the left session, and then runs the reconstruction algorithm $\VSS_\Recon$ to obtain the puzzle solutions $x_1||\dots||x_t$ (it aborts if reconstruction is unsuccessful).  
    \item \label[Step]{algo:G'i:G''i:step:3}
    It then samples an uniform index $s \pick [t]$ and commits to the value $(s||x_s)$ (i.e., it uses the $s$-th puzzle solution obtained from the previous step) in the left \Cref{prot:bbnmc:extcom}. 

    In more detail, it prepares $n$ views $\Set{\msf{cv}^{(2)}_i}_{i \in [n]}$, corresponding to an MitH execution for the $(n+1, 2k)$-$\VSS_\Share$ of the message $(s||x_s)$. It commits to each $\msf{cv}^{(2)}_i$ ($i \in [n]$) independently in parallel, using $\ExtCom$, as the \Cref{prot:bbnmc:extcom} message of the left session.

    \item \label[Step]{algo:G'i:G''i:step:4}
	In \Cref{prot:bbnmc:puzzle-sol-reveal} of the left session, $\mcal{G}'_i$ checks the extracted $x_s$ is indeed the $s$-th puzzle solution as revealed by $\mcal{M}_\secpar$. If not, it aborts. 

    \item \label[Step]{algo:G'i:G''i:step:5}
    In \Cref{prot:bbnmc:C-consis-com} of the left session, it uses $(s\|x_s)$ as the `effective input' in the virtual execution of $F^C_\msf{consis}$. This is possible because $x_s$ is indeed the $s$-th puzzle solution, and thus $s\|x_s$ serves as a valid witness for the `second clause' of $F^C_\msf{consis}$.

    \item 
    All other steps are carried out as in $\mcal{G}_i$.
    
\end{enumerate}

\para{Machine $\mcal{G}_i''{(1^\secpar,  \msf{pref}')}$:} For each $i \in [\tilde{t}]$, $\mcal{G}_i''{(1^\secpar,  \msf{pref}')}$ works similarly to $\mcal{G}_i'$ except that $\mcal{G}_i''$ no longer runs the extractor $\SimExt_\ExtCom(1^\secpar, 1^{\epsilon_1^{-1}})$ in the left \Cref{bbnmc:hard-puzzle:rv-2}, and thus does not need the error parameter $\epsilon_1$ anymore. 
\end{AlgorithmBox}

\para{At a High Level.} we prove \Cref{pq:lem:bound:Gi} by contradiction. Assuming \Cref{pq:lem:bound:Gi} is false, we will derive the desired contradiction using the machine $\mcal{G}''_1$. In particular, we will prove an upper-bound and a lower-bound for the probability related to the committed value in the right \Cref{prot:bbnmc:extcom} in $\mcal{G}''_1$. We will show that these two bounds indeed contradict to each other, thus finishing the proof of \Cref{pq:lem:bound:Gi}.

In the following, we first present the upper-bound in \Cref{pq:lem:K1dprime:upperbound} and the lower-bound in \Cref{pq:claim:G1dprime:lowerbound} without a proof, and show how to derive the desired contradiction if these bounds hold. We then focus on establishing these bounds in \Cref{sec:G''1:upperbound,sec:G''1:lowerbound} respectively.

\para{The Bounds.} The upper-bound is for the probability of the event that the value $\tilde{\alpha}$ committed by $\mcal{M}_\secpar$ in the right \Cref{prot:bbnmc:extcom} in $\mcal{G}_1''$ is `valid', i.e., it is either the message committed to initially on the right (namely, $\msf{val}(\tilde{\tau})$), or a legitimate puzzle solution (i.e., is among the strings $(\tilde{x}_1,\dots,\tilde{x}_{\tilde{t}})$). We capture this formally in the following \Cref{pq:lem:K1dprime:upperbound}, and prove it in \Cref{sec:G''1:upperbound}. We remark that \Cref{pq:lem:K1dprime:upperbound} does not rely on the assumption (for contradiction) that \Cref{pq:lem:bound:Gi} is false.
    
\begin{lemma}[Upper Bound]\label{pq:lem:K1dprime:upperbound}
For the same parameter settings as in \Cref{pq:lem:bound:Gi}, it holds that
$$\Pr[\big(\tilde{\alpha} \in \Set{\msf{val}(\tilde{\tau})} \cup \Set{\tilde{x}_j}_{j \in [\tilde{t}]}\big) \wedge \big(\tilde{d} = \top\big)~:~(\OUT,\tilde{d}) \gets \mcal{G}_1''{(1^\secpar,  \msf{pref}')}] \leq p^{\msf{Sim}}_{\msf{pref'}}[\epsilon_1] + \epsilon_1 +\negl(\secpar).$$
\end{lemma}

The lower bound is for the probability of the event that  the value $\tilde{\alpha}$ committed by $\mcal{M}_\secpar$ in the right \Cref{prot:bbnmc:extcom} in $\mcal{G}_1''$ is actually a puzzle solution (i.e., is among the strings $(\tilde{x}_1,\dots,\tilde{x}_{\tilde{t}})$).  This is formally stated as the following \Cref{pq:claim:G1dprime:lowerbound}. We present its proof in \Cref{sec:G''1:lowerbound}. We remark that \Cref{pq:claim:G1dprime:lowerbound} relies on the assumption (for contradiction) that \Cref{pq:lem:bound:Gi} is false.
\begin{lemma}[Lower bound]\label{pq:claim:G1dprime:lowerbound}
    Assume that \Cref{pq:lem:bound:Gi} is false. Then, for the same parameter settings as in \Cref{pq:lem:bound:Gi}, it holds that 
    $$
    \forall i \in [\tilde{t}],~\Pr[\big( \tilde{\alpha} = \tilde{x}_i \big)\wedge (\tilde{d} = \top): (\OUT,\tilde{d}) \gets \mcal{G}_1''(1^\secpar,  \msf{pref}')] \geq \frac{p^{\msf{Sim}}_{\msf{pref'}}[\epsilon_1] - {2\epsilon_1}- \epsilon'}{t} -{\epsilon_1} -\negl(\secpar).
    $$
\end{lemma}

\para{The Final Contradiction.} Assume \Cref{pq:lem:bound:Gi} is false. Using \Cref{pq:lem:K1dprime:upperbound,pq:claim:G1dprime:lowerbound}, we derive the desired contradiction in the following. All the probabilities below are taken over  $(\OUT, \tilde{d}) \la \mcal{G}''_1(1^\secpar, \msf{pref}')$, which we omit for notation succinctness.



\begin{align*}
\Pr[\big(\tilde{\alpha} \in \Set{\msf{val}(\tilde{\tau})} \cup \Set{\tilde{x}_j}_{j \in [\tilde{t}]}\big) \wedge \big(\tilde{d} = \top\big)] 
& = 
\Pr[\big(\tilde{\alpha} = \msf{val}(\tilde{\tau})\big) \wedge \big(\tilde{d} = \top\big) ] + 
\sum_{i =1 }^{\tilde{t}} \Pr[\big(\tilde{\alpha} = x_i\big) \wedge \big(\tilde{d} = \top\big)] \\
& \ge  
\sum_{i =1 }^{\tilde{t}} \Pr[\big(\tilde{\alpha} = x_i\big) \wedge \big(\tilde{d} = \top\big)] \\
&\ge 
\tilde{t} \cdot 
\bigg(
\frac{p^{\msf{Sim}}_{\msf{pref}'}[\epsilon_1]-2\epsilon_1 - \epsilon'}{t} - \epsilon_1- \negl(\secpar) 
\bigg)\numberthis \label[Inequality]{pq:eq:bound:Ki:final-contradiction:1}\\
& = 
\tilde{t} \cdot \frac{1}{t}\cdot \big( p^{\msf{Sim}}_{\msf{pref}'}[\epsilon_1]-(t+2)\epsilon_1 - \epsilon' \big) - \negl(\secpar)\\ 
& \ge 
\frac{t+1}{t} \cdot \big(p^{\msf{Sim}}_{\msf{pref}'}[\epsilon_1]-(t+2)\epsilon_1 - \epsilon'\big) - \negl(\secpar) \numberthis \label[Inequality]{pq:eq:bound:Ki:final-contradiction:2}\\
& = 
\frac{t+1}{t} \cdot \bigg( p^{\msf{Sim}}_{\msf{pref}'}[\epsilon_1]-(t+2)\epsilon_1 - \epsilon' - \frac{t}{t+1} \cdot\epsilon_1\bigg) + \epsilon_1 - \negl(\secpar)\\
& =
\frac{t+1}{t} \cdot \bigg( p^{\msf{Sim}}_{\msf{pref}'}[\epsilon_1]-\frac{t^2+4t+2}{t+1}\cdot\epsilon_1 - \epsilon' \bigg) + \epsilon_1 - \negl(\secpar)\\
& = 
\frac{t+1}{t} \cdot \big(p^{\msf{Sim}}_{\msf{pref}'}[\epsilon_1]- 2\epsilon'\big)+ \epsilon_1 - \negl(\secpar) \numberthis \label{pq:eq:bound:Ki:final-contradiction:3}\\
& = 
p_{\msf{pref}}^{\msf{Sim}}[\epsilon_1]+ \epsilon_1 + \bigg(\frac{p^{\msf{Sim}}_{\msf{pref}'}[\epsilon_1]}{t} - 2\epsilon' - \frac{2\epsilon'}{t}\bigg) - \negl(\secpar) \\
& \ge p^{\msf{Sim}}_{\msf{pref}'}[\epsilon_1]+ \epsilon_1 + \frac{5t^2 - t -1}{5t^3}\cdot \epsilon - \negl(\secpar), \numberthis \label[Inequality]{pq:eq:bound:Ki:final-contradiction:4}\\
\end{align*}
where \Cref{pq:eq:bound:Ki:final-contradiction:1} follows from \Cref{pq:claim:G1dprime:lowerbound}, \Cref{pq:eq:bound:Ki:final-contradiction:2} follows from the assumption that $\tilde{t} \ge t+1$, \Cref{pq:eq:bound:Ki:final-contradiction:3} follows from our parameter setting $\epsilon_1(\secpar)=\frac{t+1}{t^2+4t+2}\cdot \epsilon'(\secpar)$,  
and \Cref{pq:eq:bound:Ki:final-contradiction:4} follows from the assumption that $p^{\msf{Sim}}_{\msf{pref}'}[\epsilon_1] \ge \epsilon(\secpar)$ and our parameter setting $\epsilon'(\secpar) = \frac{\epsilon(\secpar)}{10t^2}$.

Recall that $t$ is the tag taking values from $[n]$ with $n$ being a polynomial of $\secpar$. Also recall that $\epsilon(\secpar)$ is an inverse polynomial on $\secpar$. Therefore, \Cref{pq:eq:bound:Ki:final-contradiction:4} can be written as:
$$\Pr[\big(\tilde{\alpha} \in \Set{\msf{val}(\tilde{\tau})} \cup \Set{\tilde{x}_j}_{j \in [\tilde{t}]}\big) \wedge \big(\tilde{d} = \top\big):(\OUT,\tilde{d}) \gets \mcal{G}_1''(1^\secpar,\msf{pref'})] \ge p^{\msf{Sim}}_{\msf{pref}'}[\epsilon_1] + \epsilon_1(\secpar)+ \frac{1}{\poly(\secpar)} - \negl(\secpar),$$
which contradicts the upper-bound shown in \Cref{pq:lem:K1dprime:upperbound}, yielding the desired contradiction.

This concludes the proof of \Cref{pq:lem:bound:Gi}.

\subsection{The Upper Bound} 
\label{sec:G''1:upperbound} 
In this part, we present the proof for \Cref{pq:lem:K1dprime:upperbound}.


We start by recalling from \Cref{eq:def:p-pref} that by definition: 
\begin{equation}\label{G''1:upperbound:proof:eq:1}
\Pr[\tilde{d} = \top : (\OUT,\tilde{d}) \gets \mcal{G}_1{(1^\secpar, 1^{\epsilon_1^{-1}},  \msf{pref}')}] = p^{\msf{Sim}}_{\msf{pref'}}{[\epsilon_1]}.
\end{equation}

We will use the machines $\mcal{G}_1'$ and $\mcal{G}_1''$ (see \Cref{algo:G'i:G''i}). First, recall that compared with $\mcal{G}_1$, $\mcal{G}'_1$ performs a brute-force computation to learn $s \| x_s$ (as per \Cref{algo:G'i:G''i:step:2}); It commits to $s\|x_s$ in \Cref{prot:bbnmc:extcom} of the left session (as per \Cref{algo:G'i:G''i:step:3}) and use $s\|x_s$ as the witness to perform the proof of consistency in \Cref{prot:bbnmc:PoC} of the left session (as per \Cref{algo:G'i:G''i:step:5}). In other words, what $\mcal{G}_1'$ does is simply to change the witness committed in \Cref{prot:bbnmc:extcom} in the commit-and-prove protocol consisting of \Cref{prot:bbnmc:extcom} and \Cref{prot:bbnmc:PoC}; Note that that `witness' used by $\mcal{G}'_1$ (i.e., $s\|x_s$) still satisfies the target predicate $F^C_{\msf{consis}}$. This will not be noticed by $\mcal{M}_\secpar$ as the commit-and-prove protocol is witness-indistinguishable\footnote{This can be proven formally using standard techniques and thus we omit the details. A formal proof can be found in, e.g., \cite[Section 6.5]{C:CCLY22}.}. Also note that the brute-force computation performed by $\mcal{G}'_1$ does not affect the computational indistinguishability between $\mcal{G}'_1$ and $\mcal{G}_1$, as that step happens before the beginning of \Cref{prot:bbnmc:extcom} and can be treated as non-uniform advice when invoking the witness-indistinguishability of the commit-and-prove protocol. This argument implies the following:
\begin{align*}
& \bigg|\Pr[\tilde{d} = \top : (\OUT,\tilde{d}) \gets \mcal{G}_1{(1^\secpar, 1^{\epsilon_1^{-1}},  \msf{pref}')}] \\
&\hspace{3em} - \Pr[\tilde{d} = \top : (\OUT,\tilde{d}) \gets \mcal{G}_1'{(1^\secpar, 1^{\epsilon_1^{-1}},  \msf{pref}')}]\bigg| \leq \negl(\secpar). \numberthis \label[Inequality]{G''1:upperbound:proof:eq:2}
\end{align*}
\Cref{G''1:upperbound:proof:eq:1} and \Cref{G''1:upperbound:proof:eq:2} together imply the following:
\begin{equation}\label[Inequality]{pq:eq:G1prime:ub:success}
    \Pr[\tilde{d} = \top : (\OUT,\tilde{d}) \gets \mcal{G}_1'{(1^\secpar, 1^{\epsilon_1^{-1}},  \msf{pref}')}] \leq p^{\msf{Sim}}_{\msf{pref'}}[\epsilon_1] +\negl(\secpar)
\end{equation}

Next, notice that the difference between $\mcal{G}'_1$ and $\mcal{G}''_1$ is that the latter stops running machine $\SimExt_\ExtCom(1^\secpar, 1^{\epsilon_1^{-1}})$ (see \Cref{algo:G'i:G''i}). Thus, $\mcal{G}'_1$ and $\mcal{G}''_1$ are at most $\epsilon_1$-far. Therefore, \Cref{pq:eq:G1prime:ub:success} implies the following:
\begin{equation}\label[Inequality]{pq:eq:G1dprime:ub:success}
    \Pr[\tilde{d} = \top : (\OUT,\tilde{d}) \gets \mcal{G}_1''{(1^\secpar,  \msf{pref}')}] \leq p^{\msf{Sim}}_{\msf{pref'}}[\epsilon_1] + \epsilon_1(\secpar) +\negl(\secpar)
\end{equation}
\Cref{pq:eq:G1dprime:ub:success} immediately implies the inequality in \Cref{pq:lem:K1dprime:upperbound}.

This completes the proof of \Cref{pq:lem:K1dprime:upperbound}.

\subsection{The Lower Bound}
\label{sec:G''1:lowerbound}
In this part, we present the proof for \Cref{pq:claim:G1dprime:lowerbound}.

\para{Assumption for Contradiction.} As mentioned earlier, this proof (in particular, in the proof of \Cref{pq:claim:Gi:lb:comval}) will make use of the negation of \Cref{pq:lem:bound:Gi}, which is for the sake of contradiction. That is, we assume for contradiction that: Under the parameter setting of \Cref{pq:lem:bound:Gi}, it holds that
\begin{equation}\label[Inequality]{eq:negation:pq:lem:bound:Gi}
 \forall i \in [\tilde{t}],\ \Pr[\big(\tilde{\alpha} = \msf{val}( \tilde{\tau}) \big) \wedge \big(\tilde{d} = \top\big) ~:~(\OUT,\tilde{d}) \gets \mcal{G}_i{(1^\secpar, 1^{\epsilon_1^{-1}},  \msf{pref}')}] < \epsilon'(\secpar)
\end{equation}


\para{Preparatory Claims.} Before showing this bound, we will first consider three preparatory claims (i.e., \Cref{pq:claim:bound:Gi,pq:claim:Gi:lb:comval,pq:claim:Gi:vss-hide}) regarding the machine $\mcal{G}_i$'s. They will help us to bound the probability regarding $\mcal{G}''_1$ as in \Cref{pq:claim:G1dprime:lowerbound} eventually.

\begin{MyClaim}\label{pq:claim:bound:Gi} 
For the same parameter settings as in \Cref{pq:claim:G1dprime:lowerbound}, it holds that 
$$\forall i \in [\tilde{t}], ~\Pr[\tilde{d} = \top ~:~ (\OUT, \tilde{d}) \gets \mcal{G}_i{(1^\secpar, 1^{\epsilon_1^{-1}},  \msf{pref}')}] \ge p^{\msf{Sim}}_{\msf{pref'}}[\epsilon_1] - 2\epsilon_1(\secpar) -\negl(\secpar).$$
\end{MyClaim} 

\begin{proof} 

This lemma again makes use of the machines $\mcal{G}_1'$ and $\mcal{G}_1''$ defined in  \Cref{algo:G'i:G''i}.

We first claim that
\begin{align*}
\forall i \in [\tilde{t}],~ & \bigg|\Pr[\tilde{d} = \top : (\OUT,\tilde{d}) \gets \mcal{G}_i{(1^\secpar, 1^{\epsilon_1^{-1}},  \msf{pref}')}] \\
&\hspace{3em} - \Pr[\tilde{d} = \top : (\OUT,\tilde{d}) \gets \mcal{G}_i'{(1^\secpar, 1^{\epsilon_1^{-1}},  \msf{pref}')}]\bigg| \leq \negl(\secpar). \numberthis \label[Inequality]{G''i:upperbound:proof:eq:2}
\end{align*} 
Indeed, we have already shown \Cref{G''i:upperbound:proof:eq:2} for the case $i = 1$ (see \Cref{G''1:upperbound:proof:eq:2}). \Cref{G''i:upperbound:proof:eq:2} follows from the same argument as we presented for \Cref{G''1:upperbound:proof:eq:2}. Thus, we omit the details.

Similarly, we also have the following `all-$i$' version of \Cref{pq:eq:G1dprime:ub:success}:
\begin{align*}
 \forall i \in [\tilde{t}],~& \bigg|\Pr[\tilde{d} = \top : (\OUT,\tilde{d}) \gets \mcal{G}'_i{(1^\secpar, 1^{\epsilon_1^{-1}},  \msf{pref}')}] \\
&\hspace{3em} - \Pr[\tilde{d} = \top : (\OUT,\tilde{d}) \gets \mcal{G}_i''{(1^\secpar,  \msf{pref}')}]\bigg| \leq \epsilon_1(\secpar)  + \negl(\secpar). \numberthis \label[Inequality]{pq:eq:Gi:prime:ub:success}
\end{align*} 

Next, using a similar (non-uniform) argument as for \Cref{G''1:upperbound:proof:eq:2} over the commit-and-prove protocol consisting of the hard puzzle setup step (i.e., \Cref{bbnmc:hard-puzzle:puzzle-setup}) of the right session, we can establish the following
\begin{align*}
 \forall i \in [\tilde{t}],~& \bigg|\Pr[\tilde{d} = \top : (\OUT,\tilde{d}) \gets \mcal{G}''_i{(1^\secpar,  \msf{pref}')}] \\
&\hspace{3em} - \Pr[\tilde{d} = \top : (\OUT,\tilde{d}) \gets \mcal{G}_1''{(1^\secpar,  \msf{pref}')}]\bigg| \leq \negl(\secpar). \numberthis \label[Inequality]{pq:eq:G1:Gi:dprime:ub:success}
\end{align*}

Using \Cref{G''i:upperbound:proof:eq:2,pq:eq:Gi:prime:ub:success,pq:eq:G1:Gi:dprime:ub:success} by setting $i = 1$, we obtain 
\begin{equation}\label[Inequality]{eq:bound:G''1:by:p-pref}
\Pr[\tilde{d} = \top : (\OUT,\tilde{d}) \gets \mcal{G}_1''{(1^\secpar,  \msf{pref}')}]\bigg| \ge p^{\msf{Sim}}_{\msf{pref'}}[\epsilon_1] - \epsilon_1(\secpar)  - \negl(\secpar),
\end{equation}
where recall the definition of $p^{\msf{Sim}}_{\msf{pref'}}[\epsilon_1]$ from \Cref{eq:def:p-pref}.

Finally, we have the following for all $i \in [\tilde{t}]$:
\begin{align*}
& \Pr[\tilde{d} = \top : (\OUT, \tilde{d}) \gets \mcal{G}_i{(1^\secpar, 1^{\epsilon_1^{-1}},  \msf{pref}')}] \\ 
\ge~ & 
\Pr[\tilde{d} = \top : (\OUT, \tilde{d}) \gets \mcal{G}'_i{(1^\secpar, 1^{\epsilon_1^{-1}},  \msf{pref}')}] -\negl(\secpar) \numberthis \label[Inequality]{pq:claim:bound:Gi:final:eq:1} \\
\ge~ & 
\Pr[\tilde{d} = \top : (\OUT, \tilde{d}) \gets \mcal{G}''_i{(1^\secpar, \msf{pref}')}] - \epsilon_1(\secpar) - \negl(\secpar) \numberthis \label[Inequality]{pq:claim:bound:Gi:final:eq:2} \\
\ge~ & 
\Pr[\tilde{d} = \top : (\OUT, \tilde{d}) \gets \mcal{G}''_1{(1^\secpar,   \msf{pref}')}] - \epsilon_1(\secpar) - \negl(\secpar) \numberthis \label[Inequality]{pq:claim:bound:Gi:final:eq:3} \\
\ge~ & 
p^{\msf{Sim}}_{\msf{pref'}}[\epsilon_1] - 2\epsilon_1(\secpar) - \negl(\secpar) \numberthis \label[Inequality]{pq:claim:bound:Gi:final:eq:4}
,\end{align*}
where \Cref{pq:claim:bound:Gi:final:eq:1} follows from \Cref{G''i:upperbound:proof:eq:2}, \Cref{pq:claim:bound:Gi:final:eq:2} follows from \Cref{pq:eq:Gi:prime:ub:success}, \Cref{pq:claim:bound:Gi:final:eq:3} follows from \Cref{pq:eq:G1:Gi:dprime:ub:success}, and \Cref{pq:claim:bound:Gi:final:eq:4} follows from \Cref{eq:bound:G''1:by:p-pref}.

    This concludes the proof of \Cref{pq:claim:bound:Gi}. 

\end{proof}

\begin{MyClaim}\label{pq:claim:Gi:lb:comval}
Assume that \Cref{pq:lem:bound:Gi} is false. For the same parameter settings as in \Cref{pq:claim:G1dprime:lowerbound}, it holds that
$$\forall i \in  [\tilde{t}],~\Pr[\big(\tilde{\alpha} \in \Set{\tilde{x}_j}_{j \in [\tilde{t}]}\big) \wedge \big(\tilde{d} = \top \big):(\OUT,\tilde{d}) \gets \mcal{G}_i{(1^\secpar, 1^{\epsilon_1^{-1}},  \msf{pref}')}] \geq p^{\msf{Sim}}_{\msf{pref'}}[\epsilon_1] - 2\epsilon_1(\secpar) - \epsilon'(\secpar) - \negl(\secpar).$$
\end{MyClaim}

\begin{proof}
In the following, we will fix an arbitrary $i \in [\tilde{t}]$ and describe events only within $\mcal{G}_i$, and omit the rider that the variables in question are generated from $\mcal{G}_i(1^\secpar, 1^{\epsilon_1^{-1}},  \msf{pref}')$ for notation convenience. 

First, it follows from \Cref{pq:claim:Gi:lb:soundness} and \Cref{pq:claim:bound:Gi} that
\begin{equation}\label[Inequality]{pq:claim:Gi:lb:comval:proof:eq:1}
\Pr[\big(\tilde{\alpha} \in \Set{\msf{val}(\tilde{\tau})} \cup \Set{\tilde{x}_j}_{j \in [\tilde{t}]}\big) \wedge \big( \tilde{d} = \top\big)] \ge  p^{\msf{Sim}}_{\msf{pref'}}[\epsilon_1] - 2\epsilon_1(\secpar) -\negl(\secpar).
\end{equation}
\Cref{pq:claim:Gi:lb:comval:proof:eq:1} and \Cref{eq:negation:pq:lem:bound:Gi} (i.e., our assumption for contradiction) together immediately imply
\begin{equation}
\Pr[\big(\tilde{\alpha} \in  \Set{\tilde{x}_j}_{j \in [\tilde{t}]}\big) \wedge \big( \tilde{d} = \top\big)] \ge  p^{\msf{Sim}}_{\msf{pref'}}[\epsilon_1] - 2\epsilon_1(\secpar) - \epsilon'(\secpar) -\negl(\secpar).
\end{equation}

This completes the proof of \Cref{pq:claim:Gi:lb:comval}.

\end{proof}


\begin{MyClaim}\label{pq:claim:Gi:vss-hide}
For the same parameter settings as in \Cref{pq:claim:G1dprime:lowerbound}, it holds that
$$\forall i \in [\tilde{t}],~\Pr[\big(\tilde{\alpha} \in \Set{\tilde{x}_j}_{j \in [\tilde{t}] \setminus \Set{i}}\big) \wedge (\tilde{d} = \top): (\OUT,\tilde{d}) \gets \mcal{G}_i(1^\secpar, 1^{\epsilon_1^{-1}},  \msf{pref}')] \leq \negl(\secpar) ,$$
\end{MyClaim}

\begin{proof}
We first define a new machine $\mcal{G}^*_i$ as follows.

\begin{AlgorithmBox}[label={machine:Gi*}]{Machine \textnormal{$\mcal{G}^*_i(1^\secpar, 1^{\epsilon_1^{-1}},  1^{\epsilon^{*-1}}, \msf{pref}')$}} 
This machine takes an additional parameter $\epsilon^*$ than $\mcal{G}_i$. On input $(1^\secpar, 1^{\epsilon_1^{-1}},  1^{\epsilon^{*-1}}, \msf{pref}')$, $\mcal{G}^*_i$ is identical to $\mcal{G}_i(1^\secpar, 1^{\epsilon_1^{-1}}, \msf{pref}')$ except for the following difference:
\begin{itemize}
\item
In \Cref{bbnmc:hard-puzzle:com-ch} of the right session, it invokes the the extractor for $\ExtCom$ with simulation error set to $\epsilon^*$, to extracts the set $\tilde{\eta}$ committed by $\mcal{M}$.

\item
In \Cref{bbnmc:hard-puzzle:rv-3} of the right session, it prepares the shares $\Set{\tilde{\msf{rv}}^{(3)}_j}_{j\in [n]}$ differently. First, recall that each $\tilde{\msf{rv}}^{(3)}_j$ is the output of the MitH party $P_j$ using input $\tilde{\msf{rv}}^{(1)}_j\|\tilde{\msf{rv}}^{(2)}_j$ to compute the ideal functionality $F^R_{\msf{consis}}$. $\mcal{G}^*_i$ modifies the input to $P_j$'s as follows: 
\begin{itemize}
\item
It defines a new vector $(\tilde{x}^*_1, \ldots, \tilde{x}^*_{\tilde{t}})$, where $\tilde{x}^*_i$ is equal to the $\tilde{x}_i$ sampled in \Cref{bbnmc:hard-puzzle:rv-1}, but $\tilde{x}^*_j = 0^\secpar$ for all $j \ne i$.
\item
It creates a new set of VSS shares $\Set{\tilde{\msf{rv}}^{*(1)}_j}_{j \in [n]}$ for the new $(\tilde{x}^*_1, \ldots, \tilde{x}^*_{\tilde{t}})$ satisfying the requirement that $\tilde{\msf{rv}}^{*(1)}_j = \tilde{\msf{rv}}^{(1)}_j$ for all $j \in \tilde{\eta}$. That is, it takes the shares $\Set{\tilde{\msf{rv}}^{(1)}_j}_{j \in \tilde{\eta}}$, which is the VSS shares for the original vector $(\tilde{x}_1, \ldots, \tilde{x}_{\tilde{t}})$, and computes a new set of `the remainder' shares $\Set{\tilde{\msf{rv}}^{*(1)}_j}_{j \in [n]\setminus \tilde{\eta}}$, such that the new set $\Set{\tilde{\msf{rv}}^{*(1)}}_{i \in [n]\setminus \tilde{\eta}} \cup \Set{\tilde{\msf{rv}}^{(1)}}_{i \in \tilde{\eta}}$ constitutes VSS sharing of the new vector $(\tilde{x}^*_1, \ldots, \tilde{x}^*_{\tilde{t}})$. We remark that this is possible because any $k$ shares of a $(n+1, k)$-VSS scheme contain no information of the underlying secret; Thus, any $k$ shares of some secret can be `extended' to $n$ shares that constitute a VSS of a different secret. Indeed, the VSS scheme we use (from \cite{EC:CDDHR99}) satisfies this property.
\end{itemize}
With the above, $\mcal{G}^*_i$ prepares $\Set{\tilde{\msf{rv}}^{(3)}_j}_{j\in [n]}$ by running the $(n,k)$-MitH execution with party $P_j$ ($\forall j \in [n]$) using $\tilde{\msf{rv}}^{*(1)}_j\|\tilde{\msf{rv}}^{(2)}_j$ as its input.
\item
It finishes the remaining execute in the same manner as $\mcal{G}_i$.
\end{itemize}
\end{AlgorithmBox}

We first claim that for any noticeable $\epsilon^*(\secpar)$ and any $i \in [\tilde{t}]$, it holds that 
 \begin{align*}
& \Pr[\tilde{\alpha} \in \Set{\tilde{x}_j}_{j \in [\tilde{t}] \setminus \Set{i}} \wedge (\tilde{d} = \top): (\OUT,\tilde{d}) \gets \mcal{G}_i(1^\secpar, 1^{\epsilon_1^{-1}},  \msf{pref}')] \\ 
~ \cind_{\epsilon^*}& 
\Pr[\tilde{\alpha} \in \Set{\tilde{x}_j}_{j \in [\tilde{t}] \setminus \Set{i}} \wedge (\tilde{d} = \top): (\OUT,\tilde{d}) \gets \mcal{G}^*_i(1^\secpar, 1^{\epsilon_1^{-1}},1^{\epsilon^{*-1}},  \msf{pref}')]. \numberthis \label{eq:ind:Gi:G*i}
\end{align*}
To see \Cref{eq:ind:Gi:G*i}, note that the only difference between $\mcal{G}_i$ and $\mcal{G}^*_i$ is how the shares $\msf{rv}^{(3)}_j$'s are generated (modulo the extractor with $\epsilon^*$-simulation that $\mcal{G}^*_i$ invokes in \Cref{bbnmc:hard-puzzle:com-ch}, which is already taken into account by the symbol $\cind_{\epsilon^*}$). First, we remark that the new $\msf{rv}^{*(1)}_j$'s generated by $\mcal{G}^*_i$ still satisfy the predicate $F^R_{\msf{consis}}$ in \Cref{bbnmc:hard-puzzle:rv-3}. This is because these $\msf{rv}^{*(1)}_j$'s will reconstruct to $(\tilde{x}^*_1, \ldots, \tilde{x}^*_{\tilde{t}})$, where  $\tilde{x}^*_i$ does equal to $\tilde{x}_i$, to which the (unchanged) $\msf{rv}^{(2)}_j$'s will reconstruct. Thus, if we compare the $\Set{\msf{rv}^{(3)}_j}_{j\in \tilde{\eta}}$ shares between $\mcal{G}_i$ and $\mcal{G}^*_i$, they are the views of parties $P_j$ ($j\in \tilde{\eta}$) resulted from different inputs that lead to the same output for $F^R_{\msf{consis}}$. By the $(n,k)$-privacy of the underlying MPC, we know that any $k$ shares (i.e., those in set $\tilde{\eta}$) does not reveal the input of other parties $P_j$ for $j \in [n]\setminus \tilde{\eta}$. Therefore, the view of $\mcal{M}_\secpar$ is computationally indistinguishable between $\mcal{G}_i$ and $\mcal{G}^*_i$ (modulo the error $\epsilon^*$ accounting for the extractor with $\epsilon^*$-simulation invoked by $\mcal{G}^*_1$ in \Cref{bbnmc:hard-puzzle:com-ch}). 

This seems to already establish \Cref{eq:ind:Gi:G*i}. But we remark that there is a caveat: what we have shown so far is about $\mcal{M}_\secpar$ view. But the event in \Cref{eq:ind:Gi:G*i} is about the committed value $\tilde{\alpha}$. To compensate for that, note that the committed $\tilde{\alpha}$ can be efficiently extracted using the extractor $\SimExt$ for the $\ExtCom$ in \Cref{prot:bbnmc:extcom} of the right session, with an arbitrarily small noticeable simulation error $\epsilon$. Thus, the indistinguishability of $\mcal{M}_\secpar$ does translate to the event regarding the committed $\tilde{\alpha}$. This finises the proof of \Cref{eq:ind:Gi:G*i}.

With \Cref{eq:ind:Gi:G*i} in hand (where note that the $\epsilon^*$ can be made arbitrarily small), to prove \Cref{pq:claim:Gi:vss-hide}, it suffices to prove the following regarding machine $\mcal{G^*}$: for all $i \in [\tilde{t}]$ and all $\epsilon^*$, it holds that 
\begin{equation}\label[Inequality]{eq:convert:Gi:G*i}
\Pr[\tilde{\alpha} \in \Set{\tilde{x}_j}_{j \in [\tilde{t}] \setminus \Set{i}} \wedge (\tilde{d} = \top): (\OUT,\tilde{d}) \gets \mcal{G}^*_i(1^\secpar, 1^{\epsilon_1^{-1}}, 1^{\epsilon^{*-1}},  \msf{pref}')] \leq \negl(\secpar).
\end{equation}

In the following, we establish \Cref{eq:convert:Gi:G*i} by reducing it to the VSS hiding game shown in \Cref{chall:vss:hide}.

Assume for the sake of contradiction that \Cref{eq:convert:Gi:G*i} does not hold. Namely, there exists an $i \in [\tilde{t}]$ and a inverse polynomial quantity $\nu(\secpar)$ such that for infinitely many $\secpar \in \Naturals$, it holds that
 $$\Pr[\tilde{\alpha} \in \Set{\tilde{x}_j}_{j \in [\tilde{t}] \setminus \Set{i}} \wedge (\tilde{d} = \top): (\OUT,\tilde{d}) \gets \mcal{G}^*_i(1^\secpar, 1^{\epsilon_1^{-1}}, 1^{\epsilon^{*-1}},  \msf{pref}')] > \nu(\secpar).$$ 
 We show there exists an adversary $\Adv$ that wins the VSS hiding game shown in \Cref{chall:vss:hide}. $\Adv$ works as follows: 
\begin{enumerate}
\item 
$\Adv$ picks two tuples of messages $\Set{\tilde{x}_j}_{j \in [\tilde{t}]}$ and $\Set{\tilde{x}'_j}_{j \in [\tilde{t}]}$ that are distinct in every entry other than $i$ (i.e., $\tilde{x}_j \neq \tilde{x}'_j ~\forall j \in [\tilde{t}] \setminus \Set{i}$), but it holds that $\tilde{x}_i = \tilde{x}'_i$. Externally, it sends $m_0 \coloneqq \Set{\tilde{x}_j}_{j \in [\tilde{t}] }$ and $m_1 \coloneqq \Set{\tilde{x}'_j}_{j \in [\tilde{t}]}$ to the challenger $\msf{Ch}$ for the VSS hiding game, as \Cref{chall:vss:hide:step:1} in \Cref{chall:vss:hide}. 

\item \label[Step]{algo:G*i:A:against:VSS:hiding:step:2}
Internally $\Adv$ starts executing $\mcal{G}^*_i(1^\secpar,1^{\epsilon_1^{-1}}, 1^{\epsilon^{*-1}},\msf{pref'})$. It proceeds with this execution till the end of \Cref{bbnmc:hard-puzzle:com-ch} of the right session. Note that by definition of $\mcal{G}^*_i$ (see \Cref{machine:Gi*}), the set $\tilde{\eta}$ has already been extracted at this moment. $\Adv$ then commit to $\tilde{\eta}$ using $\ExtCom$ to the external challenger $\msf{Ch}$ as \Cref{chall:vss:hide:step:2} in \Cref{chall:vss:hide}.

\item \label[Step]{algo:G*i:A:against:VSS:hiding:step:3}
The challenger next sends commitments as \Cref{chall:vss:hide:step:3} in \Cref{chall:vss:hide}. $\Adv$ forwards these commitments to $\mcal{M}_\secpar$ (at \Cref{bbnmc:hard-puzzle:rv-1} on the right).

\item \label[Step]{algo:G*i:A:against:VSS:hiding:step:4}
$\Adv$ then sends the $\tilde{\eta}$ together with the decommitment information w.r.t.\ its $\ExtCom$ made in \Cref{algo:G*i:A:against:VSS:hiding:step:2}, as the \Cref{chall:vss:hide:step:4} message in \Cref{chall:vss:hide}.

\item \label[Step]{algo:G*i:A:against:VSS:hiding:step:5}
Then, $\msf{Ch}$ will send the shares $\Set{\msf{v}_j}_{j \in \tilde{\eta}}$ to $\Adv$, as the \Cref{chall:vss:hide:step:5} message in \Cref{chall:vss:hide}. $\Adv$ records this values.

\item \label[Step]{algo:G*i:A:against:VSS:hiding:step:6}
Internally, $\Adv$ continues the execute as $\mcal{G}^*_i$, until the beginning of \Cref{bbnmc:hard-puzzle:rv-3}. It executes \Cref{bbnmc:hard-puzzle:rv-3} in the following manner (which is also identical to $\mcal{G}^*_i$ by renaming some variables as explained below):
\begin{itemize}

 \item
 It first prepares the shares $\Set{\msf{rv}^{*(1)}_j}_{j\in [n]}$ that constitute a VSS of $(\tilde{x}^*_1, \ldots, \tilde{x}^*_{\tilde{t}})$ where  $\tilde{x}^*_i = \tilde{x}_i $ but $\tilde{x}^*_j = 0^\secpar$ for all $j \ne i$, and these shares satisfy the requirement that $\msf{rv}^{*(1)}_j = \msf{v}_j$ for $j\in \tilde{\eta}$, as we explain in the description of $\mcal{G}^*_i$ with $\Set{\msf{v}_j}_{j\in \tilde{\eta}}$ playing the role of $\Set{\msf{rv}^{(1)}_j}_{j\in \tilde{\eta}}$ in the description in \Cref{machine:Gi*}.

\item \label[Step]{algo:G*i:A:against:VSS:hiding:step:8} 
$\mcal{G}^*_i$ then prepares $\Set{\tilde{\msf{rv}}^{(3)}_j}_{j\in [n]}$ by running the $(n,k)$-MitH execution with party $P_j$ ($\forall j \in [n]$) using $\tilde{\msf{rv}}^{*(1)}_j\|\tilde{\msf{rv}}^{(2)}_j$ as its input.
\end{itemize} 
It is worth noting that in this step, $\Adv$ does not make use of the shares $\Set{\msf{v}_j}_{j\in [n]\setminus \tilde{\eta}}$ (which are anyway only know to the external $\msf{Ch}$ but not to $\Adv$). $\Adv$ only uses the shares $\Set{\msf{v}_j}_{j\in  \tilde{\eta}}$ that is revealed by $\msf{Ch}$ in \Cref{algo:G*i:A:against:VSS:hiding:step:5}.

\item \label[Step]{algo:G*i:A:against:VSS:hiding:step:9}
$\Adv$ then internally finish the remaining execution in the same manner as $\mcal{G}^*_i$, with only one difference---In \Cref{prot:bbnmc:extcom} of the right session, $\Adv$ invokes the extractor $\SimExt_\ExtCom$ with error parameter $\epsilon_{\Adv}$ to get an extracted value $\tilde{v}$. It parses $\tilde{v}$ as $j||a$. If $a = \tilde{x}_j$, $\Adv$ halts and output $b' = 0$; If $a = \tilde{x}'_j$, $\Adv$ halts and output $b'=1$; If neither of the cases happen, $\Adv$ halts and output a random bit $b'$
\end{enumerate}
By the above description, it is not hard to see that up to \Cref{algo:G*i:A:against:VSS:hiding:step:8}, the internal execution of $\Adv$ perfectly emulates the view of $\mcal{M}_\secpar$ in game $\mcal{G}^*_i$. If $\mcal{M}_\secpar$ indeed commits to an $\tilde{\alpha} \in \Set{\tilde{x}_j}_{j \in [\tilde{t}] \setminus \Set{i}}$ (and $\tilde{d} = \top$) with probability $\nu(\secpar)$, $\Adv$ in \Cref{algo:G*i:A:against:VSS:hiding:step:9} will extract a $\tilde{v} \in \Set{\tilde{x}_j}_{j \in [\tilde{t}] \setminus \Set{i}}$ (or $\tilde{v} \in \Set{\tilde{x}'_j}_{j \in [\tilde{t}] \setminus \Set{i}}$, depending on the $\msf{Ch}$ uses $m_0$ or $m_1$) with probability at least $\nu(\secpar) - \epsilon_\Adv(\secpar)$. Therefore, the advantage of $\Adv$ in the VSS hiding game is at least:
$$\big(\nu(\secpar) - \epsilon_\Adv(\secpar)\big) \cdot 1 + \big(1 - (\nu(\secpar) - \epsilon_\Adv(\secpar)) \big)\cdot \frac{1}{2} = \frac{1}{2} + \frac{\nu(\secpar) - \epsilon_\Adv(\secpar)}{2}.$$
Note that we can set $\epsilon_\Adv(\secpar)$ to be an arbitrarily small noticeable function. By setting $\epsilon_\Adv(\secpar) \coloneqq \frac{\nu(\secpar)}{2}$, the above lower bound becomes $\frac{1}{2} + \frac{\nu(\secpar)}{4}$. Since $\frac{\nu(\secpar)}{4}$ is still a noticeable function, this contradicts \Cref{lem:game:VSS:hiding}, breaking the VSS hiding game.



This concludes the proof of \Cref{pq:claim:Gi:vss-hide}.

\end{proof}

\para{Finishing the Proof of \Cref{pq:claim:G1dprime:lowerbound}.} With the above preparatory \Cref{pq:claim:bound:Gi,pq:claim:Gi:lb:comval,pq:claim:Gi:vss-hide}, we now proceed to finish the proof of \Cref{pq:claim:G1dprime:lowerbound}.

\subpara{From $\mcal{G}_i$ to $\mcal{G}'_i$.} We start by comparing machine $\mcal{G}_i$ and $\mcal{G}'_i$ (see \Cref{algo:G'i:G''i}). Note that $\mcal{G}_i'$ differs from $\mcal{G}_i$ in that it performs brute-force computation to extract the puzzle solutions from \Cref{bbnmc:hard-puzzle:rv-1} of the left session (see \Cref{algo:G'i:G''i:step:2} of \Cref{algo:G'i:G''i}), while $\mcal{G}_i$ extract a puzzle solution $(j\|x_j)$ using the extractability from $\ExtCom$. Note that the $(j\|x_j)$ extracted by $\mcal{G}_i$ must be among the $t$ real solutions (which are all extracted by $\mcal{G}'_i$ using brute force). Also, recall form \Cref{algo:G'i:G''i} that $\mcal{G}'_i$ picks a random $(s\|x_s)$ to finish the reminder execution as in $\mcal{G}_i$. Thus, in the case where $j=s$ (i.e., $\mcal{G}'_i$ happens to guess the same $(j\|x_j)$ as extracted by $\mcal{G}_i$), then the games $\mcal{G}_i$ and $\mcal{G}_i'$ are {\em identical}. Moreover, since $\mcal{G}'_i$ guesses $s$ uniformly at random from $[t]$, the event $j = s$ happens with probability at least $1/t$. Therefore, the following holds:
\begin{align*}
\forall i \in [\tilde{t}],~ & \Pr[\big(\tilde{\alpha} = \tilde{x}_i \big) \wedge (\tilde{d} = \top): (\OUT,\tilde{d}) \gets \mcal{G}'_i{(1^\secpar, 1^{\epsilon_1^{-1}}, \msf{pref'})}] \\ 
 \geq & ~
\frac{\Pr[\big( \tilde{\alpha} = \tilde{x}_i \big) \wedge (\tilde{d} = \top): (\OUT,\tilde{d}) \gets \mcal{G}_i{(1^\secpar, 1^{\epsilon_1^{-1}}, \msf{pref'})}]}{t}. \numberthis \label[Inequality]{proof:lower-bound:final:eq:1}
\end{align*}
On the other hand, notice that \Cref{pq:claim:Gi:lb:comval} and \Cref{pq:claim:Gi:vss-hide}, we conclude that
\begin{equation}\label[Inequality]{pq:eq:Gi:comval:xi}
    \forall i \in [\tilde{t}],~\Pr[\big(\tilde{\alpha} = \tilde{x}_i\big) \wedge \big(\tilde{d} = \top\big): (\OUT,\tilde{d}) \gets \mcal{G}_i{(1^\secpar, 1^{\epsilon_1^{-1}}, \msf{pref'})}] \geq p^{\msf{Sim}}_{\msf{pref'}}[\epsilon_1] - 2\epsilon_1(\secpar) - \epsilon'(\secpar)- \negl(\secpar).
\end{equation}
\Cref{proof:lower-bound:final:eq:1,pq:eq:Gi:comval:xi} together imply the following: 
\begin{equation}\label[Inequality]{pq:eq:Giprime:lb}
\forall i \in [\tilde{t}],~\Pr[\big( \tilde{\alpha} = \tilde{x}_i \big)\wedge (\tilde{d} = \top): (\OUT,\tilde{d}) \gets \mcal{G}_i'(1^\secpar, 1^{\epsilon_1^{-1}}, \msf{pref'})] \geq  \frac{p^{\msf{Sim}}_{\msf{pref'}}[\epsilon_1] - 2\epsilon_1(\secpar) - \epsilon'(\secpar)}{t} -\negl(\secpar).
\end{equation}

\subpara{From $\mcal{G}'_i$ to $\mcal{G}''_i$.} Next, note that the machines $\mcal{G}_i'$ and $\mcal{G}_i''$ only differ in that $\mcal{G}_i''$ no longer invokes $\SimExt_\ExtCom(1^\secpar, 1^{\epsilon_1^{-1}})$ to extract from \Cref{bbnmc:hard-puzzle:rv-2} on the left (see \Cref{algo:G'i:G''i}). As a consequence, it holds that 
\begin{equation}\label[Inequality]{pq:eq:Gidprime:lowerbound}
\forall i \in [\tilde{t}],~\Pr[\big( \tilde{\alpha} = \tilde{x}_i \big)\wedge (\tilde{d} = \top): (\OUT,\tilde{d}) \gets \mcal{G}_i''(1^\secpar, \msf{pref'})] \geq \frac{p^{\msf{Sim}}_{\msf{pref'}}[\epsilon_1] - 2\epsilon_1(\secpar) - \epsilon'(\secpar)}{t} -\epsilon_1 -\negl(\secpar).
\end{equation}

\subpara{From $\mcal{G}''_i$ to $\mcal{G}''_1$.} We first note that to finish our current proof of \Cref{pq:claim:G1dprime:lowerbound}, it suffices to show the following inequality
 \begin{align*}
        \forall i \in [\tilde{t}],~ & \bigg|  \Pr[\big( \tilde{\alpha} = \tilde{x}_i \big)\wedge (\tilde{d} = \top): (\OUT,\tilde{d}) \gets \mcal{G}_i''(1^\secpar, \msf{pref'})] \\
         & \hspace{3em}- \Pr[\big( \tilde{\alpha} = \tilde{x}_i \big)\wedge (\tilde{d} = \top): (\OUT,\tilde{d}) \gets \mcal{G}_1''(1^\secpar,\msf{pref'})] \bigg| 
         \leq \negl(\secpar), \numberthis \label[Inequality]{pq:eq:Gidprime:compare}
\end{align*} 
because \Cref{pq:eq:Gidprime:lowerbound,pq:eq:Gidprime:compare} together imply \Cref{pq:claim:G1dprime:lowerbound} immediately. Thus, the only thing left is to prove \Cref{pq:eq:Gidprime:compare}.

\begin{proof}[Proof of \Cref{pq:eq:Gidprime:compare}]
    
    For the sake of contradiction, assume that there exist an $i \in [\tilde{t}]$ and an inverse polynomial $\kappa(\secpar)$ such that for infinitely many $\secpar\in \Naturals$, it holds that 
    \begin{align*}
        & \bigg|  \Pr[\big( \tilde{\alpha} = \tilde{x}_i \big)\wedge (\tilde{d} = \top): (\OUT,\tilde{d}) \gets \mcal{G}_i''(1^\secpar,\msf{pref'})] \\
         &\hspace{3em} - \Pr[\big( \tilde{\alpha} = \tilde{x}_i \big)\wedge (\tilde{d} = \top): (\OUT,\tilde{d}) \gets \mcal{G}_1''(1^\secpar,\msf{pref'})] \bigg| 
         > \kappa(\secpar) \numberthis \label[Inequality]{pq:eq:Gidprime:contra}
    \end{align*} 

    We next introduce a new machine $\mcal{\hat{G}}_i$ for this proof. 
\begin{AlgorithmBox}[label={algo:G:hat}]{Machine \textnormal{$\mcal{\hat{G}}_i(1^\secpar, 1^{\hat{\epsilon}^{-1}},\msf{pref'})$}}

    \para{Machine $\mcal{\hat{G}}_i(1^\secpar, 1^{\hat{\epsilon}^{-1}},\msf{pref'})$:} For each $i \in [\tilde{t}]$, this machine works similar to $\mcal{G}_i''(1^\secpar, \msf{pref'})$ (see \Cref{algo:G'i:G''i}), except that
    \begin{itemize}
    \item
    In \Cref{prot:bbnmc:extcom} of the right session, instead of using the honest receiver's algorithm,  $\mcal{\hat{G}}_i$ invokes $\SimExt_\ExtCom(1^\secpar, 1^{\hat{\epsilon}^{-1}})$ to extract a value $\tilde{v}$, which is supposed to be the value committed by $\mcal{M}_\secpar$ in the right \Cref{prot:bbnmc:extcom}.

    In more detail, the shares $\Set{\tilde{\msf{cv}}^{(2)}_i}_{i \in [n]}$ are committed by $\mcal{M}_\secpar$ using independent $\ExtCom$ in parallel in the right \Cref{prot:bbnmc:extcom}. $\mcal{G}_1$ will extract all of these shares using the parallel extractability of $\ExtCom$ with error parameter $\hat{\epsilon}$ (as per \Cref{def:epsilon-sim-ext-com:parallel}), and compute $\tilde{v} \coloneqq \VSS_{\Recon}(\tilde{\msf{cv}}^{(2)}_1, \ldots, \tilde{\msf{cv}}^{(2)}_n)$.

    \end{itemize} 
\end{AlgorithmBox}
  
    We start by comparing the value $\tilde{\alpha}$ committed to in \Cref{prot:bbnmc:extcom} on the right in $\mcal{G}_i''$ and the value $\tilde{v}$ extracted by $\SimExt_\ExtCom$ in \Cref{prot:bbnmc:extcom} on the right within $\mcal{\hat{G}}_i$. Similar to before, we can base this comparison on the simulation-extraction guarantee of $\SimExt_\ExtCom(1^\secpar, 1^{\hat{\epsilon}^{-1}})$, which implies that for any noticeable $\hat{\epsilon}$, it holds that
    \begin{align*}
        \forall i,j \in [\tilde{t}],~ & \bigg| \Pr[\big( \tilde{\alpha} = \tilde{x}_j \big)\wedge (\tilde{d} = \top): (\OUT,\tilde{d}) \gets \mcal{G}_i''(1^\secpar, \msf{pref'})] \\
         & \hspace{3em} - 
         \Pr[\big( \tilde{v} = \tilde{x}_j \big)\wedge (\tilde{d} = \top): (\OUT, \tilde{d}) \gets \mcal{\hat{G}}_i(1^\secpar, 1^{\hat{\epsilon}^{-1}}, \msf{pref'})] \bigg| 
         \leq \hat{\epsilon}(\secpar). \numberthis \label[Inequality]{pq:eq:Kidprime:compare}
    \end{align*} 

Next, using a similar (non-uniform) argument as for \Cref{pq:eq:G1:Gi:dprime:ub:success} over the commit-and-prove protocol consisting of the hard puzzle setup step (i.e., \Cref{bbnmc:hard-puzzle:puzzle-setup}) of the right session, we can establish the following: for any noticeable $\hat{\epsilon}(\secpar)$, it holds that
    \begin{align*}
        \forall i,j \in [\tilde{t}],~& \bigg| \Pr[\big( \tilde{v} = \tilde{x}_j \big)\wedge (\tilde{d} = \top): (\OUT, \tilde{d}) \gets \mcal{\hat{G}}_i(1^\secpar, 1^{\hat{\epsilon}^{-1}}, \msf{pref'})] \\
         &\hspace{3em} - \Pr[\big( \tilde{v} = \tilde{x}_j \big)\wedge (\tilde{d} = \top): (\OUT, \tilde{d}) \gets \mcal{\hat{G}}_1(1^\secpar, 1^{\hat{\epsilon}^{-1}}, \msf{pref'})] \bigg| 
         \leq \negl(\secpar). \numberthis \label[Inequality]{pq:eq:K1dprime:compare}
    \end{align*} 
It then follows from \Cref{pq:eq:K1dprime:compare} and \Cref{pq:eq:Kidprime:compare} that for any noticeable $\hat{\epsilon}(\secpar)$, it holds that
    \begin{align*}
        \forall i,j \in [\tilde{t}],~& \bigg| \Pr[\big( \tilde{\alpha} = \tilde{x}_j \big)\wedge (\tilde{d} = \top): (\OUT,\tilde{d}) \gets \mcal{G}_i''(1^\secpar, \msf{pref'})] \\
         &\hspace{3em} - \Pr[\big( \tilde{v} = \tilde{x}_j \big)\wedge (\tilde{d} = \top): (\OUT, \tilde{d}) \gets \mcal{\hat{G}}_1(1^\secpar, 1^{\hat{\epsilon}^{-1}}, \msf{pref'})] \bigg| 
         \leq \hat{\epsilon}(\secpar) +  \negl(\secpar). \numberthis \label[Inequality]{pq:eq:K1dprime:compare:2}
    \end{align*} 
Setting $j = i$ in \Cref{pq:eq:K1dprime:compare:2} implies that for any noticeable $\hat{\epsilon}(\secpar)$, it holds that
    \begin{align*}
        \forall i \in [\tilde{t}],~& \bigg| \Pr[\big( \tilde{\alpha} = \tilde{x}_i \big)\wedge (\tilde{d} = \top): (\OUT,\tilde{d}) \gets \mcal{G}_i''(1^\secpar, \msf{pref'})] \\
         &\hspace{3em} - \Pr[\big( \tilde{v} = \tilde{x}_i \big)\wedge (\tilde{d} = \top): (\OUT, \tilde{d}) \gets \mcal{\hat{G}}_1(1^\secpar, 1^{\hat{\epsilon}^{-1}}, \msf{pref'})] \bigg| 
         \leq \hat{\epsilon}(\secpar) +  \negl(\secpar). \numberthis \label[Inequality]{pq:eq:K1dprime:compare:3}
    \end{align*} 
Setting $i = 1$ (and then renaming $j$ to $i$) in \Cref{pq:eq:K1dprime:compare:2} implies that for any noticeable $\hat{\epsilon}(\secpar)$, it holds that
    \begin{align*}
        \forall i \in [\tilde{t}],~& \bigg| \Pr[\big( \tilde{\alpha} = \tilde{x}_i \big)\wedge (\tilde{d} = \top): (\OUT,\tilde{d}) \gets \mcal{G}_1''(1^\secpar, \msf{pref'})] \\
         &\hspace{3em} - \Pr[\big( \tilde{v} = \tilde{x}_i \big)\wedge (\tilde{d} = \top): (\OUT, \tilde{d}) \gets \mcal{\hat{G}}_1(1^\secpar, 1^{\hat{\epsilon}^{-1}}, \msf{pref'})] \bigg| 
         \leq \hat{\epsilon}(\secpar) +  \negl(\secpar). \numberthis \label[Inequality]{pq:eq:K1dprime:compare:4}
    \end{align*} 
Combining \Cref{pq:eq:K1dprime:compare:3,pq:eq:K1dprime:compare:4} implies that for any noticeable $\hat{\epsilon}(\secpar)$, it holds that
    \begin{align*}
        \forall i \in [\tilde{t}],~& \bigg| \Pr[\big( \tilde{\alpha} = \tilde{x}_i \big)\wedge (\tilde{d} = \top): (\OUT,\tilde{d}) \gets \mcal{G}_i''(1^\secpar, \msf{pref'})] \\
         &\hspace{3em} - \Pr[\big( \tilde{\alpha} = \tilde{x}_i \big)\wedge (\tilde{d} = \top): (\OUT,\tilde{d}) \gets \mcal{G}_1''(1^\secpar, \msf{pref'})] \bigg| 
         \leq 2\cdot\hat{\epsilon}(\secpar) +  \negl(\secpar). \numberthis \label[Inequality]{pq:eq:K1dprime:compare:5}
    \end{align*}
By setting $\hat{\epsilon}(\secpar) \coloneqq \frac{\kappa(\secpar)}{4}$ in \Cref{pq:eq:K1dprime:compare:5}, we obtain a contradiction to \Cref{pq:eq:Gidprime:contra}.

This concludes the proof of \Cref{pq:eq:Gidprime:compare}.

\end{proof}



This eventually concludes our proof for \Cref{pq:claim:G1dprime:lowerbound}. 
    .

%% file: sections/SimExt-lemma.tex
\section{Simulation-Extractor \textnormal{$\SimExt$}: 1-1 Settings}
\label{sec:simext:1-1}

\subsection{Noisy Simulation-Extraction Lemma}
\label{sec:noisy-sim-ext}

\begin{lemma}[Noisy Simulatable-Extraction Lemma]
\label{lem:Noisy-SimExt}
Let $\mcal{G}$ be a QPT algorithm that takes the security parameter $1^\secpar$, an error parameter $1^{\gamma^{-1}}$, a quantum state $\rho$, {and a classical string $z$} as input,  and outputs  $d\in \{\top,\bot\}$ and a quantum state $\rho_\out$. 

Suppose that there exists a QPT algorithm $\mcal{K}$ (referred to as the simulation-less extractor) that takes as input the security parameter $1^\secpar$, two error parameters $1^{\gamma^{-1}}$ and $1^{\zeta^{-1}}$, a quantum state $\rho$, {and a classical string $z$}, and outputs $s\in \bit^{\poly(\secpar)}\cup \{\bot\}$   
satisfying the following w.r.t.\ some sequence of classical strings {$\{s^*_{z}\}_{z\in \bit^*}$.}
\takashi{The index of $s^*$ is changed from $\secpar$ to $z$. (Imagine that $s^*$ is the committed message and $z$ is the partial transcript. Then this should be a more natural formalization.)}

\begin{enumerate}
 \item  \label{item:s_star_or_bot}
    For any $\secpar$,  $\rho_\secpar$,  $z_\secpar$, and any noticeable functions $\gamma(\secpar)$ and $\zeta(\secpar)$, it holds that  

$$\Pr[s \notin \Set{s^*_{z_\secpar}, \bot}~:~s \la \mcal{K}(1^\secpar,1^{\gamma^{-1}}, 1^{\zeta^{-1}}, \rho_\secpar,{z_\secpar})]\le \zeta(\secpar) +\negl(\secpar).$$

    \item \label{item:gamma_delta} \takashi{I modified the statement to match \Cref{lem:Simultaneous-SimExt}.}
 For any noticeable function $\gamma(\secpar)$, there exists a noticeable function $\delta(\secpar)$, 
 which is efficiently computable from $\gamma(\secpar)$, so that the following requirement is satisfied: For 
 any noticeable function $\zeta(\secpar)$ and
 any sequence $\{\rho_\secpar,{z_\secpar}\}_{\secpar\in\mathbb{N}}$ of polynomial-size quantum states and classical strings, 
 if 
$$
\Pr[d=\top ~:~ (d,\rho_\out) \leftarrow \mcal{G}(1^\secpar,1^{\gamma^{-1}},\rho_\secpar,{z_\secpar})]\geq  8\gamma(\secpar), 
$$  
then 
$$
\Pr[s =s^*_{{z_\secpar}}~:~ s\la \mcal{K}(1^\secpar,1^{\gamma^{-1}}, 1^{\zeta^{-1}}, \rho_\secpar,{z_\secpar})]\geq   \delta(\secpar)-\zeta(\secpar)-\negl(\secpar).
$$
\end{enumerate}
Then, there exists a QPT algorithm $\SimExt$ such that for any noticeable function $\epsilon=\epsilon(\secpar)$, there exists a noticeable function $\gamma=\gamma(\secpar)\le \epsilon(\secpar)$ that is efficiently computable from $\epsilon$ and satisfies the following:
For any sequence $\{\rho_\secpar,{z_\secpar}\}_{\secpar\in\mathbb{N}}$ of polynomial-size quantum states and classical strings,  
$$
\{\SimExt(1^\secpar,1^{\epsilon^{-1}},\rho_\secpar,{z_\secpar})\}_{\secpar \in \Naturals}
~\statind_{\epsilon}~ 
\{(\rho_\out,\Gamma_d(s^*_{{z_\secpar}}))~:~(d,\rho_\out)\leftarrow \mcal{G}(1^\secpar,1^{\gamma^{-1}},\rho_\secpar,{z_{\secpar}})\}_{\secpar \in \Naturals},
$$
where  $
\Gamma_d(s^*_{{z_\secpar}})\defeq 
\begin{cases}
s^*_{{z_\secpar}} & \text{if}~ d=\top \\
\bot & \text{otherwise}
\end{cases}
$.
\end{lemma}
\takashi{This is almost identical to (the updated version of) the proof of Lemma 20 in LPY.
The full proof takes 7 pages, but I guess the difference will be just a few words. Please check if the following proof sketch is okay.
}

\begin{proof}[Proof sketch]
    Since the proof is almost identical to that of \cite[Lemma 20]{arXiv:LPY23}, we only describe the differences.\footnote{\cite{arXiv:LPY23} is the full version of \cite{FOCS:LPY23} on arXiv.} 
    There are the following two differences in the statement:
    \begin{itemize}
    \item We introduce an additional error parameter $\zeta$, which gives an upper bound of the probability that $\mcal{K}$ outputs $s\notin \{s^*_{z_\secpar},\bot\}$. In \cite[Lemma 20]{arXiv:LPY23}, the probability was assumed to be $0$.
    \item The lower bound of $\mcal{G}'s$ success probability in \Cref{item:gamma_delta} is $8\gamma(\secpar)$ instead of $\gamma(\secpar)$. 
    \end{itemize}
The second point can be easily dealt with by simply replacing $\gamma$ with $8\gamma$ in the original proof.
The first point introduces an additional noticeable error polynomially related to $\zeta$ in the simulation for the case of $b=\top$.
Since $\zeta$ can be chosen to be an arbitrarily small noticeable function, we can manage the additional error by appropriately setting the parameters. 

Below, we give more concrete explanation for the readers who are familiar with the proof of \cite[Lemma 20]{arXiv:LPY23}. 
We only need to modify the proof of \cite[Lemma 26]{arXiv:LPY23}, which claims that the simulation for the case $b=\top$ works.  
The first difference causes an error probability $\zeta+\negl(\secpar)$ in \cite[Claims 29 and 30]{arXiv:LPY23}, which eventually causes an error $\zeta^{1/2}+\negl(\secpar)$ in \cite[Eq. (84)]{arXiv:LPY23} where the square root appears due to the gentle measurement lemma.
As a result, \cite[Eq. (84)]{arXiv:LPY23} should be replaced with  $\left(12(8\gamma)^{1/2}+2\nu^{1/2}\right)^{1/2}+\zeta^{1/2}+\negl(\secpar)$ instead of $\left(12\gamma^{1/2}+2\nu^{1/2}\right)^{1/2}$.
(Note that $\gamma$ is replaced with $8\gamma$ to deal with the second point as explained above.) 
It suffices to set 
$
\gamma:=\frac{1}{8}\left(\frac{\epsilon}{10}\right)^4$, 
$\nu:=\left(\frac{\epsilon}{4}\right)^4$, and 
$\zeta:=\left(\frac{\epsilon}{2}\right)^2$ so that $\left(12(8\gamma)^{1/2}+2\nu^{1/2}\right)^{1/2}+\zeta^{1/2}<\epsilon$.
\end{proof}

%% file: sections/SE-1-1.tex

\subsection{Converting $\mcal{K}$ to \textnormal{$\SimExt$}}
\label{sec:sim-less-to-sim:1-1}

In this part, we build the simulation-extractor $\SimExt$ as required by \Cref{lem:simext:closeness}. This eventually finishes the proof of \Cref{lem:simext:closeness}, which is the only left piece in the proof of 1-1 non-malleability of \Cref{protocol:BB-NMCom}.

The existence of the desired $\SimExt$ relies on \Cref{pq:lem:small-tag:proof:se:proof:K,lem:Noisy-SimExt} that we established previously. Roughly speaking, we will use \Cref{lem:Noisy-SimExt} to convert the simulation-less extractor $\mcal{K}$ from \Cref{pq:lem:small-tag:proof:se:proof:K} to the desired $\SimExt$ satisfying the stipulated property in \Cref{lem:simext:closeness}. However, we remark that this must be done with proper choice of the parameters. In the following, we show how this can be done.

We first define machines $\mcal{K}'$ and $\mcal{G}'$, which are ``wrappers'' for the machines $\mcal{K}$ and $\mcal{G}_1$ in \Cref{pq:lem:small-tag:proof:se:proof:K}. These machines will help us set parameters properly so we can invoke \Cref{lem:Noisy-SimExt}:

\subpara{Machine $\mcal{G}'$:} it takes as input $(1^\secpar, 1^{\gamma^{-1}}, \msf{pref}')$ and proceeds as follows:
\begin{enumerate}
\item
Set $\epsilon\coloneqq 8\gamma$.
\item
Compute $\epsilon_1$ from $\epsilon$. Note that this can be done because \Cref{pq:lem:small-tag:proof:se:proof:K} stipulates the there is a noticeable $\epsilon_1 \le \epsilon$ that is efficiently computable from $\epsilon$.
\item
Run machine $\mcal{G}_1(1^\secpar, 1^{\epsilon_1^{-1}}, \msf{pref}')$ (as per \Cref{pq:lem:small-tag:proof:se:proof:K}) and output whatever it outputs.
\end{enumerate}

\subpara{Machine $\mcal{K}'$:} it takes as input $(1^\secpar,1^{\gamma^{-1}}, 1^{\zeta^{-1}},\msf{pref}')$ and proceeds as follows:
\begin{enumerate}
\item
Set $\epsilon\coloneqq 8\gamma$.
\item
Compute $\epsilon_1$ from $\epsilon$. Note that this can be done because \Cref{pq:lem:small-tag:proof:se:proof:K} stipulates the there is a noticeable $\epsilon_1 \le \epsilon$ that is efficiently computable from $\epsilon$.
\item
Set $\epsilon_2 \coloneqq \zeta$.
\item
Run machine $\mcal{K}(1^\secpar, 1^{\epsilon_1^{-1}}, 1^{\epsilon_2^{-1}}, \msf{pref}')$ (as per \Cref{pq:lem:small-tag:proof:se:proof:K}) and output whatever it outputs.
\end{enumerate}
In the following, we invoke \Cref{lem:Noisy-SimExt} with $\mcal{G}'$, $\mcal{K}'$, $(\msf{st}_{\mcal{M}}, \msf{st}_R, \tau,\eta,\msf{VI}_{\eta})$, $\tilde{\tau}$, and $\msf{val}(\tilde{\tau})$  playing the role of $\mcal{G}$,  $\mcal{K}$, $\rho_\secpar$, $z_\secpar$, and $s^*_{z_\secpar}$ respectively in \Cref{lem:Noisy-SimExt}. To do that, we first prove the  $\mcal{G}'$ and $\mcal{K}'$ indeed satisfy the conditions \Cref{item:s_star_or_bot,item:gamma_delta} in \Cref{lem:Noisy-SimExt}. 

\para{For \Cref{item:s_star_or_bot} in \Cref{lem:Noisy-SimExt}.} First, by \Cref{pq:lem:small-tag:proof:se:proof:K}, we know that the machine $\mcal{K}$ when invoked with parameters $(1^\secpar, 1^{\epsilon_1^{-1}}, 1^{\epsilon_2^{-1}}, \msf{pref}')$ outputs $\msf{val}(\tilde{\tau})$ with probability at most $\epsilon_2(\secpar) + \negl(\secpar)$. Since $\epsilon_2$ is set to $\zeta$ in machine $\mcal{K}'$, this implies that $\mcal{K}'$ outputs $\msf{val}(\tilde{\tau})$ with probability at most $\zeta(\secpar) + \negl(\secpar)$, satisfying \Cref{item:s_star_or_bot} in \Cref{lem:Noisy-SimExt}.

\para{For \Cref{item:gamma_delta} in \Cref{lem:Noisy-SimExt}.} First, we claim that if $\mcal{G}'(1^\secpar, 1^{\gamma^{-1}}, \msf{pref}')$ output $\tilde{d}  = \top$ with probability $8\gamma$, then it must hold that $p_\msf{pref}^{\msf{Sim}}[\epsilon_1] \ge \epsilon$, with the $\epsilon_1$ defined in $\mcal{G}'$. To see that, first notice that  
\begin{align*}
\Pr[\tilde{d} = \top ~:~ (\OUT, \tilde{d}) \la \mcal{G}'(1^\secpar, 1^{\gamma^{-1}}, \msf{pref}')] =
\Pr[\tilde{d} = \top ~:~ (\OUT, \tilde{d}) \la \mcal{G}_1(1^\secpar, 1^{\epsilon_1^{-1}}, \msf{pref}')] 
\end{align*}
where $\epsilon_1$ is defined in the description of $\mcal{G}'$. Thus, if the LHS of the above equation is greater than $8\gamma$, then it must hold that 
\begin{align*}
p_\msf{pref}^{\msf{Sim}}[\epsilon_1] = \Pr[\tilde{d} = \top ~:~ (\OUT, \tilde{d}) \la \mcal{G}_1(1^\secpar, 1^{\epsilon_1^{-1}}, \msf{pref}')] \ge 8\gamma = \epsilon. 
\end{align*}

Next, recall from \Cref{pq:lem:small-tag:proof:se:proof:K} that under the condition of $p_\msf{pref}^{\msf{Sim}}[\epsilon_1] \ge \epsilon$, it must hold that 
$$\Pr[\Val = \msf{val}(\tilde{\tau}) : \Val \gets \mcal{K}{(1^\secpar, 1^{\epsilon_1^{-1}},1^{\epsilon_2^{-1}}, \msf{pref}')}] \ge {\frac{\epsilon'(\secpar)-\epsilon_2(\secpar)}{\tilde{t}}},$$
where  $\epsilon' = \frac{\epsilon(\secpar)}{10t^2}$. Then, by definition of $\mcal{K}'$, it must hold that 
\begin{align*}
\Pr[\Val = \msf{val}(\tilde{\tau}) : \Val \gets \mcal{K}'{(1^\secpar, 1^{\gamma^{-1}},1^{\zeta^{-1}}, \msf{pref}')}] &= \Pr[\Val = \msf{val}(\tilde{\tau}) : \Val \gets \mcal{K}{(1^\secpar, 1^{\epsilon_1^{-1}},1^{\epsilon_2^{-1}}, \msf{pref}')}] \\ 
&\ge 
{\frac{\epsilon'(\secpar)-\epsilon_2(\secpar)}{\tilde{t}}}\\
& =
\delta(\secpar) - \frac{\zeta(\secpar)}{\tilde{t}}  \\ 
& \ge 
\delta(\secpar) - \zeta(\secpar) -\negl(\secpar),
\end{align*}
where we $\delta(\secpar) \coloneqq \frac{\epsilon'(\secpar)}{\tilde{t}}$ with $\epsilon' = \frac{\epsilon(\secpar)}{10t^2}$. (Also note that $\epsilon_2 = \zeta$ by definition of $\mcal{K}'$.)

The above shows that the \Cref{item:gamma_delta} in \Cref{lem:Noisy-SimExt} is satisfied.

\para{Invoking \Cref{lem:Noisy-SimExt}.} Since $\mcal{K}'$ and $\mcal{G}'$ satisfy the conditions in \Cref{lem:Noisy-SimExt}, we can invoke it to claim the existence of a machine $\SimExt$ such that
 for any noticeable $\epsilon(\secpar)$, there exists a noticeable $\gamma(\secpar) \le \epsilon(\secpar)$ such that
$$\{\SimExt(1^\secpar,1^{\epsilon^{-1}},\msf{pref}')\}_{\secpar \in \Naturals}
~\statind_{\epsilon}~ 
\{(\OUT, \msf{val}_{\tilde{d}}(\tilde{\tau}))~:~(\OUT,  \tilde{d})\leftarrow \mcal{G}'(1^\secpar,1^{\gamma^{-1}},\msf{pref}')\}_{\secpar \in \Naturals}.$$

Finally, recall that $\mcal{G}'(1^\secpar,1^{\gamma^{-1}},\msf{pref}')$ is identical to machine $\mcal{G}_1(1^\secpar,1^{\epsilon_1^{-1}},\msf{pref}')$ where $\epsilon_1 \le \epsilon (=8\gamma)$ is efficiently computable from $\gamma$ as in the description of $\mcal{G}'$. Thus, the above implies that: 
For any noticeable $\epsilon(\secpar)$, there exists a noticeable $\epsilon'(\secpar) \le 8 \epsilon(\secpar)$ that is efficiently computable from $\epsilon(\secpar)$, such that
$$\{\SimExt(1^\secpar,1^{\epsilon^{-1}},\msf{pref}')\}_{\secpar \in \Naturals}
~\statind_{\epsilon}~ 
\{(\OUT, \msf{val}_{\tilde{d}}(\tilde{\tau}))~:~(\OUT, \tilde{d})\leftarrow \mcal{G}_1(1^\secpar,1^{\epsilon'^{-1}},\msf{pref}')\}_{\secpar \in \Naturals},$$
which is exactly \Cref{lem:simext:closeness}.

%% file: sections/two-sided.tex

\section{Post-Quantum Non-Malleable Commitments: One-to-One and Two-sided}
\label{sec:two-sided:main-body}

In this section, we show how to remove the `one-sided' restriction from \Cref{protocol:BB-NMCom}. 
 
Recall that our proof for the non-malleability of \Cref{protocol:BB-NMCom} works only if $t<\tilde{t}$. However, this is not guaranteed in the real main-in-the-middle attack---the adversary can of course use a smaller tag in the right session. Fortunately, this problem can be addressed by the so-called `two-slot' technique proposed by Pass and Rosen \cite{STOC:PasRos05}. The idea is to create a situation where no matter how the MIM adversary $\mcal{M}$ schedules the messages, there is always a `slot' for which the `$t<\tilde{t}$' condition holds; As long as this is true, non-malleability can be proven using the same techniques as we did for \Cref{protocol:BB-NMCom}.

To do that, first observe that the only place where \Cref{protocol:BB-NMCom} makes use of the tag $t$ is \Cref{bbnmc:hard-puzzle:puzzle-setup} {\bf Hard Puzzle Setup}: The receiver is required to setup a $t$-solution hard puzzle where $t$ is determined by the tag. Of course, \Cref{prot:bbnmc:puzzle-sol-reveal,prot:bbnmc:PoC} also depend on $t$ but that is rather a consequence of \Cref{bbnmc:hard-puzzle:puzzle-setup} using a $t$-solution hard puzzle. 

This observation allows us to instantiate the \cite{STOC:PasRos05} technique for \Cref{protocol:BB-NMCom} as follows. We view \Cref{bbnmc:hard-puzzle:puzzle-setup} as a `slot' in \cite{STOC:PasRos05} terminology. We ask the receiver to repeat this `slot' twice sequentially, using $t$ and $(T-t)$ as their respective tag, where recall that $T$ is the upper-bound for the size of tag space and is a polynomial on the security parameter $\secpar$. That is,
\begin{itemize}
 \item
 {\bf Slot-A:}
$R$ first executes \Cref{bbnmc:hard-puzzle:puzzle-setup} as it is, setting a $t$-solution hard puzzle;
\item
{\bf Slot-B:}
$R$ then executes \Cref{bbnmc:hard-puzzle:puzzle-setup} again, but using $(T-t)$ in place of $t$ in the first execution. This sets a $(T-t)$-solution hard puzzle. 
 \end{itemize} 
We also modify \Cref{prot:bbnmc:puzzle-sol-reveal,prot:bbnmc:PoC} as follows:
\begin{itemize}
 \item
In \Cref{prot:bbnmc:puzzle-sol-reveal}, $R$ reveals the solutions to {\em both} the $t$ solutions w.r.t.\ {\bf Slot-A} {\em and} the $(T-t)$ solutions w.r.t.\ {\bf Slot-B};
\item
In \Cref{prot:bbnmc:PoC}, we change the trapdoor statement from `$C$ manages to commit to a puzzle solution in \Cref{prot:bbnmc:extcom}' to `$C$ manages to commit to a puzzle solution {\em either} for {\bf Slot-A} {\em or} for {\bf Slot-B} in \Cref{prot:bbnmc:extcom}'.
 \end{itemize} 
By the above design, it is easy to see that one of the following case must happen no matter how $\mcal{M}$ sets the tags $t$ and $\tilde{t}$:
\begin{enumerate}
\item
{$t = \tilde{t}$:} This is the trivial case that is already ruled out by the definition of non-malleability.
\item
{$t < \tilde{t}$:} In this case, non-malleability follows by applying the same argument as we did for \Cref{protocol:BB-NMCom} to {\bf Slot-A}. 
\item
{$t > \tilde{t}$:} In this case, it must hold that $(T-t) < (T-\tilde{t})$. In other words, the tag for the left {\bf Slot-B} is smaller than the tag for the right {\bf Slot-B}. Therefore,  non-malleability follows by applying the same argument as we did for \Cref{protocol:BB-NMCom} to {\bf Slot-B}. 
\end{enumerate}
Therefore, the modified protocol is non-malleable without the `one-sided' restriction.

We remark that the same technique has been employed by \cite{FOCS:LPY23} to remove the `one-sided' restriction in their original protocol as well. Our application does not encounter any new challenges compared with the same step in \cite{FOCS:LPY23}.  Thus, we omit the proof details and only present the formal description of this updated protocol in \Cref{sec:two-sided:full}.

We summarize the result of this section as the following theorem.

\begin{theorem}\label{thm:two-sided:non-malleability}
  Assuming the existence of post-quantum one-way functions, there exists (i.e., \Cref{protocol:BB-NMCom:two-sided}) a black-box, constant-round construction of 1-1 (two-sided) post-quantum non-malleable commitments (as per \Cref{def:NMCom:pq} with $k=1$) in the synchronous setting,  supporting tag space $[T]$ with $T(\secpar)$ being any polynomial in the security parameter $\secpar$.
\end{theorem} 

%% file: sections/simultaneous_extraction.tex

\newcommand{\projimp}{\mathsf{ProjImp}}
\newcommand{\API}{\mathsf{API}}
\newcommand{\shiftdis}[1]{\Delta_{\mathsf{Shift}}^{#1}}
\newcommand{\repair}{\mathsf{Repair}}
\newcommand{\repairA}{\mcal{A}\text{-}\mathsf{Repair}}

\section{Simultaneous Extraction Lemma}

So far, we have obtained a post-quantum non-malleable commitment in the 1-1 MIM setting. Recall that our final goal is to obtain a construction secure in the more demanding 1-many MIM setting. Jupping ahead, we will manage to show (in \Cref{sec:BB-NMCom:one-many}) that the same protocol \Cref{protocol:BB-NMCom} (more accurately, its two-sided version \Cref{protocol:BB-NMCom:two-sided}), without any modifications, is indeed already secure in the 1-many MIM setting. However, this or course requires a different security proof. 

In this section, our develop a `simultaneous extraction lemma' (\Cref{lem:Simultaneous-SimExt}). This lemma will play a crucial role later when we upgrade the security proof for \Cref{protocol:BB-NMCom} (or \Cref{protocol:BB-NMCom:two-sided}) to the 1-many setting in \Cref{sec:BB-NMCom:one-many}.

\subsection{Lemma Statement}

\begin{lemma}[Simultaneous Extraction Lemma]
\label{lem:Simultaneous-SimExt}
Let $\mcal{V}$ be a QPT algorithm that takes the security parameter $1^\secpar$, an error parameter $1^{\gamma^{-1}}$, 
a quantum state $\rho$, {and a classical string $z$}  
as input,  and outputs  $d\in \{\top,\bot\}$. 
\takashi{
This is essentially the same as $\mcal{G}$ as in the simulatable extraction lemma. 
I changed the name and syntax since the quantum output $\rho_\out$ is redundant in this lemma.
But it also makse sense to use the same $\mcal{G}$ for consistency.
}

Suppose that for $i\in [n]$, 
there exists a QPT algorithm $\mcal{K}_i$ (referred to as the extractor) that takes as input the security parameter $1^\secpar$, two error parameters $1^{\gamma^{-1}}$ and $1^{\zeta^{-1}}$, 
a quantum state $\rho$ 
and outputs $s\in \bit^{\poly(\secpar)}\cup \{\bot\}$   
satisfying the following w.r.t.\ some sequence of classical strings {$\{s^*_{z,i}\}_{z\in \bit^*,i\in[n]}$.} \takashi{This notation may be slightly weird since $n$ may be a function of $\secpar$.} 
\begin{itemize}
 \item    \label{item:simultaneous_s_star_or_bot} 
    {\bf Assumption 1:} For any $\secpar$,  $\rho_\secpar$,  $z_\secpar$, $i\in [n]$, and any noticeable functions $\gamma(\secpar)$ and $\zeta(\secpar)$, it holds that  

$$\Pr[s \notin \Set{s^*_{z_\secpar,i}, \bot}~:~s \la \mcal{K}_i(1^\secpar,1^{\gamma^{-1}}, 1^{\zeta^{-1}}, \rho_\secpar,{z_\secpar})]\le \zeta(\secpar) +\negl(\secpar).$$

    \item \label{item:simultaneous_gamma_delta}
    {\bf Assumption 2:} For any noticeable function $\gamma(\secpar)$, there exists a noticeable function  $\delta(\secpar)$, 
 which is efficiently computable from $\gamma(\secpar)$, so that the following requirement is satisfied: 
 For any noticeable function $\zeta(\secpar)$
 and any sequence $\{\rho_\secpar,{z_\secpar}\}_{\secpar\in\mathbb{N}}$ of polynomial-size quantum states and classical strings and $i\in[n]$, 
 if 
$$
\Pr[d=\top ~:~ d \leftarrow \mcal{V}(1^\secpar,1^{\gamma^{-1}},\rho_\secpar,{z_\secpar})]\geq  \gamma(\secpar), 
$$  
then 
$$
\Pr[s =s^*_{{z_\secpar,i}}~:~ s\la \mcal{K}_i(1^\secpar,1^{\gamma^{-1}}, 1^{\zeta^{-1}}, \rho_\secpar,{z_\secpar})]\geq   \delta(\secpar)-\zeta(\secpar)-\negl(\secpar).
$$
\end{itemize}
Then, there exists a QPT algorithm $\mcal{K}$ that satisfies the following:  
\begin{enumerate}
 \item  \label[Property]{item:simultaneous_conclusion_s_star_or_bot}
    For any $\secpar$,  $\rho_\secpar$,  $z_\secpar$,  and any noticeable functions $\gamma(\secpar)$ and $\zeta(\secpar)$, it holds that  

$$\Pr[\bar{s} \notin \Set{s^*_{z_\secpar,1}||s^*_{z_\secpar,2}...||s^*_{z_\secpar,n}, \bot}~:~\bar{s} \la \mcal{K}(1^\secpar,1^{\gamma^{-1}}, 1^{\zeta^{-1}}, \rho_\secpar,{z_\secpar})]\le \zeta(\secpar) +\negl(\secpar).$$

    \item \label[Property]{item:simultaneous_conclusion_gamma_delta}
  For any noticeable function $\gamma(\secpar)$,  there exists a noticeable function  $\delta'(\secpar)$,  
 which is efficiently computable from $\gamma(\secpar)$, so that 
 the following requirement is satisfied: For 
 any noticeable funntion $\zeta(\secpar)$ and 
 any sequence $\{\rho_\secpar,{z_\secpar}\}_{\secpar\in\mathbb{N}}$ of polynomial-size quantum states and classical strings, 
 if 
$$
\Pr[d=\top ~:~ d \leftarrow \mcal{V}(1^\secpar,1^{\gamma^{-1}},\rho_\secpar,{z_\secpar})]\geq  8\gamma(\secpar), 
$$  
then 
$$
\Pr[\bar{s} =s^*_{z_\secpar,1}||s^*_{z_\secpar,2}...||s^*_{z_\secpar,n}~:~ \bar{s}\la \mcal{K}(1^\secpar,1^{\gamma^{-1}}, 1^{\zeta^{-1}}, \rho_\secpar,{z_\secpar})]\geq   \delta'(\secpar)-\zeta(\secpar)-\negl(\secpar).
$$
\takashi{Indeed, $8\gamma(\secpar)$ and $\delta'(\secpar)$ can be made arbitrarily close to $\gamma(\secpar)$.}
\end{enumerate}
\end{lemma}

\subsection{Preparation}
We take several tools and lemmas from \cite{TCC:Zhandry20,FOCS:CMSZ21} and give slight extensions of them.   
\begin{definition}[Projective Implementation \cite{TCC:Zhandry20}]
Let $\mcal{M}=(M_0,M_1)$ be a binary outcome POVM. Let $\mcal{E}=\{E_p\}_{p\in S}$ be a projective measurement indexed by $p\in S$ for some finite subset $S$ of $[0,1]$.\footnote{In \cite{TCC:Zhandry20}, $\mcal{E}$ is labeled by a distribution $D$. The definition here is identical to theirs if we interpret $p$ as a distribution that takes $1$ with probability $p$ and otherwise takes $0$.} Consider the following experiment:
\begin{enumerate}
\item Apply the measurement $\mcal{E}$ to obtain $p\in S$. 
\item Output $1$ with probability $p$ and output $0$ with probability $1-p$.  
\end{enumerate}
We say that $\mcal{E}$ is a projective implementation of $\mcal{M}$ if for any initial state, the above experiment yields the identical distribution to that obtained by applying the POVM $\mcal{M}$.  
\end{definition}

\begin{lemma}[{\cite[Lemma 3.3]{TCC:Zhandry20}}]
Any binary outcome POVM $\mcal{M}$ has a unique projective implementation. 
\end{lemma} \xiao{It appears to me that \cite[Lemma 3.3]{TCC:Zhandry20} only claims the existence for projective implementation for {\em commutative} $\mcal{M}$. Are you suggesting that the commutative condition is redundant? Or you actually mean that ``If a binary outcome POVM $\mcal{M}$ has a  projective implementation, then it must be unique''?}
\takashi{I'm relying on the fact that any binary-outcome POVM commute. This is implicitly used in \cite{TCC:Zhandry20}. For example, see the paragraph starting from "In our case,..." in page 7.} \xiao{I see. You're right!}

For a binary outcome POVM $\mcal{M}$, 
we write $\projimp(\mcal{M})$ to mean its projective implementation.

\begin{definition}[Shift Distance~\cite{TCC:Zhandry20}]\label{def:shift_distance}
For two distributions $D_0,D_1$, with cumulative density functions $f_0,f_1$, respectively, 
the shift distance with parameter $\epsilon$
is defined as
\begin{align*}
\shiftdis{\epsilon}(D_0,D_1) \coloneqq \sup_{x\in \mathbb{R}}\min_{y\in [f_1(x-\epsilon),f_1(x+\epsilon)]}|f_0(x)-y|.
\end{align*}
For two real-valued measurements $\mcal{M}$ and $\mcal{N}$ over the same quantum system, the shift distance between $\mcal{M}$ and $\mcal{N}$ with parameter $\epsilon$ is
\[
\shiftdis{\epsilon}(\mcal{M},\mcal{N}) \coloneqq  \sup_{\ket{\psi}}\shiftdis{\epsilon}(\mcal{M}(\ket{\psi}),\mcal{N}(\ket{\psi})).
\]
\end{definition}

By the definition, we can see the following: If $\shiftdis{\epsilon}(\mcal{M},\mcal{N})\le \eta$, then for any state $\ket{\psi}$ and $x \in \mathbb{R}$, 
\begin{align*}
\Pr[\mcal{M}(\ket{\psi}) \le x] & \le \Pr[\mcal{N}(\ket{\psi})\le x + \epsilon] + \eta,&& \Pr[\mcal{M}(\ket{\psi}) \ge x]  \le \Pr[\mcal{N}(\ket{\psi})\ge x - \epsilon] + \eta,\\
\Pr[\mcal{N}(\ket{\psi})\le x] & \le \Pr[\mcal{M}(\ket{\psi}\le x + \epsilon] + \eta,&& \Pr[\mcal{N}(\ket{\psi})\ge x]  \le \Pr[\mcal{M}(\ket{\psi}\ge x - \epsilon] + \eta.
\end{align*}

\begin{definition}[Almost Projective Measurements~\cite{TCC:Zhandry20}]
A real-valued measurement $\mcal{M}=(M_i)_{i\in I}$ is $(\epsilon,\eta)$-almost projective if the following is true: for any quantum state $\ket{\psi}$, apply $\mcal{M}$ twice in a row to $\ket{\psi}$, obtaining outcomes $x,y$. Then $\Pr[|x-y|\le \epsilon]\ge 1-\eta$. 
\end{definition}

\xiao{@Takashi: Just for my understanding: It seems \cite{TCC:Zhandry20} and \cite{FOCS:CMSZ21} works with the original \cite[Theorem 6.2]{TCC:Zhandry20}. However we need this variant \Cref{lem:API}. I want to learn the reason behind this difference. (I saw your remarks after this lemma. But I'd like to discuss more about it.)}

\xiao{@Takashi: Also, why doesn't \Cref{lem:API} follow directly from \cite[Lemma 4.9]{FOCS:CMSZ21}?}
\takashi{
As I explained in the first item of the remark, 
the difference is that they focus on a mixture of projective measurements.
I remark that the second point of the remark was already implicitly dealt with in \cite{FOCS:CMSZ21} (see \cite[Remark 4.8(a)]{FOCS:CMSZ21}).}

The following is a variant of \cite[Theorem 6.2]{TCC:Zhandry20}.  
\begin{lemma}\label{lem:API}
For any binary-outcome POVM $\mcal{M}=(M_0,M_1)$ and reals $0<\epsilon,\eta<1$, there is a real-valued measurement $\API_{\mcal{M}}^{\epsilon,\eta}$ that satisfies the following: 
\begin{enumerate}
\item \label{item:API_shiftdis}
$\shiftdis{\epsilon}(\API_{\mcal{M}}^{\epsilon,\eta},\projimp(\mcal{M}))\le \eta$.
\item  \label{item:API_almost_projective}
$\API_{\mcal{M}}^{\epsilon,\eta}$ is $(\epsilon,\eta)$-almost projective.
\item \label{item:API_run_time}
The run time of $\API_{\mcal{M}}^{\epsilon,\eta}$ is $T_{\mcal{M}}\cdot \poly(\epsilon^{-1},\log (\eta^{-1}))$, where $T_{\mcal{M}}$ is the run time of the POVM $\mcal{M}$. 
\end{enumerate}
\end{lemma}

There are the following two differences from the original statement of \cite[Theorem 6.2]{TCC:Zhandry20}.
\begin{enumerate}
\item We consider general binary-outcome POVM whereas they focuses on a special case called  ``mixture of projective measurements." \xiao{@Takashi: Why cannot we view our case of binary-outcome POVM  as a special case of "mixture of projective measurements"? For example, in \cite[Section 6]{TCC:Zhandry20} notation, if the index set $\mcal{I} = \Set{1}$, then doesn't it correspond to our case?}\takashi{No, because $\mathcal{M}$ may not be a projective measurement.
Note that if $\mathcal{M}$ is a projective measurement, its projective implementation is trivial since it is projective from the beginning.
}

\item We require the run time of $\API_{\mcal{M}}^{\epsilon,\eta}$ is  $T_{\mcal{M}}\cdot \poly(\epsilon^{-1},\log (\eta^{-1}))$ whereas they require it only for the \emph{expected} run time. \xiao{@Takashi: did you actually strengthen Zhandry? It seems that Zhandry only claimed strict QPT for $p$ far from 0 or 1.}
\takashi{I don't think he made any formal claim about strict QPT. If you are talking about \cite[Remark 6.4]{TCC:Zhandry20}, I believe this means that 1. if $p\in [1/4,3/4]$, achieving strict QPT is straightforward, and 2. we can reduce the general case to the case of $p\in [1/4,3/4]$ by introducing "dummy projections". The proof of \Cref{lem:API} exactly follows this idea. }
\end{enumerate}
For the first difference, we observe that the original proof can be easily extended to general binary-outcome POVM by using Jordan's lemma.  
The second difference can be resolved by using an idea of ``scaling down" as sketched in \cite[Remark 6.4]{TCC:Zhandry20}.

For completeness, we prove \Cref{lem:API}. We note that the proof is based on the proof of \cite[Theorem 6.2]{TCC:Zhandry20} and we often repeat very similar arguments to theirs. 

\begin{proof}[Proof of \Cref{lem:API}]
First, we construct $\tilde{\API}_{\mcal{M}}^{\epsilon,\eta}$ that satisfies the requirements if $\projimp(\mcal{M})$ is supported by $p\in [1/4,3/4]$, i.e., for any state $\rho$, it holds that 
$$\Pr[\frac{1}{4} \le p \le \frac{3}{4} ~:~p\gets \projimp(\mcal{M})(\rho)]=1.$$ 
Looking ahead, this assumption is used to make sure that $\tilde{\API}_{\mcal{M}}^{\epsilon,\eta}$ runs in strict QPT (rather than expected QPT as in \cite[Theorem 6.2]{TCC:Zhandry20}). 
At the end of the proof, we modify it to $\API_{\mcal{M}}^{\epsilon,\eta}$ that works for any binary-outcome measurement. 

Suppose that $\projimp(\mcal{M})$ is supported by $p\in [1/4,3/4]$. 
Let $\regX$ be a quantum register for states on which $\mcal{M}=(M_0,M_1)$ acts.
Let $U$ be a purification of $\mcal{M}$ on $\regX$ and an ancilla register $\regY$. That is, we define the unitary $U$ in such a way that for any state $\rho_\regX$ on $\regX$ and $b\in \bit$, we have 
\begin{align*}
    \Tr(M_b \rho_\regX)=\Tr(U^\dagger(\ket{b}\bra{b}\otimes I) U (\rho_\regX \otimes \ket{0}\bra{0}_{\regY}))
\end{align*}
where $\ket{b}\bra{b}\otimes I$ means the operator that projects the first qubit of $\regX$ onto $\ket{b}$. 
We define two projectors $\Pi_0$ and $\Pi_1$ over $\regX$ and $\regY$ as:
\begin{align*}
\Pi_0 \coloneqq I_\regX \otimes \ket{0^n}\bra{0^n}_\regY,~~~\Pi_1 \coloneqq U^\dagger(\ket{1}\bra{1}\otimes I) U
\end{align*}
where $n$ is the number of qubits in $\regY$. 
By applying Jordan's lemma to $\Pi_0$ and $\Pi_1$, we can see that there is an orthogonal decomposition of the Hilbert space over $\regX$ and $\regY$ into two-dimensional subspaces $\{S_j\}_j$ that satisfies the following:\footnote{In general, there may also appear one-dimensional subspaces. However, by our assumption that $\projimp(\mcal{M})$ is supported by $p\in [1/4,3/4]$, 
all eigenvalues of $\Pi_0 \Pi_1 \Pi_0$ belongs to $[1/4,3/4]$, and thus one-dimensional subspaces do not appear in our case. 
}
For each two-dimensional subspace $S_j$, there exist two orthonormal bases $(\ket{\alpha_j},\ket{\alpha_j^{\bot}})$ and $(\ket{\beta_j},\ket{\beta_j^{\bot}})$ of $S_j$ such that 
\begin{align*}
    \Pi_0\ket{\alpha_j}=\ket{\alpha_j},~~~ \Pi_0\ket{\alpha_j^{\bot}}=0,\\
    \Pi_1\ket{\beta_j}=\ket{\beta_j},~~~ \Pi_1\ket{\beta_j^{\bot}}=0.
\end{align*}
Moreover, if we let 
\begin{align*}
    p_j\defeq \bra{\alpha_j}\Pi_1 \ket{\alpha_j},
\end{align*}
then we have $1/4\le p_i \le 3/4$ and
\begin{align*}
\ket{\alpha_j}=\sqrt{p_j}\ket{\beta_j}+\sqrt{1-p_j}\ket{\beta_j^{\bot}},~~~
\ket{\beta_j}=\sqrt{p_j}\ket{\alpha_j}+\sqrt{1-p_j}\ket{\alpha_j^{\bot}}.
\end{align*} 
In particular, this implies that 
\begin{align}
\label{eq:transition_alpha_beta}
\begin{split}
\Pi_1\ket{\alpha_j}=\sqrt{p_j}\ket{\beta_j},~~~ &(I-\Pi_1)\ket{\alpha_j}=\sqrt{1-p_j}\ket{\beta_j^\bot},\\
\Pi_1\ket{\alpha_j^\bot}=\sqrt{1-p_j}\ket{\beta_j},~~~ &(I-\Pi_1)\ket{\alpha_j^\bot}=\sqrt{p_j}\ket{\beta_j^\bot},\\
\Pi_0\ket{\beta_j}=\sqrt{p_j}\ket{\alpha_j},~~~ &(I-\Pi_0)\ket{\beta_j}=\sqrt{1-p_j}\ket{\alpha_j^\bot},\\
\Pi_0\ket{\beta_j^\bot}=\sqrt{1-p_j}\ket{\alpha_j},~~~ &(I-\Pi_0)\ket{\beta_j}=\sqrt{p_j}\ket{\alpha_j^\bot}
\end{split}
\end{align}
Since $\Pi_0\ket{\alpha_j}=\ket{\alpha_j}$, we can write $\ket{\alpha_j}=\ket{\alpha'_j}_{\regX}\ket{0}_{\regY}$ for each $j$. 
For each $p\in [1/4,3/4]$, 
we define a projector $E_p$ on $\regX$ as 
$$
E_p \coloneqq \sum_{j:p_j=p} \ket{\alpha'_j}\bra{\alpha'_j}. 
$$
Then one can see that $\mcal{E}=\{E_p\}_{p\in S}$ is the projective implementation of $\mcal{M}$ where \xiao{@Takashi: in terms of writing, I think this part might need some ``guidance sentence'' to let the reader know why you're building $E_p$. If I understand it correctly, you are constructing the ``unique'' projective implementation $\projimp(\mcal{M})$? The explicit form of this $\projimp(\mcal{M})$ you provide here will be used in later part of the proof?}
\takashi{Yes. So strictly speaking, we should have proven that $\mcal{E}$ satisfies the definition of projective implementation, but I thought this was almost obvious.}

\begin{takashienv}{The reason why $\mathcal{E}$ is the projective implementation}
    {Here is a brief explanation:
Given any state $\ket{\psi}$, we can decompose the state as
\[
\ket{\psi}\ket{0^n}=\sum_{j} c_j \ket{\alpha_j}.
\]
We have 
\[
\Tr(M_1\ket{\psi}\bra{\psi})=\|\Pi_1\ket{\psi_1}\|^2
=\|\sum_{j} c_j \sqrt{p_j} \ket{\beta_j}\|^2
=\sum_{j} c_j^2 p_j
\]
where the first equality follows from the definition of $\Pi_1$, the second from $\Pi_1\ket{\alpha_j}=\sqrt{p_j} \ket{\beta_j}$, 
and the third from the orthogonality of $\ket{\beta_j}$'s and $\|\ket{\beta_j}\|=1$.So applying the POVM $(M_0,M_1)$ on $\ket{\psi}$ results in the outcome $1$ with probability  $\sum_{j} c_j^2 p_j$.

On the other hand, if we first apply $\mathcal{E}$ on $\ket{\psi}$, then we get an outcome $p$ with probability 
\[
\sum_{j:p_{j}=p}c_{j}^2.
\]
Thus, if we output $1$ with probability $p$, the overall probability of outputting $1$ is
\[
\sum_{p}(\sum_{j:p_{j}=p}c_{j}^2)p
=\sum_{j}c_j^2 p_j
\]
where the first sum is taken over all $p$ such that $p=p_j$ for some $j$. 
So the probability of outputting $1$ is the same as that by the POVM $(M_0,M_1)$ for any initial state. (I focused on pure states, but this immediately extends to the mixed states by decomposing it into a mix of pure states.) This means that $\mathcal{E}$ is the projective implementation of $(M_0,M_1)$. 
}
\end{takashienv}

$$S \coloneqq \{p\in [1/4,3/4] ~:~ \exists j~\text{s.t.}~p_j=p\}.$$ 

We describe the algorithm $\tilde{\API}_{\mcal{M}}^{\epsilon,\eta}$ on register $\regX$ below:
\begin{enumerate}
\item Prepare and initialize the register $\regY$ to the all-zero state.
\item Initialize a classical list $L=(0)$.
\item \label{step:main_loop}
Repeat the following ``main loop"  
for $i=1,2, \ldots, T$, where $T\coloneqq \lceil  \ln(6/\eta)/\epsilon^2\rceil$: 
\begin{enumerate}
\item Apply the projective measurement $(I-\Pi_1,\Pi_1)$, obtaining an outcome $b_{2i-1}$, and append $b_{2i-1}$ to the end of $L$. 
\item Apply the projective measurement $(\Pi_0,I-\Pi_0)$, obtaining an outcome $b_{2i}$, and append $b_{2i}$ to the end of $L$. 
\end{enumerate}
\item Let $t$ be the number of bit flips in the sequence $L=(0,b_1,b_2,...,b_{2T})$, and let $\tilde{p} \coloneqq t/2T$. 
\item \label[Step]{step:API_recover}
If $b_{2T}=1$, repeat the ``main loop" until the first time $b_{2i}=0$ or it is repeated $T'=\lceil\log_{5/8}(\eta/3)\rceil$ times.   
We say that it fails if $b_{2i}=0$ does not occur within $T'$ times repetition. 
\item Discard  $\regY$ and output $\tilde{p}$. 
\end{enumerate} 
The run time requirement of \Cref{item:API_run_time} is clear from the description. We next establish one by one. First, we remark that  $\tilde{\API}_{\mcal{M}}^{\epsilon,\eta}$ just applies projective measurements $(I-\Pi_1, \Pi_1)$ and $(\Pi_0,I-\Pi_0)$  on registers $\regX,\regY$. Therefore, when proving \Cref{item:API_shiftdis,item:API_almost_projective}, we can analyze each subspace separately. \xiao{@Takashi: just for my understanding: Though I believe this claim, I never go through the rigorous reasoning to prove it myself. I may need to discuss with you about it, especially for the case where there are multiple subspaces sharing the same $p_j$.}
\takashi{I don't think there's any issue in that case, but it might be better to write down the details. I was just too lazy to do so. It seems that Mark's proof writes a little bit more details on how it works. (His proof is essentially identical to ours except for how to define the decomposition into subspaces.)
} 

\begin{takashienv}{More details on why we can analyze each subspace separately.}
 We would be able to argue this formally as follows:
For any state $\ket{\psi}$, we decompose the state as
\[
\sum_{j} c_j \ket{\alpha'_j}.
\]
\begin{itemize}
\item Let $D$ be the distribution of the outcome of $\tilde{\API}_{\mcal{M}}^{\epsilon,\eta}(\ket{\psi})$ and $D_j$ be the distribution of outcome of $\tilde{\API}_{\mcal{M}}^{\epsilon,\eta}(\ket{\alpha'_j})$. 
Then $D$ can be written as a convex sum of $D_j$:
\[
D=\sum_{j} |c_j|^2 D_j.
\]
(A formal proof of this fact may need a tedious calculation, but this seems obvious to me from the fact that "different subspaces do not interfere each other" during the execution of $\tilde{\API}_{\mcal{M}}^{\epsilon,\eta}$.)
\item Let $D'$ be the distribution of outcome of $\mathsf{ProjImp}(\mathcal{M})(\ket{\psi})$ and $D'_j$ be the distribution whose density function concentrates on $p_j$. Then, by the fact that $\mathcal{E}$ is the projective implementation of $\mathcal{M}$,  we have 
\[
D'=\sum_{j} |c_j|^2 D'_j.
\]
\end{itemize}
Then we have 
\[
\shiftdis{\epsilon}(D,D')\le
\sum_{j}|c_j|^2\shiftdis{\epsilon}(D_j,D'_j)
\le 
\sum_{j}|c_j|^2 \eta =\eta
\]
where the first inequality follows from the definition of shift distance and the second inequality follows from the analysis of the case where the initial state is $\ket{\alpha'_j}$.  
This completes the proof of Item 1. 

For Item 2, let $D''$ be the distribution of $(p,p')$ obtained by a consecutive execution of $\tilde{\API}_{\mcal{M}}^{\epsilon,\eta}$ on the initial state $\ket{\psi}$. 
Let $D''_j$ be the similar distribution for the initial state $\ket{\alpha'_j}$. 
Then, 
\[
D''=\sum_{j}c_j D''_j.
\]
(Again, rely on the non-interference among different subspaces.)
Then, 
\[
\Pr[|p-p'|> \epsilon:(p,p')\leftarrow D'']
\le 
\sum_{j}|c_j|^2\Pr[|p-p'|> \epsilon:(p,p')\leftarrow D''_j]
\le 
\sum_{j}|c_j|^2 \eta =\eta
\]
where the first inequality is obvious and the second inequality follows from the analysis for the case where the initial state is $\ket{\alpha'_j}$.  
\end{takashienv}
That is, we can focus on the case where the initial state is $\ket{\alpha'_j}$ for some $j$. 

\subpara{Proving \Cref{item:API_shiftdis} of \Cref{lem:API}.}  Note that $\projimp(\mcal{M})$ on  $\ket{\alpha'_j}$ 
results in $p_j$ with probability $1$. \xiao{@Takashi: I guess this is one of the places where you utilize the explicit form of $\projimp(\mcal{M})$ you provided previously?}  \takashi{Yes, and I beleieve this is the only place.}
By \Cref{eq:transition_alpha_beta}, we can see that the list $L$ obtained by applying $\tilde{\API}_{\mcal{M}}^{\epsilon,\eta}$ on $\ket{\alpha'_j}$  is according to the following distribution:
\begin{itemize}
\item Let $K$ be a list of $2T$ independent coin flips with expected value $p_j$.
\item Set $L_i$ be the parity of the first $i$ bits of $K$.
\end{itemize}
Then $t=2T\tilde{p}$ is the number of $1$s in $K$. Thus, by Hoeffding's bound, we have 
$$
\Pr[|p_j-\tilde{p}|\ge \epsilon/2]\le 2e^{-2(2T)(\epsilon/2)^2}\le \eta/3<\eta
$$
where we used $T\ge \ln(6/\eta)/\epsilon^2$. 
This implies 
$\shiftdis{\epsilon}(\tilde{\API}_{\mcal{M}}^{\epsilon,\eta},\projimp(\mcal{M}))\le \eta$, finishing the proof of \Cref{item:API_shiftdis}. 

\subpara{Proving \Cref{item:API_almost_projective} of \Cref{lem:API}.}  Suppose that we sequentially run $\tilde{\API}^{\epsilon,\eta}_{\mcal{M}}$ twice on the initial state $\ket{\alpha'_j}$.  
Let $\tilde{p}_0$ and $\tilde{p}_1$ be the measurement outcome of the first and second application, respectively. 
If the first application of $\tilde{\API}^{\epsilon,\eta}_{\mcal{M}}$ does not fail, then the state in $\regX$ goes back to $\ket{\alpha'_j}$ at the end of the first application. 
Thus, by repeating a similar analysis to the above, we can see that 
$$
\Pr[|\tilde{p}_b-p_j|\ge \epsilon/2]\le \eta/3
$$
for $b\in \bit$ 
conditioned on that the first application does not fail. 
Moreover, each trial in \Cref{step:API_recover} of the description of $\tilde{\API}^{\epsilon,\eta}_{\mcal{M}}$  
succeeds with probability $2p_j(1-p_j)\ge 3/8$ \xiao{@Takashi: Why's that the success probability for each trial is $2p_j(1-p_j)$? Oh, I see, you actually meant to say ``succeeds with probability at least  $2p_j(1-p_j)$''?} 
\takashi{I menat that the probability is exactly equal to $2p_j(1-p_j)$ when the initial state is $\ket{\alpha'_j}$. 
Note that the probability of observing a "bit flip" in one step is $p_j$. Here, we are interested in the probability of observing a "bit flip" (from $1$ to $0$) in $2$ steps. This can happen in either way of "flip -> non-flip" or "non-flip -> flip". So its probability is $p_j(1-p_j)+(1-p_j)p_j=2p_j(1-p_j).$
}
where we used $p_j\in [1/4,3/4]$, and thus the probability of failure is at most $(1-3/8)^{T'}\le \eta/3$  
where we used $T'\ge \log_{5/8}(\eta/3)$. 
Combining the above, we have 
$$
\Pr[|\tilde{p}_0-\tilde{p}_1|\ge \epsilon]\le 
\eta/3+\eta/3+\eta/3=
\eta,
$$
which implies \Cref{item:API_almost_projective}. 
This finishes the proof of \Cref{lem:API} for the case where $\projimp(\mcal{M})$ is supported by $p\in [1/4,3/4]$. 

\para{For the General Case of $p \in [0,1]$.} Finally, we extend it to general binary-outcome POVMs. 
For any binary-outcome POVM $\mcal{M}=(M_0,M_1)$,  let  $\mcal{M}' \coloneqq (\frac{I}{4}+\frac{M_0}{2},\frac{I}{4}+\frac{M_1}{2})$. 
That is, $\mcal{M}'$ corresponds to the process that either outputs a uniformly random bit or applies $\mcal{M}$  
with probability $1/2$ for each.   
Let $\mcal{E}=\{E_p\}_{p\in S}$ be the projective implementation of $\mcal{M}$. Then it is easy to see that 
the projective implementation of $\mcal{M}'$ is 
$\mcal{E'}=\{E'_{p'}\}_{p'\in S'}$
where $E'_{p'} \coloneqq E_{2p'-1/2}$ and $S' \coloneqq \{p': 2p'-1/2\in S\}$. 
For any $p'\in S'$, 
since 
$2p'-1/2\in [0,1]$, we have $p'\in [1/4,3/4]$.  
Thus, $\projimp(\mcal{M}')$ is supported by $p'\in [1/4,3/4]$ and $\tilde{\API}$ is applicable for $\mcal{M}'$. 
Based on this observation, we construct $\API_\calM^{\epsilon,\eta}$ as follows:
\begin{enumerate}
\item Apply $\tilde{\API}_{\calM'}^{\epsilon/2,\eta}$, obtaining an outcome $p'$.
\item Output $p \coloneqq 2p'-1/2$.
\end{enumerate}
Then the properties of $\tilde{\API}_{\calM'}^{\epsilon/2,\eta}$ which we showed above are directly translated into those of $\API_\calM^{\epsilon,\eta}$, which concludes the proof of \Cref{lem:API}. 

\end{proof}

\begin{lemma}[{\cite[Lemma 4.10]{FOCS:CMSZ21}}]\label{lem:repair}
Let $\mcal{N}$ be an $(\epsilon,\eta)$-almost projective measurement on a Hilbert space $\mcal{H}$, and $\mcal{P}=(P_0,P_1)$ be a binary-outcome projective measurement on $\mcal{H}$. 
Then there is a quantum algorithm $\mathsf{Repair}^{\mcal{N},\mcal{P}}$ on $\mcal{H}$ satisfying the following: 
\begin{itemize}
\item For a positive integer $T$, consider the following procedure $\mathsf{RepairExpt}^{\mcal{N},\mcal{P}}(1^T)$ on $\mcal{H}$: 
\begin{enumerate}
\item Apply $\mcal{N}$, obtaining outcome $p$; 
\item Apply $\mcal{P}$, obtaining outcome $b$;
\item Apply $\mathsf{Repair}^{\mcal{N},\mcal{P}}(1^T,b,p)$.
\item Output $p$.
\end{enumerate}
Then $\mathsf{RepairExpt}^{\mcal{N},\mcal{P}}(1^T)$ is $(2\epsilon,2(\eta+1/T)+4\sqrt{\eta})$-almost projective.
\item The expected run time of  $\mathsf{Repair}^{\mcal{N},\mcal{P}}(1^T,b,p)$ is at most $(T_{\mcal{N}}+T_{\mcal{P}})\cdot (4T\sqrt{\eta}+3)$ where  $T_{\mcal{N}}$ and $T_{\mcal{P}}$ are run times of $\mcal{
N}$ and $\mcal{P}$, respectively. \takashi{We may need to carefully clarify what is meant by an expected run time of a quantum algorithm. 
This is briefly mentioned in the Preliminaries of \cite{FOCS:CMSZ21}
}
\end{itemize}
\end{lemma}

We show the following corollary.
\begin{corollary}\label{cor:repair}
Let $\mcal{N}$ be an $(\epsilon,\eta)$-almost projective measurement on a Hilbert space $\mcal{H}$, and $\mcal{A}$ be a quantum algorithm that takes a quantum state in $\mcal{H}$ as input and outputs a classical string satisfying the following:
There are some classical string $s^*$ and $0\le \zeta \le 1$ such that for any state $\rho$, 
$$\Pr[s \notin \Set{s^*, \bot}~:~s \la \mcal{A}(\rho)]\le \zeta.$$
Then for any positive integer $T$ and $p\in[0,1]$, there is a measurement $\repairA(1^T,p)$ 
satisfying the following: 
\begin{itemize}
\item For any $T$, $p$, and any state $\rho$ in $\mcal{H}$, if we apply $\repairA(1^T,p)$ on $\rho$, then the distribution of the measurement outcome is identical to that of $\mcal{A}(\rho)$. 
\item For a positive integer $T$, consider the following procedure $\mathsf{RepairExpt}^{\mcal{N},\repairA}(1^T)$ on $\mcal{H}$: 
\begin{enumerate}
\item Apply $\mcal{N}$, obtaining outcome $p$; 
\item Apply $\repairA(1^T,p)$, obtaining outcome $s$; 
\item Output $p$.
\end{enumerate}
Then $\mathsf{RepairExpt}^{\mcal{N},\repairA}(1^T)$ is $\big(2\epsilon,2(\eta+1/T)+4\sqrt{\eta}+\sqrt{\zeta}\big)$-almost projective. 
\item The expected run time of  
$\repairA(1^T,p)$
is at most $(T_{\mcal{N}}+T_{\mcal{A}})\cdot (4T\sqrt{\eta}+3)$ where  $T_{\mcal{N}}$ and $T_{\mcal{A}}$ are run times of $\mcal{
N}$ and $\mcal{A}$, respectively. 
\end{itemize}
\end{corollary}
\begin{proof}
Intuitively, $\repairA$ first runs $\mcal{A}$ and then applies the repair procedure of \Cref{lem:repair}. A formal proof is given below. 

We can describe $\mcal{A}$ by using a unitary $U$ over the input register $\reginp$, output register $\regS$, and working register $\regW$ as follows:
\begin{description}
\item $\mcal{A}(\rho):$ Set $\rho$ in $\reginp$, initialize $\regS$ and $\regW$ to be all-zero states, apply $U$, measure $\regS$, and output the outcome $s$. 
\end{description}
We define a binary projective measurement $\mcal{P}=(P_0,P_1)$ on $\reginp$, $\regS$, and $\regW$ as 
$$
P_1 \coloneqq U^\dagger (\sum_{s\neq \bot}\ket{s}\bra{s})_{\regS} U
$$
and $P_0 \coloneqq I-P_1$. 
We apply \Cref{lem:repair} for $\mcal{N}$ and $\mcal{P}$ to get $\mathsf{Repair}^{\mcal{N},\mcal{P}}$ satisfying the requirements of \Cref{lem:repair}.\footnote{
Strictly speaking, $\mcal{N}$ is a POVM on $\reginp$ but $\mcal{P}$ is a projector on  $(\reginp,\regS,\regW)$ and thus  \Cref{lem:repair} is not directly applicable.  
We abuse the notation to simply write $\mcal{N}$ to mean its trivial extension  to registers $(\reginp,\regS,\regW)$ that does not touch $(\regS,\regW)$.
} By using it, we construct $\repairA(1^T,p)$ on $\reginp$ as follows:
\begin{enumerate}
\item Initialize $\regS$ and $\regW$ to all-zero states. 
\item Apply $\mcal{P}$, obtaining an outcome $b$.
\item \label[Step]{item:measure_s} If $b=0$, then set $s \coloneqq \bot$. If $b=1$, then apply $U$, measure $\regS$ to obtain $s$, and apply $U^\dagger$.
\item Apply $\mathsf{Repair}^{\mcal{N},\mcal{P}}(1^T,b,p)$.
\item \label[Step]{item:output_s}
Output $s$ as the measurement outcome. 
\end{enumerate}
It is clear from the construction that the distribution of $s$ obtained by applying $\repairA(p)$ on $\rho$ is identical to the distribution of $\mcal{A}(\rho)$. 
The requirement about the run time directly follows from that of \Cref{lem:repair}. 
Below, we show that $\mathsf{RepairExpt}^{\mcal{N},\repairA}(1^T)$ is $(2\epsilon,2(\eta+1/T)+4\sqrt{\eta}+\sqrt{\zeta})$-almost projective. 

Let $\repairA'(p)$ be a quantum process that works similarly to $\repairA(p)$ except that \Cref{item:measure_s,item:output_s} are removed. 
Then, it is not hard to see that $\mathsf{RepairExpt}^{\mcal{N},\repairA'}(1^T)$ is identical to 
$\mathsf{RepairExpt}^{\mcal{N},\mcal{P}}(1^T)$, and thus it is $(2\epsilon,2(\eta+1/T)+4\sqrt{\eta})$-almost projective by \Cref{lem:repair}. 
Moreover, we observe that the measurement in \Cref{item:measure_s} of $\repairA(p)$ for the case of $b=1$ yields a fixed value $s^*$ with probability except for $\zeta$ by the assumption about $\mcal{A}$. Thus, by the gentle measurement lemma \cite[Lemma 2.2]{DBLP:journals/toc/Aaronson05}, the trace distance between the states before and after the step is at most $\sqrt{\zeta}$.    
This implies that $\mathsf{RepairExpt}^{\mcal{N},\repairA}(1^T)$  is $(2\epsilon,2(\eta+1/T)+4\sqrt{\eta}+\sqrt{\zeta})$-almost projective.

This finishes the proof of \Cref{cor:repair}.
\end{proof}

\subsection{Proof of \Cref{lem:Simultaneous-SimExt}}

Let $\mcal{M}_{\secpar,\gamma,z}$ be the binary-outcome POVM corresponding to $\mcal{V}(1^\secpar,1^{\gamma^{-1}},\cdot,z)$. That is, it is defined in such a way that 
$
\Pr[\mcal{M}_{\secpar,\gamma,z}(\rho)=1]=
\Pr[\mcal{V}(1^\secpar,1^{\gamma^{-1}},\rho,z)=\top]$
for any state $\rho$.
Let $\API_{\mcal{M}_{\secpar,\gamma,z}}^{\epsilon,\eta}$ be the $(\epsilon,\eta)$-almost projective measurement
as given in \Cref{lem:API}.
For each $i,\secpar,\gamma,\zeta,\rho,z,\epsilon,\eta$, we apply \Cref{cor:repair} to 
the $(\epsilon,\eta)$-almost projective measurement $\API_{\mcal{M}_{\secpar,\gamma,z}}^{\epsilon,\eta}$ and
the algorithm $\mcal{K}_i(1^\secpar,1^{\gamma^{-1}},1^{\zeta^{-1}},\cdot,z)$, 
and we denote the corresponding repairing measurement by $\mcal{K}_i(1^\secpar,1^{\gamma^{-1}},1^{\zeta^{-1}},\cdot,z)\text{-}\repair$
and the corresponding repairing experiment by $\mathsf{RepairExpt}_{\secpar,\gamma,\zeta,z}^{\epsilon,\eta}(1^T)$.\footnote{If we strictly follow the notation in \Cref{cor:repair}, then  the experiment should be written as $\mathsf{RepairExpt}^{\API_{\mcal{M}_{\secpar,\gamma,z}}^{\epsilon,\eta},\mcal{K}_i(1^\secpar,1^{\gamma^{-1}},1^{\zeta^{-1}},\cdot,z)\text{-}\repair}(1^T)$, but we simply write $\mathsf{RepairExpt}_{\secpar,\gamma,\zeta,z}^{\epsilon,\eta}(1^T)$ for brevity.}
By the assumption about $\mcal{K}_i$ and \Cref{cor:repair}, $\mathsf{RepairExpt}_{\secpar,\gamma,\zeta,z}^{\epsilon,\eta}(1^T)$ 
is $\big(2\epsilon,2(\eta+1/T)+4\sqrt{\eta}+\sqrt{\zeta}+\negl(\secpar)\big)$-almost projective. \xiao{@Takashi: According to the statement of the current \Cref{lem:Simultaneous-SimExt}, the $\mcal{K}_i$'s probability is bounded by $\zeta + \negl$. However, the \Cref{cor:repair} does not take care of the $\negl$ term. Do we need to say something about it?}
\takashi{You are right. I added a negligible term.}


We first construct \emph{expected} QPT algorithm $\mcal{K}$ that satisfies the requirements, after which we argue that we can modify it to be \emph{strict} QPT by truncation. 

The \emph{expected} QPT algorithm $\mcal{K}$ is described as follows:
\begin{description}
\item $\mcal{K}(1^\secpar,1^{\gamma^{-1}},1^{\zeta^{-1}},\rho,z)$: 
Do the following:
\begin{enumerate}
\item \label{item:set_delta}
Compute $\delta$ as in {\bf Assumption 2} of \Cref{lem:Simultaneous-SimExt} from the given $\gamma$. 
\item 
Take an positive integer $T'$ in such a way that 
$(1-\delta/3)^{T'}\le \gamma/n$ holds. (For example, $T'=O(\delta^{-1} \log \gamma^{-1} \log n)$ suffices).
\item 
Set parameters as follows:
\begin{align*}
T & \coloneqq \lceil 6nT'\gamma^{-1} \rceil \\ 
\epsilon
& \coloneqq  \min\{2\gamma/(2nT'+1),\gamma/4\} \\ 
\eta
& \coloneqq  \left( \gamma/(18nT')\right)^2 \\ 
\zeta' 
& \coloneqq  \min \{\zeta/(nT'),\left( \gamma/(3nT')\right)^2,\delta/2\}
\end{align*} 

\item \label[Step]{item:check_initial}
Apply $\API_{\mcal{M}_{\secpar,\gamma,z}}^{\epsilon,\eta}$ on $\reginp$ to obtain an outcome $\tilde{p}$. 
 If $\tilde{p}<4\gamma-\epsilon$, output $\bot$ and halt.     
\item For $i=1,2,...,n$, do the following:
\begin{enumerate}
\item For $j=1,2,...,T' $, do the following  \label[Step]{item:K_loop_j}
\begin{enumerate}
\item Apply $\API_{\mcal{M}_{\secpar,\gamma,z}}^{\epsilon,\eta}$ to obtain an outcome $\tilde{p}_{i,j}$.

\item \label[Step]{item:check_prob}
If $\tilde{p}_{i,j}<\tilde{p}_{i,j-1}-2\epsilon$, output $\bot$ and halt, 
where when $i=1$ and $j=1$, $\tilde{p}_{i,j-1} \coloneqq \tilde{p}$ and when  $i\ge 2$ and $j=1$, $\tilde{p}_{i,j-1} \coloneqq \tilde{p}_{i-1,T'}$.  

\item \label[Step]{item:apply_K_i}
Apply $\mcal{K}_i(1^\secpar,1^{\gamma^{-1}},1^{\zeta'^{-1}},\cdot,z)\text{-}\repair^{\epsilon,\eta}(1^T,\tilde{p}_{i,j})$ to obtain an outcome $s_{i,j}$. 

\item \label[Step]{item:check_s}
If $s_{i,j}\ne \bot$, set $s_i \coloneqq s_{i,j}$, break the inner loop, and proceed to the outer loop for $i+1$. 
\end{enumerate}

\item \label[Step]{item:check_time_out}
If $s_{i,j}=\bot$ for all $j\in[T']$, output $\bot$ and halt.  
\end{enumerate}

\item Output $s_{1}||s_{2}||...||s_{n}$.
\end{enumerate}
\end{description}
We can see that $\mcal{K}$ runs in expected QPT 
by \Cref{lem:API} and \Cref{cor:repair}. 

\subpara{Proving \Cref{item:simultaneous_conclusion_s_star_or_bot} of \Cref{item:simultaneous_gamma_delta}.} We observe that whenever  $\mcal{K}$  does not output $\bot$, 
for each $i\in [n]$, 
$s_i$ is a non-$\bot$ value obtained by $\mcal{K}_i(1^\secpar,1^{\gamma^{-1}},1^{\zeta'^{-1}},\cdot,z)\text{-}\repair^{\epsilon,\eta}(\tilde{p}_{i,j})$. By \Cref{cor:repair}, its distribution is identical to the output distribution of $\mcal{K}_i(1^\secpar,1^{\gamma^{-1}},1^{\zeta'^{-1}},\cdot,z)$. 
\xiao{ @Takashi: 
I have confusion regarding this step: I think \Cref{cor:repair} means that (correct me if I'm wrong): if we run $\repairA$ on the ``original'' state $\rho$, then it leads to the same distribution as $\mathcal{A}(\rho)$. However, I'm not sure whether the ``same distribution'' claim holds when we compare $\mcal{A}(\rho)$ with $\repairA(\rho')$, {\em where $\rho'$ is the result we obtained after applying $\mcal{N}$ to the original $\rho$}. Do we need the $\mcal{A}(\rho) = \repairA(\rho')$ condition here to make your argument go through? 
}

\xiao{Oh, I think I see it now. You are talking about {\bf Assumption 1}, which holds for any $\rho$. So what I concerned does not matter here. Is this correct?}

\takashi{Yes!}

By {\bf Assumption 1} of \Cref{lem:Simultaneous-SimExt}, it outputs non-$\bot$ value other than $s^*_{z,i}$ with probability at most $\zeta'+\negl(\secpar)$. 
Since we apply it at most $T'$ times, the probability that it ever occurs is at most $T' (\zeta'+\negl(\secpar))$. 
By taking union bound over all $i\in [n]$, the probability that it occurs for some $i\in [n]$ is at most $n T' (\zeta'+\negl(\secpar))\le \zeta+\negl(\secpar)$. This finishes the proof of \Cref{item:simultaneous_conclusion_s_star_or_bot}. 

\subpara{Proving  \Cref{item:simultaneous_conclusion_gamma_delta} of \Cref{item:simultaneous_gamma_delta}.} 
Suppose that $\rho$ and $z$ satisfy the requirement of \Cref{item:simultaneous_conclusion_gamma_delta}, i.e., we have
\begin{align}\label{eq:_assumption_V}
\Pr[d=\top ~:~ d \leftarrow \mcal{V}(1^\secpar,1^{\gamma^{-1}},\rho,{z})]\geq  8\gamma. 
\end{align}  
We define the following events in the execution of $\mcal{K}(1^\secpar,1^{\gamma^{-1}},1^{\zeta^{-1}},\rho,z)$: 
\begin{itemize}
\item $\mathsf{Bad}_1$: The event that $\mcal{K}$ returns $\bot$ in   \Cref{item:check_initial}.
\item $\mathsf{Bad}_2$: The event that $\mcal{K}$ returns $\bot$ in \Cref{item:check_prob} for some $i,j$.    
\item $\mathsf{Bad}_3$: The event that $\mcal{K}$ returns $\bot$ in  \Cref{item:check_time_out} for some $i$.  
\end{itemize}  
Note that we have 
\begin{align}\label{eq:bot_prob}
\Pr[\mcal{K}(1^\secpar,1^{\gamma^{-1}},1^{\zeta^{-1}},\rho,z)=\bot]=\Pr[\mathsf{Bad}_1]+\Pr[\mathsf{Bad}_2]+\Pr[\mathsf{Bad}_3].
\end{align}
Below (\Cref{lem:bad_1,lem:bad_2,lem:bad_3}), we upper bound each term in the RHS of \Cref{eq:bot_prob}.

\begin{lemma}\label{lem:bad_1}
$\Pr[\mathsf{Bad}_1]\le 1-4\gamma+\eta$
\end{lemma}
\begin{proof}[Proof of \Cref{lem:bad_1}]
\Cref{eq:_assumption_V} implies
$$
\Pr[\mcal{M}_{\secpar,\gamma,z}(\rho)=1]\geq 8\gamma. 
$$
By the definition of $\projimp(\mcal{M}_{\secpar,\gamma,z})$ and an averaging argument, we have 
$$
\Pr[p\ge 4\gamma :p\gets \projimp(\mcal{M}_{\secpar,\gamma,z})(\rho)]\ge 4\gamma.
$$
By \Cref{item:API_shiftdis} of \Cref{lem:API}, we have 
$$
\Pr[\tilde{p}\ge 4\gamma- \epsilon :\tilde{p}\gets \API_{\mcal{M}_{\secpar,\gamma,z}}^{\epsilon,\eta}(\rho)]\ge 4\gamma-\eta.
$$
This completes the proof of \Cref{lem:bad_1}.

\end{proof}

\begin{lemma}\label{lem:bad_2}
$\Pr[\mathsf{Bad}_2]\le \gamma+\negl(\secpar)$. 
\end{lemma}
\begin{proof}[Proof of \Cref{lem:bad_2}]
Since $\API_{\mcal{M}_{\secpar,\gamma,z}}^{\epsilon,\eta}$ is $(\epsilon,\eta)$-almost projective, we have 
$$\Pr[\tilde{p}_{1,1}<\tilde{p}-2\epsilon]\le \eta.$$ 
Note that the loop done in \Cref{item:K_loop_j} of $\mcal{K}(1^\secpar,1^{\gamma^{-1}},1^{\zeta^{-1}},\rho,z)$ is identical to $\mathsf{RepairExpt}_{\secpar,\gamma,\zeta',z}^{\epsilon,\eta}(1^T)$ except for an additional check in \Cref{item:check_prob}.  
Since $\mathsf{RepairExpt}_{\secpar,\gamma,\zeta',z}^{\epsilon,\eta}(1^T)$ 
is $\big(2\epsilon,2(\eta+1/T)+4\sqrt{\eta}+\sqrt{\zeta'}+\negl(\secpar)\big)$-almost projective by \Cref{cor:repair}, it holds for each 
$(i,j)\ne (1,1)$ that
$$\Pr[\tilde{p}_{i,j}<\tilde{p}_{i,j-1}-2\epsilon]\le 2(\eta+1/T)+4\sqrt{\eta}+\sqrt{\zeta'}+\negl(\secpar)\le \gamma/(nT')+\negl(\secpar)$$
where we used $\eta\le \left( \gamma/(18nT')\right)^2$, $T=\lceil 6nT'\gamma^{-1} \rceil$, and $\zeta'\le \left( \gamma/(3nT')\right)^2$.

Noting that 
$$\eta\le \left( \gamma/(18nT')\right)^2\le \gamma/(nT'),$$ 
the union bound over all $(i,j)\in [n]\times [T']$
gives \Cref{lem:bad_2}.

\end{proof}

\begin{lemma}\label{lem:bad_3}
$\Pr[\mathsf{Bad}_3]\le \gamma$.
\end{lemma}
\begin{proof}[Proof of \Cref{lem:bad_3}]
   For each $i,j$, let $\rho_{i,j}$ be the state just before applying 
$\mcal{K}_i(1^\secpar,1^{\gamma^{-1}},1^{\zeta'^{-1}},\cdot,z)\text{-}\repair^{\epsilon,\eta}(\tilde{p}_{i,j})$ in
 \Cref{item:apply_K_i}. 
 Note that whenever \Cref{item:apply_K_i} is invoked, either of $\mathsf{\Bad}_1$ or $\mathsf{Bad}_2$ has not occurred by that point, which implies $\tilde{p}_{i,j}\ge 4\gamma-(2nT'+1)\epsilon \ge 2\gamma$ where we used $\epsilon\le 2\gamma/(2nT'+1)$.  
 Since $\API_{\mcal{M}_{\secpar,\gamma,z}}^{\epsilon,\eta}$ is $(\epsilon,\eta)$-almost projective by \Cref{item:API_almost_projective} of \Cref{lem:API}, 
 $$
 \Pr[
 \tilde{p}'_{i,j}\ge 2\gamma- \epsilon
 :
 \tilde{p}'_{i,j}\gets \API_{\mcal{M}_{\secpar,\gamma,z}}^{\epsilon,\eta}(\rho_{i,j})]\ge 1-\eta.
 $$
 Since we have $\shiftdis{\epsilon}(\API_{\mcal{M}_{\secpar,\gamma,z}}^{\epsilon,\eta},\projimp(\mcal{M}_{\secpar,\gamma,z}))\le \eta$ by \Cref{item:API_shiftdis} of \Cref{lem:API}, we have  
  $$
 \Pr[
 p_{i,j}\ge 2\gamma- 2\epsilon
 :
 p_{i,j}\gets \projimp(\mcal{M}_{\secpar,\gamma,z})(\rho_{i,j})]\ge 1-2\eta.
 $$
 This implies 
 $$
 \Pr[d=\top ~:~ d \leftarrow \mcal{V}(1^\secpar,1^{\gamma^{-1}},\rho_{i,j},{z})]\geq (1-2\eta)(2\gamma- 2\epsilon)
 \ge \gamma
 $$
 where we used $\eta\le \left( \gamma/(18nT')\right)^2\le \gamma/4$ and $\epsilon\le \gamma/4$. 
Thus, by {\bf Assumption 2} of \Cref{lem:Simultaneous-SimExt}, 
we have 
$$
\Pr[s_i =s^*_{{z,i}}~:~ s_i\la \mcal{K}_i(1^\secpar,1^{\gamma^{-1}}, 1^{\zeta'^{-1}}, \rho_{i,j},{z})]\geq   \delta-\zeta'-\negl(\secpar) \ge \delta/3
$$
for sufficiently large $\secpar$ 
where we used $\zeta'\le \delta/2$. 
Noting that $\mcal{K}_i(1^\secpar,1^{\gamma^{-1}},1^{\zeta'^{-1}},\cdot,z)\text{-}\repair^{\epsilon,\eta}(1^T,\tilde{p}_{i,j})$ on $\rho_{i,j}$ yields the identical distribution as  $\mcal{K}_i(1^\secpar,1^{\gamma^{-1}}, 1^{\zeta'^{-1}}, \rho_{i,j},{z})$ by \Cref{cor:repair} \xiao{@Takashi: I have the same confusion as the last time you invoked \Cref{cor:repair}: $\rho_{i,j}$ is different from $\rho$; does it require $\rho_{i,j}=\rho$ to apply the first item of \Cref{cor:repair}? If I understand your writing correctly, this is not a real issue here either --- it is just a typo that you should have written $\rho_{i, j}$ in place of $\rho$ in the input ot $\mcal{K}_i$?}\takashi{Yes, you are right, I fixed it.}, for each $i,j$, the probability of breaking the inner loop in \Cref{item:check_s} is at least $\delta/3$. Thus, for each $i$, the probability that this does not happen for all $j\in [T']$ is at most $(1-\delta/3)^{T'}\le \gamma/n$. \xiao{@Takashi: Just for my own understanding: is it true that if I'm willing to set $T'$ large enough, I can actually make this bad-3 probability negligible?}\takashi{I believe so.} 
By taking the union bound over all $i\in [n]$, \Cref{lem:bad_3} holds.

\end{proof}

Combining \Cref{eq:bot_prob} and \Cref{lem:bad_1,lem:bad_2,lem:bad_3}, we have 
\begin{align*}
\Pr[\mcal{K}(1^\secpar,1^{\gamma^{-1}},1^{\zeta^{-1}},\rho,z)=\bot]\le 1-4\gamma+ \eta+ \gamma + \gamma+\negl(\secpar)\le 1-\gamma+\negl(\secpar)
\end{align*}
where we used $\eta \le \left( \gamma/(18nT')\right)^2\le \gamma$. 
Combined with \Cref{item:simultaneous_conclusion_s_star_or_bot} of \Cref{lem:Simultaneous-SimExt} which is already proven, we have 
\begin{align*}
\Pr[\mcal{K}(1^\secpar,1^{\gamma^{-1}},1^{\zeta^{-1}},\rho,z)=s_{z,1}^*||...||s_{z,n}^*]\ge \gamma-\zeta-\negl(\secpar). 
\end{align*} 
This finishes the proof that $\mcal{K}$ satisfies  \Cref{item:simultaneous_conclusion_gamma_delta} of \Cref{lem:Simultaneous-SimExt}.
\xiao{@Takashi: It appears to me that the current proof (so far) does not mention $\delta'$ at all? If I understand it correct, you mean that $\delta' = \gamma$ in the expected QPT case, and $\delta' = \gamma/2$ in the strictly QPT case?}
\takashi{Yes.}

\para{On Strictly QPT.} Finally, we argue how to convert $\mcal{K}$ into a \emph{strict} QPT one. 
For some polynomial $C(\secpar)$, 
suppose that we modify $\mcal{K}$ so that if it runs $C(\secpar)$ times longer 
than its expected run time, then it immediately outputs $\bot$ and halts. 
Then $\mcal{K}$ now runs in strict QPT. This modification does not affect \Cref{item:simultaneous_conclusion_s_star_or_bot} since $\mcal{K}$ only outputs $\bot$ in the case of the time out. 
By Markov's inequality, the time out occurs with probability at most $C(\secpar)^{-1}$, which may decrease the probability in \Cref{item:simultaneous_conclusion_gamma_delta} by at most $C(\secpar)^{-1}$. Thus, if we set $C(\secpar)$ in such a way that $C(\secpar)^{-1}\le \delta'(\secpar)/2$, then  \Cref{item:simultaneous_conclusion_gamma_delta} is still satisfied if we replace $\delta'(\secpar)$ with $\delta'(\secpar)/2$.  

This completes the proof of \Cref{lem:Simultaneous-SimExt}. 
\takashi{Please check if this argument is okay.} \xiao{I just finished reading. It looks good modulo the comments I left.}

%% file: sections/PQNMC-one-many.tex

\section{Post-Quantum Non-Malleable Commitments: One-Many}
\label{sec:BB-NMCom:one-many}

In this section, we turn to the construction of a black-box commitment scheme that satisfies the weak {\em one-many} definition of post-quantum non-malleability (as described in \Cref{def:NMCom:weak:pq}). The construction for this setting is identical to \Cref{protocol:BB-NMCom} (more accurately, its `two-sided' version described in \Cref{sec:two-sided:main-body} and \Cref{protocol:BB-NMCom:two-sided}) given in \Cref{sec:BB-NMC:construction}, and thus uses the same component primitives. In effect, we show that the security of \Cref{protocol:BB-NMCom} extends to the one-many case as well. This is stated formally below. 
\begin{theorem}\label{thm:pqmnc:1-many}
For any polynomial $N(\secpar)$ in the security parameter, \Cref{protocol:BB-NMCom:two-sided} is a black-box, constant-round construction of a 1-$N$ post-quantum weakly non-malleable commitment (as per \Cref{def:NMCom:weak:pq}) in the synchronous setting, supporting tag space $[T]$ with $T(\secpar)$ being any polynomial in $\secpar$.
\end{theorem}

Note that in \Cref{thm:pqmnc:1-many}, we have referred to the number of right sessions in the non-malleability game by $N$, while this parameter was denoted by $k$ in \Cref{protocol:BB-NMCom}. We rename this for clarity, since $k$ is already used to denote an important quantity in our construction.

To prove \Cref{thm:pqmnc:1-many}, note that completeness and hiding can be argued similar as for \Cref{protocol:BB-NMCom}. We focus on showing 1-many non-malleability in the following.

\subsection{Proof Overview}

Technically, we need to prove weak 1-many non-malleability for \Cref{protocol:BB-NMCom:two-sided}. But to simplify the presentation, we only  focus on the `one-sided' version of it (i.e., \Cref{protocol:BB-NMCom}) in the following. This is because we can use the same `two-slots' trick as explained in \Cref{sec:two-sided:main-body} to lift the proof to the `two-sided' setting as well.

While the setting is now different for the 1-many case, and our formalism for handling this case has to change to account for this, we wish to emphasize that our {\em core strategy} and indeed our intuition for this proof remains essentially the same! Intuitively, the fundamental idea behind our proof is to design experiments that allow us to reduce non-malleability to the {\em hiding} property of the initial commitment on the left. Now while the 1-many setting involves multiple parallel right sessions, there is only a single session on the left, and so our proof strategy and technique remains largely unchanged---At a high level, we seek to extract a `trapdoor' from \Cref{bbnmc:hard-puzzle:rv-2} of the left session to then decouple the main body of the left session from the commitment made in the prefix phase (i.e., \Cref{bbnmc:init-com}), and subsequently we extract the value from the \Cref{prot:bbnmc:extcom} $\ExtCom$ of the right session and show that this must be the committed value $\tilde{m}$ in the right session with sufficiently high probability. 

This helps us better understand what remains unchanged (in the 1-many case) from our earlier proof for \Cref{protocol:BB-NMCom}, and what needs new treatment. The key thing that changes here is the extraction step, which is now more involved. Note that to establish 1-many non-malleability, we must obtain the tuple of values committed by $\mcal{M}$ {\em across} all the $N$ right sessions (and show this must not depend on the left session commitment). So in our proof, we must extract the right side committed value as before, but now we must do so {\em simultaneously} from all the parallel sessions with sufficiently high probability. It is not clear that our technique so far extends directly to this case; indeed, this extension to the {\em simultaeneous extraction} case turns out to be nontrivial and will be our focus here. 

More technically, we will arrange our task as follows:
\begin{itemize}
\item First, we describe how to capture the reduction from non-malleability to the VSS hiding game \Cref{chall:vss:hide} by modifying the MIM experiment. This is essentially identical to what is presented in \Cref{sec:pq-nmc:1-1:proof:reduction-to-hiding,bbnmc:proof:fin}. We will sketch the argument while mostly focusing on the syntactic differences arising from the 1-many setting. 
\item 
We state the key lemma formalizing the existence of the simulator-extractor (i.e., the analog of $\SimExt$ from \Cref{bbnmc:proof:fin}) in the 1-many setting. This is used to complete the proof of non-malleability. 
\item 
We then show how to build an {\em instanced} version of the simulation-less extractor $\mcal{K}$ from \Cref{pq:lem:small-tag:proof:se:proof:K}: {\em  For each session $j \in [N]$ on the right}, we design an extractor $\mcal{K}^{(j)}$ that works essentially the same way as $\mcal{K}$ in the 1-1 setting (i.e., in \Cref{pq:lem:small-tag:proof:se:proof:K}). 
\item 
Finally, we show how to get a full-fledged simulation-extractor from these $\mcal{K}^{(j)}$'s using our simultaneous extraction lemma given in \Cref{lem:Simultaneous-SimExt}. 
\end{itemize}  

\subsection{Reduction to VSS Hiding Game} 

We start by describing the analogs of the key experiments introduced in \Cref{sec:pq-nmc:1-1:proof:reduction-to-hiding}. We begin by defining the real man-in-the-middle experiment. 

\para{Game $H^{\mcal{M}_\secpar}(\secpar,m,N,\rho_\secpar)$:\label{pq:gameH:1-many:description}} Analogous to \Cref{pq:gameH:description}, this is just the real MIM interaction, now in the 1-$N$ setting. The game now takes the number of parallel right sessions $N$ as input. We continue to denote the left committer by $C$. But for clarity, we now refer to the various receivers on the right by $R^{(1)},\dots,R^{(N)}$ respectively. Recall that in the 1-$N$ man-in-the-middle experiment, these receivers function honestly and independently.

The output of this game is again denoted by $\Output_{H^{\mcal{M}_\secpar}}(\secpar, m, \rho_\secpar)$ and consists of the following parts:
\begin{enumerate}
\item
$\OUT$: The (quantum) output of $\mcal{M}$ at the end of this game;
\item For each $j \in [N]$, 
$\tilde{\tau}^{(j)}$: The commitment transcript sent by $\mcal{M}$ to $R^{(j)}$ in the \Cref{bbnmc:init-com} of the $j$-th right session; 
\item Also for each $j \in [N]$, 
$\tilde{d}^{(j)} \in \Set{\top, \bot}$: The output of the honest receiver $R^{(j)}$ in the $j$-th right session, indicating if $\mcal{M}_\secpar$'s commitment in the $j$-th right session is accepted ($\tilde{d}^{(j)} = \top$) or not ($\tilde{d}^{(j)} = \bot$).
\end{enumerate} 


We will also need to refer to the value committed in the right sessions. Toward that, we define the function $\Gamma_{\{\tilde{d}^{(j)}\}_{j=1}^{k}}(\cdot)$ in exactly the style given in \Cref{def:NMCom:weak:pq}.



Specifically, let $\msf{val}^{(j)}(\tilde{\tau}) = \msf{val}^{(j)}(\tilde{\tau}^{(j)})$ denote the value committed in the $j$-th right session by $\mcal{M}$ (note that this is well defined by the statistical binding property). 

Thus, to prove 1-$N$ non-malleability as per \Cref{def:NMCom:weak:pq}, we need only establish the following:
\begin{align*}
&\big\{\Gamma_{\{\tilde{d}_0^{(j)}\}_{j=1}^{N}}\big(\OUT_0, \Set{\msf{val}^{(j)}(\tilde{\tau_0}^{(j)})}_{j \in [N]}\big): (\OUT_0, \Set{\tilde{\tau}^{(j)}_0, \tilde{d}^{(j)}_0}_{j \in [N]}) \gets H^{\mcal{M}_\secpar}(\secpar,m_0, N, \rho_\secpar) \big\} \\
\cind ~& 
\big\{\Gamma_{\{\tilde{d}_1^{(j)}\}_{j=1}^{N}}\big(\OUT_1, \Set{\msf{val}^{(j)}(\tilde{\tau}_1^{(j)})}_{j \in [N]}\big): (\OUT_1, \Set{\tilde{\tau}^{(j)}_1, \tilde{d}^{(j)}_1}_{j \in [N]}) \gets H^{\mcal{M}_\secpar}(\secpar, m_1, N, \rho_\secpar) \big\} \numberthis \label{pq:eq:classical:1-many:H:m-0:m-1},
\end{align*}
where both ensembles are indexed by $\secpar \in \Naturals$ and $(m_0, m_1) \in \bits^{\ell(\secpar)} \times \bits^{\ell(\secpar)}$.

We next turn to defining the machine $\tilde{H}^{\mcal{M}_\secpar}$ in the 1-$N$ setting, which is the analog of \Cref{pq:game:Htil} in \Cref{sec:pq-nmc:1-1:proof:reduction-to-hiding}.

\begin{AlgorithmBox}[label={pq:game:Htil:1-many}]{Game \textnormal{$\tilde{H}^{\mcal{M}_\secpar}(\secpar, \epsilon,m,N,\rho_\secpar)$} in $1$-$N$ Setting}

{\bf Input:} This takes as input the same parameters $\secpar$, $\rho_\secpar$, $m$, and $N$ as for $H^{\mcal{M}_\secpar}$. It additionally takes as input a noticeable function $\epsilon(\cdot)$. 

It proceeds as follows:
\begin{enumerate}
    \item \label[Step]{pq:game:Htil:1-many:pref}
    {\bf (Prefix phase.)} 
    This proceeds as follows. 
    \begin{enumerate}
        \item \label[Step]{pq:game:Htil:1-many:pref:1}
        Sample a random size-$k$ subset $\eta \subset [n]$.
        \item \label[Step]{pq:game:Htil:1-many:pref:2}
       	Execute $H^{\mcal{M}_\secpar}(m,N,\secpar,\rho_\secpar)$ until the end of \Cref{bbnmc:init-com}. At the moment, it already receives the \Cref{bbnmc:init-com} commitment made by the left-session honest committer $C$. It performs brute-force computation to obtain from $C$'s commitment the committed shares $\msf{cv}_i$ and their decommitment information for $i \in \eta$. We denote these values as $\msf{VI}_\eta \coloneqq \Set{(\msf{cv}_i, \msf{decom}_i}_{i \in \eta}$. 
    \end{enumerate}
    \subpara{Notation:}  Let $\msf{st}_{\mcal{M}}$ denote the state of $\mcal{M}$ at the end of \Cref{bbnmc:init-com}; Let $\msf{st}_C$  (and \ $\Set{\msf{st}_R^{(j)}}_{j \in [N]}$) denote the state of the honest committer (and\ receivers) at the end of \Cref{bbnmc:init-com}; 
    Let $\Set{\tilde{\tau}^{(j)}}_{j \in [N]}$ denote the commitments sent by $\mcal{M}$ in \Cref{bbnmc:init-com} in all the $N$ right sessions. We denote the tuple $(\msf{st}_{\mcal{M}}, \Set{\msf{st}_R^{(j)}}_{j \in [N]}, \tau,\Set{\tilde{\tau}^{(j)}}_{j \in [N]})$ as $\msf{pref}^{1:N}$. We use the following nation to express the execution of this {\bf Prefix phase}:
    \begin{equation}
    \msf{pref'}^{1:N}\coloneqq(\msf{pref}^{1:N}, \eta, \msf{VI}_\eta) \la \tilde{H}_{\msf{pref}}^{\mcal{M}_\secpar}(\secpar,m,N, \rho_\secpar). 
    \end{equation}
    We remark that this prefix generation step is independent of the error parameter $\epsilon$.
  
    \item \label[Step]{pq:game:Htil:body} 
    {\bf (Remainder phase.)} This involves the following steps: 
    \begin{enumerate}
        \item 
        $\tilde{H}^{\mcal{M}_\secpar}$ invokes the $\mcal{G}_1(1^\secpar, 1^{\epsilon^{-1}}, \msf{pref}^{1:N}, \eta, \msf{VI}_\eta))$ defined in \Cref{pq:machine:g1:1-many} (where $\msf{pref}^{1:N}$ is now as defined above in the 1-$N$ setting), which now outputs a tuple $(\OUT, \Set{\tilde{d}^{(j)}}_{j \in [N]})$. 
        \item 
        $\tilde{H}^{\mcal{M}_\secpar}$ outputs $(\OUT, \Set{\tilde{\tau}^{(j)}}_{j \in [N]},\Set{\tilde{d}^{(j)}}_{j \in [N]})$. 
    \end{enumerate}
\end{enumerate}
\end{AlgorithmBox}


We now describe the subprocedure $\mcal{G}_1(\cdot)$ adapted to the 1-$N$ setting, which is the analog of \Cref{pq:machine:g1} in \Cref{sec:pq-nmc:1-1:proof:reduction-to-hiding}.

\begin{AlgorithmBox}[label={pq:machine:g1:1-many}]{Machine \textnormal{$\mcal{G}_1(1^\secpar, 1^{\epsilon^{-1}}, \msf{pref}^{1:N}, \eta, \msf{VI}_\eta)$} in $1$-$N$ Setting}
Machine $\mcal{G}_1(1^\secpar, 1^{\epsilon^{-1}}, \msf{pref}^{1:N}, \eta, \msf{VI}_\eta)$ works in the same manner as the $\mcal{G}_1(1^\secpar, 1^{\epsilon^{-1}}, \msf{pref}, \eta, \msf{VI}_\eta)$ defined in \Cref{pq:machine:g1} but in the 1-$N$ setting. This is even no need to give a full description of the current $\mcal{G}_1(1^\secpar, 1^{\epsilon^{-1}}, \msf{pref}^{1:N}, \eta, \msf{VI}_\eta)$, because it has literally identical {\em syntax} as the $\mcal{G}_1(1^\secpar, 1^{\epsilon^{-1}}, \msf{pref}, \eta, \msf{VI}_\eta)$ defined in \Cref{pq:machine:g1}---All the $\mcal{G}_1(1^\secpar, 1^{\epsilon^{-1}}, \msf{pref}, \eta, \msf{VI}_\eta)$ does is to make some modifications on the left session; Here in the 1-$N$ setting, we also has only a single left session. So, what our current $\mcal{G}_1(1^\secpar, 1^{\epsilon^{-1}}, \msf{pref}^{1:N}, \eta, \msf{VI}_\eta)$ does is to do the same thing on the left session as \Cref{pq:machine:g1} and follow the honest receivers' algorithm on the right sessions.

The only point that deserves a remark is the format of the output of the current $\mcal{G}_1$ in the 1-$N$ setting, which we describe as follows:
\begin{itemize}

    \item It finally outputs the values $(\OUT, \Set{\tilde{d}^{(j)}}_{j \in [N]})$, where again $\OUT$ is $\mcal{M}$'s final output and $\Set{\tilde{d}^{(j)}}_{j \in [N]}$ are the  final decisions by the honest $R^{(j)}$s in the $N$ right session.
\end{itemize}
\end{AlgorithmBox}



As in  \Cref{sec:pq-nmc:1-1:proof:reduction-to-hiding}, one can again show that the outputs of $H$ and $\tilde{H}$ defined above are computationally indistinguishable (i.e., an analog of \Cref{lem:Htil:similarity}). We next turn our attention to the machine $\tilde{G}$ in the 1-$N$ setting that makes use of our {\em simultaeneous} simulator-extractor. This machine is the analog of \Cref{pq:game:Gtil} in  \Cref{sec:pq-nmc:1-1:proof:reduction-to-hiding}.

\begin{AlgorithmBox}[label={pq:game:Gtil:1-many}]{Game \textnormal{$\tilde{G}^{\mcal{M}_\secpar}(\secpar, m, N,\rho_\secpar,\epsilon)$}}
This proceeds in two phases as well: 
\begin{enumerate}

    \item \label[Step]{pq:game:Gtil:1-many:pref}
    {\bf (Prefix phase.)} This is identical to {the prefix phase of \Cref{pq:game:Htil}, i.e., it computes $(\msf{pref}^{1:N}, \eta, \msf{VI}_\eta) \la \tilde{H}_{\msf{pref}}^{\mcal{M}_\secpar}(\secpar, m, N,\rho_\secpar)$.}

    \item {\bf Remainder phase:}\label[Step]{pq:game:Gtil:1-many:body} This involves the following steps: 
    \begin{itemize} 
        \item It invokes a machine $\SimExt$, which is guaranteed to exist by the following \Cref{lem:simext:closeness:1-many}: $\SimExt$ takes in as input a tuple $(1^\secpar, 1^{\epsilon^{-1}}, \msf{pref}^{1:N}, \eta, \msf{VI}_\eta)$ and outputs $(\msf{OUT}, \Set{\msf{Val}^{(j)}}_{j \in [N]})$. 
        \item  $\tilde{G}^{\mcal{M}_\secpar}$ outputs $(\msf{OUT},\Set{\msf{Val}^{(j)}}_{j \in [N]})$ as its own output.
    \end{itemize}
\end{enumerate}
\end{AlgorithmBox}


With this, we can turn to the main lemma: comparing executions of $\tilde{H}$ and $\tilde{G}$. Note that in the latter game we also obtain the committed values in the right sessions. Then \Cref{lem:simext:closeness:1-many} that is analogous to \Cref{lem:simext:closeness} shows that these executions yield outputs that are close up to a controllable error parameter: 

\begin{lemma}[1-many Simulation-Extractor]\label{lem:simext:closeness:1-many}
    Let $\mcal{G}_1(\cdot)$ be the efficient procedure defined in \Cref{pq:machine:g1:1-many}. For any polynomial $N(\secpar)$ in the security parameter $\secpar$, there exists a simulation-extractor $\SimExt$ such that for any $(\msf{pref}^{1:N},\eta,\msf{VI}_\eta)$ in the support of $\tilde{H}^{\mcal{M}_\secpar}_\msf{pref}$, and for any noticeable $\epsilon(\secpar)$, there is a noticeable $\epsilon'(\secpar) \le 8\epsilon(\secpar)$ that is efficiently computable from $\epsilon(\secpar)$ such that the following holds:
    \begin{align*}
        &
        \big\{(\msf{OUT}, \Set{\msf{Val}^{(j)}}_{j \in [N]}): (\msf{OUT}, \Set{\msf{Val}^{(j)}}_{j \in [N]}) \gets \SimExt(1^\secpar, 1^{\epsilon^{-1}},\msf{pref}^{1:N},\eta,\msf{VI}_\eta) \big\}
        \\
        \cind_{\epsilon} 
        ~&
        \big\{\Gamma_{\{\tilde{d}^{(j)}\}_{j=1}^{N}}\big(\OUT, \Set{\msf{val}^{(j)}_{\tilde{d}}(\tilde{\tau})}_{j \in [N]}\big): (\OUT, \Set{\tilde{d}^{(j)}}_{j \in [N]}) \gets \mcal{G}_1(1^\secpar, 1^{\epsilon'^{-1}}, \msf{pref}^{1:N},\eta,\msf{VI}_\eta) \big\}.
    \end{align*}
\end{lemma} 

The remainder of the proof to non-malleability can be shown just as in the earlier setting (via the reduction to the VSS hiding game as done in \Cref{bbnmc:proof:fin}). We omit the details to avoid repetition. Instead, we will focus on the key task of {\em building this simulator-extractor $\SimExt$}. Our approach will be the same as before---Namely, for each right session $j \in [N]$, we will first describe a base extractor $\mcal{K}^{(j)}$ that is essentially the extractor $\mcal{K}$ from \Cref{pq:lem:small-tag:proof:se:proof:K} but localized to session $j$, and then show (in \Cref{sec:sim-ext:1-many:final}) how to use these `localized' versions of $\mcal{K}$ to obtain the 1-$N$ simulation-extractor $\mcal{SE}$ as required by \Cref{lem:simext:closeness:1-many}.

\subsection{Localized Simulation-Less Extractors $\mcal{K}^{(j)}$}

Recall that we defined the shorthand $\msf{pref'}^{1:N} \coloneqq (\msf{pref}^{1:N},\eta,\msf{VI}_{\eta})$. Also, we define the following quantity $p^{\msf{Sim}}_{\msf{pref'}^{1:N}}[\epsilon_1]$ that will be important in the formal statement of $\mcal{K}^{(j)}$s. It is the 1-$N$ analog of the quantity $p^{\msf{Sim}}_{\msf{pref}'}[\epsilon_1]$ defined in \Cref{eq:def:p-pref}. 
\begin{equation}\label{eq:def:p-pref:1-many}
p^{\msf{Sim}}_{\msf{pref'}^{1:N}}[\epsilon_1] \coloneqq \Pr[\wedge_{j \in [N]} \big(\tilde{d}^{(j)} = \top \big) : (\OUT,\Set{\tilde{d}^{(j)}}_{j \in [N]}) \gets \mcal{G}_1(1^\secpar, 1^{\epsilon_1^{-1}}, \msf{pref}'^{1:N})].
\end{equation}

The following lemma states the formal guarantee provided by these localized base extractors $\mcal{K}^{(j)}$. It is the 1-$N$ analog of \cref{pq:lem:small-tag:proof:se:proof:K}.  

\begin{lemma}[Localized Simulation-less Extraction]\label{pq:lem:small-tag:proof:se:proof:K:1-many}
For any polynomial $N(\secpar)$, let $\tilde{H}^{\mcal{M}_\secpar}_{\msf{pref}}(\secpar, m, N, \rho_\secpar)$ be as defined in \Cref{pq:game:Htil:1-many}. There exist QPT machines $\Set{\mcal{K}^{(j)}}_{j \in [N]}$ such that for any noticeable $\epsilon(\secpar)$, there is a noticeable $\epsilon_1(\secpar) \le \epsilon(\secpar)$ that can be efficiently computed from $\epsilon$, such that for any noticeable $\epsilon_2(\secpar)$ and any tuple $\msf{pref'}^{1:N} = (\msf{st}_{\mcal{M}}, \Set{\msf{st}_{R^{(j)}}}_{j \in [N]}, \tau,\Set{\tilde{\tau}^{(j)}}_{j \in [N]},\eta,\msf{VI}_{\eta})$ in the support of $\tilde{H}^{\mcal{M}_\secpar}_{\msf{pref}}(\secpar, m, N, \rho_\secpar)$, the following conditions hold:
\begin{enumerate}
\item \label[Property]{pq:property:small-tag:proof:se:proof:K:syntax:1-many}
{\bf (Almost Uniqueness:)} For each $j \in [N]$, $\mcal{K}^{(j)}$ takes as input $(1^\secpar, 1^{\epsilon_1^{-1}}, 1^{\epsilon_2^{-1}}, \msf{pref}'^{1:N})$. It outputs a value $\msf{Val}^{(j)} \in \bits^{\ell(\secpar)} \cup \Set{\bot}$ that satisfies
$$\Pr[\msf{Val}^{(j)} \notin \Set{\msf{val}(\tilde{\tau}^{(j)}) , \bot} ~:~\msf{Val}^{(j)} \ra \mcal{K}^{(j)}{(1^\secpar, 1^{\epsilon_1^{-1}}, 1^{\epsilon_2^{-1}}, \msf{pref}'^{1:N})}] \leq {\epsilon_2(\secpar) + \negl(\secpar)}.$$

\item \label[Property]{pq:property:small-tag:proof:se:proof:K:1-many}
{\bf (Extraction:)} If $p^{\msf{Sim}}_{\msf{pref'}^{1:N}}[\epsilon_1] \ge \epsilon(\secpar)$, then for each $j \in [N]$ it holds that
$$\Pr[\Val^{(j)} = \msf{val}(\tilde{\tau}^{(j)}) : \Val^{(j)} \gets \mcal{K}^{(j)}{(1^\secpar, 1^{\epsilon_1^{-1}},1^{\epsilon_2^{-1}}, \msf{pref}'^{1:N})}] \ge {\frac{\epsilon'(\secpar)-\epsilon_2(\secpar)}{\tilde{t}}},$$
where $p^{\msf{Sim}}_{\msf{pref}'^{1:N}}[\epsilon_1]$ is defined in \Cref{eq:def:p-pref:1-many} and  $\epsilon'(\secpar) \coloneqq \frac{\epsilon(\secpar)}{10t^2}$. 
\end{enumerate}
\end{lemma}


This statement of the simulation-less extractor(s) in the one-many setting looks quite different from the one appearing in \Cref{pq:lem:small-tag:proof:se:proof:K}, and at first glance, seems to be significant extension of the latter. At the very least, there are more moving parts with the multiple parallel sessions on the right. Further consideration however reveals that this is not quite the case. The trick to this is to see that since the receivers $R^{(1)},\dots R^{(N)}$ in the right sessions operate {\em independently}, and act honestly (unless we extract in that session), we can simply treat them as standard interactions of the commitment. 

In particular, when invoking this extractor on a particular session $j \in [N]$, we can treat the other right sessions as {\em context}: in more detail, given the 1-many man-in-the-middle adversary $\mcal{M}$, we can come up with a new adversary $\mcal{M}^{(j)}$ that runs the other parallel sessions internally in a one-many MIM interaction with $\mcal{M}$ and then treats the $j$-th session as the {\em sole} right session in a one-one MIM interaction. We can then apply \Cref{pq:lem:small-tag:proof:se:proof:K} to derive these guarantees for $K^{(j)}$ (where the MIM adversary is $\mcal{M}^{(j)}$). Armed with this reasoning, we can see that \Cref{pq:lem:small-tag:proof:se:proof:K:1-many} readily follows from \Cref{pq:lem:small-tag:proof:se:proof:K}. 

This also allows us to set parameters the same way as in \Cref{pq:lem:small-tag:proof:se:proof:K}. Namely, we work by first fixing a noticeable $\epsilon(\cdot)$ and then set $\epsilon_1(\secpar)\defeq \frac{t+1}{t^2+4t+2}\cdot \epsilon'(\secpar)$ with $ \epsilon'(\secpar) \coloneqq \frac{\epsilon(\secpar)}{10t^2}$.

\subsection{Simulation-Extractor $\mcal{SE}$: 1-Many Setting}
\label{sec:sim-ext:1-many:final}

To finish the proof of \Cref{lem:simext:closeness:1-many} (and thus the proof of weak 1-many non-malleability), there are two things left now. First, we will build a {\em simultaneous} simulation-less extractor $\mcal{K}$ that extracts the $\tilde{m}^{(j)}$'s in all the right sessions. As long as we have such an $\mcal{K}$, we can re-use the noisy simulation extraction lemma (i.e., \Cref{lem:Noisy-SimExt}) to upgrade it to the desired simulation-extractor $\mcal{SE}$ as required by \Cref{lem:simext:closeness:1-many}. In the following, we elaborate on these two steps.

\para{Simultaneous Simulation-less Extractor $\mcal{K}$.} Such a $\mcal{K}$ can be built by applying the simultaneous extraction lemma we developed in \Cref{lem:Simultaneous-SimExt} to the localized simulation-less extractors $\mcal{K}^{(j)}$'s. For that, we first need to prove that these $\mcal{K}^{(j)}$'s in \Cref{pq:lem:small-tag:proof:se:proof:K:1-many} indeed satisfied the prerequisites in \Cref{lem:Simultaneous-SimExt}. Similar as in \Cref{sec:sim-less-to-sim:1-1}, we will not show it directly with the $\mcal{K}^{(j)}$'s. Some `wrapper' machines need to be defined to make the parameters match. Fortunately, this step is almost identical to what we did in \Cref{sec:sim-less-to-sim:1-1}. We will be able to use almost the same parameter settings.  

\subpara{Machine $\mcal{G}'$:} it takes as input $(1^\secpar, 1^{\gamma^{-1}}, \msf{pref}'^{1:N})$ and proceeds as follows:
\begin{enumerate}
\item
Set $\epsilon\coloneqq \gamma$.
\item
Compute $\epsilon_1$ from $\epsilon$. Note that this can be done because \Cref{pq:lem:small-tag:proof:se:proof:K:1-many} stipulates the there is a noticeable $\epsilon_1 \le \epsilon$ that is efficiently computable from $\epsilon$.
\item
Run machine $\mcal{G}_1(1^\secpar, 1^{\epsilon_1^{-1}}, \msf{pref}'^{1:N})$ (as per \Cref{pq:lem:small-tag:proof:se:proof:K:1-many}) and output whatever it outputs.
\end{enumerate}

\subpara{Machine $\mcal{K}'^{(j)}$ ($j \in [\tilde{t}]$):} it takes as input $(1^\secpar,1^{\gamma^{-1}}, 1^{\zeta^{-1}},\msf{pref}'^{1:N})$ and proceeds as follows:
\begin{enumerate}
\item
Set $\epsilon\coloneqq \gamma$.
\item
Compute $\epsilon_1$ from $\epsilon$. Note that this can be done because \Cref{pq:lem:small-tag:proof:se:proof:K:1-many} stipulates the there is a noticeable $\epsilon_1 \le \epsilon$ that is efficiently computable from $\epsilon$.
\item
Set $\epsilon_2 \coloneqq \zeta$.
\item
Run machine $\mcal{K}^{(j)}(1^\secpar, 1^{\epsilon_1^{-1}}, 1^{\epsilon_2^{-1}}, \msf{pref}'^{1:N})$ (as per \Cref{pq:lem:small-tag:proof:se:proof:K:1-many}) and output whatever it outputs.
\end{enumerate}

Compare the above machines with the $\mcal{G}'$ and $\mcal{K}'$ defined in \Cref{sec:sim-less-to-sim:1-1}, the only differences regarding the parameters is that we set $\epsilon_1 = \gamma$ directly, instead of $\epsilon_1 = 8\gamma$. That is because the RHS of the {\bf Assumption 2} of \Cref{lem:Simultaneous-SimExt} is $\gamma(\secpar)$ ( instead of $8\gamma(\secpar)$ in the RHS of \Cref{item:gamma_delta} in \Cref{lem:Noisy-SimExt}).

Now, if we treat the above $\mcal{G}'$ as machine $\mcal{V}$ in \Cref{lem:Simultaneous-SimExt}, treat the above $\Set{\mcal{K}'^{(j)}}_{j \in [N]}$ as the machines $\Set{\mcal{K}_i}_{i \in [n]}$ (i.e., $n = N$), and set $\delta(\secpar) \coloneqq \frac{\epsilon'(\secpar)}{\tilde{t}}$, then it is straightforward to see that that {\bf Assumption 1} and {\bf Assumption 2} of \Cref{lem:Simultaneous-SimExt} are satisfied.\footnote{Similar as in \Cref{sec:sim-less-to-sim:1-1}, $\msf{pref}'^{1:N}$, $\Set{\tilde{\tau}^{(j)}}_{j\in [N]}$, and $\Set{\msf{val}(\tilde{\tau}^{(j)})}_{j \in [N]}$  play the role of, $\rho_\secpar$, $z_\secpar$, and $\Set{s^*_{z_\secpar, j}}_{j \in [N]}$ respectively.} Thus, \Cref{lem:Simultaneous-SimExt} implies the desired {\em simultaneous} simulation-less extractor $\mcal{K}$ that is able to extract all the committed values in the $N$ right sessions.

\para{1-Many Simulation-Extractor $\mcal{SE}$.} Finally, simply observe that \Cref{item:simultaneous_conclusion_s_star_or_bot,item:simultaneous_conclusion_gamma_delta} in \Cref{lem:Simultaneous-SimExt} is exactly the prerequisites of the noisy simulation-extraction lemma (i.e., \Cref{lem:Noisy-SimExt}). Thus, a straightforward application of \Cref{lem:Noisy-SimExt} to the machines $\mcal{K}$ and $\mcal{V}$ (with them being the $\mcal{K}$ and $\mcal{G}$ in \Cref{lem:Noisy-SimExt}) implies our desired simulation-extraction $\mcal{SE}$ in the 1-$N$ setting. This finishes the proof of \Cref{lem:simext:closeness:1-many} , and thus in turns finishes our proof of weak 1-$N$ non-malleability (i.e., \Cref{def:NMCom:weak:pq}). 

Note that limitation to weak 1-many non-malleability is somewhat inherent to our approach: the simultaeneous simulator-extractor we define has a requirement on the machine that corresponds to $\mcal{G}'$ --- namely, it should output $\top$ with a certain noticeable probability. In our interpretation, this output corresponds to the {\em conjunction} of all the verifier decisions (accept/reject) in the right sessions. This condition translates to assuming that our MIM adversary completes every right session successfully with a reasonably large probability, and so our treatment can only consider adversaries that obey this constraint --- and not ones, for example, that {\em always} abort in certain right sessions. We thus eschew showing standard 1-many non-malleability; and as we shall later see, weak one-many non-malleability suffices for the applications we have in mind for our commitment. 

%% file: sections/parallelOT.tex
\section{Post-Quantum Multi-Party Parallel OT}
\label{sec:parallel-OT}

In this section, we describe a black-box, constant-round protocol implementing the $n$-party 
OT functionality (as defined in \Cref{fig:F-OT}) w.r.t.\ the $\epsilon$-simulatable PQ-MPC security notion (as per \Cref{def:mpc}).  We state the definition we will be able to realize and work-with in \Cref{def:eps-mal-par-OT}.


\begin{FigureBox}[label={fig:F-OT}]{The Ideal Functionality $\mathcal{F}^n_{\text{OT}}$}
The functionality $\mathcal{F}^n_{\text{OT}}$ is specified by the number of distinct senders $S_i$ and receivers $R_j$ (for $i,j \in [n]$). It acts as follows: 

\para{Sender's Message:} $\mathcal{F}^n_{\text{OT}}$ receives $(\algo{send},i,j,s_0^i,s_1^i)$ from a sender $S_i$ (which also specifies a purported receiver $R_j$). It ignores this message if $i=j$. Otherwise, it records this quintuple, and ignores subsequent messages that have the same initial triple $(\algo{send},i,j)$. 

\para{Receiver's Message:} $\mathcal{F}^n_{\text{OT}}$ receives $(\algo{receive},i,j,b)$ from a receiver $R_j$. 
It responds with $(\algo{open},i,j,s_b^i)$ to the receiver $R_j$.  
    
\end{FigureBox}

\begin{definition}[Post-Quantum Multi-Party OT with $\epsilon$-Simulation]\label{def:eps-mal-par-OT}
A protocol $\Prot$ is called a malicious  multiparty secure oblivious transfer protocol if for every polynomial $n \coloneqq n(\secpar)$, 
$\Prot$ 
is a post-quantum ($\epsilon$-simulatable) MPC protocol for $\mathcal{F}^n_{\text{OT}}$ precisely as stated in \Cref{def:mpc}. 
\end{definition}

Below we will describe a constant-round, black-box protocol and prove that it satisfies \Cref{def:eps-mal-par-OT}. Our approach will in fact be to first describe a {\em 2-party} OT protocol $\Prot$ running in {\em constant rounds}, and then show that the {\em parallel repetition} of $\Prot$ satisfies the requirements of \Cref{def:eps-mal-par-OT}.

More precisely for a 2-party protocol $\Prot$, let us define the {\em n-fold parallel repetition} as follows: consider $n$ entities where each entity would like to participate in an OT session {\em both} as a sender and as a receiver against all the other entities in parallel. In other words, for each $i\neq j \in [n]$, we consider two OT sessions between parties $P_i$ and $P_j$ where in one $P_i$ plays the role of sender and $P_j$ that of the receiver, and vice versa in the other.  
This comprises $2 \cdot \binom{n}{2}$ parallel independent executions of $\Prot$.  We will show that the $n$-fold 
parallel repetition of $\Prot$ (denoted by $\Prot^n$) is a post-quantum ($\epsilon$-simualatable) MPC protocol for $\mathcal{F}^n_{\text{OT}}$. 

We note that sequential composition would serve the same purpose were we not constrained by the constant-round requirement. Relying on parallel composition, however, we get the desired constant-round OT protocol since $\Prot^n$ has the same round complexity as $\Prot$.


\subsection{Building Blocks}
\label{sec:parallelOT:building-blocks}
Before diving into the main construction of the parallel malicious secure OT protocol, we need some building blocks. 

\para{Malicious-Sender Secure OT:} The first component we require is a constant-round OT protocol with a simpler or weaker property: namely, with security against {\em malicious senders} and {\em semi-honest receivers}. Additionally, we require that the associated simulator for proving security be straight-line. Fortunately, such schemes are known to be easily obtainable from any of the following: 
 post-quantum dense cryptosystems, post-quantum linearly homomorphic PKE, or post-quantum lossy PKE (see, e.g., \cite{TCC:CDMW09,FOCS:Wee10}). These schemes are black-box constructions and also work in the post-quantum setting (due to straightline security proofs).  


\begin{theorem}[\cite{TCC:CDMW09,FOCS:Wee10}]
    There exist two-round and black-box post-quantum OT schemes with indistinguishability security against honest receivers and simulation security against malicious QPT senders, based on (any of) post-quantum 
    dense cryptosystems, linearly homomorphic public key encryption, or lossy public key encryption. 

\end{theorem}

We will denote such a protocol by $\Gamma$, where the sender uses as inputs two strings $(s_0,s_1)$ and receiver uses as input a bit $r$. For technical reasons we will also refer in our construction to the receiver's private random tape in the protocol, which we denote as a string $\tau$ of length $t(\secpar)$ that is a polynomial in the security parameter.

\para{Post-Quantum Extractable Commitment:} We make use of the post-quantum parallelly extractable commitment scheme $\ExtCom$ with $\epsilon$-simulation (as per \Cref{def:epsilon-sim-ext-com:parallel}), which can be built in black-box from any post-quantum OWFs (see \Cref{lem:parallel-extcom:CCLY}).


\para{1-Many Weak Non-Malleable Commitment:} The final component we require is a constant-round, post-quantum, {\em 1-many weakly} non-malleable commitment scheme that is also parallel $\epsilon$-simulation extractable. 
To make our overall OT construction fully black-box, we also require this construction to be fully black-box. Fortunately, such a construction is available to us from \Cref{sec:BB-NMCom:one-many}. 

We note that the extractablity property mentioned above is easily observed due to the intrinsic execution of $\ExtCom$ in \Cref{prot:bbnmc:extcom} of \Cref{protocol:BB-NMCom}, for which we can invoke the associated $\SimExt_\ExtCom$ (it is easy to see that this step also commmits to the initial committed value with overwhelming probability). Indeed, such an observation was made in \cite{FOCS:LPY23}. We denote this protocol by $\ENMC$.


\subsection{Construction}

Our construction is given below in \Cref{prot:mal-OT}.


\begin{ProtocolBox}[label={prot:mal-OT}]{The parallel malicious secure OT scheme $\Prot$}

{\bf Parameters:} The security parameter is denoted by $\secpar$. Other parameters will be specified by polynomials in $\secpar$ unless otherwise specified. 

{\bf Receiver's input:} A bit $r \in \bits$ 

{\bf Sender's input:} Strings $s_0,s_1 \gets \bits^\ell$

The protocol proceeds as follows:

\begin{enumerate}

\item\label[Step]{prot:parOT:cointoss} {\bf Phase I: Random Tape Coin Tossing}
\begin{enumerate}
    \item\label[Step]{prot:parOT:cointoss:rec-sample} The receiver samples $2\secpar$ uniform random strings $(r_1^R,\tau_1^R),\dots,(r_{2\secpar}^R,\tau_{2\secpar}^R)$ of length $t+1$ corresponding to samples of the receiver's input bit and randomness for $\Gamma$.
    \item\label[Step]{prot:parOT:cointoss:rec-extcom} The receiver then runs $2\secpar$ parallel executions of $\ExtCom$ with the sender, where the receiver commits to the values $(r_1^R,\tau_1^R),\dots,(r_{2\secpar}^R,\tau_{2\secpar}^R)$ independently.

    \item\label[Step]{prot:parOT:cointoss:com-sample} The sender then samples $2\secpar$ uniform random strings $(r_1^S,\tau_1^S),\dots,(r_{2\secpar}^S,\tau_{2\secpar}^S)$ of its own and sends these back to the receiver. 
    \item\label[Step]{prot:parOT:cointoss:mix} The receiver sets $r_i= r_i^R \oplus r_i^S$ and $\tau_i = \tau_i^R \oplus \tau_i^S$ for $i \in [2\secpar]$. 
\end{enumerate}

\item\label[Step]{prot:parOT:baseOT}{\bf Phase II: Base OT Execution}
\begin{enumerate}
    \item\label[Step]{prot:parOT:baseOT:com-sample} The sender samples $2\secpar$ pairs of uniform random strings $(s_1^0,s_1^1),\dots,(s_{2\secpar}^0,s_{2\secpar}^1)$. 
    \item\label[Step]{prot:parOT:baseOT:OT} The sender and receiver then execute $2\secpar$ parallel executions of $\Gamma$. For the $i$th execution, the sender uses the inputs $(s_i^0,s_i^1)$ and the receiver uses input $r_i$ and randomness $\tau_i$ (for each $i \in [2\secpar]$). 
\end{enumerate}

\item\label[Step]{prot:parOT:cut-n-choose}{\bf Phase III: Cut \& Choose}
\begin{enumerate}
    \item\label[Step]{prot:parOT:cut-n-choose:enmc} The sender samples an uniform random string $q_S \pick \bits^\secpar$. It then runs an execution of $\ENMC$ with the receiver to commit to the string $q_S$. 
    \item\label[Step]{prot:parOT:cut-n-choose:resp} The receiver then samples an uniform string $q_R \pick \bits^\secpar$ and sends this back to the sender. 
    \item\label[Step]{prot:parOT:cut-n-choose:rec-sample} The sender provides the {\em decommitment} of its commitment to $q_S$ to the receiver.  
    
    \item\label[Step]{prot:parOT:cut-n-choose:mix}
     Both parties then compute $q = q_S \oplus q_R$. They also compute the description of a subset $Q \subset [2\secpar]$ of size $\secpar$ using the following correspondence: $Q= \Set{2i-q_i}_{i=1}^\secpar$ where $q_i$ is the $i$th bit of $q$. More descriptively, we imagine the previous $2\secpar$ executions of $\Gamma$ to be laid out in $\secpar$ pairs (of adjacent executions). Then $Q$ marks the subset of executions to be `opened', including the first or second execution in each of the $\secpar$ pairs depending on whether $q_i$ is $0$ or $1$ (so $Q$ has exactly one member in each pair). 
     
    
    \item\label[Step]{prot:parOT:cut-n-choose:decom} 
    For each $i \in Q$, the receiver {\em decommits} its phase I commitment to $(r_i^R,\tau_i^R)$. 
    \item \label[Step]{prot:parOT:cut-n-choose:consis-check}The sender then computes $(r_i,\tau_i)$ for all such sessions $i \in Q$. It next checks that $(r_i,\tau_i)$ is consistent with the receiver's messages in the $i$th parallel session of $\Gamma$ in phase II. The sender aborts if this is not the case. 
\end{enumerate}

\item\label[Step]{prot:parOT:combine}{\bf Phase IV: OT Combiner}
\begin{enumerate}
    \item\label[Step]{prot:parOT:combine:mask} For every $j \notin Q$, the receiver computes $\alpha_j = r \oplus r_j$ (recall $r$ is its original input bit) and sends the list $\Set{\alpha_j}_{j \notin Q}$ to the sender. 
    \item\label[Step]{prot:parOT:combine:send} The sender then computes $\sigma_0 =  s_0 \oplus (\bigoplus_{j \notin Q} s_j^{\alpha_j})$ and $\sigma_1 =  s_1 \oplus (\bigoplus_{j \notin Q} s_j^{1-\alpha_j})$. It sends $(\sigma_0,\sigma_1)$ to the receiver. 
    \item\label[Step]{prot:parOT:open} Finally, the receiver computes and outputs the string $s_r = \sigma_r \oplus (\bigoplus_{j \notin Q} s_j^{r_j})$. 
\end{enumerate}

\end{enumerate}
    
\end{ProtocolBox}

\begin{remark}
    We note here that we can use an $\ENMC$ scheme as described above in lieu of $\ExtCom$ in \Cref{prot:parOT:cointoss} without losing anything in terms of functionality (indeed, earlier works do exactly this). We use the extractable commitment $\ExtCom$ separately for modularity and to invite a clearer examination of how the separate components and assumptions are used in our construction and what role they play in security. 
\end{remark}

\subsection{Security}

The correctness of \Cref{prot:mal-OT} is straightforward. We turn to proving security for this scheme. This is captured formally by the following theorem. 

\begin{theorem}\label{thm:mal-OT}
    Let $\secpar$ denote the security parameter. Let $\Gamma$, $\ExtCom$, and $\ENMC$ be as described in \Cref{sec:parallelOT:building-blocks}. 
    Then, the $n$-fold parallel execution of \Cref{prot:mal-OT} (i.e., an execution among $n$ parties, where each pair of parties run two independent parallel instances with reversed roles of sender and receiver) realizes \Cref{def:eps-mal-par-OT}, for any polynomial $n=n(\secpar)$.   
\end{theorem}


\begin{AlgorithmBox}[label={fig:OT-Sim}]{The simulator $\Sim$ for the parallel malicious secure OT protocol}

\para{Description:} The simulator $\Sim$ enjoys black-box access to the possibly quantum adversary $\Adv$. The simulator may interact using quantum communication with the adversary occasionally (indeed, this is required to carry out the quantum analog of rewinding). It also takes in as input the security parameter as $1^\secpar$, and the error parameter as $1^{1/\epsilon}$. Thus its runtime is $\poly(\secpar,\epsilon^{-1})$. 

\para{Operation:} Note that the simulation is for an $n$-fold parallel execution of $\Prot$, where $\Adv$ may corrupt a single party in each session of $\Prot$. In each session, $\Sim$ will act according to which of the parties $\Adv$ corrupts. Its operation is detailed below. 

\subpara{Sender Corruption:} $\Sim$ takes over the receiver operation for the session. It acts as follows:  
\begin{itemize}
    \item {\bf Preamble:} It starts by sampling an uniform string $q \pick \bits^\secpar$ and computing the associated subset $Q \subset [2\secpar]$.
    \item {\bf Phase I:} For each $i \in Q$, $\Sim$ acts just as the honest receiver for this phase, committing to uniformly sampled $(r_i^R,\tau_i^R)$. However, for each $i \notin Q$, $\Sim$ commits to the strings $(0,0^t)$ (i.e., sets $r_i,\tau_i$ to be the zero strings of appropriate length).  
    \item {\bf Phase II:} For each $i \in Q$, $\Sim$ plays out the execution of $\Gamma$ honestly with the above sampled $(r_i^R,\tau_i^R)$. However, for each $i \notin Q$, it $\Sim$ runs the simulator $\Sim_\Gamma$ for $\Gamma$ to simulate the view of $\Adv$ in this phase, and also extracts $\Adv$'s inputs in this phase, namely the strings $\Set{(s_i^0,s_i^1)}_{i \notin Q}$. 
    \item {\bf Phase III:} $\Sim$ uses the weak parallel $\epsilon$-simulator-extractor for $\ENMC$ to extract the value $q_S$ from the sender-side non-malleable commitment. Note that such an extractor returns $\bot$ if even one of the component executions was declared invalid by the $\ENMC$ receiver. In such a scenario, the simulator calls for the ideal functionality $\mathcal{F}^n_{\text{OT}}$ to abort the {\em entire} execution, and halts its own operation, outputting the adversary's state. 
    Otherwise, it then sets $q_R = q \oplus q_S$ and sends this back to the sender. Next, when the corrupted sender opens its commitment to $q_S$, $\Sim$ opens its own commitments to $\Set{(r_i^R,\tau_i^R)}_{i \in Q}$. 
    \item {\bf Phase IV:} $\Sim$ sends uniformly chosen bits $\Set{\alpha_j}_{j \notin Q}$ to the corrupted sender, which sends strings $(\sigma_0,\sigma_1)$ back in turn. $\Sim$ then computes $s_0 =  \sigma_0 \oplus (\bigoplus_{j \notin Q} s_j^{\alpha_j})$ and $s_1 =  \sigma_1 \oplus (\bigoplus_{j \notin Q} s_j^{1-\alpha_j})$ (recall that it extracted $\Set{(s_i^0,s_i^1)}_{i \notin Q}$ earlier in {\bf Phase II}). $\Sim$ then sends the functionality $\Func^n_\algo{OT}$ for the relevant session with inputs $(s_0,s_1)$. This completes its execution for this session. 
\end{itemize}

\para{Receiver corruption:} $\Sim$ takes over the sender operation for this session. It proceeds as follows: 
\begin{itemize}
    \item {\bf Phase I:} For $i \in [2\secpar]$, $\Sim$ 
    \begin{itemize}
    \item
     uses the $\epsilon$-simulator-extractor for $\ExtCom$ to extract the values $(r_i^R,\tau_i^R)$ committed by the corrupted receiver, 
     \item
      samples uniformly random strings $(r_i,\tau_i)$, and 
      \item
      sets $r_i^S = r_i \oplus r_i^R$, $\tau_i^S = \tau_i \oplus \tau_i^R$ and sends these values to the corrupted receiver. 
    \end{itemize}
    \item {\bf Phase II:} $\Sim$ acts exactly as the honest sender does in this phase. 
    \item {\bf Phase III:} $\Sim$ acts exactly as the honest sender does in this phase.
    \item {\bf Phase IV:} $\Sim$ computes an index $j^* \notin Q$ such that the values $(r_{j^*},\tau_{j^*})$ are consistent with the messages the corrupted receiver sent in the $j^*$th execution of $\Gamma$ in {\bf Phase II} (note that to perform this check, $\Sim$ crucially needs the values it extracted in {\bf Phase I}). \underline{If there exists no such session, $\Sim$ outputs a special symbol $\algo{Fail}$ and halts immediately.} If the execution continues, it next receives the values $\Set{\alpha_j}_{j \notin Q}$ from the corrupted receiver, and computes $r=\alpha_{j^*} \oplus r_{j^*}$ and queries $\Func^n_\algo{OT}$ for the appropriate session with input $r$. Upon receiving a reply $s_r$ from $\Func^n_\algo{OT}$, $\Sim$ then computes the values $(\sigma_0,\sigma_1)$ as follows: 
    \begin{itemize}
        \item If $r=0$, then it sets $\sigma_0 =  s_0 \oplus (\bigoplus_{j \notin Q} s_j^{\alpha_j})$ and samples an uniform $\sigma_1 \pick \bits^\ell$. 
        \item If $r=1$, then it samples an uniform $\sigma_0 \pick \bits^\ell$ and sets $\sigma_1 =  s_1 \oplus (\bigoplus_{j \notin Q} s_j^{1-\alpha_j})$. 
    \end{itemize}
\end{itemize}
\end{AlgorithmBox}

\begin{proof}
    Our proof relies on a simulator for the $n$-fold parallel execution of the scheme $\Prot$, which we can denote by $\Prot^n$. Our simulator $\Sim$ for this protocol is described in \Cref{fig:OT-Sim}. It is easily seen that this simulator runs in (quantum) polynomial time. 

    Our proof, very broadly, rests primarily on two claims \Cref{malOT:lem:fail,lem:malOT:cond}. The first is that the special abort condition $\algo{Fail}$ specified in the description of $\Sim$ is triggered with at most negligible probability. The second claim says that in the event that $\Sim$ does manage to not trigger $\algo{Fail}$, it goes on to furnish a viable simulation of the execution $\Prot^n$. The logic of the proof is thus fairly straightforward, and we now turn to formalizing these claims and their respective justifications. 

    \begin{lemma}\label{malOT:lem:fail}
        Denote the adversary for $\Prot^n$ by $\Adv$. Recall that $\Sim = \Sim^\Adv$ is the simulator for $\Prot^n$ (i.e., the procedure that produces $\IDEAL_{\mathcal{F}^n_{\text{OT}}, \Sim^\Adv}(\secpar, \vb{x}, \rho_\secpar)$). Then we have $$\Pr[\algo{Fail} \gets \Sim] \leq \negl(\secpar)$$
    \end{lemma}

    \begin{proof}
        A similar proof already appears in \cite{FOCS:Wee10,STOC:Goyal11}. The proof is in fact almost identical, with two notable differences: (1) we need to take care of the $\epsilon$ simulation error; (2) while \cite{FOCS:Wee10} defines non-malleability w.r.t.\ extractability and \cite{STOC:Goyal11} defines non-malleability w.r.t.\ replacement, we do not need such adjustments since we have fully many-many non-malleability.

        Assume that there is in fact an adversary $\Adv$ for $\Prot^n$ that runs in polynomial time, and also is such that $\Pr[\algo{Fail} \gets \Sim] = \nu(\secpar)$ where $\nu(\cdot)$ is non-negligible. Assume that $\Sim$ outputs $\algo{Fail}$ in a specific session $k \in [n']$ (where $n'\coloneqq 2 \cdot\binom{n}{2}$). 
        
        We begin by examining exactly when $\Sim$ outputs $\algo{Fail}$ during simulation. From the description of $\Sim$, this happens only when the receiver is corrupted in session $k$, {\em and} there is {\em no} sub-index $j^* \in [2\secpar]$ which is (i) {\em not} opened in the cut and choose phase, and (ii) $\Adv$ behaves honestly in the session of $\Gamma$ corresponding to $j^*$. In fact, with further consideration, we can infer that the following must also happen: 
        \begin{itemize}
            \item For each pair in the $2\secpar$ executions, $\Adv$ behaves honestly in {\em exactly} one execution in the pair: if it did cheat in both, then it cannot succeed in opening either execution in the pair during the cut and choose execution; if indeed it did not cheat in either, we have nothing to worry about and can set $j^*$ to be the index of the unopened execution. Consequently, there is an {\em unique} bit for every pair of (sub-)executions of $\Gamma$, and therefore a {\em unique} string $q^* \in \bits^\secpar$ across the $2\secpar$ executions of $\Gamma$, which when picked in the cut and choose phase will allow $\Adv$ to cheat. 
            \item Further, $\Adv$ must ensure in \Cref{prot:parOT:cut-n-choose} that $q^*$ is indeed the result, i.e., it sends $q_R$ such that $q_S \oplus q_R = q^*$. 
        \end{itemize}
        
        We will use this observation to attack the non-malleability of $\ENMC$ and eventually derive a contradiction. To begin, we consider certain hybrids relating to \Cref{prot:parOT:cut-n-choose} of $\Prot^n$. We define the view or output of a hybrid to be the adversary's view of the $\Prot^n$ of the hybrid execution (therefore, corresponding to the output variable $\IDEAL$ for the simulator).

        \para{Hybrid $H_0$:} This is simply the simulated execution of $\Prot^n$. For clearer contrast with subsequent hybrids, we make note of the index $k$ identified as above. Further, we use   $K' \subset [n']$ to denote the {\em indices of the sessions in which $\Adv$ corrupts the sender} (note that $k$ and $K'$ are random variables). In particular, we emphasize that $H_0$ aborts the entire execution if any of the $\ENMC$ executions for  \Cref{prot:parOT:cut-n-choose:enmc} in the OT sessions in $K'$ are not accepted by the corresponding receiver (see {\bf Phase-III} for sender corruption in \Cref{fig:OT-Sim}). Further, the hybrid also records the values $\Set{q_S^{k'}}_{k' \in K'}$ that it extracts in \Cref{prot:parOT:cut-n-choose:enmc}. 

        \begin{remark}
            In $H_0$, we have that the following condition holds by the above analysis: with probability $\nu$, the receiver picks $q^k_R$ such that $q^k_S \oplus q^k_R = q^*$.
        \end{remark}

        \para{Hybrid $H_1$:} This hybrid only differs from $H_0$ in the following operation: instead of using the simulator-extractor for $\ENMC$ to extract the values $\Set{q_S^{k'}}_{k' \in K'}$ from \Cref{prot:parOT:cut-n-choose:enmc}, it instead extracts them by running a {\em brute force attack} on the transcript of the \Cref{prot:parOT:cut-n-choose:enmc} $\ENMC$ execution (recall that the hybrid then requires these values to finish its execution of \Cref{prot:parOT:combine}). If however any of the $\ENMC$ sessions in $K'$ corresponding to \Cref{prot:parOT:cut-n-choose:enmc} are not completed successfully, the extracted values for {\em all} $\ENMC$ sessions in $K'$ are set to be $\bot$. 
        
        Additionally, we make the following syntactic change for \Cref{prot:parOT:cut-n-choose:resp} in session $k$: instead of the hybrid sampling $q^k_S$ directly, it instead samples an uniform value ${q'}_S^k \pick \bits^\secpar$, and then sets $q_S^k = {q'}_S^k$.


        \subpara{$\Output(H_0) \sind_\epsilon \Output(H_1)$:} A cursory examination reveals that the actual view of the adversary remains identical in both hybrids till \Cref{prot:parOT:cut-n-choose}. The extracted values $\Set{q_S^{k'}}_{k' \in K'}$, and the resulting adversary state, are $\epsilon$-indistinguishable in $H_0$ and $H_1$ by the (weak-parallel) $\epsilon$-simulation-extraction guarantees of $\ENMC$ (as per \Cref{def:epsilon-sim-ext-com:parallel}), and  \Cref{prot:parOT:combine} is otherwise unaffected (crucially, note that both hybrids suspend the entire execution whenever any $\ENMC$ within OT sessions in $K'$ is not accepted by the corresponding receiver, and the `extracted' values are set to $\bot$ in such cases). Indistinguishability of the entire view of $\Adv$ follows. 

        \begin{remark}
            We note from the above indistinguishability  condition that in particular, the receiver in session $k$ continues to pick $q^k_R$ such that $q^k_S \oplus q^k_R = q^*$ with probability at least $\nu-\epsilon$. 
        \end{remark}

        \para{Hybrid $H_2$:} This hybrid samples an uniform ${q'}^k_S$ as in $H_1$, but sets $q_S^k = 0^\secpar$ instead of ${q'}_S^k$. All other steps remain the same as in $H_1$. We note here that the operation of $H_2$ is efficient {\em but} for the brute force extraction from \Cref{prot:parOT:cut-n-choose:enmc}. 

        \subpara{$\Output(H_1) \cind \Output(H_2)$:} At first glance it may appear that this should be guaranteed directly by the hiding guarantee of the $\ENMC$ in \Cref{prot:parOT:cut-n-choose:enmc}. This is however not the case: the reason is that while we have so far focused only on the session $k$, in truth the hybrid is also following the simulation strategy for corrupted {\em senders} in the sessions in $K'$. Note that in these sessions, the simulator itself is extracting the sender side $\ENMC$ values {\em using brute force} (starting from $H_1$)! This happens in parallel to the change we would make in $H_2$, and hence we cannot appeal to the computational hiding property. 
        
        What we can do however is appeal to the {\em weak 1-many non-malleability} property of $\ENMC$. As pointed out, the executions of $\ENMC$ resemble that of a 1-many man-in-the-middle attack, where the one in session $k$ is the `left' session, and those in the sessions in $K'$ form the `right' sessions. Note that such a reduction, argued naively, will still inherit the brute-force extraction issue (whereas non-malleability is still a computational property). However, one can maneuver around this difficulty using the design of the non-malleability challenge: the extracted values $\Set{q_S^{k'}}_{k' \in K'}$ (required by the hybrid to complete the execution and manufacture its output) are not extracted by the adversary itself --- instead, the challenger does so itself and then presents them to the distinguisher. Then, if all the right sessions are successfully completed, we allow the distinguisher (using a standard nonuniformity argument) to `resurrect' the hybrid and complete the execution. Since the only brute force step by the hybrid is actually carried out by the challenger, this makes the adversary-distinguisher pair efficient, giving us a valid reduction to weak 1-many non-malleability.      

        Note that {\em weak} 1-many non-malleability indeed suffices in this setting, because of the deliberate construction of our hybrids. In more detail, observe that at the end of \Cref{prot:parOT:cut-n-choose:enmc}, $H_1$ has the corresponding information to exactly reconstruct $\Gamma_{\{d_j\}_{j=1}^{k}}(\msf{mim}[k]^{\mcal{M}_\secpar}_{\langle C, R \rangle}({q'}_S^k, \rho_\secpar))$, whereas $H_2$ has the information to exactly reconstruct $\Gamma_{\{d_j\}_{j=1}^{k}}(\msf{mim}[k]^{\mcal{M}_\secpar}_{\langle C, R \rangle}(0^\secpar, \rho_\secpar))$. In other words, any cases where the adversary chooses to selectively not complete some of the right $\ENMC$ sessions are discarded by the hybrids --- and so only cases where all right sessions are successfully completed are considered. This ensures that we can then successfully rely on weak one-may non-malleability to argue similarity.    
        
        We proceed to formally reduce this claim to the weak 1-many nonmalleability of $\ENMC$ as defined in \Cref{def:NMCom:weak:pq}. In more detail,  assume there is a distinguisher $D$ that can distinguish (with non-negligible advantage $\kappa(\secpar)$) between the outputs of $H_2$ and $H_1$. 
        We describe a valid man-in-the-middle adversary and distinguisher pair $(\tilde{A},\tilde{D})$ for the weak 1-many non-malleability game. We make use of the original adversary $\Adv$ and the assumed distinguisher $D$. The description of $\tilde{A}$ and $\tilde{D}$ are given below. 

        The adversary $\tilde{\Adv}$ works as follows: 
        \begin{itemize}
            \item Internally, it runs the hybrid $H_2$ with $\Adv$ embedded in the execution. It continues this up to the start of \Cref{prot:parOT:cut-n-choose}. 
            \item Next, it externally begins participating in a 1-many challenge for $\ENMC$, with challenge messages $0^\secpar$ and ${q'}_S^k$ (uniformly sampling the latter). 
            \item It forwards the challenger's sender (or `left-side') messages as the \Cref{prot:parOT:cut-n-choose:enmc} sender messages in session $k$ of $\Prot$, and forwards out the receiver's session $k$ replies out to the challenger as its own left-side receiver messages. Similarly, it cross-forwards the $\ENMC$ interactions for \Cref{prot:parOT:cut-n-choose:enmc} in the sessions in $K'$ to act as the `right-side' $\ENMC$ interactions in the external challenge. 
            \item At the end of \Cref{prot:parOT:cut-n-choose:enmc}, it records its view and the adversary's state as its output, and halts. Note that the adversary' state may be quantum, but this is okay as the MIM adversary for post-quantum non-malleability is allowed to output quantum information. 
        \end{itemize}

        The distinguisher $\tilde{D}$ works as follows: 
        \begin{itemize}
            \item $\tilde{D}$ receives the output of $\tilde{A}$ from the challenger, along with the left-side committed values ${q'}_S^k$. It then reconstructs the operation of the hybrid up to the start of \Cref{prot:parOT:combine} (along with the appropriate state of $\Adv$ at this stage).  
            \item From the 1-many non-malleability challenger, it then receives the values committed by $\tilde{\Adv}$ in the right side sessions, namely $\Set{q_S^{k'}}_{k' \in K'}$. 
            \item It then completes the execution of the hybrid and feeds the resulting hybrid output to $D$. It then outputs whatever $D$ does. 
        \end{itemize}

        It is easy to see that the operation of $\tilde{A}$ and $\tilde{D}$ follow exactly the execution of the hybrid $H_1$ if the $\ENMC$ challenger commits to $q'^k_S$ in the left side commitment, and conversely these follow exactly the execution of $H_2$ when the $\ENMC$ challenger commits to $0^\secpar$. Consequently, within $\tilde{D}$, the view fed to the distinguisher $D$ comes either from $H_1$ or $H_2$. Therefore $\tilde{D}$ succeeds in winning the weak 1-many nonmalleability challenge whenever $D$ distinguishes successfully within these hybrids. Thus $(\tilde{A},\tilde{D})$ win the weak 1-many non-malleability challenge with non-negligible probability $\kappa$, which contradicts the security of $\ENMC$. We conclude that there is no such efficient distinguisher $D$ that manages to distinguish between $H_1$ and $H_2$ with non-negligible probability. 

        Finally, in the hybrid $H_2$, we obtain a contradiction. Set $\epsilon = \nu/2$. From the above, we can conclude that even in $H_2$, we have that the receiver in session $k$ continues to pick $q^k_R$ such that $q'^k_S \oplus q^k_R = q^*$ with non-negligible probability $\nu/2$. Recall that $q'^k_S$ is sampled uniformly by $H_2$ and not used in its internal execution of $\Prot^n$. Therefore, $q'^k_S$ is not seen by $\Adv$. This is a clear contradiction, since an uniformly sampled string of $\secpar$ bits that is absent from the view of $\Adv$ has entropy $\secpar$ from $\Adv$'s point of view; and so even a quantum (or even unbounded) machine can predict this string with probability $\frac{1}{2^\secpar}$, which is negligible.

        This concludes the proof of \Cref{malOT:lem:fail}.

    \end{proof}


We turn now to the second claim. 

\begin{lemma}\label{lem:malOT:cond}
    Conditioned on the event $E$ where $\Sim$ does {\em not} output $\algo{Fail}$, $\Sim$ produces a valid simulation of $\Prot^n$ with respect to $\Adv$. Namely, given that $E$ occurs, we have 
    $$\REAL_{\Prot^n, \Adv}(\secpar, \vb{x}, \rho_\secpar) \cind_\epsilon \IDEAL_{\mathcal{F}^n_{\text{OT}}, \Sim^\Adv}(\secpar, \vb{x}, \rho_\secpar)$$
\end{lemma}

\begin{proof}
    This is essentially the same argument that is used in previous work \cite{FOCS:Wee10,STOC:IKLP06}. We refer the reader to these for a full proof, and include a sketch of the proof for the sake of completeness. 
    
    The case for simulating for a corrupted sender is straightforward, relying directly on the malicious sender security of $\Gamma$. The only thing of note here is that since the simulator relies on the simulator-extractor for $\ENMC$, it inherits an $\epsilon$ simulation error for the portion of the view from \Cref{prot:parOT:cut-n-choose:enmc} onwards. This is reflected in the final simulation guarantee stated in the lemma.  
    
    We focus instead on the corrupted receiver case. Here we observe that if $\Sim$ does not output $\algo{Fail}$, then everything in the first three phases is essentially identical to how an honest sender acts, and it is only in the values $(\sigma_0,\sigma_1)$ that the simulator's distribution changes. A distinguisher for the real and simulated view can then be used to attack the honest receiver security of $\Gamma$. 

    The attack is somewhat subtle, but it essentially uses the following observation: the `hidden' value $\sigma_{1-r}$ has the `correct' distribution in the real execution, while in the simulated execution it is uniformly sampled. Thus a distinguisher for the real and simulated executions will be biased towards identifying an execution as `real' when the hidden value has the correct distribution. The actual reduction makes an initial guess as to the hidden value and performs a sort of retroactive check, and uses the distinguisher's output cleverly to gain advantage in the honest receiver security game for $\Gamma$. 

    Of course, to make this reduction meaningful, care has to be taken to make sure that the internal $\Gamma$ sub-execution within $\Prot$ is indeed honest. This is essentially the function of the session $j^*$ tracked by the simulator: for this session of $\Gamma$, we are guaranteed that the adversary uses honest inputs - as the simulator extracts all the receiver inputs initially and aborts if this is not the case. 
    
    Note that again we have to account for the $\epsilon$-simulation guarantee from the extraction step in phase I, but this can be handled by a standard `funneling' argument (i.e., start by assuming an distinguisher with a certain distinguishing advantage, and set $\epsilon$ to be significantly smaller to still derive a contradiction). 
     
This concludes the proof of \Cref{lem:malOT:cond}.

\end{proof}

Finally, we tie together these claims to obtain the stated result. Since $\algo{Fail}$ is a simulator specific abort message, it is very easy to distinguish real and simulated executions whenever $\algo{Fail}$ is output. Now by \Cref{lem:malOT:cond}, the maximum possible distinguishing advantage for a computationally bounded distinguisher (which we denoted by $\Delta_c$) is $\epsilon$ whenever $E$ occurs. Formally, we define $\Delta_c(X,Y) := \text{sup}_{D \in\text{PPT}}|\Pr[D(X)=1] - \Pr[D(Y)=1]|$. For brevity we will denote $\REAL_{\Prot^n, \Adv}(\secpar, \vb{x}, \rho_\secpar)$ by $\REAL$ and $\IDEAL_{\mathcal{F}^n_{\text{OT}}, \Sim^\Adv}(\secpar, \vb{x}, \rho_\secpar)$ by $\IDEAL$ below. 

We begin with the following decomposition 
\begin{align*}
    \Delta_c(\REAL,\IDEAL) = &[\Delta_c(\REAL,\IDEAL)|E]\cdot\Pr[E] \\ & + [\Delta_c(\REAL,\IDEAL)|\bar{E}]\cdot\Pr[\bar{E}]
\end{align*}
By \Cref{malOT:lem:fail}, we can write the RHS as 
\begin{align*}
     = &[\Delta_{D^*}(\REAL,\IDEAL)|E]\cdot(1-\negl(\secpar)) \\ & + [\Delta_{D^*}(\REAL,\IDEAL)|\bar{E}]\cdot\negl(\secpar),
\end{align*}
which is of course simply $$ \leq \epsilon\cdot(1-\negl(\secpar) ) + 1\cdot\negl(\secpar)$$ Which yields $$\Delta_c(\REAL,\IDEAL)\leq \epsilon' = \epsilon + \negl(\secpar)$$ Readjusting $\epsilon \rightarrow \epsilon/2$ gives us the stated result.

This concludes the proof of \Cref{thm:mal-OT}.

\end{proof}

%% file: sections/MPC.tex
\section{Post-Quantum $\epsilon$-Simulatable MPC}
\label{sec:MPC}

In this section we will describe and show security of a black-box, constant-round $\epsilon$-MPC protocol. In fact we have gathered essentially all the ingredients needed for this task. The sole remaining component is the black-box compiler given in \cite{C:IshPraSah08}. Their protocol is a constant-round black-box MPC protocol, albeit with {\em UC security} and additionally assuming {\em ideal} OT channels. Here we will argue that this protocol when initialized with our malicious parallel OT protocol\footnote{Indeed, this observation has been employed in the classical setting \cite{FOCS:Wee10,STOC:Goyal11}.}, will give us an MPC protocol with all the desired properties. We capture this in the following lemma. 

\begin{lemma}
    The MPC protocol described in \cite{C:IshPraSah08}, instantiated with the OT construction given in \Cref{prot:mal-OT}  (in lieu of an ideal OT functionality) is a \emph{black-box}, {\em constant-round}, \emph{post-quantum} $\epsilon$-simulatable MPC protocol as defined in \Cref{def:mpc}. 
\end{lemma}

\begin{proof}
    The construction simply involves instantiating the MPC protocol from \cite{C:IshPraSah08} using our OT scheme, as is already stated. We refer to their protocol as IPS for convenience. 
    
    We begin with a brief overview of the IPS MPC protocol. This involves composing {\em two} MPC protocols (titled the {\em inner} and {\em outer} protocols) in a specific fashion. The outer protocol uses the so-called {\em client-server} model, which involves parties called servers that have no input of their own but carry out the majority of the computation in the protocol. The key stratagem devised in IPS is to {\em emulate} the function of these servers distributedly using the inner MPC protocol. To ensure honesty, the protocol uses a mechanism introduced in IPS known as {\em watchlists}, which ensure that each party is able to monitor some emulated servers. 

    Thus, in the running of the IPS protocol, there are various OT calls that are of two kinds. The first kind is used to initialize the watchlist mechanism, and this can be performed at the start of the protocol. The second kind is in the operation of the inner OT protocol that is used to emulate the servers (the inner protocol is in the OT-hybrid model, and needs to make calls to the ideal OT functionality). 

    We make the following two observations about the IPS protocol. These are easily verifiable from the descriptions present in \cite{C:IshPraSah08}. The first is that the watchlist setup can be initialized with an $n$-party OT functionality (this is observed in their work), i.e., $2\cdot\binom{n}{2}$ OT calls in totally where each pair of parties run two OT calls with reversed role of sender and receiver. So we can use an $n$-fold parallel execution \Cref{prot:mal-OT} in the beginning that suffices to setup the watchlists. The second is a {\em randomized OT trick} that can be used to `prepone' the OT executions required by the inner protocol (this is also observed in their work). This modification is also needed for security. The idea is to basically initially perform $n$-fold OT executions with {\em random} values for the senders and receivers. Subsequently, the sender can send appropriately offset values (that encode OT inputs of its choice) to the receiver and the latter can recover its intended message from this. 

    This is to say that using this trick, the inner OT calls can also be pushed to the beginning of the protocol where we perform an $n$-fold parallel execution of \Cref{prot:mal-OT} (for sufficiently long sender inputs). Therefore, we can complete an execution of the IPS protocol by beginning with two $n$-fold parallel (randomized) OT executions, for the watchlist and for the inner MPC protocol respectively. Subsequently, we proceed with the IPS protocol, setting up the watchlists and then executing the composed MPC protocol. Everytime the inner protocol would make an OT call, we use the random OT transformation and consume a predefined portion of the initial parallel randomized OT call to perform the actual OT interaction in the protocol. \xiao{check this part with rohit}
    
    
    Next we will argue why this achieves the desired security guarantee - security is already somewhat apparent and straightforward to establish, and we limit ourselves to addressing the more prominent concerns in this regard. We treat these in turn. 

    \para{Constant-Round:} The first thing to determine is simply whether the composed protocol is still constant round. While the total number of atomic OT calls made in the IPS protocol does depend on the number of parties (and hence grows polynomially with $\secpar$), these can be \emph{batched} into a single parallel OT execution and shunted to the start of the protocol as described above. Now the IPS protocol itself is constant round (this includes the interactions made due to the randomized OT trick). In turn, our OT protocol from \Cref{prot:mal-OT} also runs in constant rounds. The resulting protocol is therefore constant rounds. Indeed, this exact pipeline has been used in previous work on black-box MPC protocols to get protocols with {\em constant round overhead} (over the parallel OT part) in the classical standalone setting (see \cite{FOCS:Wee10,STOC:Goyal11}). 

    \para{Security:} As pointed out, the security of the IPS protocol when initialized with a parallel OT protocol has been noted and employed in previous work (\cite{FOCS:Wee10,STOC:Goyal11}). Our setting however presents two new challenges that are not present in the more standard setting, and we tackle them in turn. 

    \subpara{Post-Quantum security:} A simple examination of the IPS security proof and that of our OT protocol reveals that both of these are \emph{black-box} and also \emph{quantum compatible} - namely, they enjoy straightline simulation and are not reliant on classical rewinding. 
     The combined security proof for the composed MPC protocol inherits these properties. 

    \subpara{Sequential composition:} A detail we have elided so far is that our OT is limited to $\varepsilon$-simulatability. This can affect the hybrids where we \emph{sequentially} simulate various parallel OT executions in the protocol. 
     Fortunately, this exact kind of post-quantum sequential composition guarantee for $\varepsilon$-simulation has been shown in the work of \cite[Section 7.2]{C:CCLY22}. 
     
\end{proof}
 

%% file: sections/ExtCom-from-OT.tex
\section{Post-Quantum Equivocal Commitments} \takashi{I separated this section because it would be confusing if we talk about equivocal commitments after introducing batch commitments since equivocality is only defined for stand-alone commitments and not for batch commitments.}
Here, we define equivocal commitments and 
construct an $\omega(1)$-round equivocal commitment scheme from OWFs. Looking ahead, this is used as a building block for constructing fully-simulatable (rather than $\epsilon$-simulatable) extractable (batch) commitments in \Cref{sec:ExtCom-from-OT}.  

\subsection{Definition}
The definition of equivocality is given below.  
We stress that we require the simulation error to be negligible since this is used for achieving extractability with negligible simulation errors. 
\begin{definition}[PQ-EqCom]\label{def:eqcom}
A post-quantum commitment scheme $\langle C, R\rangle$ (as per \Cref{def:com}) is {\em equivocal} if there exists a QPT algorithm $\SimEq=(\SimEq_0,\SimEq_1)$ (called the simulation equivocator) such that for any non-uniform QPT $R^*(\rho)$ and any polynomial $\ell(\cdot)$, 
\begin{align*}
   &\left\{
   (\ST_{R^*},\decom):
   \begin{array}{ll}
   (\ST_{R^*},\ST_{\SimEq})\gets \SimEq_0^{R^*(\rho)}(1^\secpar)\\
   \decom\gets \SimEq_1(\ST_{\SimEq},m)
   \end{array}
   \right\}_{\secpar\in \mathbb{N},m\in \bits^{\ell(\secpar)}}\\
   \cind
    &\left\{
    (\ST_{R^*},\decom):
    \begin{array}{ll}
    (\tau,\ST_{C},\ST_{R^*})\gets \langle C(m),R^*(\rho) \rangle(1^\secpar)\\
    \decom \gets C(\ST_C)
    \end{array}
    \right\}_{\secpar\in \mathbb{N},m\in \bits^{\ell(\secpar)}}
\end{align*}
\end{definition} 

The main theorem we prove in this section is the following: 
\begin{theorem}\label{thm:eqcom}
Assuming the existence of post-quantum OWFs, there is a black-box construction of $\omega(1)$-round equivocal commitment schemes.\footnote{Formally speaking, this means that for any time-constructible function $r(\cdot)=\omega(\secpar)$, 
there is a $r(\secpar)$-round equivocal commitment scheme. We use a similar convention throughout the paper.}  \takashi{Are you okay with this convention?}
\end{theorem}

\subsection{Construction with Noticeable Binding Error}
For the ease of presentation, we start by constructing a \emph{constant-round} equivocal commitment scheme $\wEqCom$ that has a noticeable binding error. 
Later, we argue that $\omega(1)$-times sequential repetition of $\wEqCom$ achieves negligible soundness error while preserving the equivocality. The construction is based on the ideas of \cite{STOC:Kilian88,FOCS:Kilian94}, which have been used in many later works, e.g., \cite{TCC:PasWee09,C:BCKM21b}.  
 The scheme $\wEqCom$ is described in \Cref{protocol:wEqCom} where it
 makes black-box use of a constant-round statistically-binding and computationally-hiding commitment scheme $\Com$ (e.g., Naor's commitment).  

\begin{ProtocolBox}[label={protocol:wEqCom}]{Equivocal Commitments with Noticeable Soundness Error $\wEqCom$}
{\bf Parameters:} Let $k=\Theta(\log \secpar)$ be a positive integer.

{\bf Inputs:} Both parties receive $\secpar$ as the common input. The committer in addition gets a string $m\in \bits^{\ell(\secpar)}$ as its private input where $\ell(\cdot)$ is a polynomial.
For each $i\in [\ell]$, $m_i$ denotes the $i$-th bit of $m$.

\para{Commit Stage:} 
\begin{enumerate}
\item 
$C$ picks uniformly random bits $r_{i,j}$ 
and defines
\begin{align*}
\begin{pmatrix*}[r]
s^{00}_{i,j} & ~~~s^{01}_{i,j} \\
s^{10}_{i,j} & ~~~s^{11}_{i,j} \\
\end{pmatrix*}
=
\begin{pmatrix}
r_{i,j} & ~~~m_i\oplus r_{i,j} \\
r_{i,j} & ~~~m_i\oplus r_{i,j} \\
\end{pmatrix}
\end{align*}
for 
$i\in [\ell]$ and $j\in [k]$. 
\item \label[Step]{wEqCom:Com}
$C$ commits to $\{s_{i,j}^{ab}\}_{i\in [\ell],j\in[k],(a,b)\in\bits^2}$ using $\Com$ in a bit-by-bit manner in parallel.  
Let $\{\tau_{i,j}^{ab}\}_{i\in [\ell],j\in[k],(a,b)\in\bits^2}$ be the corresponding transcripts. 
 \item $R$ randomly picks uniformly random bits $c_j$ for $j\in [k]$ and sends them to $C$.
 \item $C$ reveals $s_{i,j}^{0c_j}$ and $s_{i,j}^{1c_j}$ along with the corresponding decommit information w.r.t. $\Com$ in \Cref{wEqCom:Com} to $R$ for $i\in [\ell]$ and $j\in [k]$. 
\item $R$ accepts if all the decommit information are valid and $s_{i,j}^{0c_j}=s_{i,j}^{1c_j}$ for all  $i\in [\ell]$ and $j\in [k]$ and otherwise rejects. 
\end{enumerate}

{\bf Decommit Stage:} 
\begin{enumerate}
\item $C$ picks uniformly random bits $d_{i,j}$ and reveals $m$ and 
$s_{i,j}^{d_{i,j}(1-c_j)}$ along with the corresponding decommit information w.r.t. $\Com$ in \Cref{wEqCom:Com} of Commit Stage to $R$ for $i\in [\ell]$ and $j\in [k]$. 
\item  $R$ accepts if all the decommit information are valid and $s_{i,j}^{d_{i,j}(1-c_j)}=m_i\oplus s_{i,j}^{d_{i,j}c_j}$ for all  $i\in [\ell]$ and $j\in [k]$ and otherwise rejects. (Note that $s_{i,j}^{d_{i,j}c_j}$ is already revealed in the commit stage.)
\end{enumerate}
\end{ProtocolBox}
\begin{theorem}
The scheme $\wEqCom$ (\Cref{protocol:wEqCom}) is constant-round and satisfies computational hiding, equivocality (as per \Cref{def:eqcom}), and $(2^{-k}+\negl(\secpar))$-statistical binding, which is defined similarly to statistical binding (as per \Cref{def:com}) except that we only require the malicious committer's winning probability to be at most $2^{-k}+\negl(\secpar)$ instead of $\negl(\secpar)$. 
\end{theorem}
\begin{proof}
It is clear from the description that it is constant-round. 
Since computatioanl hiding immediately follows from equivocality, we prove $(2^{-k}+\negl(\secpar))$-statistical binding and equivocality below.

\para{$(2^{-k}+\negl(\secpar))$-Statistical Binding.}
In an execution of the protocol between unbounded-time malicious committer $C^*$ and honest receiver $R$,  
let $\Bad$ be the event that any of $\tau_{i,j}^{ab}$ can be decommitted to more than one messages.  By statistical binding of $\Com$, $\Bad$ occurs with a negligible probability. Below, we assume that $\Bad$ does not occur. 
For $i\in [\ell]$, let $\inconsistent_i$ be the event that for all $j\in [k]$, there exists $b_{i,j}$ such that $\val(\tau_{i,j}^{0b_{i,j}})\neq \val(\tau_{i,j}^{1b_{i,j}})$. 
When $\inconsistent_i$ occurs, then $C^*$ can pass the verification by $R$ in the commit stage only if $c_j=1-b_{i,j}$ for all $j\in [k]$. Since $c_j$ is uniformly random, this occurs with probability $2^{-k}$. Thus, whenever $C^*$ passes the verification in the commit stage, then neither of $\Bad$ or $\inconsistent_i$ for any $i\in [k]$ occurs except for probability $2^{-k}+\negl(\secpar)$. 
When neither of $\Bad$ or $\inconsistent_i$ occurs, there is $j^*_i$ such that $\val(\tau_{i,j^*_i}^{00})= \val(\tau_{i,j^*_i}^{10})$ and $\val(\tau_{i,j^*_i}^{01})= \val(\tau_{i,j^*_i}^{11})$, in which case the $i$-th bit can be only decommitted to $\val(\tau_{i,j^*_i}^{00})\oplus \val(\tau_{i,j^*_i}^{01})=\val(\tau_{i,j^*_i}^{10})\oplus \val(\tau_{i,j^*_i}^{11})$.\footnote{We define $\bot\oplus \beta = \bot$ for $\beta\in\{0,1,\bot\}$.} 
Thus, except for probability $2^{-k}+\negl(\secpar)$, all the bits can be decommitted to either of $0$ or $1$.   
This means that it satisfies $(2^{-k}+\negl(\secpar))$-statistical binding. 

\para{Equivocality.}
The proof strategy is similar to that in the security proof of the equivocality compiler of \cite{C:BCKM21b}, which in turn is based on quantum zero-knowledge proofs by Watrous \cite{SIAM:Watrous09}. 
We first construct a weaker simulation equivocator that guesses the challenges and works only when the guess is correct. 
Then we compile it to the full-fledged simulation equivocator (as per \Cref{def:eqcom}) by using Watrous' rewinding lemma (\Cref{lem:Watrous}).

First, we consider the following algorithm $\sfQ$:

\smallskip
\noindent
$\sfQ^{R^*(\rho)}(1^\secpar)$:  
\begin{enumerate}
\item Randomly pick bits $c'_j$ for $j\in[k]$ and $e_{i,j}$ and $r_{i,j}$  for $i\in[\ell]$ and $j\in [k]$.
\item 
Define
\begin{align*}
\begin{pmatrix*}[r]
s^{0c'_j}_{i,j} & ~~~s^{0(1-c'_j)}_{i,j} \\
s^{1c'_j}_{i,j} & ~~~s^{1(1-c'_j)}_{i,j} \\
\end{pmatrix*}
=
\begin{pmatrix}
r_{i,j} & ~~~e_{i,j} \\
r_{i,j} & ~~~1-e_{i,j} \\
\end{pmatrix}
\end{align*}
for 
$i\in [\ell]$ and $j\in [k]$. 
\item 
Commit to $\{s_{i,j}^{ab}\}_{i\in [\ell],j\in[k],(a,b)\in\bits^2}$ using $\Com$ in a bit-by-bit manner in parallel.  
Let $\{\tau_{i,j}^{ab}\}_{i\in [\ell],j\in[k],(a,b)\in\bits^2}$ be the corresponding transcripts. 
Send them to the malicious receiver $R^*$. 
 \item Receive bits $\{c_j\}_{j\in [k]}$ from $R^*$.  
 \item If $c_j\ne c'_j$ for some $j\in[k]$, output $\beta=1$ and immediately halt. 
 Otherwise, proceed to the next step. 
 \item Reveal $s_{i,j}^{0c_j}=s_{i,j}^{1c_j}=r_{i,j}$ along with the corresponding decommit information w.r.t. $\Com$ to $R$ for $i\in [\ell]$ and $j\in [k]$. 
 \item Define $\decom_{i,j}^{\eta}$ to be the decommit information for $\tau_{i,j}^{(e_{i,j}\oplus \eta)(1-c_j)}$ w.r.t. $\Com$ for $i\in [\ell]$, $j\in [k]$, and $\eta\in\bits$.  Note that the committed message in $\tau_{i,j}^{(e_{i,j}\oplus \eta)(1-c_j)}$ is $\eta$. 
 \item Output a quantum state $\sigma$ that consists of the final state $\ST_{R^*}$ of $R^*$ and $\{\decom_{i,j}^{\eta}\}_{i\in[\ell],j\in[k],\eta\in\bits}$ 
 along with a bit $\beta=0$. 
\end{enumerate}
Let $\sfQ_\rho^{0}$ be the distribution of $\sigma$ output by $\sfQ^{R^*(\rho)}(1^\secpar)$ conditioned on that $\beta=0$. Let $p(\rho)$ be the probability that  $\sfQ^{R^*(\rho)}(1^\secpar)$ returns $\beta=0$. By  computational hiding of $\Com$, it is easy to see that we have 
$$
|p(\rho)-2^{-k}|\le \negl(\secpar)
$$
for any quantum advice $\rho$. 
Thus, by applying Watrous' rewinding lemma (\Cref{lem:Watrous}) with $p_0=2^{-k}-\negl(\secpar)$, $q=2^{-k}$, $\gamma=\negl(\secpar)$, and $T=\lfloor
\frac{\log (1/\gamma)}{4p_0(1-p_0)}\rfloor$, we obtain a QPT algorithm $\tilde{\sfQ}$ that makes black-box use of $R^*(\rho)$ such that 
\begin{align} \label{eq:TD_Q_and_R}
\TD(\sfQ_\rho^{0},\tilde{\sfQ}^{R^*(\rho)}(1^\secpar))\le 4\sqrt{\gamma}\frac{\log (1/\gamma)}{p_0(1-p_0)}=\negl(\secpar)
\end{align}
where we used $k=\Theta(\log \secpar)$ and thus $p_0=2^{-k}-\negl(\secpar)=1/\poly(\secpar)$. 
We remark that $\tilde{\sfQ}$ plays the role of $\sfR$ in \Cref{lem:Watrous}. We changed the notation to avoid confusion with the malicious receiver $R^*$. We also remark that $\tilde{\sfQ}$ makes black-box use of $R^*(\rho)$ since it makes black-box use of $\sfQ$, which in turn makes black-box use of $R^*(\rho)$.

We are now ready to describe the simulation equivocator $\SimEq=(\SimEq_0,\SimEq_1)$:

\smallskip
\noindent
$\SimEq_0^{R^*(\rho)}(1^\secpar)$:  
\begin{enumerate}
\item Run $\tilde{\sfQ}^{R^*(\rho)}(1^\secpar)$ to obtain $\ST_{R^*}$ and $\{\decom_{i,j}^{\eta}\}_{i\in[\ell],j\in[k],\eta\in\bits}$.  
\item Output $\ST_{R^*}$ and $\ST_{\SimEq}:=\{\decom_{i,j}^{\eta}\}_{i\in[\ell],j\in[k],\eta\in\bits}$. 
\end{enumerate}

\smallskip
\noindent
$\SimEq_1(\ST_{\SimEq},m)$:  
\begin{enumerate}
\item Parse $\ST_{\SimEq}=\{\decom_{i,j}^{\eta}\}_{i\in[\ell],j\in[k],\eta\in\bits}$. 
\item Let $m_i$ be the $i$-th bit of $m$ for $i\in[\ell]$.  
\item Output $\decom:=\{\decom_{i,j}^{m_i}\}_{i\in[\ell],j\in[k]}$. 
\end{enumerate}

Let $\bar{\SimEq}_0^{R^*(\rho)}(1^\secpar)$ be a not necessarily QPT algorithm that works similarly to $\SimEq_0^{R^*(\rho)}(1^\secpar)$ except that it samples $\ST_{R^*}$ and $\{\decom_{i,j}^{\eta}\}_{i\in[\ell],j\in[k],\eta\in\bits}$ from $\sfQ_\rho^0$.  
By \Cref{eq:TD_Q_and_R}, we have 
\begin{align*}
   &\left\{
   (\ST_{R^*},\decom):
   \begin{array}{ll}
   (\ST_{R^*},\ST_{\SimEq})\gets \SimEq_0^{R^*(\rho)}(1^\secpar)\\
   \decom\gets \SimEq_1(\ST_{\SimEq},m)
   \end{array}
   \right\}_{\secpar\in \mathbb{N},m\in \bits^{\ell(\secpar)}}\\
   \sind
   &\left\{
   (\ST_{R^*},\decom):
   \begin{array}{ll}
   (\ST_{R^*},\ST_{\SimEq})\gets \bar{\SimEq}_0^{R^*(\rho)}(1^\secpar)\\
   \decom\gets \SimEq_1(\ST_{\SimEq},m)
   \end{array}
   \right\}_{\secpar\in \mathbb{N},m\in \bits^{\ell(\secpar)}}.
\end{align*}
Moreover, by computational hiding of $\Com$, it is easy to show that 
\begin{align*}
   &\left\{
   (\ST_{R^*},\decom):
   \begin{array}{ll}
   (\ST_{R^*},\ST_{\SimEq})\gets \bar{\SimEq}_0^{R^*(\rho)}(1^\secpar)\\
   \decom\gets \SimEq_1(\ST_{\SimEq},m)
   \end{array}
   \right\}_{\secpar\in \mathbb{N},m\in \bits^{\ell(\secpar)}}\\
   \cind
    &\left\{
    (\ST_{R^*},\decom):
    \begin{array}{ll}
    (\tau,\ST_{C},\ST_{R^*})\gets \langle C(m),R^*(\rho) \rangle(1^\secpar)\\
    \decom \gets C(\ST_C)
    \end{array}
    \right\}_{\secpar\in \mathbb{N},m\in \bits^{\ell(\secpar)}}.
\end{align*}
Combining the above, the proof of equivocality is completed.
\end{proof}

\subsection{Reducing Binding Error}
We show that sequential repetition of $\wEqCom$ (\Cref{protocol:wEqCom}) reduces the binding error to be negligible while preserving equivocality. 
\begin{ProtocolBox}[label={protocol:EqCom}]{Equivocal Commitments $\EqCom$}
{\bf Parameters:} Let $n=\omega(1)$ be a positive integer.

{\bf Inputs:} Both parties receive $\secpar$ as the common input. The committer in addition gets a string $m\in \bits^{\ell(\secpar)}$ as its private input where $\ell(\cdot)$ is a polynomial.

\para{Commit Stage:} 
\begin{enumerate}
\item 
$C$ commits to $m$ using $\wEqCom$ $n$ times in a sequential manner. 
\end{enumerate}
{\bf Decommit Stage:} 
\begin{enumerate}
\item $C$ reveals $m$ along with the corresponding decommit information w.r.t. all the $n$ executions of $\wEqCom$. 
\item  $R$ accepts if all the decommit information are valid  and otherwise rejects. 
\end{enumerate}
\end{ProtocolBox}
\begin{theorem}\label{thm:security-eqcom}
The scheme $\EqCom$ (\Cref{protocol:EqCom}) satisfies computational hiding, equivocality (as per \Cref{def:eqcom}), and statistical binding. 
\end{theorem}
\begin{proof}
Statistical binding follows from $(2^{-k}+\negl(\secpar))$-statistical binding of $\wEqCom$ noting that the binding error is exponentially reduced by sequential repetition  and $(2^{-k}+\negl(\secpar))^n=\negl(\secpar)$ when $k=\Theta(\log \secpar)$ and $n=\omega(1)$. 
Equivocality immediately follows from that of $\wEqCom$ noting that equivocality is preserved under sequential composition. Indeed, this can be shown by a straightforward hybrid argument (see e.g., \cite{C:BCKM21b}). 
Computational hiding immediately follows from equivocality.  
\end{proof}

Since $\EqCom$ runs in $\omega(1)$ rounds makes black-box use of OWFs, \Cref{thm:security-eqcom} implies \Cref{thm:eqcom}. 

\section{Post-Quantum Extractable Batch Commitments}
\label{sec:ExtCom-from-OT}

\subsection{Definitions}
\takashi{TODO: add an intuition for the definition.}

\takashi{I'm not sure if "batch" is the right wording here. This may give a wrong impression that there is some efficiency gain.
Another option is to use "parallel", but I'm wondering if it's confusing because this is different from parallel extractable commitments defined in \cref{def:epsilon-sim-ext-com:parallel}. Any idea?
}

\takashi{Another idea that comes to my mind is to regard it as a commitments with "local decommitments" capturing the inituituion that we can decommit each component of the message locally (without affecting secrecy of other parts).}

\xiao{I think what you said is known in the literature as ``vector commitments''. The only difference is when people say vector commitments, it usually means a succinct (and thus computationally binding) com where statistical binding holds for some specific positions (but not all positions). Thus, even ``vector commitments'' is not good for us. I guess the best name would be ``commitment with local decommitment''. But this seems to long and we need to modify the current notations. In summary, I'm ok with the name ``batch''.}

\begin{definition}[Post-Quantum Batch Commitments]\label{def:bcom}
A {\em post-quantum batch commitment scheme}  $\langle C, R \rangle$ is a classical interactive protocol between interactive \PPT machines $C$ and $R$. Let $\vb{m}=(m_1,...,m_n)\in \bits^{\ell(\secpar)\times n(\secpar)}$ (where $\ell(\cdot)$ and $n(\secpar)$ are some polynomials) be a sequence of messages that $C$ wants to commit to. The protocol consists of the following stages:
\begin{itemize}
\item
{\bf Commit Stage:} $C(\vb{m})$ and $R$ interact with each other to generate a transcript (which is also called a commitment) denoted by $\tau$,\footnote{That is, we regard the whole transcript as a commitment.} 
$C$'s state $\ST_{C}$, and 
$R$'s output $b_{\mathrm{com}}\in\Set{\bot,\top}$ indicating acceptance $(i.e., b_{\mathrm{com}}=\top)$ 
or rejection $(i.e., b_{\mathrm{com}}=\bot)$.
 We denote this execution by $(\tau,\ST_{C},b_{\mathrm{com}}) \gets \langle C(m), R \rangle(1^\secpar)$. When $C$ is honest, $\ST_C$ is classical, but when we consider a malicious quantum committer $C^*(\rho)$, we allow it to generate any quantum state $\ST_{C^*}$. 
 Similarly, a malicious quantum receiver $R^*(\rho)$ can output any quantum state, which we denote by $\OUT_{R^*}$ instead of $b_{\mathrm{com}}$. 
\item
{\bf Decommit Stage:}
$C$ generates a sequence of decommitments $\decom=(\decom_1,...,\decom_n)$ from $\ST_C$.
We denote this procedure by $\decom \gets C(\ST_C)$. 
Then it sends a sequence of messages $\vb{m}=(m_1,...,m_n)$ and the sequence of decommitments $\decom=(\decom_1,...,\decom_n)$ to $R$. 
For each $i\in [n]$, 
$R$ runs a deterministic verification procedure $b_{\mathrm{dec},i} \gets \Verify_i(\tau,m_i,\decom_i)$ 
where $b_{\mathrm{dec},i}=\top$ and $b_{\mathrm{dec},i}=\bot$  indicate acceptance and rejection on the $i$-th bit, respectively. 
W.l.o.g., we assume that $R$ always rejects (i.e., $\Verify_i(\tau,\cdot,\cdot) = \bot$ for all $i\in [n]$) whenever $b_\mathrm{com} = \bot$. (Note that w.l.o.g., $\tau$ can include $b_\mathrm{com}$ because we can always modify the protocol to ask $R$ to send $b_\mathrm{com}$ as the last round message.)
\end{itemize}

The scheme satisfies the following requirements:
\begin{enumerate}
\item
{\bf (Completeness.)} For any polynomials $\ell:\mathbb{N} \rightarrow \mathbb{N}$ 
and $n:\mathbb{N} \rightarrow \mathbb{N}$, any $\vb{m} \in \bits^{\ell(\secpar)\times n(\secpar)}$, and any $i\in [n]$, it holds that
\begin{equation*}
\Pr[
b_{\mathrm{com}}=b_{\mathrm{dec},i}=\top
: 
\begin{array}{l}
(\tau, \ST_{C}, b_{\mathrm{com}}) \gets \langle C(\vb{m}),R \rangle(1^\secpar) \\
(\decom_1,...,\decom_n) \gets C(\ST_C)\\
 b_{\mathrm{dec},i}\gets \Verify_i(\tau,m_i,\decom_i)
\end{array}
] = 1.
\end{equation*}

\item
{\bf (Statistically binding.)} For any unbounded-time committer $C^*$, the following holds: 
\begin{align*}
    \Pr[
    \begin{array}{l}
    \exists~i\in[n],m_0,m_1,\decom_0,\decom_1,~s.t.~m_0\neq m_1 ~\land\\
     \Verify_i(\tau,m_0,\decom_0)=\Verify_i(\tau,m_1,\decom_1)=\top
    \end{array}
    :(\tau,\ST_{C^*},b_{\mathrm{com}}) \gets \langle C^*, R \rangle(1^\secpar)]=\negl(\secpar).
\end{align*}

\item
{\bf (Computationally Hiding.)} For any non-uniform QPT receiver $R^*$ and any polynomials $\ell : \mathbb{N} \rightarrow \mathbb{N}$
and 
$n : \mathbb{N} \rightarrow \mathbb{N}$
, the following holds:
\begin{align*}
\big\{ \OUT_{R^*}\langle C(\vb{m}_0),R^* \rangle(1^\secpar),\{\decom_{i'}\}_{i'\ne i}\big\}_{\secpar \in \mathbb{N},~i\in[n(\secpar)], ~\vb{m}_0, \vb{m}_1 \in \Delta_i}\\
\cind~ \big\{ \OUT_{R^*}\langle C(\vb{m}_1),R^* \rangle(1^\secpar),\{\decom_{i'}\}_{i'\ne i}\big\}_{\secpar \in \mathbb{N},~i\in[n(\secpar)],  ~\vb{m}_0, \vb{m}_1 \in \Delta_i},
\end{align*}
where 
$$
\Delta_i:=\{(\vb{m}_0=(m_{0,1},...,m_{0,n}), \vb{m}_1=(m_{1,1},...,m_{1,n}))\in (\bits^{\ell(\secpar)})^2
: \forall i'\in[n]\setminus \{i\}~m_{0,i'}=m_{1,i'}\}
,$$ 
$\OUT_{R^*}\langle C(\vb{m}_b),R^* \rangle(1^\secpar)$ $(b \in \bits)$ denotes the output of $R^*$ at the end of the commit stage, 
and $(\decom_1,...,\decom_n)$ denotes the sequence of decommitments generated by $C$ in the decommit stage.
\end{enumerate}
\end{definition}
\begin{remark}
By a straightforward hybrid argument, the above computational hiding property implies the following: 
\begin{itemize}
\item
{\bf (Computationally Hiding w.r.t. Subsets.)} For any non-uniform QPT receiver $R^*$ and any polynomials $\ell : \mathbb{N} \rightarrow \mathbb{N}$
and 
$n : \mathbb{N} \rightarrow \mathbb{N}$
, the following holds:
\begin{align*}
\big\{ \OUT_{R^*}\langle C(\vb{m}_0),R^* \rangle(1^\secpar),\{\decom_{i'}\}_{i'\notin S}\big\}_{\secpar \in \mathbb{N},~S\subseteq[n(\secpar)], ~\vb{m}_0, \vb{m}_1 \in \Delta_S}\\
\cind~ \big\{ \OUT_{R^*}\langle C(\vb{m}_1),R^* \rangle(1^\secpar),\{\decom_{i'}\}_{i'\notin S}\big\}_{\secpar \in \mathbb{N},~S\subseteq [n(\secpar)],  ~\vb{m}_0, \vb{m}_1 \in \Delta_S},
\end{align*}
where 
$$\Delta_S:=\{(\vb{m}_0=(m_{0,1},...,m_{0,n}), \vb{m}_1=(m_{1,1},...,m_{1,n}))\in (\bits^{\ell(\secpar)})^2
: \forall i'\in[n]\setminus S~m_{0,i'}=m_{1,i'}\}
,$$ 
$\OUT_{R^*}\langle C(\vb{m}_b),R^* \rangle(1^\secpar)$ $(b \in \bits)$ denotes the output of $R^*$ at the end of the commit stage, 
and $(\decom_1,...,\decom_n)$ denotes the sequence of decommitments generated by $C$ in the decommit stage.
\end{itemize}
\end{remark}

Similarly to the stand-alone setting (\Cref{def:com-val}), we define committed values for batch commitments as follows.  
\begin{definition}[Committed Values for Batch Commitments]\label{def:bcom-val}
For a statistically binding batch commitment scheme $\langle C, R \rangle$ (as per \Cref{def:bcom}), we define the value function as follows:
\begin{equation*}
    \val_i(\tau)\defeq 
    \begin{cases}
    m_i&\text{~if~}\exists\text{~unique~}m_i\text{~s.t.~}\exists~\decom_i, \Verify_i(\tau,m_i,\decom_i)=1\\
    \bot &\text{otherwise}
    \end{cases},
\end{equation*}
where $\Verify_i$ is as defined in \Cref{def:bcom}.
\end{definition}

Then we define extractability for batch commitments. This is a natural extension of that in the stand-alone setting (\Cref{def:epsilon-sim-ext-com:strong}) but there is a crucial difference that we require simulation with negligible errors instead of $\epsilon$-simulation because the purpose of this section is to achieve negligible simulation errors at the cost of sacrificing round complexity. 

\begin{definition}[PQ-ExtBCom]\label{def:sim-ext-bcom:strong}
A post-quantum batch commitment scheme $\langle C, R\rangle$ (as per \Cref{def:bcom}) is {\em extractable} if there exists a QPT algorithm $\SimExt$ (called the simulation extractor) such that for any non-uniform QPT $C^*(\rho)$, 
\begin{equation*}
\big\{ \SimExt^{C^*(\rho)}(1^\secpar) \big\}_\secpar
\cind 
\big\{(\{\val_i(\tau)\}_{i\in [n]}, \ST_{C^*}):(\tau,\ST_{C^*},b_{\mathrm{com}}) \gets \langle C^*(\rho), R \rangle(1^\secpar)\big\}_\secpar,  
\end{equation*}
where $\val_i(\tau)$ is the value committed by $C^*$ as defined in \Cref{def:bcom-val}. 
\end{definition}
\begin{remark}\label{rem:comp_vs_stat_ind}
 We remark that we only require computational indistinguishability between the simulated and real execution while \Cref{def:epsilon-sim-ext-com:strong} requires statistical indistinguishability (with a noticeable simulation error). This is because computational indistinguishability is sufficient for our purpose.   \takashi{More explanations may be useful.}
\end{remark}
\begin{remark} 
One may find a conceptual similarity between 
extractable batch commitments (\Cref{def:sim-ext-bcom:strong}) and
parallelly extractable commitments (\Cref{def:epsilon-sim-ext-com:parallel}). 
Indeed, if a (stand-alone) commitment scheme (as per \Cref{def:com}) satisfies the negligible simulation error version of parallel extractability (as per \Cref{def:epsilon-sim-ext-com:parallel}), then its parallel composition is an extractable batch commitment scheme  (as per \Cref{def:sim-ext-bcom:strong}). However, we do not know how to construct a commitment scheme that satisfies the negligible simulation error version of parallel extractability. This is why we introduced extractable batch commitments.   \takashi{I guess this remark is very confusing. We may improve the explanation later.}
\end{remark}

As an intermediate tool towards constructing extractable batch commitments (as per \Cref{def:sim-ext-bcom:strong}), we introduce a weaker security notion which we call \emph{extractability with over-extraction}.
Intuitively, it is similar to the full-fledged extractability  (\Cref{def:sim-ext-bcom:strong}) except that we allow the simulation extractor to extract non-$\bot$ messages even if the transcript is invalid (i.e., there is no accepting decommitment). 
\begin{definition}[PQ-ExtBCom with Over-extraction]\label{def:sim-ext-bcom:over}
A post-quantum batch commitment scheme $\langle C, R\rangle$ (as per \Cref{def:bcom}) is {\em extractable with over-extraction} if there exists a QPT algorithm $\SimExtO$ (called the simulation extractor with over-extraction) such that for any non-uniform QPT $C^*(\rho)$, 
\begin{equation*}
\big\{ (\tau,\ST_{C^*}):(\tau,\ST_{C^*},\{m_{\ext,i}\}_{i\in[n]})\gets \SimExtO^{C^*(\rho)}(1^\secpar) \big\}_\secpar
\cind
\big\{(\tau, \ST_{C^*}):(\tau,\ST_{C^*},b_{\mathrm{com}}) \gets \langle C^*(\rho), R \rangle(1^\secpar)\big\}_\secpar,  
\end{equation*}
and 
\begin{equation*}
    \Pr[\exists i\in[n]~s.t.~\val_i(\tau)\notin\{m_{\ext,i},\bot\}]\le \negl(\secpar)
\end{equation*}
where 
$(\tau,\ST_{C^*},\{m_{\ext,i}\}_{i\in[n]})\gets \SimExtO^{C^*(\rho)}(1^\secpar)$ and $\val_i(\com)$ is the value committed by $C^*$ as defined in \Cref{def:bcom-val}.  \takashi{Strictly speaking, this may not be weaker than \Cref{def:sim-ext-bcom:strong} since the second condition may not follow from \Cref{def:sim-ext-bcom:strong}?}
\end{definition}
\begin{remark}
Extractability with over-extraction is conceptually similar to weak extractability defined in \cite{C:CCLY22} in the sense that both only require the extracted message be correct only when the transcript is valid. However, the crucial difference is that extractability with over-extraction requires that simulation of the committer's state be indistinguishable from the real one even when the transcript is invalid whereas weak extractability only requires it when  the transcript is valid.
\end{remark}

\subsection{Extractable Batch Commitments with Over-extraction}\label{sec:construction_OverExtBCom} 
Our construction of an extractable batch commitment scheme with over-extraction is described in \Cref{protocol:OverExtBCom}. It makes black-box use of the following building blocks:
\begin{enumerate}
\item 
A constant-round statistically-binding, computationally-hiding commitment $\Com$, (e.g., Naor's commitment). 
\item A parallel oblivious transfer protocol $\OT$ that satisfies $\epsilon$-simulation security against malicious receivers and indistinguishability-based security against malicious senders.  \takashi{Should we need to give a definition? } 
Assuming $t(\secpar)$-round semi-honest OT,  \cite{C:CCLY22} gives a black-box and $O(t(\secpar))$-round construction of such a protocol.\footnote{Their construction is shown to satisfy $\epsilon$-simulation security for both malicious senders and malicious receivers, and  $\epsilon$-simulation security immediately implies indistinguishability-based security.} 
\takashi{Maybe $(t(\secpar)+O(1))$-round? Need to recheck.}

\item An equivocal commitment scheme $\EqCom$ (as per \Cref{def:eqcom}). 
A $\omega(1)$-round construction of it is known assuming only black-box access to post-quantum secure OWFs (\Cref{thm:eqcom}).  
\end{enumerate}
\begin{ProtocolBox}[label={protocol:OverExtBCom}]{Extractable Batch Commitments with Over-Extraction $\OverExtBCom$}
{\bf Parameters:} Let $k=\omega(\log \secpar)$ be a positive integer. We use $k$ to mean the number of parallel sessions in $\OT$. 

{\bf Inputs:} Both parties receive $\secpar$ as the common input. The committer in addition gets a sequence of strings $\vb{m}=(m_1,...,m_n)\in \bits^{\ell(\secpar)\times n(\secpar)}$ as its private input where $\ell(\cdot)$ and $n(\cdot)$ are polynomials.

\para{Commit Stage:} 
\begin{enumerate}
\item \label[Step]{OverExtBCom:Com} 
$C$ commits to $m_i$ using $\Com$ for all $i\in[n]$ in parallel. Let $\com_i$ be the transcript of the $i$-th execution. 
\item 
For $i\in[n]$, $C$ generates $2k$-out-of-$2k$ XOR secret sharing $\{s^{b}_{i,j}\}_{j\in [k],b\in \bit}$ of $m_i$. 
That is, they are uniformly random under the constraint that $\bigoplus_{j\in[k],b\in\bit}s^{b}_{i,j}=m_i$. \item \label[Step]{OverExtBCom:OT} 
$C$ and $R$ execute 
$n$-parallel executions of $\OT$.\footnote{Note that $\OT$ itself is a $k$-parallel OT, and thus $nk$-parallel executions are happening in total.} 
We refer to the $i$-th execution by $\OT_i$ where 
$C$ uses
$\{s^{0}_{i,j},s^{1}_{i,j}\}_{j\in[k],b\in\bits}$ as input and $R$ uses independently and uniformly random bits $\{r_{i,j}\}_{j\in [k]}$ as input. 
\item \label[Step]{OverExtBCom:CF}
$C$ and $R$ now engage in the following coin-flipping subprotocol as detailed below.  
\begin{enumerate}
\item \label[Step]{OverExtBCom:EqCom}
$R$ samples a random string $\theta_R\gets \bit^{nk}$ and commits to it using $\EqCom$.   
\item \label[Step]{OverExtBCom:pick_theta_C}
$C$ samples a random string $\theta_C\gets \bit^{nk}$ and sends it to $R$.  
\item \label[Step]{OverExtBCom:EqCom-open}
$R$ sends to $C$ the value $\theta_R$ together with the corresponding decommitment information w.r.t. 
the $\EqCom$ in \Cref{OverExtBCom:EqCom}. Now, $C$ and $R$ agree on a random string $\theta \coloneqq \theta_R \xor \theta_C\in \bit^{nk}$. 
Interpret $\theta$ as a family $\{t_{i,j}\}_{i\in[n],j\in[k]}$ of bits.  
That is, let $t_{i,j}$ be the $(i-1)k+j$-th bit of $\theta$ for $i\in [n]$ and $j\in [k]$. 
\end{enumerate}
\item \label[Step]{OverExtBCom:reveal_s}
$C$ sends $s_{i,j}^{t_{i,j}}$ to $R$ for $i\in [n]$ and $j\in [k]$.  
\item $R$ never rejects in the commit stage, i.e., it always outputs $b_{\mathrm{com}}=\top$.  
$C$ sets the randomness used in the commit stage as $\ST_{C}$ and keep it for the decommit stage. 
\end{enumerate}

{\bf Decommit Stage:} 
\begin{enumerate}
\item 
For $i\in[n]$, 
$C$ defines $\decom_i$ to be a string consisting of the decommit information of $\com_i$ w.r.t. $\Com$ in \Cref{OverExtBCom:Com} of the Commit Stage, $\{s^{0}_{i,j},s^{1}_{i,j}\}_{j\in[k],b\in\bits}$, and the sender's randomness used in $\OT_i$ in \Cref{OverExtBCom:OT} of the Commit Stage.
\item $C$ sends $\vb{m}=(m_1,...,m_n)$ along with $\decom=(\decom_1,...,\decom_n)$ to $R$.
\item 
For each $i\in [n]$, $R$ runs the verification procedure $\Verify_i$ that accepts if 
\begin{enumerate}
\item the decommitment of $\com_i$ is valid w.r.t. the committed message $m_i$,
\item $\bigoplus_{j\in[k],b\in\bit}s^{b}_{i,j}=m_i$,
\item the revealed randomness for $\OT_i$ is consistent to the transcript of $\OT_i$ with input $\{s^{0}_{i,j},s^{1}_{i,j}\}_{j\in[k],b\in\bits}$, and
\item the string sent in \Cref{OverExtBCom:reveal_s} of the Commit Stage is consistent to $\{s^{0}_{i,j},s^{1}_{i,j}\}_{j\in[k],b\in\bits}$, i.e., it is equal to $s_{i,j}^{t_{i,j}}$ where $t_{i,j}$ is the bit generated in \Cref{OverExtBCom:CF} of the Commit Stage. 
\end{enumerate}
$R$ accepts if $\Verify_i$ accepts for all $i\in[n]$ and 
otherwise rejects. 
\end{enumerate}
\end{ProtocolBox}


\para{Security.} The security of \Cref{protocol:OverExtBCom} is stated as the following theorem. 
\begin{theorem}\label{thm:OverExtBCom}
 Assuming the existence of $t(\secpar)$-round semi-honest OT protocols, there exists (i.e., \Cref{protocol:OverExtBCom}) a black-box, $O(t(\secpar))+\omega(1)$-round construction of  batch commitments 
 that satisfies statistical binding, computational hiding (as per \Cref{def:bcom}), and extractability with over-extraction (as per \Cref{def:sim-ext-bcom:over}). 
\end{theorem}   
\begin{proof}
Statistical binding property immediately follows from that of $\Com$. Below, we show computational hiding and extractability with over-extraction. 

\para{Computational Hiding.}
Let $R^*(\rho)$ be any malicious QPT receiver.
Fix $i^*\in [\ell]$ and pair of sequence of messages $\vb{m}_0$ and $\vb{m}_1$ that differ only on the $i^*$-th component.
We consider the following hybrids for $b\in \bit$ and noticeable function $\epsilon$.

\para{Hybrid $H_b$:} This hybrid simulates execution between $C$ with input $\vb{m}_b$ and $R^*(\rho)$ and outputs $(\OUT_{R^*}\langle C(\vb{m}_b),R^* \rangle(1^\secpar),\{\decom_{i}\}_{i\ne i^*})$ 
where $(\decom_1,...,\decom_n)$ denotes the sequence of decommitments generated by $C$ in the decommit stage.

\para{Hybrid $H^\epsilon_b$:} This hybrid is identical to $H_b$, except that the execution of $\OT_{i^*}$ in \Cref{OverExtBCom:OT} of the commit stage is replaced with its $\epsilon$-simulation. 

\subpara{$\Output_{H_b} \cind_\epsilon \Output_{H^\epsilon_b}$:} This follows directly from $\epsilon$-simulation security of $\OT$ against malicious receivers. 

\subpara{$\Output_{H^\epsilon_0} \sind \Output_{H^\epsilon_1}$:}
In $H^\epsilon_b$, let $\{r^*_{i^*,j}\}_{j\in[k]}$ be the receiver's input to the ideal functionality of parallel OT provided by the simulator. Then the ideal functionality returns $\{s_{i^*,j}^{r_{i^*,j}}\}_{i^*\in[n],j\in[k]}$ to the simulator where $\{s_{i^*,j}^{0},s_{i^*,j}^{1}\}_{j\in[k]}$ is the honest committer's input to $\OT_{i^*}$ generated according to the description of the protocol. Especially, no information of $\{s_{i^*,j}^{1-r_{i^*,j}}\}_{j\in[k]}$ is used until this point. Since $\{s_{i^*,j}^{b}\}_{j\in[k],b\in\bit}$ is a $2k$-out-of-$2k$ XOR secret sharing of $m_{b,i^*}$, which is the $i^*$-th component of $\vb{m}_b$, no information of $m_{b,i^*}$ is revealed in $H_b^\epsilon$ unless all the remaining shares $\{s_{i^*,j}^{1-r_{i^*,j}}\}_{j\in[k]}$ are revealed at later stages, which happens only if $t_{i^*,j}=1-r_{i^*,j}$ for all $j\in[k]$ where $\{t_{i,j}\}_{i\in [n],j\in[k]}$ is the result of coin-flipping in \Cref{OverExtBCom:CF} of the commit stage. However, by computational hiding of $\EqCom$, the malicious receiver can cause only a negligible bias on the distribution of $\{t_{i,j}\}_{i\in [n],j\in[k]}$. Thus, the probability that $t_{i^*,j}=1-r_{i^*,j}$ for all $j\in[k]$ is at most $2^{-k}+\negl(\secpar)=\negl(\secpar)$. Thus, with probability $1-\negl(\secpar)$, $m_{b,i^*}$ remains information-theoretically hidden. This implies $\Output_{H^\epsilon_0} \sind \Output_{H^\epsilon_1}$. 

Combining the above, we obtain 
$\Output_{H_0} \cind_{2\epsilon} \Output_{H_1}$. Since this holds for any noticeable function $\epsilon$, this implies $\Output_{H_0} \cind \Output_{H_1}$, which means that the protocol satisfies computational hiding.

\para{Extractability with Over-extraction.} 
Let $\SimEq=(\SimEq_0,\SimEq_1)$ be the simulation equivocator for $\EqCom$. We construct the simulation extractor with over-extraction $\SimExtO$ as follows: 

\smallskip
\noindent
$\SimExtO^{C^*(\rho)}(1^\secpar)$: 
\begin{enumerate}
\item Interact with $C^*$ in \Cref{OverExtBCom:Com} playing the role of $R$. 
\item Execute $n$-parallel executions of $\OT$ in \Cref{OverExtBCom:OT}  with $C^*$ where $\SimExtO$ plays the role of the honest receiver of $\OT$ that uses a independently and uniformly random bits $\{r_{i,j}\}_{j\in[k]}$ as input in $\OT_i$ for $i\in [n]$.  Let $\{s^*_{i,j}\}_{j\in[k]}$ be the receiver's outcome of $\OT_i$ for $i\in [n]$. 
\item 
Let $\rho'$ be the internal state of $C^*$ at the end of \Cref{OverExtBCom:OT}. 
Run $(\rho'',\ST_{\SimEq})\gets \SimEq_0^{C^*(\rho')}(1^\secpar)$. (Recall that $C^*$ plays the role of receiver for $\EqCom$.)   
\item Resume $C^*$ from \Cref{OverExtBCom:pick_theta_C} with its internal state $\rho''$ to obtain $\theta_C$. 
\item   
Let $\bar{r}\in \bits^{nk}$ be the string whose $(i-1)k+j$-th bit is $1-r_{i,j}$ for $i\in[n]$ and $j\in[k]$ and set $\theta_R\coloneqq \bar{r}\oplus \theta_C$. 
\item Run $\EqCom.\decom\gets \SimEq_1(\ST_{\SimEq},\theta_R)$ and sends $\theta_R$ and $\EqCom.\decom$ to $C^*$. 
Note that $t_{i,j}$ is now programmed to be $1-r_{i,j}$. 
\item Receive $s_{i,j}^{t_{i,j}}=s_{i,j}^{1-r_{i,j}}$ from $C^*$. Note that it obtains all the shares $\{s_{i,j}^b\}_{i\in[n],j\in[k],b\in\bits}$ at this point. 
\item Compute $m_{\ext,i}=\bigoplus_{j\in[k],b\in\bit}s_{i,j}^b$ for $i \in [n]$.
\item Output the transcript $\tau$, the final state $\ST_{C^*}$ of $C^*$, and $\{m_{\ext,i}\}_{i\in[n]}$. 
\end{enumerate}

First, it is easy to see that 
\begin{equation*}
    \Pr[\exists i\in[n]~s.t.~\val_i(\tau)\notin\{m_{\ext,i},\bot\}]=0
\end{equation*}
where 
$(\tau,\ST_{C^*},\{m_{\ext,i}\}_{i\in[n]})\gets \SimExtO^{C^*(\rho)}(1^\secpar)$. 
To see this, suppose that $\val_i(\tau)\neq \bot$. In this case, there must exist secret sharing $\{s_{i,j}^b\}_{j\in[k],b\in\bits}$ of $m_i$ that is consistent to the transcript. By the perfect correctness of $\OT$, \takashi{Is it okay to assume that $\OT$ satisfies perfect correctness? Otherwise, more involved arguments would be needed.
} these shares are obtained by $\SimExtO$ and thus $m_{\ext,i}=m_i$. 

Below, we prove 
\begin{equation} \label{eq:ind_SE_over}
\big\{ (\tau,\ST_{C^*}):(\tau,\ST_{C^*},\{m_{\ext,i}\}_{i\in[n]})\gets \SimExtO^{C^*(\rho)}(1^\secpar) \big\}_\secpar
\cind
\big\{(\tau, \ST_{C^*}):(\tau,\ST_{C^*},b_{\mathrm{com}}) \gets \langle C^*(\rho), R \rangle(1^\secpar)\big\}_\secpar.  
\end{equation}
We consider the following hybrids. 

\para{Hybrid $H_0$:} This hybrid executes  $(\tau,\ST_{C^*},b_{\mathrm{com}}) \gets \langle C^*(\rho), R \rangle(1^\secpar)$ and outputs $(\tau, \ST_{C^*})$. 

\para{Hybrid $H_1$:} This hybrid is identical to the previous one except that the commitment by $R$ of $\EqCom$ in \Cref{OverExtBCom:EqCom} is generated by $\SimEq_0$, which is decomitted to $\theta_R$ by $\SimEq_1$ in \Cref{OverExtBCom:EqCom-open}. Note that $\theta_R$ is just a uniformly random string that is independent of $\{r_{i,j}\}_{i\in[n],j\in[k]}$ in this hybrid. 

\subpara{$\Output_{H_0} \cind \Output_{H_1}$:} This follows directly from equivocality of $\EqCom$

\para{Hybrid $H_2$:} This hybrid is identical to the previous one except that $R$ uses $0^k$ as input of $\OT_i$ for all $i\in[n]$ in \Cref{OverExtBCom:OT}. 

\subpara{$\Output_{H_1} \cind \Output_{H_2}$:} This follows directly from indistinguishability-based security of $\OT$ against malicious senders. 

\para{Hybrid $H_3$:} This hybrid is identical to the previous one except that 
$\theta_R$ is set as $\theta_R=\bar{r}\oplus \theta_C$ where $\bar{r}$ is as defined in the description of $\SimExtO$. Note that $\theta_R$ can depend on $\theta_C$ since $\theta_R$ is not used in   \Cref{OverExtBCom:EqCom} due to the modification made in $H_1$. 

\subpara{$\Output_{H_2} \idind \Output_{H_3}$:} This follows directly from the observation that $u_1$ is a independently and uniformly random string in both hybrids noting that no information of $r$ is used in \Cref{OverExtBCom:OT} due to the modification made on $H_2$. 

\para{Hybrid $H_4$:} This hybrid is identical to the previous one except that $R$ uses $\{r_{i,j}\}_{j\in[k]}$ as input of $\OT_i$ for all $i\in[n]$ in \Cref{OverExtBCom:OT}. 

\subpara{$\Output_{H_3} \cind \Output_{H_4}$:} This follows directly from indistinguishability-based security of $\OT$ against malicious senders. 

Now, we can see that $H_4$ just runs  $(\tau,\ST_{C^*}):(\tau,\ST_{C^*},\{m_{\ext,i}\}_{i\in[n]})\gets \SimExtO^{C^*(\rho)}(1^\secpar)$  and outputs $(\tau, \ST_{C^*})$. 
Combining the above, we obtain 
\Cref{eq:ind_SE_over}. 
This completes the proof of extractability with over-extraction.
\end{proof}

\subsection{Removing  Over-extraction}
Next, we give a compiler that upgrades extractable batch commitments with over-extraction (as per \Cref{def:sim-ext-bcom:over})
into one with full-fledged extractability without over-extraction (as per \Cref{def:sim-ext-bcom:strong}).  
It is based on the cut-and-choose technique that is very similar to the one used for upgrading ``weak" extractable commitments into ``strong" one in \cite{C:CCLY22}.\footnote{We omit the definitions of weak and strong extractability in \cite{C:CCLY22} since this is not needed for our purpose.} 
Our construction of an extractable batch commitment scheme with over-extraction is described in \Cref{protocol:ExtBCom}. It makes black-box use of the following building blocks:
\begin{enumerate}
\item
The extractable batch commitment scheme $\OverExtBCom$ with over-extraction given in \Cref{protocol:OverExtBCom}, which in turn makes black-box use of any OTs.
Note that it is $(O(t(\secpar))+\omega(1))$-round if the assumed OT is $t(\secpar)$-round (\Cref{{thm:OverExtBCom}}).
\item A commitment scheme $\epsilonExtCom$ that satisfies statistical binding, computational hiding, and extractability with $\epsilon$-simulation (as per \Cref{def:epsilon-sim-ext-com:strong}). 
Constant-round and black-box construction of such a scheme based on OWFs is given in \cite{C:CCLY22}.\footnote{In fact, we only need a weaker security called ``weak extractability with $\epsilon$-simulation" in \cite{C:CCLY22}. 
} 
\item
An $(n+1,k)$-perfectly verifiable secret sharing scheme $\VSS = (\VSS_{\Share}, \VSS_{\Recon})$ (as per \Cref{def:VSS}). We require that $k$ is a constant fraction of $n$ such that  $k \le n/3$. There are known constructions (without any computational assumptions) satisfying these properties \cite{STOC:BenGolWig88,EC:CDDHR99}.
\end{enumerate}
\takashi{I wanted to use $n$ as the length of $\bf{m}$ as previous sections, but I found that there is a notational collision with the parameter for VSS, so I used $N$ instead. Any idea to improve the readability?}

\begin{ProtocolBox}[label={protocol:ExtBCom}]{Extractable Batch Commitment $\ExtBCom$}
\para{Parameters.}
Let $n(\SecPar)$ be a polynomial on $\SecPar$. Let $k$ be a constant fraction of $n$ such that $k \le n/3$.

\para{Input:}
Both the committer $C$ and the receiver $R$ get security parameter $1^\SecPar$ as the common input. $C$ in addition gets a sequence of strings $\vb{m}=(m_1,...,m_\mlen) \in \bits^{\ell(\SecPar)\times \mlen(\secpar)}$ as his private input, where $\ell(\cdot)$ and $\mlen(\cdot)$ are polynomials.\footnote{We use $\mlen$ instead of $n$ to mean the number of committed messages of $\OverExtBCom$ to avoid notational collision with the parameter for $\VSS$.} 

\para{Commit Stage:}
\begin{enumerate}[topsep=0pt,itemsep=0pt]
\item \label[Step]{ExtBCom:Init}
For $i\in [\mlen]$, $C$ prepares $n$ views $\Set{\msf{v}_{i,j}}_{j \in [n]}$, corresponding to an MitH execution for the $(n+1, k)$-$\VSS_\Share$ of the message $m_i$ (see \Cref{rmk:mpc-in-the-head-vss} for details). 
\item \label[Step]{ExtBCom:com_to_views} 
For $i\in[\mlen]$, 
$C$ and $R$ involve an execution of $\OverExtBCom$ where $C$ commits to $\vb{v}\coloneqq \{\msf{v}_{i,j}\}_{i\in[\mlen],j\in [n]}\in \bit^{\ell\times Nn}$. 
\item \label[Step]{ExtBCom:CF}
$C$ and $R$ engage in the following coin-flipping subprotocol as detailed below.  
\begin{enumerate}
\item \label[Step]{ExtBcom:com_to_theta_R}
$R$ samples a random string $\theta_R$ of proper length and commits to it using $\epsilonExtCom$.   
\item \label[Step]{ExtBcom:send_theta_C}
$C$ samples a random string $\theta_C$ of proper length and sends it to $R$.  
\item \label[Step]{ExtBcom:reveal_theta_R}
$R$ sends to $C$ the value $\theta_R$ together with the corresponding decommitment information w.r.t.
the $\epsilonExtCom$ in \Cref{ExtBcom:com_to_theta_R}. Now, $C$ and $R$ agree on a random string $\theta \coloneqq \theta_R \xor \theta_C$. By a proper choice of length, the string $\theta$ it can be interpreted as specifying $N$ size-$k$ random subsets of $[n]$. We write $(\eta_1,...,\eta_N)$ to mean these subsets. 
\end{enumerate} 
\item \label[Step]{ExtBCom:decommit_some_shares}
For $i\in [\mlen]$, $C$ decommits to the VSS shares in the set $\eta_i$, i.e. it sends $\Set{\msf{v}_{i,j}}_{j \in \eta_i}$ along with the corresponding decommitment information w.r.t.\ $\OverExtBCom$ in \Cref{ExtBCom:com_to_views}. 
\item \label[Step]{ExtBCom_commit_stage_check}
$R$ checks the following conditions:
\begin{enumerate}[topsep=0pt,itemsep=0pt]
\item
All the decommitments in \Cref{ExtBCom:decommit_some_shares} are valid; {\bf and}
\item
for any $i\in[\mlen]$ and $j,j'\in \eta_i$, views $(\view_{i,j}, \view_{i,j'})$ are consistent (as per \Cref{def:view-consistency} and \Cref{rmk:VSS:view-consistency}) w.r.t.\ the $\VSS_\Share$ execution in \Cref{ExtBCom:Init}. 
\end{enumerate}
If all the checks pass, $R$ accepts (i.e., outputs $b_{\mathrm{com}}=\top$); otherwise, $R$ rejects (i.e., outputs $b_{\mathrm{com}}=\bot$).
\end{enumerate}
\para{Decommit Stage:}
\begin{enumerate}[topsep=0pt,itemsep=0pt]
\item \label[Step]{ExtBCom:decommit_all_shares}
For $i\in [\mlen]$, $C$ defines $\decom_i$ to be a string consisting of $\Set{\view_{i,j}}_{j\in [n]}$ together with all the corresponding decommitment information w.r.t.\ $\OverExtBCom$ in \Cref{ExtBCom:com_to_views} of the Commit Stage. 
$C$ sends $\vb{m}=(m_1,...,m_\mlen)$ and $\decom=(\decom_1,...,\decom_\mlen)$. 
\item  \label{ExtBCom_decommit_verf}
For each $i\in [\mlen]$, $R$ runs the verification procedure $\Verify_i$ that works as follows: 
\begin{enumerate}
\item Construct $\Set{\view'_{i,j}}_{j\in [n]}$ as follows: in \Cref{ExtBCom:decommit_all_shares} of the Decommit Stage, if the decommitment to $\view_{i,j}$ is valid, $R$ sets $\view'_{i,j} \coloneqq \view_{i,j}$; otherwise, $R$ sets $\view'_{i,j} \coloneqq \bot$. 
\item Accept if $m_i = \VSS_\Recon(\view'_{i,1}, \ldots, \view'_{i,n})$ and otherwise reject. 
\end{enumerate}
$R$ accepts if $\Verify_i$ accepts for all $i\in[\mlen]$ and otherwise rejects.  
\end{enumerate}
\end{ProtocolBox}

\begin{theorem}\label{thm:ExtBCom}
 Assuming the existence of $t(\secpar)$-round semi-honest OT protocols, there exists (i.e., \Cref{protocol:ExtBCom}) a black-box, $O(t(\secpar))+\omega(1)$-round construction of  batch commitments 
 that satisfies statistical binding, computational hiding (as per \Cref{def:bcom}), and extractability (as per \Cref{def:sim-ext-bcom:strong}). 
\end{theorem}   
\begin{proof}
Statistical binding property immediately follows from that of $\OverExtBCom$. Below, we show computational hiding and extractability. 

\para{Computational Hiding.}
Let $R^*(\rho)$ be any malicious QPT receiver.
Fix $i^*\in [\ell]$ and pair of sequence of messages $\vb{m}_0$ and $\vb{m}_1$ that differ only on the $i^*$-th component. 
We consider the following hybrids for $b\in \bit$ and noticeable function $\epsilon$.

\para{Hybrid $H_b$:} This hybrid simulates execution between $C$ with input $\vb{m}_b$ and $R^*(\rho)$ and outputs $(\OUT_{R^*}\langle C(\vb{m}_b),R^* \rangle(1^\secpar),\{\decom_{i}\}_{i\ne i^*})$ 
where $(\decom_1,...,\decom_n)$ denotes the sequence of decommitments generated by $C$ in the decommit stage.

\para{Hybrid $H^\epsilon_b$:} This hybrid is identical to $H_b$, except for the following changes: It takes size-$k$ random subsets $\eta_i\subseteq [n]$ for all $i\in [\mlen]$ at the beginning. 
Then it runs the $\epsilon$-simulation extractor for $\epsilonExtCom$ to extract 
$\theta_R$ while simulating the state of $R^*$ in \Cref{ExtBcom:com_to_theta_R}
and defines $\theta_C$ so that $\theta_R\oplus \theta_C$ specifies the subsets $(\eta_1,...,\eta_\mlen)$ in \Cref{ExtBcom:send_theta_C}. 

\subpara{$\Output_{H_b} \cind_\epsilon \Output_{H^\epsilon_b}$:} This follows directly from extractability with $\epsilon$-simulation  of $\epsilonExtCom$ noting that the distribution of $\theta_C$ is uniformly random in both hybrids.   

\subpara{$\Output_{H^\epsilon_0} \cind \Output_{H^\epsilon_1}$:} 
Note that the subset $\theta_{i^*}$ is fixed at the beginning in these hybrids. Then we can reduce computational indistinguishability of them to computational hiding of $\OverExtBCom$ by the same argument as the security proof of the  VSS hiding game (\Cref{chall:vss:hide}). 

Combining the above, we obtain 
$\Output_{H_0} \cind_{2\epsilon} \Output_{H_1}$. Since this holds for any noticeable function $\epsilon$, this implies $\Output_{H_0} \cind \Output_{H_1}$, which means that the protocol satisfies computational hiding.

\para{Extractability.} 
Let $\SimExtO$ be the simulation extractor with over-extraction for $\SimExtO$. We construct the simulation extractor $\SimExt$ as follows: 

\smallskip
\noindent
$\SimExt^{C^*(\rho)}(1^\secpar)$: 
\begin{enumerate}
\item Run 
$(\OverExtBCom.\tau,\rho',\{\view_{\ext,i,j}\}_{i\in[\mlen],j\in[n]})\gets \SimExtO^{C^*(\rho)}(1^\secpar)$
where $\OverExtBCom.\tau$ is the simulated transcript of the execution of $\OverExtBCom$ in \Cref{ExtBCom:com_to_views}, $\rho'$ is the simulated state of $C^*$ at the end of  \Cref{ExtBCom:com_to_views}, and $\{\view_{\ext,i,j}\}_{i\in[\mlen],j\in[n]}$ is the tuple of the extracted messages. 
\item Run the rest of the commit stage while playing the role of the honest receiver $R$. 
\item Let $\ST_{C^*}$ be the state of $C^*$ at the end of the commit stage. Define $\{m_{\ext,i}\}_{i\in \mlen}$ as follows:
\begin{enumerate}
\item If $b_{\mathrm{com}}=\bot$ (i.e., $R$ rejects in \Cref{ExtBCom_commit_stage_check} of the commit stage), then set $m_{\ext,i}\coloneqq \bot$ for all $i\in [\mlen]$.
\item Otherwise, 
set $m_{\ext,i} \coloneqq \VSS_\Recon(\view_{\ext,i,1}, \ldots, \view_{\ext,i,n})$ for $i\in[\mlen]$. 
\end{enumerate}
\item Output $(\ST_{C^*},\{m_{\ext,i}\}_{i\in [\mlen]})$. 
\end{enumerate}

For $i\in [\mlen]$, let $\Good_i$ be the event that there exists $m^*_i$ such that  $m^*_{i} =\VSS_\Recon(\view'_{i,1}, \ldots, \view'_{i,n})$
for all $\{\view'_{i,j}\}_{j\in[n]}$ such that $\view'_{i,j}=\val_{i,j}(\OverExtBCom.\tau)$ or $\val_{i,j}(\OverExtBCom.\tau)=\bot$ for all $j\in[n]$ where $\OverExtBCom.\tau$ is the transcript of $\OverExtBCom$ in \Cref{ExtBCom:com_to_views} 
of the Commit Stage.  Let $\Bad_i$ be the complementary event of $\Good_i$. 
Then for any $i\in [\mlen]$, we have 
\begin{align}\label{eq:Bad_and_acc}
    \Pr[\Bad_i\land b_{\mathrm{com}}=\top]=\negl(\secpar). 
\end{align}
We omit its proof since almost identical claim is proven in \cite[Section 5.2]{C:CCLY22}. \takashi{@Xiao Can you add a brief explanation if needed?}

It is easy to see that $m_{\ext,i}=m^*_i=\val_i(\tau)$
whenever $\Good_i$ occurs where $\tau$ is the transcript of the commit stage of $\ExtBCom$.  
By the union bound,  \Cref{eq:Bad_and_acc} implies that $\Good_i$ occurs for all $i\in [\mlen]$ whenever $b_{\mathrm{com}}=\top$ except for a negligible probability.  
Thus, whenever $b_{\mathrm{com}}=\top$, $m_{\ext,i}=\val_i(\tau)$ except for a negligible probability. Combined with extractability with over-extraction (as per \Cref{def:sim-ext-bcom:over}), this directly implies extractability  (as per \Cref{def:sim-ext-bcom:strong}).
\end{proof}

\section{Black-Box Post-Quatnum ExtCom-and-Prove}
\label{sec:bb-extcom-n-prove}

\subsection{Definition}
The following definition is taken from \cite{C:CCLY22} with modifications to admit $\negl$-simulation instead of $\epsilon$-simulation for extractability and ZK. 

\begin{definition}[Simulatable ExtCom-and-Prove]
\label{def:com-n-prove}
An ExtCom-and-Prove scheme consists of a pair of protocols $\Prot_{\textsc{ECnP}} = (\algo{ExtCom}, \algo{Prove})$ executed between a pair of \PPT machines $P$ and $V$. Let $m\in \bits^{\ell(\secpar)}$ (where $\ell(\cdot)$ is some polynomial) is a message that $P$ wants to commit to. The protocol consists of the following stages (we omit the input $1^\secpar$ to $P$ and $V$): 
\begin{itemize}
\item 
{\bf Commit Stage:} 
$P(m)$ and $V$ execute $\algo{ExtCom}$, which generates a transcript (commitment) $\com$, 
$P$'s state $\ST_P$, and  
$V$'s decision $b\in \{\top,\bot\}$ indicating acceptance (i.e., $b=\top$) or rejection (i.e., $b=\bot$). We denote this execution as $(\com, \ST_P, b) \gets \langle P(m),V \rangle_\mathsf{EC}$. 
A malicious verifier is allowed to output any quantum state, which we denote by $\ST_{V^*}$ instead of $b$, and to keep the state for the prove stage. 

\item 
{\bf Decommit Stage:}\footnote{This stage is rarely executed in applications.} $P(\ST_P)$ generates a decommitment $\decom$ and sends it to $V$ along with a message $m$. $V$ accepts or rejects. 

\item 
{\bf Prove Stage:} 
Let $\phi$ be any predicate. $P(\ST_P, \phi)$ and $V(\com,\phi)$ execute $\algo{Prove}$, after which $V$ outputs $\top$ (accept) or $\bot$ (reject). We denote the execution of this stage as $b' \gets \langle P(\ST_P), V(\com) \rangle^\phi_\mathsf{Pr}$, where $b' \in \{\top,\bot\}$ is $V$'s output. 
A malicious verifier is allowed to output an arbitrary quantum state, which we denote by $\OUT_{V^*}$ instead of $b'$. 
\end{itemize}
The following requirements are satisfied:
\begin{enumerate}
\item \label[Property]{item:com-n-prove:condition:SimExt}
{\bf Security as Simulation Extractable Commitment.} The Commit Stage and Decommit Stage constitute a post-quantum commitment scheme (as per \Cref{def:com} where $P$ and $V$ play the roles of $C$ and $R$, respectively) that is 
computationally hiding, 
statistically binding,    
and extractable. Here, the extractability means the following: \takashi{I explicitly write it here because this is not defined yet.}
There exists a QPT algorithm $\SimExt$ (called the simulation extractor) such that for any non-uniform QPT $C^*(\rho)$, 
\begin{equation*}
\big\{ \SimExt^{C^*(\rho)}(1^\secpar) \big\}_\secpar
\cind
\big\{(\val(\tau), \ST_{C^*}):(\tau,\ST_{C^*},b_{\mathrm{com}}) \gets \langle C^*(\rho), R \rangle(1^\secpar)\big\}_\secpar,  
\end{equation*}
where $\val(\tau)$ is the value committed by $C^*$ as defined in \Cref{def:com-val}.\footnote{We only require computational indistinguishability rather than statistical one unlike \Cref{def:epsilon-sim-ext-com:strong}. This is inherited from \Cref{def:sim-ext-bcom:strong} (see also \Cref{rem:comp_vs_stat_ind}).} 

\item 
{\bf Completeness.} For any $m \in \bits^{\ell(\secpar)}$ and any polynomial-time computable predicate $\phi$ s.t.\ $\phi(m) = 1$, it holds that
\begin{equation}
\Pr[b=\top ~\wedge~ b' =\top : 
\begin{array}{l}
(\com,\ST_P,b) \gets \langle P(m),V \rangle_\mathsf{EC} \\
b' \gets \langle P(\ST_P), V(\com) \rangle^\phi_\mathsf{Pr}
\end{array}
] = 1.
\end{equation}

\item {\bf Soundness.} \label[Property]{item:com-n-prove:condition:soundness}
For any predicate $\phi$ and any non-uniform QPT prover $P^*(\rho)$, 
\begin{equation}
\Pr[
\begin{array}{l}
b=\top ~\wedge~ b' =\top  \\ 
\wedge~ \phi(\val_{\ExtCom}(\com))=0 
\end{array}: 
\begin{array}{l}
(\com, \ST_{P^*}, b) \gets \langle P^*(\rho),V \rangle_\mathsf{EC} \\
b' \gets \langle P^*(\ST_{P^*}), V(\com) \rangle^\phi_\mathsf{Pr}
\end{array}
] = \negl(\secpar),
\end{equation}
where $\val_{\ExtCom}(\com)$ is as defined in \Cref{def:com-val}
and we stipulate that $\phi(\bot) = 0$.

\item {\bf Zero-Knowledge.} \label[Property]{item:com-n-prove:condition:zk}
There exists a pair of QPT simulators $(\Sim_\algo{EC}, \Sim_\algo{Pr})$ such that for any $m \in \bits^{\ell(\secpar)}$, polynomial-time computable predicate $\phi$  s.t.\ $\phi(m) = 1$, any non-uniform QPT verifier $V^*(\rho)$, and any noticeable function $\epsilon(\secpar)$, the following conditions hold:
\begin{align}
& \big\{\tilde{\ST}_{V^*} : (\tilde{\ST}_{V^*}, \ST_\algo{EC})\gets\Sim_\algo{EC}^{V^*(\rho)}\big\}_{\secpar} \cind \big\{\ST_{V^*} : (\com, \ST_P, \ST_{V^*}) \gets \langle P(m),V^*(\rho) \rangle_\mathsf{EC}\big\}_\secpar \label{item:com-n-prove:ZK:condition:1}\\
&\bigg\{\tilde{\OUT}_{V^*} : 
\begin{array}{l}
(\tilde{\ST}_{V^*}, \ST_\algo{EC})\gets\Sim_\algo{EC}^{V^*(\rho)}\\
\tilde{\OUT}_{V^*}  \gets \Sim^{V^*}_\algo{Pr}(1^{\epsilon^{-1}},\tilde{\ST}_{V^*}, \ST_\algo{EC},\phi)
\end{array}
\bigg\}_{\secpar} \cind
\bigg\{ \OUT_{V^*} : 
\begin{array}{l}
(\com, \ST_P, \ST_{V^*}) \gets \langle P(m),V^*(\rho) \rangle_\mathsf{EC} \\
\OUT_{V^*} \gets \langle P(\ST_P), V^*(\ST_{V^*}) \rangle^\phi_\mathsf{Pr}
\end{array}
\bigg\}_\secpar. \label{item:com-n-prove:ZK:condition:2} 
\end{align}

We refer to $\Sim_\algo{EC}$ (resp.\ $\Sim_\algo{Pr}$) as the Commit-Stage (resp.\ Prove-Stage) simulator.
\end{enumerate}
\end{definition}

\subsection{Construction of ExtCom-and-Prove}
We construct an ExtCom-and-Prove scheme based on extractable batch commitments. The construction is almost identical to that in \cite[Section 6.5]{C:CCLY22} except that we require full-simulation security instead of $\epsilon$-simulation. 

The construction is shown in \Cref{protocol:ExtCom-n-Prove}. It makes black-box use of the following building blocks:
\begin{enumerate}
\item
The extractable batch commitment (as per \Cref{def:sim-ext-bcom:strong}) $\ExtBCom$ (\Cref{protocol:ExtBCom}), which in turn makes black-box use of any OTs. Note that it is $(O(t(\secpar))+\omega(1))$-round if the assumed OT is $t(\secpar)$-round (\Cref{{thm:ExtBCom}}).

\item An equivocal commitment scheme $\EqCom$ (as per \Cref{def:eqcom}). 
A $\omega(1)$-round construction of it is known assuming only black-box access to post-quantum secure OWFs (\Cref{thm:eqcom}).

\item
	A constant-round statistically-binding, computationally-hiding commitment $\Com$, (e.g., Naor's commitment).  
\item
	An $(n+1,k)$-perfectly verifiable secret sharing scheme $\VSS = (\VSS_{\Share}, \VSS_{\Recon})$ (as per \Cref{def:VSS}). We require that $k$ is a constant fraction of $n$ such that  $k \le n/3$. There are known constructions (without any computational assumptions) satisfying these properties \cite{STOC:BenGolWig88,EC:CDDHR99}.
\item
	A $(n,k)$-perfectly secure MPC protocol $\Prot_\textsc{mpc}$ (as per \Cref{def:MPC}).
\end{enumerate}

\begin{ProtocolBox}[label={protocol:ExtCom-n-Prove}]{ExtCom-and-Prove scheme $\Prot_\textsc{ECnP}$}
{\bf Parameter Setting:} Let $n(\SecPar)$ be a polynomial on $\SecPar$. Let $k$ be a constant fraction of $n$ such that $k \le n/3$.

\para{Input:}
Both the prover $P$ and the verifier $V$ get $1^\SecPar$ as the common input. $P$ in addition gets a string $m \in \bits^{\ell(\SecPar)}$ as his private input, where $\ell(\cdot)$ is a polynomial.

\para{Commit Stage:}
\begin{enumerate}[topsep=0pt,itemsep=0pt]
\item \label[Step]{item:ExtCom-n-Prove:commit-stage:1}
$P$ prepares $n$ views $\Set{\msf{v}_i}_{i \in [n]}$, corresponding to an MitH execution for the $(n+1, k)$-$\VSS_\Share$ of the string $x$ (see \Cref{rmk:mpc-in-the-head-vss} for details). 
\item \label[Step]{item:ExtCom-n-Prove:commit-stage:2}
$P$ and $V$ involve in $\ExtBCom$, where $P$ commits to $(\view_1,...,\view_n)$.
\end{enumerate}
\para{Decommit Stage:}
\begin{enumerate}[topsep=0pt,itemsep=0pt]
\item  \label[Step]{item:ExtCom-n-Prove:decommit-stage:1}
$P$ sends $m$ and
$\Set{\view_i}_{i\in [n]}$ together with the corresponding decommitment information w.r.t.\ the $\ExtBCom$ in \Cref{item:ExtCom-n-Prove:commit-stage:2} of the Commit Stage.
\item
$V$ checks that all the decommitments in \Cref{item:ExtCom-n-Prove:decommit-stage:1} of the Decommit Stage are valid and $m=\VSS_\Recon(\view_1, \ldots, \view_n)$. If so, it accepts and otherwise rejects. 
\end{enumerate}

\para{Prove Stage:} Both parties learn a polynomial-time computable predicate $\phi$.
\begin{enumerate}[topsep=0pt,itemsep=0pt]
\item \label[Step]{item:ExtCom-n-Prove:prove-stage:1}
$P$ prepares $n$ views $\Set{\msf{v}'_i}_{i \in [n]}$ corresponding to an $(n, k)$-MitH execution for the functionality $F_\phi$ described below, where party $P_i$ uses $\msf{v}_i$ as input. It then commits to each of these views $\msf{v'}_i$ independently in parallel using $\Com$. 
    \begin{itemize}
        \item {\bf Functionality $F_\phi$:} This collects inputs $\msf{v}_i$ from party $i$, runs $\VSS_\Recon$ on these inputs to recover a value $x$, and outputs $\phi(x)$. 
    \end{itemize} 
\item \label[Step]{ExtCom-n-Prove:CF}
$P$ and $V$ engage in the following coin-flipping subprotocol as detailed below.  
\begin{enumerate}
\item \label[Step]{ExtCom-n-Prove:com_to_theta_P}
$P$ samples a random string $\theta_P$ of proper length and commits to it using $\EqCom$. 
\item \label[Step]{ExtCom-n-Prove:send_theta_V}
$V$ samples a random string $\theta_V$ of proper length and sends it to $P$.  
\item \label[Step]{ExtCom-n-Prove:reveal_theta_P}
$P$ sends to $V$ the value $\theta_P$ together with the corresponding decommitment information w.r.t. the $\EqCom$ in \Cref{ExtCom-n-Prove:com_to_theta_P}. Now, $P$ and $V$ agree on a random string $\theta \coloneqq \theta_P \xor \theta_V$. By a proper choice of length, the string $\theta$ it can be interpreted as specifying a size-$k$ random subset $\eta\subset [n]$. 
\end{enumerate} 
\item \label[Step]{item:ExtCom-n-Prove:prove-stage:6}
$P$ sends to $V$ in {\em one round} the following messages:
\begin{enumerate}[topsep=0pt,itemsep=0pt]
\item \label[Step]{item:ExtCom-n-Prove:prove-stage:6:a}
$\Set{\view_i}_{i \in \eta}$ together with the corresponding decommitment information w.r.t.\ the $\ExtBCom$ in \Cref{item:ExtCom-n-Prove:commit-stage:2} of the Commit Stage; {\bf and}
\item \label[Step]{item:ExtCom-n-Prove:prove-stage:6:b}
$\Set{\view'_i}_{i \in \eta}$ together with the corresponding decommitment information w.r.t.\ the $\Com$ in \Cref{item:ExtCom-n-Prove:prove-stage:1} of the Prove Stage.
\end{enumerate}
\item 
$V$ checks the following conditions:
\begin{enumerate}[topsep=0pt,itemsep=0pt]
\item \label[Step]{item:ExtCom-n-Prove:prove-stage:7:a}
All the decommitments in \Cref{item:ExtCom-n-Prove:prove-stage:6:a,item:ExtCom-n-Prove:prove-stage:6:b} are valid; {\bf and}
\item \label[Step]{item:ExtCom-n-Prove:prove-stage:7:b}
for any $i\in \eta$, $\view_i$ is the prefix of $\view'_i$ ; {\bf and}
\item \label[Step]{item:ExtCom-n-Prove:prove-stage:7:c}
for any $i,j\in \eta$, views $(\view'_i, \view'_j)$ are consistent (as per \Cref{def:view-consistency} and \Cref{rmk:VSS:view-consistency}) w.r.t.\ the 
$\VSS_\Share$ execution in \Cref{item:ExtCom-n-Prove:commit-stage:1} of the Commit Stage
and the $\Prot_\textsc{mpc}$ execution as described in \Cref{item:ExtCom-n-Prove:prove-stage:1} of the Prove Stage.
\end{enumerate}
If all the checks pass, $V$ accepts; otherwise, $V$ rejects.
\end{enumerate}
\end{ProtocolBox}

\begin{theorem}\label{thm:extcom-n-prove}
Assume the existence of $t(\secpar)$-round semi-honest OTs. Then, there exists a $(O(t(\secpar))+\omega(1)$-round construction of an ExtCom-and-Prove scheme $\Prot_\textsc{ECnP}$ (i.e., \Cref{protocol:ExtCom-n-Prove}) satisfying \Cref{def:com-n-prove}. Moreover, this construction makes only black-box use of the assumed OT.
\end{theorem}

This can be proven similarly to the security proof of $\epsilon$-ExtCom-and-Prove in \cite[Section 6.5]{C:CCLY22}. The only differences from their construction are that 
\begin{enumerate}
\item we use fully-simulatable extractable batch commitments to commit to the views in \Cref{item:ExtCom-n-Prove:commit-stage:2} whereas they use $\epsilon$-simulatable parallel extractable commitments, and
\item we implement the coin-flipping subprotocol \Cref{ExtCom-n-Prove:CF} using equivocal commitments instead whereas they used $\epsilon$-simulation extractable commitments. \takashi{Originally, I wrote the coin-flipping step by using $\ExtBCom$ to look more similar to CCLY. 
But I found using $\EqCom$ will be slightly better in terms of round-complexity, and thus I did so.
}  
\end{enumerate}

The difference between full-simulation and $\epsilon$-simulation is directly connected to that we achive full-simulation instead of $\epsilon$-simulation for the resulting protocol. 
For the coin-flipping subprotocol, what we need here is \emph{one-sided} simulation security where we require simulation-based security against malicious verifiers but only require a weaker security against malicious provers that they cannot bias the result of coin-flipping.   
This can be achieved using either equivocal commitments as is done here or extractable commitments as is done in \cite[Section 6.5]{C:CCLY22}. With the above remarks in mind, it is straightforward to adapt the proof in \cite[Section 6.5]{C:CCLY22} to our setting. 
Thus, we omit the proof of \Cref{thm:extcom-n-prove}.
\takashi{Okay?}

%% file: sections/full-MPC.tex

\section{Post-Quantum Black-Box MPC with Full Simulation}
\label{sec:full-MPC}

\subsection{Black-Box PQ-2PC with Full Simulation}
\label{sec:full-MPC:2PC}
In this part, we prove the following theorem.
\begin{theorem}\label{thm:2pc:main}
Assuming the existence of a constant-round, semi-honest post-quantum OT, there exists a black-box, $\omega(1)$-round construction of post-quantum 2PC. 
\end{theorem}

To prove \Cref{thm:2pc:main}, we follow the paradigm established in earlier works, in particular \cite{EC:GLSV21,C:CCLY22}. This involves two steps.

\para{Step-1:} In \cite{C:CCLY22}, the authors first define an ideal functionality $\Func^t_\textsc{so-com}$ for ``selective-opening secure'' commitments, which is shown in \Cref{figure:functionality:so-com}. More descriptively, this is an idealization of a commitment that offers {\em selective opening} security in a {\em bounded-parallel} execution. That is, it can be used by a committer to commit to an a-priori bounded number, say a polynomial $t(\secpar)$, of strings within a single invocation; later, the receiver may specify an arbitrary subset $I \subset [t]$ of positions, and the committer must decommit to the $i$-th commitment it made, for each $i \in I$.

The intuitive benefit in having access to such a construct arises from the fact that it allows for implementations of {\em cut-and-choose} protocols which naturally involve committing to several instances of certain data and then later opening a receiver-chosen subset of these committed instances. The techniques we use to get 2PC will involve these techniques. 



\begin{FigureBox}[label={figure:functionality:so-com}]{The Ideal Functionality \textnormal{$\Func^t_\textsc{so-com}$} \cite{EC:GLSV21,C:CCLY22}}
\para{Commit Stage:} $\Func^t_\textsc{so-com}$ receives from the committer $C$ a query $\big(\algo{Commit}, sid, (m_1, \ldots, m_t)\big)$. $\Func^t_\textsc{so-com}$ records $\big(sid, (m_1, \ldots, m_t)\big)$ and sends $(\algo{Receipt}, sid)$ to the receiver $R$. $\Func^t_\textsc{so-com}$ ignores further $\algo{Commit}$ messages with the same $sid$.

\para{Decommit Stage:} $\Func^t_\textsc{so-com}$ receives from $R$ a query $(\algo{Reveal}, sid, I)$, where $I$ is a subset of $[t]$. If no $\big(sid, (m_1, \ldots, m_t)\big)$ has been recorded, $\Func^t_\textsc{so-com}$ does nothing; otherwise, it sends to $R$ the message $\big(\algo{Open}, sid, \Set{m_i}_{i\in I} \big)$. 
\end{FigureBox}



\para{Step 2:} Then, it is shown in \cite[Section 7.4]{C:CCLY22} that $\Func^t_\textsc{so-com}$ is indeed black-box 2PC-complete. Namely, \cite[Section 7.4]{C:CCLY22} shows that given a protocol $\pi$ that securely implements $\Func^t_\textsc{so-com}$ against QPT adversaries, one can construct a {\em general-purpose} 2PC protocol (i.e, computing any efficient 2-party functionality) that is secure against QPT adversaries. Moreover, the 2PC construction makes only {\em black-box} use of $\pi$ and involves only a constant multiplicative blow up in the number
of rounds (as compared to $\pi$).  

We must keep in mind the following caveat: the protocol $\pi$ as described in \cite[Section 7.4]{C:CCLY22} in fact implements $\Func^t_\textsc{so-com}$ w.r.t.\ $\epsilon$-simulation. Hence the final 2PC they obtained is also w.r.t.\ $\epsilon$-simulation. It is straightforward however to see that the same proof works w.r.t.\ standard negligibly-close simulation as well. Namely, if one starts with a $\pi$ that implements $\Func^t_\textsc{so-com}$ w.r.t.\ the standard notion of negligible-close simulation, then the resulting 2PC protocol will also be secure w.r.t.\ the standard notion of negligible-close simulation.

\para{Implementing \textnormal{$\Func^t_\textsc{so-com}$}.} From the above discussion we can see that to prove \Cref{thm:2pc:main}, it suffices to construct a $\omega(1)$-round, black-box, post-quantum protocol implementing the $\Func^t_\textsc{so-com}$ functionality. For that, we will make use of the $\omega(1)$-round ExtCom-and-Prove protocol described in \Cref{protocol:ExtBCom}. Note that it is okay to make use of this ExtCom-and-Prove protocol because this protocol makes only black-box use of a semi-honest post-quantum OT protocol, which is indeed the minimal assumption for our current goal of 2PC.

We can then conclude the proof of \Cref{thm:2pc:main} using the following lemma.

\begin{lemma}[\text{\cite[Lemma 26]{C:CCLY22}}]
\label{lem:ExtCin-and-Prove:to:SO-Com}
Assume the existence of Post-Quantum ExtCom-and-Prove (as per \Cref{def:com-n-prove}). Then, for any polynomial $t(\secpar)$, there exists a post-quantum protocol implementing $\Func^t_\textsc{so-com}$. Moreover, this construction makes only black-box use of the ExtCom-and-Prove protocol and incurs only a constant blow up in the number of rounds.
\end{lemma}
\begin{proof}
This proof is essentially identical to the proof of \cite[Lemma 26]{C:CCLY22}, relying on the extractable commit-and-prove protocol given in \Cref{protocol:ExtCom-n-Prove}. The idea is simple: the committer uses the Commit Stage of the Extcom-and-Prove protocol to commit to different messages $(m_1,\dots,m_t)$ of its choice. Next, when required to decommit to a certain subset $I \subset [t]$ of messages, the committer reveals these messages $\Set{m_i}_{i \in I}$ to the receiver and then uses to Prove Stage to prove that the revealed messages are indeed the committed ones for the appropriate positions. Note that this protocol then is purely black-box and only adds a small constant number of rounds of communication (relating to sending the revealed subset and the decommitment information) over the underlying Extcom-and-Prove protocol.   

Security against a cheating committer is obtained via the soundness of \Cref{protocol:ExtCom-n-Prove}, and that against a cheating reciever can be seen from the zero-knowledge of the same underlying protocol. 
The crucial (and in fact only) difference is that \cite[Lemma 26]{C:CCLY22} uses an underlying ExtCom-and-Prove protocol that offers $\epsilon$-simulation. This is why they only manage to obtain a protocol implementing $\Func^t_\textsc{so-com}$ w.r.t.\ $\epsilon$-simulation. In contrast, \Cref{protocol:ExtCom-n-Prove} does achieve the standard notion of negligibly close simulation. It is then easy to verify that our protocol for $\Func^t_\textsc{so-com}$ achieves the standard notion of full simulation as well, using the same proof.
\end{proof}

\subsection{Black-Box PQ-MPC with Full Simulation}
\label{sec:full-MPC:MPC}
Here we turn to the problem of obtaining a {\em fully simulatable} black-box post-quantum MPC protocol. More precisely, we show the following theorem.

\begin{theorem}\label{thm:mpc:main}
Assuming the existence of a semi-honest post-quantum OT, there exists a black-box construction of post-quantum MPC in polynomial rounds. 
\end{theorem}

This theorem follows directly from \Cref{thm:2pc:main} and \cite{C:IshPraSah08}. To start, we observe that \Cref{thm:2pc:main} provides a black-box construction of post-quantum maliciously secure and fully simulatable OT --- via the constructed 2PC protocol (recall that the latter can be made to implement {\em any} 2 party functionality). We have further seen in \Cref{sec:MPC} that the black-box compiler given in \cite{C:IshPraSah08} from OT to MPC works in the post-quantum setting as is.  

There is however a caveat: recall that the original \cite{C:IshPraSah08} result is in the OT hybrid model in the UC setting, where the OT primitive is modeled as an ideal UC functionality. Such modeling indeed allows parallel OT calls (this has been observed and discussed in \cite[Section 7]{C:CCLY22} and in \Cref{sec:MPC}). But the OT protocol we obtain from \Cref{thm:2pc:main} is only secure in the {\em standalone} setting. 

Fortunately, this does not become a problem for our application. Recall that the IPS compiler involves carrying out a certain polynomial number of OT calls or executions at the start of the protocol, which are carried out in parallel as observed above. For our purposes, we simply make required number of OT calls in sequence instead of in parallel, which adds to the round complexity of our protocol but preserves the desired order asymptotics --- it is easy to check that our overall MPC protocol still takes only polynomial rounds in $n$ (and thus also in $\secpar$).

%% file: sections/two-sided-formal.tex

\section{Full Description of the Two-Sided PQ-MNC Protocol}
\label{sec:two-sided:full}

In this section, we present the full description of the 1-1 PQ-NMC protocol without the `one-sided' restriction. As explained in \Cref{sec:two-sided:main-body}, this is obtained by applying the \cite{STOC:PasRos05} `two-slot' technique to \Cref{protocol:BB-NMCom}.

The protocol is presented in \Cref{protocol:BB-NMCom:two-sided}. It makes use of exactly the same building blocks for \Cref{protocol:BB-NMCom} (as listed at the beginning of \Cref{sec:BB-NMC:construction}).

\begin{ProtocolBox}[label={protocol:BB-NMCom:two-sided}]{(Two-Sided) PQ-NMC: Black-Box and Constant-Round}
{\bf Parameter Setting:} The tag space is defined to be $[T]$, where $T$ is a polynomial in the security parameter $\secpar$. Let $n$ be a polynomial in $\SecPar$, and $k$ be a constant fraction of $n$ such that $2k \le n/3$. 

\para{Input:}
Both the committer $C$ and the receiver $R$ get the security parameter $1^\secpar$ and a tag $t \in [T]$ as the common input; $C$ gets a string $m \in \bits^{\ell(\SecPar)}$ as its private input, where $\ell(\cdot)$ is a polynomial. 

\para{Commit Phase:}
\begin{enumerate}
\item\label[Step]{bbnmc:two-sided:init-com}
{\bf (Initial Com to $m$.)} In this stage, $C$ commits to the message with MitH.
\begin{itemize}
\item $C$ prepares $n$ views $\Set{\msf{cv}^{(1)}_i}_{i \in [n]}$, corresponding to an MitH execution for the $(n+1, 2k)$-$\VSS_\Share$ of the message $m$. $C$ commits to each $\msf{cv}^{(1)}_i$ ($i \in [n]$) independently in parallel, using Naor's commitment.
\end{itemize}

\item \label[Step]{bbnmc:two-sided:hard-puzzle:puzzle-setup:A}
{\bf (Hard-Puzzle-A.)} In this stage, $R$ sets up a $t$-solution hard puzzle. It then commits to one solution of the puzzle and proves in zero-knowledge the consistency with MitH. This corresponds to the {\bf Slot-A} as described in \Cref{sec:two-sided:main-body}.
\begin{enumerate}

\item \label[Step]{bbnmc:two-sided:hard-puzzle:A:com-ch}
$C$ samples a size-$k$ random subset $\msf{ch}_A \subseteq [n]$, and commits to it using $\ExtCom$. 

\item \label[Step]{bbnmc:two-sided:hard-puzzle:A:rv-1}
$R$ samples $t$ random strings $x^A_1, \ldots, x^A_t \pick \bits^\secpar$. $R$ prepares $n$ views $\Set{\msf{rv}^{(1, A)}_i}_{i \in [n]}$, corresponding to an MitH execution for the $(n+1, k)$-$\VSS_\Share$ of the string $x^A_1 \| \ldots \|x^A_t$. $R$ commits to each $\msf{rv}^{(1, A)}_i$ ($i \in [n]$) independently in parallel, using Naor's commitment.

\item \label[Step]{bbnmc:two-sided:hard-puzzle:A:rv-2}
$R$ prepares another $n$ views $\Set{\msf{rv}^{(2, A)}_i}_{i \in [n]}$, corresponding to an MitH execution for the $(n+1, k)$-$\VSS_\Share$ of the string $1\|x^A_1$.  $R$ commits to each $\msf{rv}^{(2,A)}_i$ ($i \in [n]$) independently in parallel, using $\ExtCom$.

\item \label[Step]{bbnmc:two-sided:hard-puzzle:A:rv-3}
$R$ then prepares another $n$ views $\Set{\msf{rv}^{(3,A)}_i}_{i \in [n]}$, corresponding to an  $(n, k)$-MitH execution of the $n$-party functionality $F^{R,A}_{\msf{consis}}$ described below, where party $P_i$ ($i\in [n]$) uses $\msf{rv}^{(1, A)}_i \| \msf{rv}^{(2, A)}_i$ as input. $R$ commits to each $\msf{rv}^{(3, A)}_i$ ($i \in [n]$) independently in parallel, using Naor's commitment.
\begin{itemize}
\item {\bf Functionality  $F^{R, A}_{\msf{consis}}$:} It collects input (and parses it as) $\msf{rv}^{(1,A)}_i \| \msf{rv}^{(2,A)}_i$ from party $i$ for each $i \in [n]$. It then runs the recovery algorithm of $\VSS$ to obtain $a_1\|\ldots\|a_t \coloneqq \VSS_\Recon(\msf{rv}^{(1, A)}_1, \ldots, \msf{rv}^{(1, A)}_n)$ and $j\|b_{j} \coloneqq \VSS_\Recon(\msf{rv}^{(2, A)}_1, \ldots, \msf{rv}^{(2, A)}_n)$. If $j \in [t]$ and $b_{j} = a_{j}$, it outputs 1 to each party; otherwise, it outputs 0 to each party.
\end{itemize}

\item \label[Step]{bbnmc:two-sided:hard-puzzle:A:decom-ch}
$C$ sends $\msf{ch}_A$ together with the decommitment information (w.r.t.\ \Cref{bbnmc:two-sided:hard-puzzle:A:com-ch}).

\item \label[Step]{bbnmc:two-sided:hard-puzzle:A:open}
$R$ sends $\Set{(\msf{rv}^{(1, A)}_i, \msf{rv}^{(2, A)}_i, \msf{rv}^{(3, A)}_i)}_{i \in \msf{ch}_A}$ together with the decommitment information (w.r.t.\ their respective commitments in \Cref{bbnmc:two-sided:hard-puzzle:A:rv-1,bbnmc:two-sided:hard-puzzle:A:rv-2,bbnmc:two-sided:hard-puzzle:A:rv-3}).

\item \label[Step]{bbnmc:two-sided:hard-puzzle:A:consistency-check}
$C$ checks the validity of the decommitment information and the consistency among the revealed views $\Set{(\msf{rv}^{(1, A)}_i, \msf{rv}^{(2, A)}_i, \msf{rv}^{(3, A)}_i)}_{i \in \msf{ch}_A}$. In particular, it checks for each $i \in \msf{ch}_A$ that $\msf{rv}^{(1, A)}_i\| \msf{rv}^{(2, A)}_i$ is the prefix of $\msf{rv}^{(3, A)}_i$. It also checks for each distinct pair $i, j\in \msf{ch}_A$ that $(\msf{rv}^{(1, A)}_i, \msf{rv}^{(1, A)}_j)$ are consistent, $(\msf{rv}^{(2, A)}_i, \msf{rv}^{(2, A)}_j)$ are consistent, and $(\msf{rv}^{(3, A)}_i, \msf{rv}^{(3, A)}_j)$ are consistent, where by `consistent' we refer to the consistency requirements as per \Cref{def:view-consistency} and \Cref{rmk:VSS:view-consistency}. It also checks for each $i\in \msf{ch}_A$ the final output of $P_i$ contained in $\msf{rv}^{(3, A)}_i$ is 1. It aborts immediately if any of the checks fail.
\end{enumerate}

\item \label[Step]{bbnmc:two-sided:hard-puzzle:puzzle-setup:B}
{\bf (Hard-Puzzle-B.)} In this stage, $R$ sets up a $(T-t)$-solution hard puzzle. It then commits to one solution of the puzzle and proves in zero-knowledge the consistency with MitH. This corresponds to the {\bf Slot-B} as described in \Cref{sec:two-sided:main-body}.
\begin{enumerate}

\item \label[Step]{bbnmc:two-sided:hard-puzzle:B:com-ch}
$C$ samples a size-$k$ random subset $\msf{ch}_B \subseteq [n]$, and commits to it using $\ExtCom$. 

\item \label[Step]{bbnmc:two-sided:hard-puzzle:B:rv-1}
$R$ samples $(T-t)$ random strings $x^B_1, \ldots, x^B_{T-t} \pick \bits^\secpar$. $R$ prepares $n$ views $\Set{\msf{rv}^{(1, B)}_i}_{i \in [n]}$, corresponding to an MitH execution for the $(n+1, k)$-$\VSS_\Share$ of the string $x^B_1 \| \ldots \|x^B_t$. $R$ commits to each $\msf{rv}^{(1, B)}_i$ ($i \in [n]$) independently in parallel, using Naor's commitment.

\item \label[Step]{bbnmc:two-sided:hard-puzzle:B:rv-2}
$R$ prepares another $n$ views $\Set{\msf{rv}^{(2, B)}_i}_{i \in [n]}$, corresponding to an MitH execution for the $(n+1, k)$-$\VSS_\Share$ of the string $1\|x^B_1$.  $R$ commits to each $\msf{rv}^{(2,B)}_i$ ($i \in [n]$) independently in parallel, using $\ExtCom$.

\item \label[Step]{bbnmc:two-sided:hard-puzzle:B:rv-3}
$R$ then prepares another $n$ views $\Set{\msf{rv}^{(3,B)}_i}_{i \in [n]}$, corresponding to an  $(n, k)$-MitH execution of the $n$-party functionality $F^{R,B}_{\msf{consis}}$ described below, where party $P_i$ ($i\in [n]$) uses $\msf{rv}^{(1, B)}_i \| \msf{rv}^{(2, B)}_i$ as input. $R$ commits to each $\msf{rv}^{(3, B)}_i$ ($i \in [n]$) independently in parallel, using Naor's commitment.
\begin{itemize}
\item {\bf Functionality  $F^{R, B}_{\msf{consis}}$:} It collects input (and parses it as) $\msf{rv}^{(1,B)}_i \| \msf{rv}^{(2,B)}_i$ from party $i$ for each $i \in [n]$. It then runs the recovery algorithm of $\VSS$ to obtain $a_1\|\ldots\|a_{T-t} \coloneqq \VSS_\Recon(\msf{rv}^{(1, B)}_1, \ldots, \msf{rv}^{(1, B)}_n)$ and $j\|b_{j} \coloneqq \VSS_\Recon(\msf{rv}^{(2, B)}_1, \ldots, \msf{rv}^{(2, B)}_n)$. If $j \in [T-t]$ and $b_{j} = a_{j}$, it outputs 1 to each party; otherwise, it outputs 0 to each party.
\end{itemize}

\item \label[Step]{bbnmc:two-sided:hard-puzzle:B:decom-ch}
$C$ sends $\msf{ch}_B$ together with the decommitment information (w.r.t.\ \Cref{bbnmc:two-sided:hard-puzzle:B:com-ch}).

\item \label[Step]{bbnmc:two-sided:hard-puzzle:B:open}
$R$ sends $\Set{(\msf{rv}^{(1, B)}_i, \msf{rv}^{(2, B)}_i, \msf{rv}^{(3, B)}_i)}_{i \in \msf{ch}_B}$ together with the decommitment information (w.r.t.\ their respective commitments in \Cref{bbnmc:two-sided:hard-puzzle:B:rv-1,bbnmc:two-sided:hard-puzzle:B:rv-2,bbnmc:two-sided:hard-puzzle:B:rv-3}).

\item
$C$ checks the validity of the decommitment information and the consistency among the revealed views $\Set{(\msf{rv}^{(1, B)}_i, \msf{rv}^{(2, B)}_i, \msf{rv}^{(3, B)}_i)}_{i \in \msf{ch}_B}$ in the same manner as in \Cref{bbnmc:two-sided:hard-puzzle:A:consistency-check}. It also checks for each $i\in \msf{ch}_B$ the final output of $P_i$ contained in $\msf{rv}^{(3, B)}_i$ is 1. It aborts immediately if any of the checks fail.
\end{enumerate}

\item \label[Step]{prot:bbnmc:two-sided:extcom}
{\bf (ExtCom to $m$.)} $C$ commits to $m$ once again with an extractable MitH.
\begin{itemize}
\item
$C$ prepares $n$ views $\Set{\msf{cv}^{(2)}_i}_{i \in [n]}$, corresponding to an MitH execution for the $(n+1, 2k)$-$\VSS_\Share$ of the message $m$. $C$ commits to each $\msf{cv}^{(2)}_i$ ($i \in [n]$) independently in parallel, using $\ExtCom$.
\end{itemize}

\item\label[Step]{prot:bbnmc:two-sided:puzzle-sol-reveal}
{\bf (Puzzle Solution Reveal.)} $R$ reveals $(x^A_1, \ldots, x^A_t)$ and $(x^B_1, \ldots, x^B_{T-t})$ by decommitting to $\Set{\msf{rv}^{(1, A)}_i}_{i \in [n]}$ and $\Set{\msf{rv}^{(1, B)}_i}_{i \in [n]}$. 

\item \label[Step]{prot:bbnmc:two-sided:PoC}
{\bf (Committer's Consistency Proof.)} This stage should be interpreted as $C$ proving consistency between its actions in \Cref{bbnmc:two-sided:init-com,prot:bbnmc:two-sided:extcom} (i.e., these two steps commit to the same value) using a WI argument, where the trapdoor statement is that: $C$ manages to commit to a puzzle solution in \Cref{prot:bbnmc:two-sided:extcom} {\em either} for {\bf Hard-Puzzle-A} {\em or} {\bf Hard-Puzzle-B}. Note that this is corresponding to the modification described in \Cref{sec:two-sided:main-body}.

This step is again conducted in MitH. Note that for the honest committer, the `effective witness' is the same message $m$ reconstructed from both $\Set{\msf{cv}^{(1)}_i}_{i \in [n]}$ and $\Set{\msf{cv}^{(2)}_i}_{i \in [n]}$, and so the virtual MPC execution in reality evaluates the `first clause' of $F^C_{\msf{consis}}$ as defined below. 
\begin{enumerate}
\item\label[Step]{prot:bbnmc:two-sided:C-consis-com}
$C$ prepares $n$ views $\Set{\msf{cv}^{(3)}_i}_{i \in [n]}$, corresponding to an $(n,2k)$-MitH execution of the $n$-party functionality $F^C_{\msf{consis}}$ described below, where party $P_i$ ($i\in [n]$) uses $\msf{cv}^{(1)}_i \| \msf{cv}^{(2)}_i$ as input. $C$ commits to each $\msf{cv}^{(3)}_i$ ($i \in [n]$) independently in parallel, using Naor's commitment.
\begin{itemize}
\item {\bf Functionality  $F^C_{\msf{consis}}$:} It collects input (and parses it as) $\msf{cv}^{(1)}_i \| \msf{cv}^{(2)}_i$ from party $i$ for each $i \in [n]$. It then runs the recovery algorithm of $\VSS$ to obtain $a \coloneqq \VSS_\Recon(\msf{cv}^{(1)}_1, \ldots, \msf{cv}^{(1)}_n)$ and $b \coloneqq \VSS_\Recon(\msf{cv}^{(2)}_1, \ldots, \msf{cv}^{(2)}_n)$. It outputs 1 to each party if
\begin{itemize}
\item
$b = a$, {\bf or}
\item
$b$ can be parsed as $j\|x'$ such that $j\in [t]$ and $x' = x^A_j$ (recall that $x^A_j$ is among the {\bf Hard-Puzzle-A} solutions revealed by Rrevealed in \Cref{prot:bbnmc:two-sided:puzzle-sol-reveal}, {\bf or}
\item
$b$ can be parsed as $j\|x'$ such that $j\in [T-t]$ and $x' = x^B_j$ (recall that $x^B_j$ is among the {\bf Hard-Puzzle-B} solutions revealed by Rrevealed in \Cref{prot:bbnmc:two-sided:puzzle-sol-reveal}.
\end{itemize} 
Otherwise, it outputs 0 to each party.
\end{itemize}

\item\label[Step]{prot:bbnmc:two-sided:cointoss}
{\bf (Trapdoor Coin-Flipping)} $C$ and $R$ then execute the {\bf Coin-Flipping Stage} of the trapdoor coin-flipping protocol shown in \Cref{prot:td:ct}, with the trapdoor predicate $\phi(\cdot)$ defined as follows
\begin{itemize}
\item
{\bf Predicate $\phi(\cdot)$:} It has the values $(x^A_1, \ldots, x^A_t)$ and $(x^B_1, \ldots, x^B_{T-t})$ hard-wired (recall that these values are revealed in \Cref{prot:bbnmc:two-sided:puzzle-sol-reveal}). On input $j\|x'$, $\phi$ outputs 1 if and only if {\em either}
\begin{itemize}
\item
$j\in [t]$ and $x' = x^A_j$ {\em or}
\item
$j\in [T-t]$ and $x' = x^B_j$.
\end{itemize}  
\end{itemize}
By the completeness of the trapdoor coin-flipping protocol (i.e., \Cref{def:com-n-prove:property:2} in \Cref{def:com-n-prove}), at the end of this step, $C$ and $R$ agree on a string $\eta$. By a proper choice of length, the string $\eta$ can be interpreted as specifying a size-$k$ random subset of $[n]$. In the following, we abuse notation by using $\eta$ to denote the corresponding size-$k$ random subset.

\item\label[Step]{prot:bbnmc:two-sided:mpc:reveal}
$C$ sends $\Set{(\msf{cv}^{(1)}_i, \msf{cv}^{(2)}_i, \msf{cv}^{(3)}_i)}_{i \in \eta}$ together with the decommitment information (w.r.t.\ their respective commitments in \Cref{bbnmc:two-sided:init-com,prot:bbnmc:two-sided:extcom,prot:bbnmc:two-sided:C-consis-com}).

\item
$R$ checks the validity of the decommitment information and the consistency among the revealed views $\Set{(\msf{cv}^{(1)}_i, \msf{cv}^{(2)}_i, \msf{cv}^{(3)}_i)}_{i \in \eta}$. 
It also checks for each $i \in \eta$ that the output of $P_i$ contained in $\msf{cv}^{(3)}_i$ is 1. It aborts if any of these checks fail.
\end{enumerate}

\end{enumerate}

\para{Decommit Stage:} 
\begin{enumerate}
\item
$C$ sends $\Set{\msf{cv}^{(1)}_i}_{i\in [n]}$ together with the decommitment information w.r.t.\ the commitments in \Cref{bbnmc:two-sided:init-com}. 
\item
$R$ checks the validity of the decommitment information and the consistency among $\Set{\msf{cv}^{(1)}_i}_{i\in [n]}$. If these checks are successful, $R$ recovers $m$ as $m \coloneqq \VSS_\Recon(\msf{cv}^{(1)}_1, \ldots, \msf{cv}^{(1)}_n)$; otherwise, $R$ rejects and output $\bot$. 
\end{enumerate}

\end{ProtocolBox}